\documentclass[a4paper]{article}
\usepackage{amsfonts,amssymb,amsmath,amsthm,cite}
\usepackage{ucs} 
\usepackage[utf8x]{inputenc}
\usepackage{graphicx}
\usepackage[english]{babel}
\usepackage{slashed}
\usepackage{textcomp}
\usepackage{comment}
\usepackage{bm}
\usepackage{titlesec}
\usepackage{mathtools}
\usepackage{etoolbox}
\patchcmd{\thebibliography}{\section*}{\section}{}{}

\evensidemargin=\oddsidemargin
\newtheorem{theorem}{Theorem}
\newtheorem{lemma}{Lemma}

\begin{document}

\vspace{10mm}
\begin{center}
\large{\textbf{Three-loop renormalization of the quantum action}\\ 
\textbf{for a four-dimensional scalar model with quartic interaction}\\
\textbf{with the usage of the background field method and a cutoff regularization}}
\end{center}
\vspace{2mm}
\begin{center}
\large{\textbf{Aleksandr V. Ivanov}}
\end{center}
\begin{center}
St. Petersburg Department of Steklov Mathematical Institute of Russian Academy of Sciences,\\ 
27 Fontanka, St. Petersburg 191023, Russia
\end{center}
\begin{center}
Leonhard Euler International Mathematical Institute in Saint Petersburg,\\ 
10 Pesochnaya nab., St. Petersburg 197022, Russia
\end{center}
\begin{center}
E-mail: regul1@mail.ru
\end{center}
\vspace{2mm}
\begin{flushright}
\large{\textbf{\textit{To the 90-th anniversary of L.D.Faddeev}}}
\end{flushright}
\vspace{10mm}

\textbf{Abstract.} The paper studies the quantum action for the four-dimensional real $\phi^4$-theory in the case of a general formulation using the background field method. The three-loop renormalization is performed with the usage of a cutoff regularization in the coordinate representation. The absence of non-local singular contributions and the correctness of the renormalization $\mathcal{R}$-operation on the example of separate three-loop diagrams are also discussed. The explicit form of the first three coefficients for the renormalization constants and for the $\beta$-function is presented. Consistency with previously known results is shown.

\vspace{2mm}

\newpage	
\tableofcontents

\newpage
\section{Introduction}
\label{29:sec:int}
One of the most widely used methods for studying models in quantum field theory \cite{9,10} is the perturbative approach. Its essence lies in the decomposition of the investigated value into an asymptotic series with respect to a small parameter. As a rule, the role of the parameter is played by the coupling constant. The need for such the technique is usually associated with a lack of mathematical understanding of the original object and with the cumbersomeness of calculations. For example, we can recall a quantum action, which is the functional integration \cite{3} of the exponential with a classical action. Such object is not fully understood, since it contains integration over a functional space and, as a result, includes open questions from a mathematical point of view. However, by decomposing some of the terms of the classical action in a series in powers of a small parameter, the problem can be reduced to calculating Gaussian integrals and, thus, circumvent some mathematical questions.

Nevertheless, when moving to the asymptotic decomposition, another snag appears: the coefficients of the series may contain divergent integrals. This is due to the fact that generalized functions that should be considered using a certain test class actually act on other generalized functions, which leads to the appearance of non-integrable functions. The answer to the question "How to get rid of divergencies?" have been given by the theory of renormalization \cite{6,7,105}. 
According to the proven recipe, it is necessary to first regularize the divergent integrals, and then subtract the singular terms by multiplying\footnote{We are talking about renormalizable and super-renormalizable theories.} the individual parts of the classical action by special renormalization constants.

Thus, regularization is an important and essential component in the study of quantum field problems using perturbative decomposition methods. There are various regularization schemes, and the choice of a specific one depends on which part of the classical action is planned to be deformed and which symmetry you want to preserve. Among the most popular types one can emphasize a cutoff regularization \cite{w6,w7,w8,Khar-2020}, dimensional one \cite{19,555}, implicit one \cite{chi-0,chi-1,chi-2}, Pauli--Villars regularization \cite{Pauli-Villars}, and one with the usage of the higher covariant derivatives \cite{Bakeyev-Slavnov}. Other examples of the use of regularization can be found in the generalized functions\footnote{The most famous example here is the Sokhotski--Plemelj theorem \cite{Soh}, which connects representations for a generalized function in two different regularizations.} theory \cite{Gelfand-1964,Vladimirov-2002}, representation theory \cite{Gel,I-1}, and the theory of integrable models \cite{I-2,I-3}.

This work is devoted to the study of a cutoff regularization in the coordinate representation using the example of three-loop calculations in the four-dimensional general real $\phi^4$-theory, see \cite{29-4,29-3}, in the Euclidean formulation using the background field method, the action density for which has the following form\footnote{The formula is shown for clarity. A detailed description is provided when describing the problem statement.}
\begin{equation*}
t^i\phi_i+
\frac{1}{2}\phi_i\Big(A^{ij}_0+M^{ij}\Big)\phi_j+\frac{g^{ijk}}{3!}\phi_i\phi_j\phi_k+
\frac{\lambda^{ijkn}}{4!}\phi_i\phi_j\phi_k\phi_n.
\end{equation*}
Meanwhile, the basic formulation will mean the case when the indices take only one value $i=j=k=n=1$, and the coefficients for odd degrees of the field are zero.
For the first time, the regularization under study was proposed in \cite{34} when studying two-loop divergences in cubic models. Since then, it has been significantly improved and applied to a number of other models \cite{Ivanov-Kharuk-2020,Ivanov-Kharuk-20222,Ivanov-Akac,Ivanov-Kharuk-2023}, including gauge ones. It is worth noting separately that this regularization has a set of interesting and non-trivial properties \cite{Ivanov-2022}. For example, it preserves the structure\footnote{It is assumed that the asymptotic series has a decomposition structure for the Laplace operator. With additional deformation of the functions included in the decomposition.} of the asymptotic expansion of the Green's function near the diagonal, it is also related to the homogenization theory, has an explicit spectral representation, and can be reformulated using higher covariant derivatives.

Using the example of the main term $R_0(x-y)$ of the asymptotic expansion of the Green's function $G(x,y)$ near the diagonal $x\sim y$, the regularization can be represented by the following deformation
\begin{equation*}
R_0^{\phantom{1}}(z)=\frac{1}{4\pi^2|z|^2}\to
R_0^\Lambda(z)=\frac{\Lambda^2\mathbf{f}\big(|z|^2\Lambda^2\big)}{4\pi^2}+\frac{1}{4\pi^2}
\begin{cases}
	\,\,\,\Lambda^2, &\mbox{if}\,\,\,|z|\leqslant1/\Lambda;\\
		|z|^{-2}, &\mbox{if}\,\,\,|z|>1/\Lambda,
\end{cases}
\end{equation*}
where $\Lambda$ is a dimensional parameter of the regularization, and $\mathbf{f}$ is a continuous function, whose support is from $[0,1]$. Therefore, in particular, $R_0^{\phantom{1}}(z)=R_0^\Lambda(z)$ for $|z|>1/\Lambda$. Schematically, such deformation is shown in Fig.\,\ref{29-picpic}. A detailed description, taking into account the deformation of the classical action, is given below.

\begin{figure}[h]
	\center{\includegraphics[width=1\linewidth]{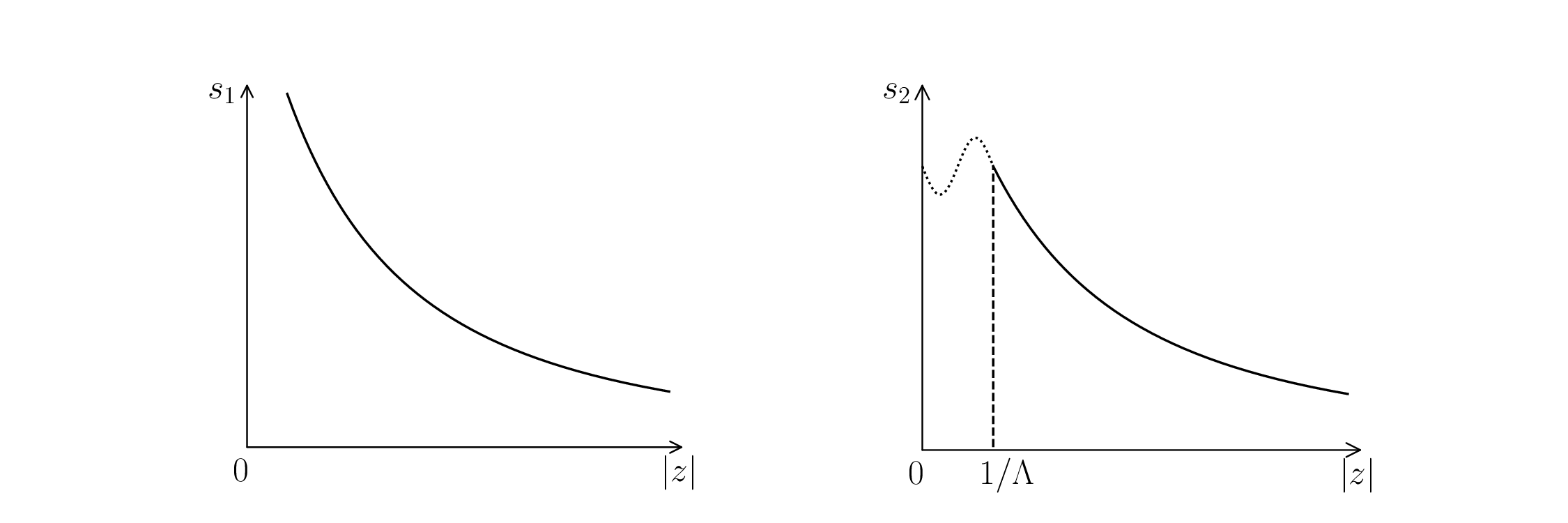}}
	\caption{Schematic representation of the deformation of the first term in the asymptotic expansion of the Green's function. Here $s_1(|z|)=R_0^{\phantom{1}}(z)$ and $s_2(|z|)=R_0^{\Lambda}(z)$ for a function $\mathbf{f}(\cdot)$.}
	\label{29-picpic}
\end{figure}

Recall that the model is formulated using the background field method, therefore, a separate important task is to check the absence of non-local singular terms. Also, in the course of research, using the example of individual diagrams, the implementation of the rules of the renormalization $\mathcal{R}$-operation is demonstrated. Thus, when considering three-loop diagrams, the local singular part is obtained by subtracting diagrams with divergences of lower order.

\vspace{3mm}
\noindent\textbf{The structure of the paper.}

\vspace{2mm}
\textbf{Section \ref{29:sec:res}} formulates two main theorems on renormalization of both the basic model and the generalized one.
Explicit answers for the first three coefficients of the renormalization constants are presented as results. The first three coefficients of the $\beta$-function are calculated, and consistency with known results is shown. It also mentions the explicit verification of the validity of the renormalization $\mathcal{R}$-operation and the absence of divergent non-local parts.

In \textbf{Section \ref{29:sec:1:state}}, the problem statement in the basic case (see above) is discussed. In addition to the basic definitions of the quantum action and elements of Feynman's diagrammatic technique, the transition from the initial effective action to the regularized one and from the regularized to the renormalized one is also analyzed. Special emphasis is placed on the process of renormalization, counter diagrams, and relations to find the coefficients for the renormalization constants. The section ends with a short list of tasks.

\textbf{Section \ref{29:sec:reg}} presents the process of introducing regularization. Starting with general considerations about the connection between the quantum action and the quantum equation of motion, a restriction on the type of regularization is derived. Next, an explicit recipe is demonstrated for how to introduce the cutoff regularization in the coordinate representation, while maintaining the fulfillment of an additional condition. Decompositions by the coupling constant for the Green's function, the determinant, and the asymptotic expansion near the diagonal for the Green's function are derived. In the last part of the section, it is shown that the regularization can be introduced by deformation of the classical action.

In \textbf{Section \ref{29:sec:contr}}, the calculation of the first two corrections (loops) for the quantum action is discussed. We derive the singular parts for the individual diagrams and the final answers for the first two coefficients of the renormalization constants.

In \textbf{Section \ref{29:sec:tri}}, three-loop diagrams are studied and the third coefficients of the renormalization constants are calculated for the basic formulation of the problem. We start the section with a number of additional auxiliary notations, part of the subsections is devoted to the careful derivation (with proofs) of singular parts for three-loop diagrams. Emphasis is also placed on the fact that in some diagrams there are nonlocal terms with singular coefficients, but in the final sum all the nonlocal parts are finite. We provide explicit ratios. At the end of the section, answers for the desired coefficients are written out, supplemented by discussions of their verification and consistency with known results.

In \textbf{Section \ref{29:sec:gen}}, we discuss a generalized formulation of the problem. The basic definitions are formulated and the relationship with the model in the basic formulation is shown. The extension of the regularization rules and renormalization is considered. Special emphasis is placed on the process of calculating the first three coefficients of the renormalization constants using the results for the studied basic model. We formulate an explicit recipe for the transition from the basic case (with a discussion of additional calculations). In the last part, the obtaining of the desired coefficients is briefly discussed, and explicit formulas for them are written out.

\textbf{Section \ref{29:sec:conc}} contains concluding remarks that were not included in the main sections. In particular, issues such as additional deformation during the regularization, extension of the results to other models, and possible simplifications in further calculations are discussed. The main results and their consistency with other types of regularization are also summarized once again. Acknowledgements are given at the end of the section.

\textbf{Section \ref{29:sec:app}} includes proofs, conclusions, and notations used in calculating the coefficients for the renormalization constants. The mentioned results are new and represent an essential part of the work, but are of a purely technical in nature, therefore, for convenience, they are outside the scope of the main text.

\section{Results}
\label{29:sec:res}
The main results of the work can be formulated in the form of two theorems. Let us start with the basic formulation of the problem, which is discussed in detail in Sections \ref{29:sec:contr} and \ref{29:sec:tri}.

\begin{theorem}\label{29-th-1}
Let the classical action of the model be described by the functional (\ref{29-1-1}), and the regularization of the quantum action (\ref{29-1-4}) is introduced by cutoff in the coordinate representation using the deformation of the quadratic form in the classical action according to the procedure from Section \ref{29:sec:reg-4}. In this connection, $\Lambda$ is the parameter of regularization, and the function $\mathbf{f}(\cdot)$ from (\ref{29-3-17}) is responsible for the deformation of the Green's function. In this case, the non-local terms are reduced, and the first three coefficients for the renormalization constants (\ref{29-1-5})--(\ref{29-1-23}), that is, for
\begin{align*}
Z_2=&~1+\hbar z_{2,1}+\hbar^2 z_{2,2}+\hbar^3 z_{2,3}+\ldots,\\
Z=&~1+\hbar z_{1}+\hbar^2 z_{2}+\hbar^3 z_{3}+\ldots,\\
Z_4=&~1+\hbar z_{4,1}+\hbar^2 z_{4,2}+\hbar^3 z_{4,3}+\ldots,
\end{align*}
can be calculated explicitly and have the following form:
\begin{align*}
z_{2,1}=&~\bar{z}_{2,1}, \\
z_1=&~\Lambda^2q_1+L\hat{\lambda}+\bar{z}_{1},\\
z_{4,1}=&~3L\hat{\lambda}+\bar{z}_{4,1},\\
z_{2,2}=&-L\hat{\lambda}^2/6+
\bar{z}_{2,2},\\
z_2=&~\Lambda^2\big(Lq_3+q_2\big)
+2L^2\hat{\lambda}^2+L\big(-\hat{\lambda}^2+\hat{\lambda}(\bar{z}_1+\bar{z}_{4,1}-2\bar{z}_{2,1})\big)
+\bar{z}_{2},\\
z_{4,2}=&~9L^2\hat{\lambda}^2+
L\big(-6\hat{\lambda}^2+6\hat{\lambda}\bar{z}_{4,1}-6\hat{\lambda}\bar{z}_{2,1}\big)
+
\bar{z}_{4,2},\\
	z_{2,3}=&-L^2\hat{\lambda}^3/2+La_1+\bar{z}_{2,3},
	\\
	z_{3}=&~\Lambda^2\big(L^2b_6+Lb_5+b_4\big)+14L^3\hat{\lambda}^3/3+L^2b_2+Lb_1+\bar{z}_{3},
	\\
	z_{4,3}=&~27L^3\hat{\lambda}^3+L^2c_2+Lc_1
	+\bar{z}_{4,3}.
\end{align*}
Here $L=\ln(\Lambda/\sigma)$, $\hat{\lambda}=\lambda/(16\pi^2)$, and the numbers "$\bar{z}$" are free parameters.
The dependence of the auxiliary coefficients and references to their explicit formulas can be represented as the following list:
\begin{align*}
q_1=&~q_1(\mathbf{f},m)\to\mbox{see (\ref{29-4-61})},\\
q_2=&~q_2(\mathbf{f},\bar{z}_{2,1},\bar{z}_{4,1},m)\to\mbox{see (\ref{29-4-251})},
\\
q_3=&~q_3(\mathbf{f},m)\to\mbox{see (\ref{29-4-251})},
\\
a_1=&~a_1(\mathbf{f},\bar{z}_{2,1},\bar{z}_{4,1})\to\mbox{see (\ref{29-7-40})},
\\
b_1=&~b_1(\mathbf{f},\bar{z}_{1},\bar{z}_{2,1},\bar{z}_{4,1},\bar{z}_{2,2},\bar{z}_{2},\bar{z}_{4,2})\to\mbox{see (\ref{29-7-42})},
\\
b_2=&~b_2(\bar{z}_{1},\bar{z}_{2,1},\bar{z}_{4,1})\to\mbox{see (\ref{29-7-43})},
\\
b_4=&~b_4(\mathbf{f},\bar{z}_{2,1},\bar{z}_{4,1},\bar{z}_{2,2},\bar{z}_{4,2},m)\to\mbox{see (\ref{29-7-45})},
\\
b_5=&~b_5(\mathbf{f},\bar{z}_{2,1},\bar{z}_{4,1},m)\to\mbox{see (\ref{29-7-37})},
\\
b_6=&~b_6(\mathbf{f},m)\to\mbox{see (\ref{29-7-38})},
\\
c_1=&~c_1(\mathbf{f},\bar{z}_{2,1},\bar{z}_{4,1},\bar{z}_{2,2},\bar{z}_{4,2})\to\mbox{see (\ref{29-7-39})},
\\
c_2=&~c_2(\bar{z}_{2,1},\bar{z}_{4,1})\to\mbox{see (\ref{29-7-391})}.
\end{align*}
\end{theorem}

It can be noted that the main logarithmic singularities for each order are written out explicitly. They do not depend on the free "$\bar{z}$"-parameters and additional deformation of the Green's function. At the same time, the main poles for $Z$ and $Z_4$ are consistent with those obtained in dimensional regularization up to the third order inclusive, see \cite{29-2,29-1}.

Let us calculate the $\beta$-function. If we enter the notation $\lambda_0=Z_4^{\phantom{1}}\lambda/Z_2^2$, then according to the general theory it is defined by equality
\begin{equation*}
\beta(\lambda_0^{\phantom{1}})=-\sigma\frac{\mathrm{d}\lambda_0}{\mathrm{d}\sigma}=
\beta_1\lambda_0^2+\beta_2\lambda_0^3+\beta_3\lambda_0^4+\ldots
\end{equation*}
Using the results obtained, the first three coefficients are calculated explicitly and have the form
\begin{align}
	\nonumber
	\beta_1=&~3,\\
	\nonumber
	\beta_2=&-\frac{17}{3},
	\\
	\label{29-15-16}
\beta_3=&~\frac{77}{4}
+12\zeta(3)
-36\alpha_1(\mathbf{f})
+17\alpha_2(\mathbf{f})
+2\alpha_6(\mathbf{f})
+18\alpha_2^2(\mathbf{f})+\\\nonumber&+
(4\pi^2)^2\big(
16\alpha_{11}(\mathbf{f})
-4\alpha_3(\mathbf{f})
-144\alpha_4(\mathbf{f})
-144\alpha_5(\mathbf{f})\big)
-12\bar{z}_{2,2}+6\bar{z}_{4,2},
\end{align}
where the expressions for numbers $\alpha_i(\mathbf{f})$ depending on the function $\mathbf{f}(\cdot)$ are given in Section \ref{29:sec:app:vch}.
It is important to note that the first two coefficients are scheme-independent and fully consistent with the known results, see \cite{chi-0,12}, for the dimensional and the implicit regularizations. The third coefficient depends on the regularization and renormalization procedure.

In addition, it is interesting to consider the values of the coefficients for the renormalization constants for some fixed parameter values. For example, if we consider all free "$\bar{z}$"-coefficients are equal to zero and $\mathbf{f}\equiv0$, then
\begin{equation*}\label{29-15-1}
z_{2,1}=0,\,\,\,
z_{1}=-\frac{2\hat{\lambda}\Lambda^2}{m^2}+L\hat{\lambda},\,\,\,
z_{4,1}=3L\hat{\lambda},
\end{equation*}
\begin{equation*}\label{29-15-13}
	z_{2,2}=-\frac{L\hat{\lambda}^2}{6},\,\,\,
	z_{2}=\frac{\Lambda^2\hat{\lambda}^2}{m^2}(-6L+1)+2L^2\hat{\lambda}^2-L\hat{\lambda}^2,\,\,\,
	z_{4,2}=9L^2\hat{\lambda}^2-6L\hat{\lambda}^2,
\end{equation*}
\begin{equation*}\label{29-15-131}
z_{2,3}=-\frac{L^2\hat{\lambda}^3}{2}+
\frac{L\hat{\lambda}^3}{12},
\end{equation*}
\begin{equation*}\label{29-15-132}
z_3=-\frac{\Lambda^2\hat{\lambda}^3}{m^2}
\bigg(
18L^2-\frac{53L}{3}+16\Theta_7+\frac{16\Theta_9}{3}+\frac{28}{9}
\bigg)+
\frac{14L^3\hat{\lambda}^3}{3}-
\frac{41L^2\hat{\lambda}^3}{6}
+\frac{65L\hat{\lambda}^3}{18},
\end{equation*}
\begin{equation*}\label{29-15-133}
z_{4,3}=27L^3\hat{\lambda}^3-
	\frac{89L^2\hat{\lambda}^3}{2}+
	L\hat{\lambda}^3\bigg(
	12\zeta(3)-144\Theta_5+\frac{611}{24}\bigg).
\end{equation*}
Here the numbers $\Theta_5$, $\Theta_7$, and $\Theta_9$ are calculated numerically and represented by formulas (\ref{29-13-15})--(\ref{29-13-152}).
In this case, the third coefficient (\ref{29-15-16}) for the $\beta$-function takes a more concise form
\begin{equation*}\label{29-15-134}
\beta_3\big|_{\mathbf{f}=0}=\frac{607}{24}+12\zeta(3)-144\Theta_5=45.907\pm 10^{-3}>0.
\end{equation*}

Let us move on to the generalized case. Given the cumbersomeness of the final expressions, the theorem will provide references to the corresponding formulas without re-writing.
\begin{theorem}\label{29-th-2}
	Let the classical action of the model be described by the functional from (\ref{29-16-1}), and the regularization of the quantum action is introduced by cutoff in the coordinate representation using the deformation of the quadratic form in the classical action by adding the functional from (\ref{29-16-7}).
In this connection, $\Lambda$ is the regularizing parameter and the function $\mathbf{f}(\cdot)$ from (\ref{29-3-17}) is responsible for the deformation of the Green's function. In this case, the non-local terms are reduced, the first three coefficients for the renormalization constants (\ref{29-16-9})--(\ref{29-16-11}) are calculated explicitly and have the form:
	\begin{itemize}
		\item the first order is represented by formulas (\ref{29-16-18})--(\ref{29-16-24}),
		\item the second order is represented by formulas (\ref{29-18-4})--(\ref{29-16-29}),
		\item the third order is represented by formulas (\ref{29-16-30})--(\ref{29-16-35}) taking into account the auxiliary notations from Sections \ref{29:sec:app:kom} and \ref{29:sec:app:kom-1}.
	\end{itemize}
\end{theorem}
It can be verified that when considering values (\ref{29-16-222}), the results of Theorem \ref{29-th-2} for the generalized model pass into the results of Theorem \ref{29-th-1} for the basic model. It can also be noted that, as in the simple case, in the generalized model, the main logarithmic divergences are scheme-independent. In particular, they do not depend on the deforming function $\mathbf{f}(\cdot)$.

Additionally, we note that in the course of research, explicit calculations have shown the validity of the $\mathcal{R}$-operation on the example of three-loop diagrams using the background field method for both the basic formulation and the generalized one.

\section{Basic problem statement}
\label{29:sec:1:state}
\subsection{Effective action}
Let us consider the four-dimensional Euclidean space $\mathbb{R}^4$, see \cite{2}, whose elements will be designated by the letters $x,y,z,u,w$. We will use Cartesian coordinates as local coordinates, and the individual components will be marked with Greek letters $\alpha,\beta,\mu,\nu\in\{1,2,3,4\}$. Despite the fact that the metric in our case is represented by the Kronecker symbol $\delta_{\mu\nu}$, we will still keep the presence of upper and lower indexes. At the same time, the Einstein convention will be used: a repeating index from above and below implies summation over all values. Next, let $m^2\geqslant0$ be the square of a mass parameter, and the coefficient $\lambda\geqslant0$ be a coupling constant and responsible for the quartic interaction. Then the classical action of the scalar $\phi^4$-model in the basic formulation can be written as
\begin{equation}\label{29-1-1}
S[\phi]=\int_{\mathbb{R}^4}\mathrm{d}^4x\,\bigg[
\frac{1}{2}\big(\partial_{x_\mu}\phi(x)\big)\big(\partial_{x^\mu}\phi(x)\big)+
\frac{m^2}{2}\phi^2(x)+\frac{\lambda}{4!}\phi^4(x)\bigg].
\end{equation}
The last formula implies that the function $\phi(\cdot)$ is smooth and has a good enough decrease at infinity, so that the integral exists. Then, integrating the first term by parts, we can represent the action as
\begin{equation}\label{29-1-2}
S[\phi]=\int_{\mathbb{R}^4}\mathrm{d}^4x\,\bigg[
\frac{1}{2}\phi(x)A(x)\phi(x)+\frac{\lambda}{4!}\phi^4(x)\bigg],
\end{equation}
where we have used the definition for the Laplace operator $A(x)$ in local coordinates
\begin{equation}\label{29-1-3}
A(x)=A_0(x)+m^2=-\partial_{x_\mu}\partial_{x^\mu}+m^2.
\end{equation}

Next, we define the quantum (effective) action, which is obtained using the following functional integration
\begin{equation}\label{29-1-4}
e^{-W/\hbar}=\int_{\mathcal{H}}\mathcal{D}\phi\,e^{-S[\phi]/\hbar},
\end{equation}
where the Planck constant $\hbar$ plays the role of the small parameter, and a functional space $\mathcal{H}$ is determined by using physical considerations\footnote{It should be noted here that the definition of the functional integral, as well as the corresponding measure, contains some open mathematical questions. Therefore, the set of functions $\mathcal{H}$ and subsequent operations with the integration domain are symbolic in nature. It is assumed that the standard properties of an ordinary integral are fulfilled for the functional integral.}. Thus, the quantum action $W$ depends on the class of functions in $\mathcal{H}$, on their behavior at infinity.

The main method of studying the quantum action is the perturbative approach, that is, decomposition by a small parameter into an asymptotic series. Such a small parameter in our case is the Planck constant $\hbar$. Let us briefly describe the transition to the asymptotic decomposition. The background field method will be used for this, see \cite{12,102,103,24,25,26,23,Ivanov-Russkikh}, the main idea of which is to shift the field
\begin{equation}\label{29-1-7}
\phi(x)\to B(x)+\sqrt{\hbar}\,\phi(x),
\end{equation}
where the function $B(\cdot)$ is called the background field and solves a quantum equation of motion\footnote{In diagrammatic language, we can say that the background field is a solution to an equation in which the sum of all strongly connected diagrams with one external free line is zero.}. As a rule, we consider the behavior at infinity of functions, which define the set $\mathcal{H}$, as boundary conditions for such equation. Thus, the dependence of the quantum action $W$ on the set $\mathcal{H}$ becomes dependent on the background field $B(\cdot)$ after the change of the variables. Therefore, the quantum action is actually the functional $W=W[B]$. Next, we will notate with the symbol $\mathcal{H}_0$ the set of functions after the mentioned transition, that is,  $\phi\in\mathcal{H}_0\leftrightarrow \phi+B/\sqrt{\hbar}\in\mathcal{H}$.

\begin{figure}[h]
\center{\includegraphics[width=0.37\linewidth]{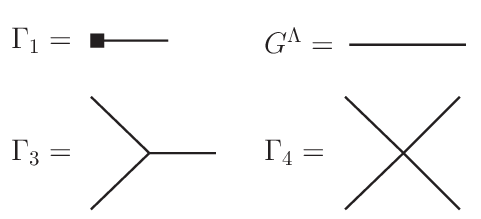}}
\caption{Elements of the diagram technique for vertices with one, three and four external free lines, as well as for the regularized Green's function $G^\Lambda$.}
\label{pic:29-1}
\end{figure}

Note that after the shift (\ref{29-1-7}), the classical action (\ref{29-1-2}) is represented as the following finite sum in powers of the Planck constant
\begin{equation}\label{29-1-8}
S\big[B+\sqrt{\hbar}\phi\big]=S[B]+\sqrt{\hbar}\,\Gamma_1[B,\phi]+\frac{\hbar}{2}\Gamma_2[B,\phi]+
\frac{\lambda\hbar^{3/2}}{3!}\Gamma_3[B,\phi]+\frac{\lambda\hbar^2}{4!}\Gamma_4[\phi],
\end{equation}
where
\begin{equation}\label{29-1-9}
\Gamma_1[B,\phi]=\int_{\mathbb{R}^4}\mathrm{d}^4x\,\phi(x)\Big(A(x)+\lambda B^2(x)/3!\Big)B(x),
\end{equation}
\begin{equation}\label{29-1-10}
\Gamma_2[B,\phi]=\int_{\mathbb{R}^4}\mathrm{d}^4x\,\phi(x)M(x)\phi(x)=
\int_{\mathbb{R}^4}\mathrm{d}^4x\,\phi(x)\Big(A(x)+\lambda B^2(x)/2\Big)\phi(x),
\end{equation}
\begin{equation}\label{29-1-11}
\Gamma_3[B,\phi]=\int_{\mathbb{R}^4}\mathrm{d}^4x\,B(x)\phi^3(x),\,\,\,
\Gamma_4[\phi]=\int_{\mathbb{R}^4}\mathrm{d}^4x\,\phi^4(x).
\end{equation}
In further work with such terms, their diagrammatic representation will be useful. In Fig. \ref{pic:29-1} the vertices with external free lines are shown for $\Gamma_1[B,\phi]$, $\Gamma_3[B,\phi]$ and $\Gamma_4[\phi]$. The functional $\Gamma_2[B,\phi]$ is the quadratic form with the differential operator $M$. As is known, the external free lines are connected by an element corresponding to the regularized kernel $G^\Lambda(\cdot\,,\cdot)$ of the inverse operator to $M$. It is indicated by a continuous line, called the propagator, and is also depicted in Fig. \ref{pic:29-1}.

Therefore, substituting decomposition (\ref{29-1-8}) into the functional integral from (\ref{29-1-4}), we obtain
\begin{multline}\label{29-1-12}
e^{-W[B]/\hbar}=\exp\bigg(-\frac{S[B]}{\hbar}-\frac{\Gamma_1[B,\delta_j]}{\sqrt{\hbar}}-
	\frac{\lambda\hbar^{1/2}}{3!}\Gamma_3[B,\delta_j]-\frac{\lambda\hbar}{4!}\Gamma_4[\delta_j]\bigg)\times\\\times
\int_{\mathcal{H}_0}\mathcal{D}\phi\,e^{-\Gamma_2[B,\phi]/2}\exp\bigg(\int_{\mathbb{R}^4}\mathrm{d}^4x\,j(x)\phi(x)\bigg)\bigg|_{j=0},
\end{multline}
where we have used the notation for the functional derivative $\delta_j$ over an auxiliary smooth field $j$ and the following relation
\begin{equation}\label{29-1-13}
F[\delta_j]\exp\bigg(\int_{\mathbb{R}^4}\mathrm{d}^4x\,j(x)\phi(x)\bigg)\bigg|_{j=0}=F[\phi]
\end{equation}
for an arbitrary functional $F[\,\cdot\,]$, which can be represented as a series in powers of the argument. In turn, the functional integral from formula (\ref{29-1-12}) is Gaussian, so the answer for it, taking into account the appropriate normalization of the measure, is written out explicitly
\begin{equation}\label{29-1-14}
\int_{\mathcal{H}_0}\mathcal{D}\phi\,e^{-\Gamma_2[B,\phi]/2}\exp\bigg(\int_{\mathbb{R}^4}\mathrm{d}^4x\,j(x)\phi(x)\bigg)=\frac{e^{g[G,j]}}{\sqrt{\det(M)}},
\end{equation}
where the functional $g[\,\cdot\,,\,\cdot\,]$ and the Green's function $G(\cdot\,,\cdot)$ are determined by the relations
\begin{equation}\label{29-1-15}
g[G,j]=\frac{1}{2}\int_{\mathbb{R}^4}\int_{\mathbb{R}^4}\mathrm{d}^4x\mathrm{d}^4y\,j(x)G(x,y)j(y),\,\,\,
M(x)G(x,y)=\delta(x-y).
\end{equation}
Of course, the Green's function is not only a solution to the last equation, but also satisfies some boundary conditions arising from the properties of the background field and the set $\mathcal{H}_0$. Their explicit form does not play an important role in further research, however, we will assume that the task of finding the Green's function is well posed.

After all these manipulations, the quantum action is written out in the form
\begin{equation}\label{29-1-16}
W[B]=S[B]-\frac{\hbar}{2}\ln\det(G)-\hbar
\exp\bigg(-\frac{\lambda\hbar^{1/2}}{3!}\Gamma_3[B,\delta_j]-\frac{\lambda\hbar}{4!}\Gamma_4[\delta_j]\bigg)e^{g[G,j]}\bigg|_{j=0}^{\mathrm{1PI}},
\end{equation}
where the symbol $\mathrm{1PI}$ means that only strongly connected\footnote{It should be noted here that after taking the logarithm of the functional integral, only connected graphs remain in the corrections. At the same time, it must be remembered that the background field $B$ is a solution of the quantum equation of motion, which reduces the sum of strongly connected diagrams with one external free line. Thus, only the strongly connected part remains in the sum.}$^{,}$\footnote{Tracking the transition from formula (\ref{29-1-12}), the logarithm $\ln$ should be written in the third term before the exponential. It has been deleted because the sign $\mathrm{1 PI}$ leaves only strongly connected terms, and the logarithm operation preserves connected ones. Therefore, the weaker operation can be omitted.} diagrams remain in the sum. It is for this reason that formula (\ref{29-1-16}) lacks the functional $\Gamma_1[B,\delta_j]$ and negative powers of the Planck constant. This leads to the fact that the quantum action contains only positive powers of the Planck constant in the asymptotic expansion and, as a result, is representable as the sum of the classical action and quantum corrections
\begin{equation*}\label{29-1-17}
W[B]=S[B]+\sum_{k=1}^{+\infty}\hbar^{k}W_k[B],
\end{equation*}
where $W_k[B]$ corresponds to the correction involving strongly connected diagrams with $k$ loops.

It is known from general theory that the object $W[B]$ is poorly defined, since each correction term $W_k[B]$ of the asymptotic series contains divergent integrals. Therefore, according to the general logic, regularization is introduced with some auxiliary parameter $\Lambda$, which leads to the convergence of integrals depending on the background field. In this case, instead of divergent integrals, we obtain terms containing singularities\footnote{They are finite, but tend to infinity when the regularization is removed. For example, the function $\Lambda^{2}$ is singular in the neighborhood of infinity by the regularizing parameter $\Lambda$.} with respect to the regularizing parameter $\Lambda$. In turn, integrals that do not depend on the background field do not require any regularization in the general case. They do not carry important information (it's just a number) and can be removed by suitable subtraction in each loop. Thus, the quantum action after an additional transformation looks like this
\begin{equation}\label{29-1-18}
W[B]\xrightarrow{\mbox{\footnotesize reg.}}W[B,\Lambda]=S[B]+\sum_{k=1}^{+\infty}\hbar^{k}
\Big(W_k[B,\Lambda]-\kappa_k\Big),
\end{equation}
where the values $\kappa_k$ subtract the divergences independent of the background field. It follows that the regularized quantum action is finite, but contains a set of terms that will tend to infinity when the regularization is removed. We need to draw attention that in this paper, regularization is understood as the deformation\footnote{Further, it will be shown that the regularization must be introduced in a special coordinated way, which allows us to maintain the connection between the action and the equation of motion.} of the Green's function
\begin{equation*}\label{29-1-19}
G(x,y)\xrightarrow{\mbox{\footnotesize reg.}}G^\Lambda(x,y).
\end{equation*}
The removal of the regularization is performed by moving to the limit $\Lambda\to+\infty$.

\subsection{Renormalization}

There is a general recipe that allows us to remove (subtract) singular terms by multiplying the components of the classical action by additional constants, which are called renormalization constants. In the case of the model under study, there should be three such constants\footnote{The expansion coefficients for constants depend on the regularizing parameter $\Lambda$, the designation of which, as a rule, will be omitted for convenience.}
\begin{align}\label{29-1-5}
Z_2&=1+\sum_{k=1}^{+\infty}\hbar^k z_{2,k}(\Lambda),\\\label{29-1-22} Z&=1+\sum_{k=1}^{+\infty}\hbar^k z_{k}(\Lambda),\\\label{29-1-23} Z_4&=1+\sum_{k=1}^{+\infty}\hbar^k z_{4,k}(\Lambda).
\end{align}
They do not depend on the spatial coordinate or the background field. Such constants are introduced into the quantum action $W[B,\Lambda]$ through the following series of substitutions
\begin{equation}\label{29-1-6}
\phi(x)\to\sqrt{Z_2^{\phantom{2}}}\phi(x),\,\,\,
m^2\to ZZ_2^{-1}m^2,\,\,\,
\lambda\to Z_4^{\phantom{2}}Z_2^{-2}\lambda
\end{equation}
in the density of the classical action (\ref{29-1-1}). Consequently, after such procedure, we obtain the third object, the renormalized regularized quantum action, 
\begin{equation}\label{29-1-20}
	W[B,\Lambda]\xrightarrow{\mbox{\footnotesize ren.}}W_{\mathrm{ren}}[B,\Lambda].
\end{equation}
The last functional no longer contains singularities, so after removing the regularization
\begin{equation*}\label{29-1-21}
W_{\mathrm{ren}}[B,+\infty]\,\,\,\mbox{is finite}.
\end{equation*}

\begin{figure}[ht]
	\center{\includegraphics[width=0.5\linewidth]{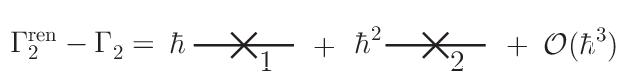}}
	\caption{Elements of the diagram technique for additional vertices (\ref{29-1-26}) and (\ref{29-1-27}) with two external free lines.}
	\label{pic:29-2}
\end{figure}

Note that the scaling of the field (\ref{29-1-6}) is performed before the shift. Therefore, when working with elements after the shift, it is necessary to multiply by $\sqrt{Z_2^{\phantom{1}}}$ both the field $\phi$ and the background field $B$. Then the renormalized action from (\ref{29-1-20}) is represented as
\begin{multline}\label{29-1-24}
	W_{\mathrm{ren}}[B,\Lambda]=\frac{Z_2}{2}S_2[B]+\frac{m^2Z}{2}S_m[B]+\frac{\lambda Z_4}{4!}S_4[B]-\bigg(\frac{\hbar}{2}\ln\det(G^\Lambda)+\hbar\,\kappa_1\bigg)-\\-\Bigg[\hbar
	\exp\bigg(\frac{1}{2}\big(\Gamma_2^{\phantom{r}}[B,\delta_j]-\Gamma_2^{\mathrm{ren}}[B,\delta_j]\big)-\frac{\lambda\hbar^{1/2}}{3!}\Gamma_3[B,\delta_j]-\frac{\lambda\hbar}{4!}\Gamma_4[\delta_j]\bigg)e^{g[G^\Lambda,j]}\bigg|_{j=0}^{\mathrm{1PI}}+\sum_{k=2}^{+\infty}\hbar^k\,\kappa_k\Bigg],
\end{multline}
where\footnote{Here it is necessary to explain the appearance of the additional term $\big(\Gamma_2^{\phantom{r}}[B,\delta_j]-\Gamma_2^{\mathrm{ren}}[B,\delta_j]\big)/2$ in the second line of (\ref{29-1-24}). The fact is that after renormalization (\ref{29-1-6}), the quadratic form contains the renormalization constants. Therefore, after calculating the Gaussian integral in the second line of formula (\ref{29-1-12}), the renormalization constants are obtained in the determinant and in the Green's function. Wanting to keep them (the determinant and the Green's function) the same, additional terms were rendered according to the rule from (\ref{29-1-13}) in the form of functional derivatives.}
\begin{equation}\label{29-1-25}
S_2[B]=\int_{\mathbb{R}^4}\mathrm{d}^4x\,\big(\partial_{x_\mu}B(x)\big)\big(\partial_{x^\mu}B(x)\big),\,\,\,
S_m[B]=\int_{\mathbb{R}^4}\mathrm{d}^4x\,B^2(x),\,\,\,
S_4[B]=\int_{\mathbb{R}^4}\mathrm{d}^4x\,B^4(x),
\end{equation}
\begin{equation}\label{29-1-26}
\Gamma_2^{\mathrm{ren}}[B,\delta_j]-\Gamma_2^{\phantom{r}}[B,\delta_j]=\sum_{k=1}^{+\infty}\hbar^k\,\mathrm{X}_k[B,\delta_j].
\end{equation}
The explicit form of the functionals from $\mathrm{X}_k[B,\delta_j]$ is presented in formula (\ref{29-1-27}) after the final description of the regularization process.
As a general rule, when new vertices appear, it is necessary to assign elements of the diagram technique to them. They are shown in Fig. \ref{pic:29-2}.

\begin{figure}[ht]
	\center{\includegraphics[width=0.55\linewidth]{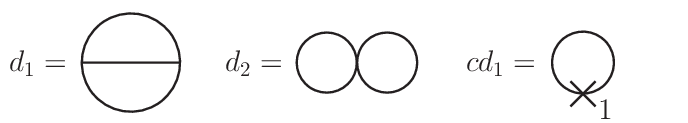}}
	\caption{The diagrams $d_1$ and $d_2$ and the counter diagram $cd_1$, giving a contribution to the two-loop term.}
	\label{pic:29-3}
\end{figure}

\begin{figure}[ht!]
	\center{\includegraphics[width=0.7\linewidth]{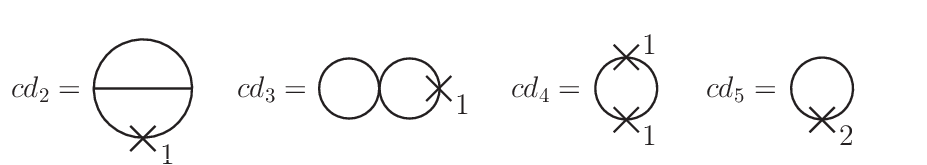}}
	\caption{The counter diagrams $cd_2$--$cd_5$ from the three-loop contribution.}
	\label{pic:29-4}
\end{figure}

\begin{figure}[ht!]
	\center{\includegraphics[width=0.57\linewidth]{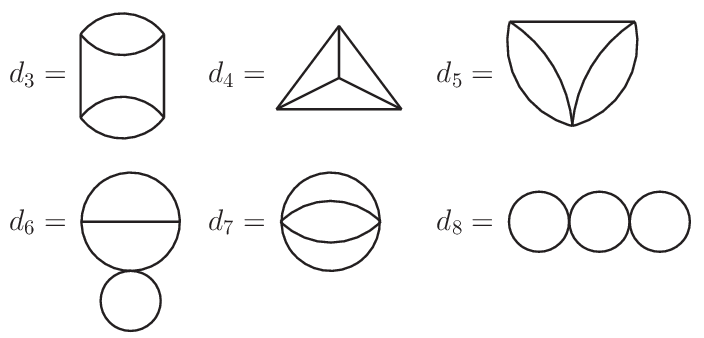}}
	\caption{The diagrams $d_3$--$d_8$ from the three-loop contribution $W_3[B,\Lambda]$.}
	\label{pic:29-5}
\end{figure}

Further, taking into account the definitions for the elements of diagram technique, the object in square brackets from formula (\ref{29-1-24}) can be rewritten in diagrammatic form, see Fig. \ref{pic:29-3}, \ref{pic:29-4}, and \ref{pic:29-5},
\begin{multline}\label{29-1-29}
\hbar\,\bigg(\frac{\lambda^2 d_1}{12}-\frac{\lambda d_2}{8}-\frac{cd_1}{2}+\kappa_2\bigg)+\hbar^2\bigg(\frac{\lambda^2z_{4,1} d_1}{6}-\frac{\lambda z_{4,1} d_2}{8}+\frac{\lambda^4 d_3}{16}+\frac{\lambda^4d_4}{24}-\\-\frac{\lambda^3d_5}{8}-\frac{\lambda^3d_6}{8}+\frac{\lambda^2d_7}{48}+\frac{\lambda^2d_8}{16}-\frac{\lambda^2cd_2}{4}+\frac{\lambda cd_3}{4}+\frac{cd_4}{4}-\frac{cd_5}{2}+\kappa_3\bigg)+\mathcal{O}\big(\hbar^3\big).
\end{multline}


The diagrams $d_i$ and the counter diagrams $cd_i$ are functionals that depend on the background field. They are also finite\footnote{The diagrams $\{d_2,d_7,d_8\}$ and the counter diagrams $\{cd_1,cd_3,cd_4,cd_5\}$ may contain divergences that do not depend on the background field. They represent the integration of a constant over $\mathbb{R}^4$. Such densities are subtracted using $\kappa_2$ and $\kappa_3$. It is due to their easy removability and lack of physical meaning that they are not taken into account.} and depend on the regularizing parameter $\Lambda$.

Decomposition (\ref{29-1-24}) taking into account (\ref{29-1-29}) is the main working object of this section. Given the fact that the renormalization constants remove (subtract) terms with singularities depending on the regularizing parameter, it can be argued that this should be true for each power of the Planck constant separately. Therefore, by studying the divergences in each individual diagram, it is possible to sum them up and obtain values for the coefficients of the renormalization constants. For each power of the Planck constant, we can obtain the corresponding linear equation
\begin{equation}\label{29-1-30}
\frac{z_{2,1}}{2}S_2[B]+\frac{m^2z_1}{2}S_m[B]+\frac{\lambda z_{4,1}}{4!}S_4[B]
\stackrel{\mathrm{s.p.}}{=}\frac{1}{2}\ln\det(G^\Lambda)+\kappa_1,
\end{equation}
\begin{equation}\label{29-1-31}
\frac{z_{2,2}}{2}S_2[B]+\frac{m^2z_2}{2}S_m[B]+\frac{\lambda z_{4,2}}{4!}S_4[B]
\stackrel{\mathrm{s.p.}}{=}\frac{\lambda^2 d_1}{12}-\frac{\lambda d_2}{8}-\frac{cd_1}{2}+\kappa_2,
\end{equation}
\begin{multline}\label{29-1-32}
\frac{z_{2,3}}{2}S_2[B]+\frac{m^2z_3}{2}S_m[B]+\frac{\lambda z_{4,3}}{4!}S_4[B]
\stackrel{\mathrm{s.p.}}{=}\frac{\lambda^2z_{4,1} d_1}{6}-\frac{\lambda z_{4,1} d_2}{8}+\\+
\frac{\lambda^4 d_3}{16}+\frac{\lambda^4d_4}{24}-\frac{\lambda^3d_5}{8}-\frac{\lambda^3d_6}{8}+\frac{\lambda^2d_7}{48}+\frac{\lambda^2d_8}{16}-\frac{\lambda^2cd_2}{4}+\frac{\lambda cd_3}{4}+\frac{cd_4}{4}-\frac{cd_5}{2}+\kappa_3,
\end{multline}
where the abbreviation $\mathrm{s.p.}$ means equality of singular terms.

\vspace{4mm}
\noindent\textbf{The main tasks for the basic formulation.} 
\begin{itemize}
	\item Study divergent parts in the diagrams $d_1$--$d_8$ and the counter diagrams $cd_1$--$cd_5$ using a cutoff regularization in coordinate representation.
	\item Check the validity of the $\mathcal{R}$-operation using the example of three-loop diagrams.
	\item Find the first three coefficients $\{z_{2,i},z_i,z_{4,i}\}_{i=1}^3$ for the renormalization constants $Z_2$, $Z$, and $Z_4$.
	\item Make sure that the non-local terms are reduced.
\end{itemize}

\section{Rules of regularization}
\label{29:sec:reg}
\subsection{General considerations}
\label{29:sec:reg-1}

In Sections \ref{29:sec:int} and \ref{29:sec:1:state}, some general information about the regularization was mentioned, but no clear recipe was provided, so we will fill this gap in this section. It is known that the Green's function can be deformed $G(x,y)\to G^\Lambda(x,y)$ in various ways. Moreover, an arbitrary "small" additional term $h^\Lambda(x,y)$ can be added to the regularized function, such that
\begin{equation}\label{29-3-1}
M(x)h^\Lambda(x,y)\to0
\end{equation}
as $\Lambda\to+\infty$ in the sense of generalized functions. In this section, we formulate an additional natural restriction, which decreases the class of possible deformations.

As it was noted earlier, the effective action $W[B]$ is a functional that depends on the background field. In its asymptotic expansion, it contains only strongly connected diagrams. This follows from the fact that the background field $B(x)$ is a solution to the quantum equation of motion $Q[B](x)=0$, in which the sum of all strongly connected diagrams with one free outer line is zero. In this case, the quantum equation of motion and the quantum action have a direct relationship: the first is obtained by the functional derivative of the second object with respect to the background field, that is
\begin{equation}\label{29-3-7}
\frac{\delta W[B]}{\delta B(x)}=Q[B](x).
\end{equation}
If such relation is broken during the regularization process $W[\,\cdot\,]\to W[\,\cdot\,,\Lambda]$ and $Q[\,\cdot\,]\to Q[\,\cdot\,,\Lambda]$, the quantum action $W[\,\cdot\,,\Lambda]$ becomes dependent not on the extreme field, that is, the solution of the quantum equation of motion 
\begin{equation*}\label{29-3-80}
	\frac{\delta W[B,\Lambda]}{\delta B(x)}=0,
\end{equation*}
but it becomes dependent on some other field $\tilde{B}$, which solves the equation $Q[\tilde{B},\Lambda]=0$.
Consequently, the desire to preserve the dependence of the effective action on the extreme field, as it was before the regularization, leads to the need to preserve the equality
\begin{equation}\label{29-3-8}
	\frac{\delta W[B,\Lambda]}{\delta B(x)}=Q[B,\Lambda](x).
\end{equation}
To fulfill the last ratio, we propose to follow\footnote{It is not claimed that the proposed procedure is the only one.} the following logic. It is clear that each diagram in the quantum action corresponds to a certain sum of diagrams in $Q[B]$, which are connected by functional differentiation with respect to the background field. Let us demand that this relationship (for each individual diagram) be fulfilled even after regularization.

Considering the fact that the diagram-forming vertices $\Gamma_3[B]$ and $\Gamma_4[B]$ do not deform\footnote{More generally, we can consider a situation where the vertices are also deformed. This option is not studied in this paper.} during the regularization, it is necessary to follow only the deformation of the Green's function. As we know, the function $G$ satisfies the following relation
\begin{equation*}\label{29-3-9}
\frac{\delta}{\delta B(z)}G(x,y)=-\lambda
G(x,z)B(z)G(z,y),
\end{equation*}
which is obtained by a variation of the second equality from (\ref{29-1-15}) with respect to $B(z)$. Therefore, the condition from (\ref{29-3-8}) will be fulfilled if the regularized Green’s function satisfies the relation
\begin{equation}\label{29-3-10}
	\frac{\delta}{\delta B(z)}G^\Lambda(x,y)=-\lambda
	G^\Lambda(x,z)B(z)G^\Lambda(z,y).
\end{equation}
It is this equality that will be an additional constraint for the regularization under study.

\subsection{Additional condition}
\label{29:sec:reg-2}

Consider the functional integral that arose in formula (\ref{29-1-12}). The formal\footnote{Since the reasoning is done before the introduction of regularization. After explaining the rules, all the arguments are easily repeated taking into account the regularization and have a clear mathematical appearance.} answer for it is presented in (\ref{29-1-14}). Therefore, using the functional derivative with respect to the field $j(x)$, the determinant and the Green's function can be written as
\begin{equation}\label{29-3-2}
\ln\det(G)=2\ln\Bigg[
\int_{\mathcal{H}_0}\mathcal{D}\phi\,e^{-\Gamma_2[0,\phi]/2}\exp\bigg(-\frac{\lambda}{4}\int_{\mathbb{R}^4}\mathrm{d}^4x\,B^2(x)\phi^2(x)\bigg)\Bigg],
\end{equation}
\begin{equation}\label{29-3-3}
G(x,y)=\sqrt{\det(M)}
\int_{\mathcal{H}_0}\mathcal{D}\phi\,e^{-\Gamma_2[0,\phi]/2}
\phi(x)\phi(y)
\exp\bigg(-\frac{\lambda}{4}\int_{\mathbb{R}^4}\mathrm{d}^4z\,B^2(z)\phi^2(z)\bigg),
\end{equation}
where the definition from (\ref{29-1-10}) has been used to rewrite the quadratic form
\begin{equation*}\label{29-3-4}
\Gamma_2[B,\phi]=\Gamma_2[0,\phi]+\frac{\lambda}{2}\int_{\mathbb{R}^4}\mathrm{d}^4x\,B^2(x)\phi^2(x).
\end{equation*}

The regularization procedure can be described as follows. Let $G_0(x-y)$ be the Green's function of the operator $A(x)$, see (\ref{29-1-3}), from the quadratic form $\Gamma_2[0,\phi]$, see (\ref{29-1-10}). Then by the regularization we will understand such a deformation of the quadratic form $\Gamma_2^{\phantom{1}}[0,\phi]\to\Gamma_2^\Lambda[0,\phi]$ by deforming the operator\footnote{Note that it is possible to deform not only $A_0$, but also $A_0+m^2$ as a whole. However, this will complicate the view of asymptotic expansions and calculations. This situation is not considered in this paper.} $A\to A_0^\Lambda+m^2$, after which the Green's function $G_0^{\phantom{1}}\to G^\Lambda_0$ satisfies the following conditions:
\begin{itemize}
	\item $A(x)G_0^\Lambda(x-y)\to\delta(x-y)$ as $\Lambda\to+\infty$ in the sense of generalized functions;
	\item there are $N>0$ and $\mu>0$, such that $G_0^\Lambda(x)$ is bounded in the domain $|x|<1/\mu$ for all finite $\Lambda>N$.
\end{itemize}
Both requirements are quite natural. Indeed, the first condition ensures the convergence of the regularized Green's function to the original one. While the second condition allows us to consider the function on the diagonal, which is necessary because diagrams with "bubbles"\footnote{That is, with densities including $G^\Lambda(x,x)$.} appear in the model (\ref{29-1-1}).

As a result, by decomposing the exponents in (\ref{29-3-2}) and (\ref{29-3-3}) in a series in powers of the field, we obtain\footnote{The right part of (\ref{29-3-3}) contains only connected diagrams, the so-called "chains". The total "vacuum" multiplier is reduced by the factor $\sqrt{\det(M)}$.} explicit, mathematically accurate formulas\footnote{Formula (\ref{29-3-6}) is an analogue of the second resolvent identity, see Theorem 4.8.2 in \cite{29-hp}.} for the regularized determinant and the Green's function
\begin{multline}\label{29-3-5}
\ln\det(G^\Lambda/G_0^\Lambda)=
\sum_{k=1}^{+\infty}\frac{(-\lambda)^k}{k2^k}
\int_{\mathbb{R}^{4\times k}}\mathrm{d}^4x_1\ldots\mathrm{d}^4x_k\,
G_0^\Lambda(x_k-x_1)B^2(x_1)G_0^\Lambda(x_1-x_2)B^2(x_2)\times\ldots\\\ldots\times G_0^\Lambda(x_{k-1}-x_k)B^2(x_k),
\end{multline}
\begin{multline}\label{29-3-6}
G^\Lambda(x,y)=G_0^\Lambda(x-y)+\sum_{k=1}^{+\infty}\frac{(-\lambda)^k}{2^k}
\int_{\mathbb{R}^{4\times k}}\mathrm{d}^4x_1\ldots\mathrm{d}^4x_k\,
G_0^\Lambda(x-x_1)B^2(x_1)G_0^\Lambda(x_1-x_2)B^2(x_2)\times\ldots\\\ldots\times G_0^\Lambda(x_{k-1}-x_k)B^2(x_k)G_0^\Lambda(x_k-y).
\end{multline}
It is not difficult to verify the validity of the following statement.
\begin{lemma}\label{29-l-1}
Taking into account all the above, function (\ref{29-3-6}) satisfies the relations from (\ref{29-3-10}) and
\begin{equation*}\label{29-3-11}
\frac{\delta}{\delta B(x)}\ln\det(G^\Lambda/G_0^\Lambda)=-\lambda B(x)G^\Lambda(x,x)
\end{equation*}
for all $x\in\mathbb{R}^4$.
\end{lemma}

Taking into account the latter statement and the fact that the transition from the regularized quantum action from (\ref{29-1-18}) to the renormalized one from (\ref{29-1-20}) is carried out by adding additional vertices that either do not depend on the background field, or in which the component with the background field is not deformed during the regularization, then it is possible to generalize relation (\ref{29-3-8}) to the renormalized quantum action
\begin{equation}\label{29-3-24}
	\frac{\delta W_{\mathrm{ren}}[B,\Lambda]}{\delta B(x)}=Q_{\mathrm{ren}}[B,\Lambda](x).
\end{equation}

\subsection{Decomposition near the diagonal}
\label{29:sec:reg-3}
In the continuation of this section, we will make some useful remarks about the deformation of the Green's function $G_0(x-y)$. As already mentioned, it satisfies the equation
\begin{equation*}\label{29-3-12}
A(x)G_0(x-y)=
\big(-\partial_{x_\mu}\partial_{x^\mu}+m^2\big)G_0(x-y)=\delta(x-y),
\end{equation*}
which is solved explicitly
\begin{equation*}\label{29-3-13}
G_0(x)=\frac{mK_1(m|x|)}{(2\pi)^2|x|},
\end{equation*}
where $K_1(\cdot)$ is the modified Bessel function of the second kind. Taking into account the properties of the special function near the diagonal, we can write out the expansion in a series
\begin{equation}\label{29-3-14}
G_0(x)=R_{0}(x)-R_{10}(x)m^2+R_{20}(x)\frac{m^4}{2}+PS_0(x)+o\big(|x|^2\big),
\end{equation}
where $PS_0(x)$ is a smooth component, and the rest of the functions are determined by the equalities
\begin{equation}\label{29-3-15}
R_0(x)=\frac{1}{4\pi^2|x|^2},\,\,\,
R_{10}(x)=-\frac{\ln\big(|x|^2\sigma^2\big)}{16\pi^2},\,\,\,
R_{20}(x)=\frac{|x|^2\big(\ln\big(|x|^2\sigma^2\big)-1\big)}{64\pi^2}.
\end{equation}
Here $\sigma>0$ is a fixed parameter\footnote{Function (\ref{29-3-14}) does not depend on the parameter $\sigma$, since $PS_0$ contains the corresponding term depending on $\sigma$ with the opposite sign. The dimensionless argument in the selected logarithm was achieved by adding and subtracting. In this case, the argument of the logarithm in $PS_0$ can be made dimensionless with the usage of the mass parameter $m$.} to make the argument dimensionless.

During the regularization of $G_0^{\phantom{1}}\to G_0^\Lambda$, each term in the decomposition (\ref{29-3-14}) undergoes a deformation, and therefore each function acquires the additional parameter $\Lambda$. The new decomposition has the form
\begin{equation}\label{29-3-16}
G_0^\Lambda(x)=R_0^\Lambda(x)-R_{10}^\Lambda(x)m^2+R_{20}^\Lambda(x)\frac{m^4}{2}+PS_0^\Lambda(x)+o\big(|x|^2\big)+o\big(\Lambda^{-2}\big).
\end{equation}
It is assumed that the non-smooth part\footnote{The smooth component can move from the $PS$-part to the $R$-functions and vice versa.} of each function with the parameter $\Lambda$ converges to the non-smooth part of the function without the parameter when the regularization is removed. Additionally, we will assume that the main term of the asymptotic behaves as follows:
\begin{equation}\label{29-3-17}
R_0^\Lambda(x)=\frac{\Lambda^2\mathbf{f}\big(|x|^2\Lambda^2\big)}{4\pi^2}+
\begin{cases}
	\Lambda^2/(4\pi^2), &\mbox{if}\,\,\,|x|\leqslant1/\Lambda;\\
\,\,\,\,\,R_0^{\phantom{1}}(x), &\mbox{if}\,\,\,|x|>1/\Lambda,
\end{cases}
\end{equation}
where the new bounded function has the properties\footnote{Note that the last limit transition will be valid, for example, when performing additional ratios $\mathbf{f}(\cdot)\in C^1(\mathbb{R}_+)$ and $\mathbf{f}^\prime(1)=0$.}
\begin{equation}\label{29-3-25}
	\mathrm{supp}\big(\mathbf{f}(\cdot)\big)\subset[0,1],\,\,\,
	\mathbf{f}(\cdot)\in C(\mathbb{R}_+)\,\,\,\mbox{and}\,\,\,
	\partial_{x_\mu}\partial_{x^\mu}\Lambda^2\mathbf{f}\big(|x-y|^2\Lambda^2\big)
	\xrightarrow{\Lambda\to+\infty}0.
\end{equation}
Here the latter relation is understood in the sense of generalized functions.


Next, we write out the asymptotics for the Green's function from (\ref{29-3-6}). It has the form
\begin{equation}\label{29-3-18}
G^\Lambda(x,y)=R_0^\Lambda(x-y)-R_{10}^\Lambda(x-y)m^2-R_{11}^\Lambda(x-y)\frac{\lambda \big(B^2(x)+B^2(y)\big)}{4}+
PS^\Lambda(x,y),
\end{equation}
where
\begin{equation}\label{29-3-22}
PS^\Lambda(x,y)=
R_{20}^\Lambda(x-y)\frac{m^4}{2}
+R_{21}^\Lambda(x-y)\frac{m^2\lambda \big(B^2(x)+B^2(y)\big)}{4}
+R_{22}^\Lambda(x-y)\frac{\lambda^2\big(B^4(x)+B^4(y)\big)}{16}+
PS_1^\Lambda(x,y),
\end{equation}
\begin{equation}\label{29-3-19}
R_{11}^\Lambda(x)=\int_{\mathrm{B}_{1/\sigma}}\mathrm{d}^4z\,R_0^\Lambda(z)R_0^\Lambda(z+x),
\end{equation}
\begin{equation}\label{29-3-20}
R_{21}^\Lambda(x)=\int_{\mathrm{B}_{1/\sigma}}\mathrm{d}^4z\,R_0^\Lambda(z)R_{10}^\Lambda(z+x)+
\int_{\mathrm{B}_{1/\sigma}}\mathrm{d}^4z\,R_{10}^\Lambda(z)R_0^\Lambda(z+x)-2\tilde{c}_2,
\end{equation}
\begin{equation}\label{29-3-21}
R_{22}^\Lambda(x)=2\int_{\mathrm{B}_{1/\sigma}\times\mathrm{B}_{1/\sigma}}\mathrm{d}^4z\mathrm{d}^4y\,R_0^\Lambda(z)
R_0^\Lambda(z+y)R_0^\Lambda(y+x)-2\tilde{c}_2,
\end{equation}
where $\sigma$ is the same auxiliary fixed parameter, and $\mathrm{B}_{1/\sigma}(x)$ is a closed ball of radius $1/\sigma$ centered at point $x$, $\mathrm{B}_{1/\sigma}\equiv\mathrm{B}_{1/\sigma}(0)$. The functions $PS^\Lambda(x,y)$ and $PS_1^\Lambda(x,y)$ are symmetric. An explicit view for $PS_1^\Lambda(x,y)$ is presented in Section \ref{29:sec:app:doc}. The constant $\tilde{c}_2$ is defined by the equality
\begin{equation}\label{29-3-27}
\tilde{c}_2=\int_{\mathrm{B}_{1/\sigma}\times\mathrm{B}_{1/\sigma}}\mathrm{d}^4z\mathrm{d}^4y\,R_0(z)
R_0(z+y)R_0(y)=\frac{1}{32\pi^2\sigma^2}
\end{equation}
and is calculated explicitly in (\ref{29-9-115}). Such a subtraction is possible because all smooth and non-local parts, as will be shown, are not included in a final singularity. At the same time, the shift subtracts a constant term and provides in (\ref{29-3-20}) and (\ref{29-3-21}) for $x=0$ an asymptotic of the form $L/\Lambda^2$, see (\ref{29-9-126}).

Also, for convenience, we introduce several auxiliary functions
\begin{equation}\label{29-3-28}
\mathrm{J}_1[B]=\int_{\mathbb{R}^2}\mathrm{d}^4x\,PS_1^{\Lambda}(x,x),\,\,\,
\mathrm{J}_2[B]=\int_{\mathbb{R}^2}\mathrm{d}^4x\,PS_1^{\Lambda}(x,x)B^2(x),
\end{equation}
\begin{equation}\label{29-3-29}
\mathrm{J}_3[B]=\int_{\mathbb{R}^2}\mathrm{d}^4x\Big(\partial_{y_\mu}\partial_{y^\mu}
PS_1^{\Lambda}(y,x)\Big)\Big|_{y=x},\,\,\,
\mathrm{J}_4[B]=\int_{\mathbb{R}^2}\mathrm{d}^4x\,PS_1^{\Lambda}(x,x)PS_1^{\Lambda}(x,x).
\end{equation}
\subsection{Final formulation}
\label{29:sec:reg-4}
Taking into account all the comments made above, it is necessary to formulate the rules of regularization in the final form. Moreover, it is important to check that the regularization is introduced by deforming the original classical action, and is not introduced arbitrarily and does not change when moving from one diagram to another. So, the initial quantum action (with divergences), taking into account the background field method, has the form
\begin{equation*}\label{29-10-1}
W[B]=-\hbar\ln\bigg(\int_{\mathcal{H}_0}\mathcal{D}\phi\,e^{-S[B+\sqrt{\hbar}\phi]/\hbar}\bigg).
\end{equation*}
Next, the deformation of the classical action is performed. To do this, we add the regularizing functional $S[\phi,\Lambda]$, which will make all the terms that depend on the background field finite. Taking into account the arguments from Section \ref{29:sec:reg-2}, it is necessary to deform the operator (\ref{29-3-1}) in the quadratic form
\begin{equation*}\label{29-10-2}
A_0^{\phantom{1}}(x)\to A_0^\Lambda(x),
\end{equation*}
so that $A_0^\Lambda(x)\to A_0^{\phantom{1}}(x)$ as $\Lambda\to+\infty$. In this case, the additional functional has the form
\begin{equation}\label{29-10-3}
S[\phi,\Lambda]=
\frac{1}{2}\int_{\mathbb{R}^4}\mathrm{d}^4x\,\phi(x)\Big(A_0^\Lambda(x)-A_0^{\phantom{1}}(x)\Big)\phi(x),
\end{equation}
and the regularized quantum action taking into account (\ref{29-1-18}) is written out in the form
\begin{equation*}\label{29-10-4}
W[B,\Lambda]=-\hbar\ln\bigg(\int_{\mathcal{H}_0}\mathcal{D}\phi\,e^{-S[B+\sqrt{\hbar}\phi]/\hbar-S[\phi,\Lambda]}\bigg)
-\sum_{k=1}^{+\infty}\hbar^k\kappa_k.
\end{equation*}
It should be noted that the so-called "long-range" models \cite{LR-1} have a similar deformation formulation, in which the quadratic operator has the form $(-\partial_{x_\mu}\partial_{x^\mu})^{\xi(\Lambda)}$, where $0<\xi(\Lambda)<1$. However, taking into account our explicit type of deformation (\ref{29-3-17}), the "long-range" models are significantly different. At the same time, both regularizations can be related to a common class in which the quadratic form is deformed.

Further, after the renormalization substitution, see (\ref{29-1-6}), the renormalized action is obtained (\ref{29-1-20}). Its explicit form is shown in formula (\ref{29-1-24}), in which it is necessary to specify the counterterms from the quadratic form (\ref{29-1-26}). They have the form
\begin{equation}\label{29-1-27}
\mathrm{X}_k[B,\phi]=z_{2,k}S_\Lambda[\phi]+m^2z_kS_m[\phi]+\frac{\lambda z_{4,k}}{2}
\int_{\mathbb{R}^4}\mathrm{d}^4x\,B^2(x)\phi^2(x),
\end{equation}
where
\begin{equation*}\label{29-10-5}
S_\Lambda[\phi]=\int_{\mathbb{R}^4}\mathrm{d}^4x\,\phi(x)A_0^\Lambda(x)\phi(x).
\end{equation*}
In conclusion of the section, we note that under the proposed conditions it is convenient to choose\footnote{Without limiting generality, since the choice is made by redefining the function $PS_1^\Lambda$, which does not contribute singular terms to the effective action.} functions in the asymptotic decomposition (\ref{29-3-18}) as follows
\begin{equation}\label{29-10-6}
R_{10}^\Lambda=R_{11}^\Lambda,\,\,\,
R_{20}^\Lambda=R_{21}^\Lambda=R_{22}^\Lambda.
\end{equation}
At the same time, the differences in notation will be preserved until the final answers, so that they can be used in generalizing regularization. For example, when the mass term is deformed. Also, saving the designations helps to more clearly understand the internal structure of the divergences and make additional intermediate checks of the results.

\section{First two corrections}
\label{29:sec:contr}
\subsection{One loop}
\label{29:sec:contr-1}
Let us consider the calculation of the one-loop contribution to the effective action (\ref{29-1-24}) containing singular terms, which have dependence on the parameter $\Lambda$. Using the explicit form of regularization (\ref{29-3-5}), we can write out the following relation
\begin{equation}\label{29-4-1}
\frac{1}{2}\ln\det(G^\Lambda/G_0^\Lambda)\stackrel{\mathrm{s.p.}}{=}-
\frac{\lambda G_0^\Lambda(0)}{4}\int_{\mathbb{R}^{4}}\mathrm{d}^4y\,B^2(y)+
\frac{\lambda^2}{16}
\bigg(\int_{\mathrm{B}_{1/\sigma}}\mathrm{d}^4y\Big(G_0^\Lambda(x)\Big)^2\bigg)
\int_{\mathbb{R}^{4}}\mathrm{d}^4y\,B^4(y),
\end{equation}
where $\sigma$ is an auxiliary fixed parameter\footnote{Here it is assumed that this parameter matches the one entered in Section \ref{29:sec:reg-3}. A different choice will only shift the singular part by a constant and, therefore, will not affect the answer, since the constant addition can be fixed arbitrarily.}. Also, we have used the fact that in the integral of the form
\begin{equation}\label{29-4-2}
\int_{\mathbb{R}^{4\times 2}}\mathrm{d}^4x_1\mathrm{d}^4x_2\,
G_0^\Lambda(x_2-x_1)B^2(x_1)G_0^\Lambda(x_1-x_2)B^2(x_2)
\end{equation}
a singular contribution follows only from the part of the integral corresponding to the domain of integration\footnote{Or using $\mathbb{R}^4\times\mathrm{B}_{1/\sigma}(x_1)$. Because of the symmetry, the choice does not play a special role.}  $\mathrm{B}_{1/\sigma}(x_2)\times\mathbb{R}^4$. Therefore, making the shift $x_1\to x_1+x_2$ and decomposing $B^2(x_1+x_2)$ into the Taylor series at the point $x_2$, we get the second term in (\ref{29-4-1}).

Let us use the decomposition from (\ref{29-3-16}), then
\begin{equation*}\label{29-4-3}
\frac{1}{2}\ln\det(G^\Lambda/G_0^\Lambda)\stackrel{\mathrm{s.p.}}{=}
-\frac{\lambda\big(R_0^\Lambda(0)-R_{10}^\Lambda(0)m^2\big)}{4}S_m[B]+\frac{\lambda^2\mathrm{A}(\sigma)}{16}S_4[B],
\end{equation*}
where
\begin{equation}\label{29-4-4}
\mathrm{A}(\sigma)=\int_{\mathrm{B}_{1/\sigma}}\mathrm{d}^4y\Big(R_0^\Lambda(x)\Big)^2
=\frac{L+\alpha_2(\mathbf{f})}{8\pi^2},
\end{equation}
see formula (\ref{29-9-10}).
Thus, substituting the last equality into relation (\ref{29-1-30}), we get
\begin{align}
	\nonumber
z_{2,1}&=\tilde{z}_{2,1}, \\\label{29-4-6}
z_1&= -\frac{\lambda R_0^\Lambda(0)}{2m^2}+\frac{\lambda R_{10}^\Lambda(0)}{2}+\tilde{z}_{1},\\\label{29-4-7}
z_{4,1}&=\frac{3\lambda}{2}\mathrm{A}(\sigma)+\tilde{z}_{4,1},
\end{align}
where $\{\tilde{z}_{2,1},\tilde{z}_{1},\tilde{z}_{4,1}\}$ are arbitrary constants ($\mathcal{O}(1)$ with respect to the $\Lambda$). Note that in a specific case, see formula (\ref{29-10-6}), we can perform the following reassignment
\begin{equation*}\label{29-4-9}
\{\tilde{z}_{2,1},\tilde{z}_{1},\tilde{z}_{4,1}\}\to
\{\bar{z}_{2,1},\bar{z}_{1},\bar{z}_{4,1}\},\,\,\,
\hat{\lambda}=\frac{\lambda}{16\pi^2},
\end{equation*}
and get the answer in a more explicit way
\begin{align}
	\nonumber
	z_{2,1}&=\bar{z}_{2,1}, \\\label{29-4-61}
	z_1&= -\frac{\Lambda^2\lambda\big(1+\mathbf{f}(0)\big)}{8\pi^2m^2}+L\hat{\lambda}+\bar{z}_{1},\\
	\nonumber
	z_{4,1}&=3L\hat{\lambda}+\bar{z}_{4,1}.
\end{align}
It is clear that using (\ref{29-9-14}) we can get the relations
\begin{equation*}\label{29-4-911}
\tilde{z}_{2,1}=\bar{z}_{2,1},\,\,\,
\tilde{z}_{1}=-\hat{\lambda}\alpha_2(\mathbf{f})+\bar{z}_{1},\,\,\,
\tilde{z}_{4,1}=-3\hat{\lambda}\alpha_2(\mathbf{f})+\bar{z}_{4,1}.
\end{equation*}

\subsection{Two loops}
\label{29:sec:contr-2}
Let us move on to calculating the two-loop contribution, which includes two diagrams $\{d_1,d_2\}$ and one counter diagram $cd_1$. Given the fact that the divergences in the coordinate representation are local\footnote{That is, they appear due to the "bad" behavior of the Green's function (\ref{29-3-3}), when the arguments are close to each other ($x\sim y$).} in nature, we will use the decomposition (\ref{29-3-18}).

Let us start with the diagram $d_1$. We use the logic of calculating the integral (\ref{29-4-2}), that is, moving to the integration domain $\mathbb{R}^4\times\mathbb{R}^4\to\mathrm{B}_{1/\sigma}(x_2)\times\mathbb{R}^4$. In this case, using the definitions from (\ref{29-1-25}), we get
\begin{align}\label{29-4-8}
d_1\stackrel{\mathrm{s.p.}}{=}&-S_2[B]\frac{\mathrm{I}_{1}(\sigma)}{8}+
S_m[B]\Big(\mathrm{I}_{4}(\sigma)-3m^2
\mathrm{I}_{2}(\sigma)\Big)-\\\label{29-4-10}
&-S_4[B]\frac{3\lambda\mathrm{I}_{3}(\sigma)}{2}+
\mathrm{J}_2[B]3\mathrm{A}(\sigma),
\end{align}
where the auxiliary integrals were written out according to the notations from (\ref{29-4-4}) and (\ref{29-8-1})--(\ref{29-8-4}).

The remaining two diagrams $d_2$ and $cd_1$ include Green's functions only with matching arguments $x=y$, so we can substitute the decomposition from (\ref{29-3-18}) and get the following answers\footnote{In the last line, the term $\mathrm{J}_4[B]$ is specially written out, despite the fact that it is not singular. This is useful in Section \ref{29:sec:tri:sootn}, so as not to write out the decomposition twice.}
\begin{align}\label{29-4-12}
	d_2-d_2\big|_{B=0}\stackrel{\mathrm{s.p.}}{=}&
	S_m[B]\bigg(-\lambda R_0^\Lambda(0)R_{11}^\Lambda(0)+\lambda m^2R_{10}^\Lambda(0)R_{11}^\Lambda(0)+\lambda m^2R_0^\Lambda(0)R_{21}^\Lambda(0)\bigg)+\\\label{29-4-13}+
	&S_4[B]\bigg(\frac{\lambda^2}{4}R_{11}^\Lambda(0)R_{11}^\Lambda(0)+\frac{\lambda^2}{4}R_0^\Lambda(0)R_{22}^\Lambda(0)\bigg)+\\\label{29-4-14}+&
		\mathrm{J}_1[B]
	\bigg(2R_0^\Lambda(0)-2m^2R_{10}^\Lambda(0)\bigg)+\\+\label{29-4-15}&
	\mathrm{J}_2[B]
	\bigg(-\lambda R_{11}^\Lambda(0)\bigg)+\mathrm{J}_4[B],
\end{align}
\begin{align}\label{29-4-16}
	cd_1-cd_1\big|_{B=0}\stackrel{\mathrm{s.p.}}{=}&
	S_m[B]\bigg(\frac{\lambda (z_{4,1}-\tilde{z}_{2,1})}{2}R_0^\Lambda(0)-\frac{m^2\lambda (z_{4,1}-\tilde{z}_{2,1})}{2}R_{10}^\Lambda(0)-\\&~~~~~~~~~~~~~~-\frac{m^2\lambda (z_1-\tilde{z}_{2,1})}{2}R_{11}^\Lambda(0)+\frac{m^4\lambda z_1}{2}R_{21}^\Lambda(0)\bigg)+\\\label{29-4-17}+
	&S_4[B]\bigg(-\frac{\lambda^2(z_{4,1}-\tilde{z}_{2,1})}{4}R_{11}^\Lambda(0)+\frac{m^2\lambda^2z_1}{8}R_{22}^\Lambda(0)\bigg)+\\\label{29-4-18}+&\mathrm{J}_1[B]
	m^2z_{1}+\mathrm{J}_2[B]
	\frac{\lambda z_{4,1}}{2}.
\end{align}
Note that when calculating the last diagram, we have used the equality
\begin{equation}\label{29-21-1}
\bigg(A_0^{\Lambda}(x)+m^2+\frac{\lambda}{2}B^2(x)\bigg)G^\Lambda(x,y)=\delta(x-y).
\end{equation}

Next, consider the linear combination of the last three diagrams from the right side of (\ref{29-1-31})
\begin{equation*}\label{29-4-19}
\frac{\lambda^2 d_1}{12}-\frac{\lambda d_2}{8}-\frac{cd_1}{2}+\kappa_2=
\frac{\lambda^2 d_1}{12}-\frac{\lambda\Big(d_2-d_2\big|_{B=0}\Big)}{8}-\frac{cd_1-cd_1\big|_{B=0}}{2}.
\end{equation*}
According to the general theory, the resulting object should contain singular contributions only in terms with $S_2$, $S_m$, and $S_4$. Non-local terms containing $PS_1^\Lambda$ should be reduced. Let us make sure that the last condition is true and write equations for the singular parts with $\mathrm{J}_1[B]$ and $\mathrm{J}_2[B]$. They have the form
\begin{equation*}\label{29-4-21}
-\frac{\lambda}{4}R_0^\Lambda(0)+\frac{m^2\lambda}{4}R_{10}^\Lambda(0)\stackrel{\mathrm{s.p.}}{=}
\frac{m^2z_{1}}{2},
\end{equation*}
\begin{equation*}\label{29-4-22}
\frac{\lambda^2}{4}\int_{\mathrm{B}_{1/\sigma}}\mathrm{d}^4x\Big(R_0^\Lambda(x)\Big)^2+
\frac{\lambda^2}{8}R_{11}^\Lambda(0)\stackrel{\mathrm{s.p.}}{=}\frac{\lambda z_{4,1}}{4}.
\end{equation*}
Given the explicit form for the coefficients (\ref{29-4-6})--(\ref{29-4-7}) from the first loop, both ratios are satisfied automatically. Next, using the equalities
\begin{equation*}\label{29-4-23}
\frac{\lambda^2}{4m^2} R_0^\Lambda(0)R_{11}^\Lambda(0)-\frac{\lambda^2} {4}R_{10}^\Lambda(0)R_{11}^\Lambda(0)+\frac{\lambda z_1}{2}R_{11}^\Lambda(0)
\stackrel{\mathrm{s.p.}}{=}\frac{\lambda\tilde{z}_1}{2}R_{11}^\Lambda(0),
\end{equation*}
\begin{equation*}\label{29-4-28}
	-\frac{\lambda^2}{4}R_0^\Lambda(0)R_{21}^\Lambda(0)
	-\frac{m^2\lambda z_1}{2}R_{21}^\Lambda(0)\stackrel{\mathrm{s.p.}}{=}0,
\end{equation*}
\begin{equation*}\label{29-4-27}
-\frac{3\lambda^2}{4}R_0^\Lambda(0)R_{22}^\Lambda(0)-\frac{3m^2\lambda z_1}{2}R_{22}^\Lambda(0)\stackrel{\mathrm{s.p.}}{=}0,
\end{equation*}
the answer for the next order of the renormalization constants is written as follows
\begin{equation*}\label{29-4-24}
z_{2,2}=-\frac{\lambda^2}{48}\mathrm{I}_{1}(\sigma)+
\tilde{z}_{2,2},
\end{equation*}
\begin{equation*}\label{29-4-25}
z_2=\frac{\lambda^2}{6m^2}\mathrm{I}_{4}(\sigma)-\frac{\lambda^2}{2}
\mathrm{I}_{2}(\sigma)
-\frac{\lambda (z_{4,1}-\tilde{z}_{2,1})}{2m^2}R_0^\Lambda(0)+\frac{\lambda (z_{4,1}-\tilde{z}_{2,1})}{2}R_{10}^\Lambda(0)+
\frac{\lambda(\tilde{z}_1-\tilde{z}_{2,1})}{2}R_{11}^\Lambda(0)+
\tilde{z}_{2},
\end{equation*}
\begin{equation*}\label{29-4-26}
z_{4,2}=-3\lambda^2\mathrm{I}_{3}(\sigma)
-\frac{3\lambda^2}{4}R_{11}^\Lambda(0)R_{11}^\Lambda(0)
+3\lambda (z_{4,1}-\tilde{z}_{2,1})R_{11}^\Lambda(0)+
\tilde{z}_{4,2},
\end{equation*}
where $\{\tilde{z}_{2,2},\tilde{z}_{2},\tilde{z}_{4,2}\}$ are arbitrary constants having the asymptotics $\mathcal{O}(1)$ with respect to the parameter $\Lambda$. Further, making the shift of the constants, we get
\begin{equation*}\label{29-4-241}
	z_{2,2}=-\frac{L\hat{\lambda}^2}{6}+
	\bar{z}_{2,2},
\end{equation*}
\begin{multline}\label{29-4-251}
	z_2=
	\frac{\Lambda^2\hat{\lambda}}{6m^2}
	\Big(-36L\hat{\lambda}\big(1+\mathbf{f}(0)\big)+16\pi^2\lambda\alpha_3(\mathbf{f})-12\big(1+\mathbf{f}(0)\big)(\bar{z}_{4,1}-\bar{z}_{2,1})\Big)+\\+2L^2\hat{\lambda}^2+L\Big(-\hat{\lambda}^2+\hat{\lambda}(\bar{z}_1+\bar{z}_{4,1}-2\bar{z}_{2,1}))\Big)
	+\bar{z}_{2},
\end{multline}
\begin{equation*}\label{29-4-261}
	z_{4,2}=9L^2\hat{\lambda}^2+
	L\Big(-6\hat{\lambda}^2+6\hat{\lambda}(\bar{z}_{4,1}-\bar{z}_{2,1})\Big)
	+
	\bar{z}_{4,2}.
\end{equation*}

\section{Three loops}\label{29:sec:tri}
\subsection{Additional definitions}\label{29:sec:tri:note}
Before calculating diagrams, it is convenient to define several auxiliary functions that will allow us to give compact explanations. Let $i_1,i_2,i_3,i_4\in\mathbb{N}\cup\{0\}$, and $\{g_j(\cdot,\cdot)\}_{j=1}^4$ is a set of symmetric functions such that the integrals below converge. Then we define the following values (multidimensional integrals)
\begin{equation*}\label{29-11-1}
\mathrm{I}_{i_1}^{i_2}\big(g_1\big)=
\int_{\mathbb{R}^{4\times2}}\mathrm{d}^4x\mathrm{d}^4y\,
B^{i_1}(x)g_1(x,y)B^{i_2}(y),
\end{equation*}
\begin{equation*}\label{29-11-2}
\mathrm{I}_{\,i_3}^{i_1i_2}\big(g_3,g_2,g_1\big)=
\int_{\mathbb{R}^{4\times3}}\mathrm{d}^4x\mathrm{d}^4y\mathrm{d}^4u\,
B^{i_1}(x)g_1(x,y)B^{i_2}(y)g_2(y,u)B^{i_3}(u)g_3(u,x),
\end{equation*}
\begin{equation*}\label{29-11-3}
\mathrm{I}_{i_3i_4}^{i_1i_2}\big(g_4,g_3,g_2,g_1\big)=
\int_{\mathbb{R}^{4\times4}}\mathrm{d}^4x\mathrm{d}^4y\mathrm{d}^4u\mathrm{d}^4z\,
B^{i_1}(x)g_1(x,y)B^{i_2}(y)g_2(y,u)
B^{i_3}(u)g_3(u,z)B^{i_4}(z)g_4(z,x).
\end{equation*}
 By permuting the variables taking into account the symmetry, it is easy to verify that the following relations are valid
\begin{equation*}\label{29-11-4}
	\mathrm{I}_{\,i_3}^{i_1i_2}\big(g_3,g_2,g_1\big)=
	\mathrm{I}_{\,i_3}^{i_2i_1}\big(g_2,g_3,g_1\big),
\end{equation*}
\begin{equation*}\label{29-11-5}
	\mathrm{I}_{i_3i_4}^{i_1i_2}\big(g_4,g_3,g_2,g_1\big)=
	\mathrm{I}_{i_4i_3}^{i_2i_1}\big(g_2,g_3,g_4,g_1\big)=
	\mathrm{I}_{i_2i_1}^{i_4i_3}\big(g_4,g_1,g_2,g_3\big).
\end{equation*}

Next, let $g(\cdot)$ be a function such that the integral below exists. Let us define the transformation $g(\cdot)\to\hat{g}(\cdot)$ as follows
\begin{equation}\label{29-8-21}
	\hat{g}(y)=\int_{\mathrm{B}_{1/\sigma}}\mathrm{d}^4x
	\Big(R_0^\Lambda(x)\Big)^2g(x+y)-\mathrm{A}(\sigma)g(y).
\end{equation}
It is seen that the integral on the right hand side diverges as $\Lambda\to+\infty$. This is due to the fact that the density of the integrand when removing regularization is proportional to $|x|^{-4}$ and, thus, is not integrable in the four-dimensional space. In turn, the regularized integral contains the logarithmic type singularity $\ln(\Lambda/\sigma)$, which is subtracted with the usage of the term with
$\mathrm{A}(\sigma)$. Thus, the function $\hat{g}(\cdot)$, which actually contains a dependence on the parameter $\Lambda$, has a finite limit when removing the regularization $\Lambda\to+\infty$.
It is easy to verify this by applying the operator $\Lambda\partial_{\Lambda}$ on both sides of the equality and taking the limit
\begin{equation*}\label{29-11-6}
\Lambda\frac{\mathrm{d}\hat{g}(y)}{\mathrm{d}\Lambda}=
\frac{1}{4\pi^4}\int_{\mathrm{B}_{1}}\mathrm{d}^4x
\big(1+\mathbf{f}\big(|x|^2\big)\big)\big(1+\mathbf{f}\big(|x|^2\big)+|x|^2\mathbf{f}^\prime\big(|x|^2\big)\big)g(x/\Lambda+y)-\frac{g(y)}{8\pi^2}
\xrightarrow{\Lambda\to+\infty}0,
\end{equation*}
where we have used formulas (\ref{29-3-17}) and (\ref{29-4-4}).
In Section \ref{29:sec:app:vch} it is shown that the functions $R_0^\Lambda(x)$, $R_{10}^\Lambda(x)$, and $R_{11}^\Lambda(x)$, can also be considered instead of $g(x)$.

\subsection{Diagram $d_3$}\label{29:sec:tri:d-3}

This section is devoted to the analysis of the divergent part for the diagram $d_3$, see Fig. \ref{pic:29-5}. Taking into account the diagram technique in Fig. \ref{pic:29-1} and the notations entered from the previous section, the diagram can be written as
\begin{equation*}\label{29-2-1}
d_3=\mathrm{I}_{11}^{11}\big(G^\Lambda G^\Lambda,G^\Lambda,G^\Lambda G^\Lambda,G^\Lambda).
\end{equation*}

Decompose the regularized Green's function $G^\Lambda(x,y)$ into in- and out-parts\footnote{According to the suggestion and terminology from \cite{Iv-Kh-23}.}, using the decomposition of the unit with the usage of the characteristic function $\chi(\cdot)$ of a set. In this case
\begin{equation*}\label{29-2-2}
G^\Lambda(x,y)=s(x,y)+b(x,y),
\end{equation*}
where
\begin{equation*}\label{29-2-3}
s(x,y)=G^\Lambda(x,y)\chi\big(|x-y|\leqslant1/\sigma\big),\,\,\,
b(x,y)=G^\Lambda(x,y)\chi\big(|x-y|>1/\sigma\big).
\end{equation*}
Here $\sigma$ is a fixed positive dimensional parameter\footnote{Here again, it is assumed that the parameter $\sigma$ matches the one entered earlier. In fact, nothing depends on this parameter at this stage, since it simply determines the division of $\mathbb{R}^4$ into two parts.} $\big([\sigma^{-1}]=[x^\mu]\big)$, large enough that the asymptotic expansion (\ref{29-3-18}) is valid near the diagonal for $G^\Lambda(x,y)$ inside the domain $|x-y|\leqslant1/\sigma$. Therefore, after substituting the described decomposition, the diagram is represented as the sum of subdiagrams
\begin{equation}\label{29-2-4}
\mathrm{I}_{11}^{11}\big(G^\Lambda G^\Lambda,G^\Lambda,G^\Lambda G^\Lambda,G^\Lambda)=
\Bigg(\prod_{i=1}^4\sum_{g_i=b,s}\Bigg)\mathrm{I}_{11}^{11}\big(g_1^2,g_2^{\phantom{1}},g_3^2,g_4^{\phantom{1}}\big),
\end{equation}
where we have used the property $s(x,y)b(x,y)=0$.

Let us analyze the diagrams separately. We will immediately identify several trivial cases that do not contain any singular contribution:
\begin{equation}\label{29-2-5}
\mathrm{I}_{11}^{11}\big(b^2,b,b^2,b\big)\stackrel{\mathrm{s.p.}}{=}0,
\end{equation}
\begin{equation}\label{29-2-6}
\mathrm{I}_{11}^{11}\big(b^2,s,b^2,b\big)\stackrel{\mathrm{s.p.}}{=}0,\,\,\,
\mathrm{I}_{11}^{11}\big(b^2,b,b^2,s\big)\stackrel{\mathrm{s.p.}}{=}0,
\end{equation}
\begin{equation}\label{29-2-7}
\mathrm{I}_{11}^{11}\big(b^2,s,b^2,s\big)\stackrel{\mathrm{s.p.}}{=}0.
\end{equation}
These equalities follow from the facts that $b(x,y)=0$ in the region $|x-y|\leqslant1/\sigma$, and the function $s(x,y)$, after removing the regularization, contains a singularity integrable in 4-dimensional space\footnote{After removing the regularization, the main term of the asymptotics for $x\sim y$ is $\big(4\pi^2|x-y|^2\big)^{-1}$. In the four-dimensional space, such function is integrable.}. Of course, in this case, the regularized functions after integration do not give singularities with respect to the parameter $\Lambda$.

Let us consider a less trivial case where the square of the function $s$ appears. There are two such contributions
\begin{equation*}\label{29-2-8}
\mathrm{I}_{11}^{11}\big(s^2,b,b^2,b\big)=\mathrm{I}_{11}^{11}\big(b^2,b,s^2,b\big),
\end{equation*}
and they are equal to each other, so we can only study the first one. It has the form
\begin{equation}\label{29-2-9}
\mathrm{I}_{11}^{11}\big(s^2,b,b^2,b\big)=
\int_{\mathbb{R}^{4\times4}}\mathrm{d}^4x\mathrm{d}^4y\mathrm{d}^4u\mathrm{d}^4z\,\Big(
B(y)s^2(y,x)B(x)b(x,u)B(u)b^2(u,z)B(z)b(z,y)\Big).
\end{equation}
Note that the function
\begin{equation}\label{29-2-13}
\rho_1(x,y)=
\int_{\mathbb{R}^{4\times2}}\mathrm{d}^4u\mathrm{d}^4z\,\Big(
B(x)b(x,u)B(u)b^2(u,z)B(z)b(z,y)B(y)\Big)
\end{equation}
exists and has a finite limit $\rho_1(y,y)$ on the diagonal, since none of the integrand components has singularities. Therefore, by subtracting and adding the last function, the integral (\ref{29-2-9}) can be rewritten as
\begin{equation}\label{29-2-14}
\mathrm{I}_{11}^{11}\big(s^2,b,b^2,b\big)=
\int_{\mathbb{R}^{4\times2}}\mathrm{d}^4x\mathrm{d}^4y\,s^2(y,x)\big(\rho_1(x,y)-\rho_1(y,y)\big)+
\int_{\mathbb{R}^{4\times2}}\mathrm{d}^4x\mathrm{d}^4y\,s^2(y,x)\rho_1(y,y).
\end{equation}
It is clear that the first term does not contain divergences after removing the regularization, since the difference neutralizes\footnote{From now on, the word "neutralizes" will be used to indicate that density has become integrable in the four-dimensional space before and after the removal of regularization.} the function $|x-y|^{-4}$. In the second integral, it should be noted that $s^2(x,y)$ can be replaced by $s_0^2(x-y)$, where
\begin{equation}\label{29-2-50}
s_0(x)=R_0^\Lambda(x)\chi\big(|x|\leqslant1/\sigma\big),
\end{equation}
since the difference $s^2(x,y)-s_0^2(x-y)$ after removing the regularization is an integrable function in the four-dimensional space. Next, making the shift  $x\to x+y$, we get the factorization
\begin{align}\label{29-2-11}
\mathrm{I}_{11}^{11}\big(s^2,b,b^2,b\big)
&\stackrel{\mathrm{s.p.}}{=}\bigg(\int_{\mathbb{R}^{4}}\mathrm{d}^4x\,s_0^2(x)\bigg)
\int_{\mathbb{R}^{4}}\mathrm{d}^4y\,\rho_1(y,y)
\\\label{29-2-21}
&\stackrel{\mathrm{s.p.}}{=}
\bigg(\int_{\mathbb{R}^{4}}\mathrm{d}^4x\,s_0^2(x)\bigg)
\int_{\mathbb{R}^{4\times3}}\mathrm{d}^4y\mathrm{d}^4u\mathrm{d}^4z\,\Big(
B^2(y)b(y,u)B(u)b^2(u,z)B(z)b(z,y)\Big),
\end{align}
which in compact notation is written out in the form
\begin{equation}\label{29-2-12}
\mathrm{I}_{11}^{11}\big(s^2,b,b^2,b\big)
\stackrel{\mathrm{s.p.}}{=}\mathrm{A}(\sigma)\mathrm{I}^{11}_{\, 2}(b,b,b^2).
\end{equation}

Consider the following set of four subdiagrams from the decomposition (\ref{29-2-4})
\begin{equation*}\label{29-2-10}
\mathrm{I}_{11}^{11}\big(s^2,b,b^2,s\big)=\mathrm{I}_{11}^{11}\big(s^2,s,b^2,b\big)=
\mathrm{I}_{11}^{11}\big(b^2,s,s^2,b\big)=\mathrm{I}_{11}^{11}\big(b^2,b,s^2,s\big),
\end{equation*}
which are also equal to each other. Of course, it is enough to study only the first of them
\begin{equation*}\label{29-2-15}
\mathrm{I}_{11}^{11}\big(s^2,b,b^2,s\big)=
\int_{\mathbb{R}^{4\times4}}\mathrm{d}^4x\mathrm{d}^4y\mathrm{d}^4u\mathrm{d}^4z\,\Big(
B(y)s^2(y,x)B(x)b(x,u)B(u)b^2(u,z)B(z)s(z,y)\Big).
\end{equation*}
At first glance, it may seem that the main divergence will be given by an integral of the form
\begin{equation*}\label{29-2-16}
\int_{\mathbb{R}^{4}}\mathrm{d}^4y\,
s(z,y)B(y)s^2(y,x),
\end{equation*}
which, generally speaking, diverges quadratically at $x\sim y$, and not logarithmically. However, it is necessary to take into account the fact that $s(z,y)$ is also included in the integral by the variable $z$. This leads to the fact that the function
\begin{equation*}\label{29-2-17}
\rho_2(x,y)=
\int_{\mathbb{R}^{4\times2}}\mathrm{d}^4u\mathrm{d}^4z\,\Big(
B(x)b(x,u)B(u)b^2(u,z)B(z)s(z,y)B(y)\Big)
\end{equation*}
has a finite limit on the diagonal $y=x$ before and after removing the regularization. Therefore, we can use addition and subtraction, as was done in formula (\ref{29-2-14}), replacing only the function $\rho_1(x,y)$ with $\rho_2(x,y)$. Next, replacing $s^2(x,y)$ with $s_0^2(x-y)$ again and making the shift $x\to x+y$, we get the factorization and the answer in the form
\begin{equation}\label{29-2-18}
\mathrm{I}_{11}^{11}\big(s^2,b,b^2,s\big)
\stackrel{\mathrm{s.p.}}{=}\bigg(\int_{\mathbb{R}^{4}}\mathrm{d}^4x\,s_0^2(x)\bigg)
\int_{\mathbb{R}^{4}}\mathrm{d}^4y\,\rho_2(y,y)=\mathrm{A}(\sigma)\mathrm{I}^{11}_{\, 2}(s,b,b^2).
\end{equation}

The above method can be used to analyze two more contributions from (\ref{29-2-4})
\begin{equation*}\label{29-2-19}
\mathrm{I}_{11}^{11}\big(s^2,s,b^2,s\big)=\mathrm{I}_{11}^{11}\big(b^2,s,s^2,s\big).
\end{equation*}
In this case, instead of the function $\rho_1(x,y)$ from (\ref{29-2-13}), we must select the function
\begin{equation*}\label{29-2-20}
\rho_3(x,y)=
\int_{\mathbb{R}^{4\times2}}\mathrm{d}^4u\mathrm{d}^4z\,\Big(
B(x)s(x,u)B(u)b^2(u,z)B(z)s(z,y)B(y)\Big).
\end{equation*}
Then, repeating the steps from (\ref{29-2-14})--(\ref{29-2-21}), we get the following answer
\begin{equation}\label{29-2-22}
\mathrm{I}_{11}^{11}\big(s^2,s,b^2,s\big)
\stackrel{\mathrm{s.p.}}{=}\bigg(\int_{\mathbb{R}^{4}}\mathrm{d}^4x\,s_0^2(x)\bigg)
\int_{\mathbb{R}^{4}}\mathrm{d}^4y\,\rho_3(y,y)=\mathrm{A}(\sigma)\mathrm{I}^{11}_{\, 2}(s,s,b^2).
\end{equation}

The remaining terms from decomposition (\ref{29-2-4}) can be represented as
\begin{equation}\label{29-2-23}
\mathrm{I}_{11}^{11}\big(s^2,G^\Lambda,s^2,G^\Lambda\big).
\end{equation}
In this case, it is convenient to make an additional re-decomposition. Indeed, representing $s^2=\big(s^2-s_0^2\big)+s_0^2$ and using the linearity of the functional from (\ref{29-2-23}) by arguments, we can rewrite it in the form
\begin{equation}\label{29-2-24}
\mathrm{I}_{11}^{11}\big(s^2,G^\Lambda,s^2,G^\Lambda\big)=
\mathrm{I}_{11}^{11}\big(s^2-s_0^2,G^\Lambda,s^2-s_0^2,G^\Lambda\big)+
2\mathrm{I}_{11}^{11}\big(s_0^2,G^\Lambda,s^2-s_0^2,G^\Lambda\big)+
\mathrm{I}_{11}^{11}\big(s_0^2,G^\Lambda,s_0^2,G^\Lambda\big).
\end{equation}
Note that the first two contributions on the right hand side contain an integral of the form
\begin{equation*}\label{29-2-25}
\rho_4(u,y)=\int_{\mathbb{R}^4}\mathrm{d}^4z\,\Big(s^2(u,z)-s_0^2(u-z)\Big)B(z)G^\Lambda(z,y),
\end{equation*}
which exists at $u\neq y$ before and after removing the regularization. Indeed, the main term of the asymptotics for $s^2(u,z)-s_0^2(u-z)$, when $u\sim z$ and $\Lambda\to+\infty$, is proportional to $\ln(|u-z|)/|u-z|^2$, while $G(z,y)$ for $z\sim y$ starts with $(4\pi^2|z-y|^2)^{-1}$. Therefore, the main term of the asymptotics for $\rho_4(u,y)$ at $u\sim y$ in the worst case can behave like $\ln^2(|u-y|)$, which is an integrable function. Next, the first two terms of (\ref{29-2-24}) are rewritten as
\begin{equation*}\label{29-2-26}
\mathrm{I}_{11}^{11}\big(s^2-s_0^2,G^\Lambda,s^2-s_0^2,G^\Lambda\big)=
\int_{\mathbb{R}^{4\times2}}\mathrm{d}^4y\mathrm{d}^4u\,B(y)\rho_4(y,u)B(u)\rho_4(u,y),
\end{equation*}
\begin{equation*}\label{29-2-27}
\mathrm{I}_{11}^{11}\big(s_0^2,G^\Lambda,s^2-s_0^2,G^\Lambda\big)=
\int_{\mathbb{R}^{4\times3}}\mathrm{d}^4y\mathrm{d}^4x\mathrm{d}^4u\,B(y)s_0^2(y-x)B(x)G^\Lambda(x,u)B(u)\rho_4(u,y).
\end{equation*}
It is clear that the first functional, taking into account the remark about the behavior of the asymptotics, does not contain terms with a singularity with respect to the parameter $\Lambda$. The second functional can be studied using the procedure from (\ref{29-2-14})--(\ref{29-2-21}), if instead of $\rho_1(x,y)$ we select
\begin{equation*}\label{29-2-28}
\rho_5(x,y)\int_{\mathbb{R}^{4}}\mathrm{d}^4u\,B(x)G^\Lambda(x,u)B(u)\rho_4(u,y)B(y).
\end{equation*}
Then the answers have the following form
\begin{equation}\label{29-2-29}
\mathrm{I}_{11}^{11}\big(s^2-s_0^2,G^\Lambda,s^2-s_0^2,G^\Lambda\big)\stackrel{\mathrm{s.p.}}{=}0,
\end{equation}
\begin{equation}\label{29-2-30}
\mathrm{I}_{11}^{11}\big(s_0^2,G^\Lambda,s^2-s_0^2,G^\Lambda\big)\stackrel{\mathrm{s.p.}}{=}
\bigg(\int_{\mathbb{R}^{4}}\mathrm{d}^4x\,s_0^2(x)\bigg)
\int_{\mathbb{R}^{4}}\mathrm{d}^4y\,\rho_5(y,y)
=\mathrm{A}(\sigma)\mathrm{I}^{11}_{\, 2}(G^\Lambda,G^\Lambda,s^2-s_0^2).
\end{equation}

Let us move on to the third term from (\ref{29-2-24}). To do this, we use another additional auxiliary partition for the regularized Green's function (\ref{29-3-18}), which can be rewritten in the equivalent symmetrized form
\begin{equation}\label{29-2-31}
G^\Lambda(x,y)=R_0^\Lambda(x-y)-
R_{10}^\Lambda(x-y)m^2-
R_{11}^\Lambda(x-y)\tilde{a}_1(x,y)+PS^\Lambda(x,y),
\end{equation}
where
\begin{equation}\label{29-2-51}
\tilde{a}_1(x,y)=\frac{\lambda}{4}\Big(B^2(x)+B^2(y)\Big).
\end{equation}
Then the studied diagram is divided into the sum of nine subdiagrams. Let us study them separately. Firstly, we consider
\begin{equation*}\label{29-2-32}
\mathrm{I}_{11}^{11}\big(s_0^2,PS^\Lambda,s_0^2,PS^\Lambda\big)=
\int_{\mathbb{R}^{4\times4}}\mathrm{d}^4x\mathrm{d}^4y\mathrm{d}^4u\mathrm{d}^4z\,\Big(
B(y)s^2_0(y-x)B(x)PS^\Lambda(x,u)B(u)s_0^2(u-z)B(z)PS^\Lambda(z,y)\Big).
\end{equation*}
Note that the function $PS^\Lambda(x,y)$ by definition has a finite limit on the diagonal $(y=x)$ before and after the removal of regularization. Following the general calculation method, we add and subtract $B(y)PS^\Lambda(y,u)$
\begin{multline*}
\int_{\mathbb{R}^{4\times4}}\mathrm{d}^4x\mathrm{d}^4y\mathrm{d}^4u\mathrm{d}^4z\,
B(y)s^2_0(y-x)\Big(B(x)PS^\Lambda(x,u)-B(y)PS^\Lambda(y,u)\Big)B(u)s_0^2(u-z)B(z)PS^\Lambda(z,y)+\\+
\int_{\mathbb{R}^{4\times4}}\mathrm{d}^4x\mathrm{d}^4y\mathrm{d}^4u\mathrm{d}^4z\,\Big(
B^2(y)s^2_0(y-x)PS^\Lambda(y,u)B(u)s_0^2(u-z)B(z)PS^\Lambda(z,y)\Big).
\end{multline*}
It is clear that in the second term, we can additionally make the shift $x\to x+y$ and get the well known factorization. In the first term, we will once again add and subtract the function $B(u)PS^\Lambda(u,y)$. This choice is due to the fact that the singularity in $s_0^2(y-x)$ was neutralized by the previous subtraction, whereas the second subtraction focused on the function $s_0^2(u-z)$. Then after additional factorization (with the shift $z\to z+u$) in one of the terms we get
\begin{multline*}
	\int_{\mathbb{R}^{4\times4}}\mathrm{d}^4x\mathrm{d}^4y\mathrm{d}^4u\mathrm{d}^4z\,
	B(y)s^2_0(y-x)\Big(B(x)PS^\Lambda(x,u)-B(y)PS^\Lambda(y,u)\Big)\times\\\times B(u)s_0^2(u-z)\Big(B(z)PS^\Lambda(z,y)-B(u)PS^\Lambda(u,y)\Big)+\\+
	\mathrm{A}(\sigma)\int_{\mathbb{R}^{4\times3}}\mathrm{d}^4x\mathrm{d}^4y\mathrm{d}^4u\,
	B(y)s^2_0(y-x)\Big(B(x)PS^\Lambda(x,u)-B(y)PS^\Lambda(y,u)\Big)B^2(u)PS^\Lambda(u,y)
	+\\+\mathrm{A}(\sigma)\mathrm{I}^{11}_{\, 2}\big(PS^\Lambda,PS^\Lambda,s_0^2\big).
\end{multline*}
Obviously, in the first term, both singular densities are neutralized. Therefore, the contribution we are interested in will be given by the second and third terms, which can be written out in a more concise form
\begin{equation}\label{29-2-35}
\mathrm{I}_{11}^{11}\big(s_0^2,PS^\Lambda,s_0^2,PS^\Lambda\big)\stackrel{\mathrm{s.p.}}{=}
2\mathrm{A}(\sigma)\mathrm{I}^{11}_{\, 2}(PS^\Lambda,PS^\Lambda,s_0^2)-\mathrm{A}^2(\sigma)\mathrm{I}^{2}_{2}\big((PS^\Lambda)^2\big).
\end{equation}
Note that the procedure described above can also be applied in the case when one of the functions $PS^\Lambda$ is replaced by one of the first three terms from the right hand side of (\ref{29-2-31}). For the sake of certainty, let us assume that $PS^\Lambda(z,y)$ has been replaced by $R_0^\Lambda(z-y)$. In this case, we can use the first subtraction again, since $PS^\Lambda(x,u)$ remains the same. And this subtraction neutralizes the singularity in $s_0^2(y-x)$. The second subtraction, already with the function $B(u)R_0^\Lambda(u-y)$ instead of $B(u)PS^\Lambda(u,y)$, can be done for the reason that the integral
\begin{equation*}\label{29-2-36}	
	\rho_6(z,u)=
\int_{\mathbb{R}^{4\times2}}\mathrm{d}^4x\mathrm{d}^4y\,
B(z)R_0^\Lambda(z-y)B(y)s^2_0(y-x)\Big(B(x)PS^\Lambda(x,u)-B(y)PS^\Lambda(y,u)\Big)B(u)
\end{equation*}
has a finite limit on the diagonal $z=u$ before and after removing the regularization. In fact, in the second time, $\rho_6(u,u)$ is added and subtracted. Therefore, taking into account the last remarks, the following relations are valid
\begin{equation}\label{29-2-52}
\mathrm{I}_{11}^{11}\big(s_0^2,PS^\Lambda,s_0^2,R_{10}^\Lambda m^2\big)\stackrel{\mathrm{s.p.}}{=}
2\mathrm{A}(\sigma)\mathrm{I}^{11}_{\, 2}(PS^\Lambda,R_{10}^\Lambda m^2,s_0^2)-\mathrm{A}^2(\sigma)\mathrm{I}^{2}_{2}(PS^\Lambda R_{10}^\Lambda m^2),
\end{equation}
\begin{equation}\label{29-2-37}
\mathrm{I}_{11}^{11}\big(s_0^2,PS^\Lambda,s_0^2,R_{11}^\Lambda \tilde{a}_1\big)\stackrel{\mathrm{s.p.}}{=}
2\mathrm{A}(\sigma)\mathrm{I}^{11}_{\, 2}(PS^\Lambda,R_{11}^\Lambda \tilde{a}_1,s_0^2)-\mathrm{A}^2(\sigma)\mathrm{I}^{2}_{2}(PS^\Lambda R_{11}^\Lambda \tilde{a}_1),
\end{equation}
\begin{equation}\label{29-2-38}
\mathrm{I}_{11}^{11}\big(s_0^2,PS^\Lambda,s_0^2,R_0^\Lambda\big)\stackrel{\mathrm{s.p.}}{=}
2\mathrm{A}(\sigma)\mathrm{I}^{11}_{\, 2}(PS^\Lambda,R_0^\Lambda,s_0^2)-\mathrm{A}^2(\sigma)\mathrm{I}^{2}_{2}(PS^\Lambda R_0^\Lambda).
\end{equation}

Let us move on to the study of the last four contributions out of nine. They can be written out in a compact form using an additional notation. Indeed, let us introduce the functions
\begin{equation}\label{29-2-41}
P_0(x)=R_0^\Lambda(x),\,\,\,P_1(x)=R_{10}^\Lambda(x),\,\,\,
P_2(x)=R_{11}^\Lambda(x),
\end{equation}
\begin{equation}\label{29-2-42}
p_0(x,y)=1,\,\,\,p_1(x,y)=-m^2,\,\,\,p_2(x,y)=-\tilde{a}_1(x,y),
\end{equation}
where the lower index is equal to the combinations we are interested in. Then the integral takes the form
\begin{multline}\label{29-2-40}
\mathrm{I}_{11}^{11}\big(s_0^2,P_ip_i,s_0^2,P_jp_j\big)=
\int_{\mathbb{R}^{4\times4}}\mathrm{d}^4x\mathrm{d}^4y\mathrm{d}^4u\mathrm{d}^4z\,
B(y)s_0^2(y,x)B(x)P_i(x-u)p_i(x,u)\times\\\times B(u)s_0^2(u,z)B(z)P_j(z-y)p_j(z,y).
\end{multline}
As before, we will use the addition and subtraction method. However, in this case, an adjustment must be introduced: only smooth components, that is, densities $p_i$, and not singular multipliers $P_i$, should be subtracted. Then the integral (\ref{29-2-40}) can be represented as the sum of four
\begin{align}\nonumber
\int_{\mathbb{R}^{4\times4}}\mathrm{d}^4x\mathrm{d}^4y\mathrm{d}^4u\mathrm{d}^4z\,
B(y)&s^2_0(y-x)P_i(x-u)\Big(B(x)p_i(x,u)-B(y)p_i(y,u)\Big)\times
\\&~~~~~~~~~~~~~~~~\times \nonumber
B(u)s_0^2(u-z)P_j(z-y)\Big(B(z)p_j(z,y)-B(u)p_j(u,y)\Big)+
\\+
\int_{\mathbb{R}^{4\times4}}\mathrm{d}^4x\mathrm{d}^4y\mathrm{d}^4u\mathrm{d}^4z\,&
B^2(y)s^2_0(y-x)P_i(x-u)p_i(y,u)\times\nonumber
\\&~~~~~~~~~~~~~~~~\times 
B(u)s_0^2(u-z)P_j(z-y)\Big(B(z)p_j(z,y)-B(u)p_j(u,y)\Big)+\label{29-2-39}
\\+
\int_{\mathbb{R}^{4\times4}}\mathrm{d}^4x\mathrm{d}^4y\mathrm{d}^4u\mathrm{d}^4z\,&\nonumber
B(y)s^2_0(y-x)P_i(x-u)\Big(B(x)p_i(x,u)-B(y)p_i(y,u)\Big)\times
\\&~~~~~~~~~~~~~~~~\times \nonumber
B^2(u)s_0^2(u-z)P_j(z-y)p_j(u,y)+
\\+
\int_{\mathbb{R}^{4\times4}}\mathrm{d}^4x\mathrm{d}^4y\mathrm{d}^4u\mathrm{d}^4z\,&\nonumber
B^2(y)s^2_0(y-x)P_i(x-u)p_i(y,u)\times
\\&~~~~~~~~~~~~~~~~\times \nonumber
B^2(u)s_0^2(u-z)P_j(z-y)p_j(u,y).
\end{align}
Here, the first term does not actually contain singularities with $\Lambda$. It can be noted that the integral
\begin{equation}\label{29-2-43}
\rho_7^i(y,u)=\int_{\mathbb{R}^{4}}\mathrm{d}^4x\,
B(y)s^2_0(y-x)P_i(x-u)\Big(B(x)p_i(x,u)-B(y)p_i(y,u)\Big)
\end{equation}
converges before and after the removal of regularization. Moreover, when $i>0$, the limit $\rho_7^i(u,u)$ exists and is finite. While the worst asymptotic term, which can appear in $\rho_7^i(y,u)$ with $y\sim u$ and $i\in\{0,1,2\}$, is proportional to $|y-u|^{-1}$ (in absolute value). This reasoning proves that the first integral in (\ref{29-2-39}) does not give divergences. 

In the second, third, and fourth integrals, we need to use the property from (\ref{29-8-21}) for regularized functions. For example, in the second integral we obtain the following chain of relations
\begin{multline}\label{29-2-44}
\int_{\mathbb{R}^{4}}\mathrm{d}^4x\,s_0^2(y-x)P_i(x-u)=
\int_{\mathbb{R}^{4}}\mathrm{d}^4x\,s_0^2(x)P_i(x+(y-u))=\\=
\int_{\mathrm{B}_{1/\sigma}}\mathrm{d}^4x\,\Big(R_0^\Lambda(x)\Big)^2P_i(x+(y-u))=
\mathrm{A}(\sigma)P_i(y-u)+\hat{P}_i(y-u).
\end{multline}
Additionally, we note that the contributions containing the functions $\rho_7^i$ and $\hat{P}_i$ will be finite after the regularization is removed, since their densities are integrable in the four-dimensional space. Indeed, in the worst case, when $i=j=0$, the modulus of the main term of the asymptotics is proportional to $|\cdot|^{-3}\ln(|\cdot|)$.
Thus, taking into account the above remarks, the intermediate result is rewritten as
\begin{align}\nonumber
	2\mathrm{A}(\sigma)\int_{\mathbb{R}^{4\times2}}\mathrm{d}^4y\mathrm{d}^4u&\,
	B^2(y)P_i(y-u)p_i(y,u)\rho_7^j(u,y)+
	\\+
	\mathrm{A}^2(\sigma)&\int_{\mathbb{R}^{4\times2}}\mathrm{d}^4y\mathrm{d}^4u\,\label{29-2-45}
	B^2(y)P_i(y-u)p_i(y,u)B^2(u)P_j(u-y)p_j(u,y)+
		\\+
	2\mathrm{A}(\sigma)&\int_{\mathbb{R}^{4\times2}}\mathrm{d}^4y\mathrm{d}^4u\,\nonumber
	B^2(y)P_i(y-u)p_i(y,u)B^2(u)\hat{P}_j(u-y)p_j(u,y)+
		\\+
	&\int_{\mathbb{R}^{4\times2}}\mathrm{d}^4y\mathrm{d}^4u\,\nonumber
	B^2(y)\hat{P}_i(y-u)p_i(y,u)B^2(u)\hat{P}_j(u-y)p_j(u,y).
\end{align}
Next, substitute the function $\rho_7^j(u,y)$ from (\ref{29-2-43}) into the first term using the relation from (\ref{29-2-44})
\begin{equation*}\label{29-2-46}
	\rho_7^j(u,y)=\int_{\mathbb{R}^{4}}\mathrm{d}^4x\,
	B(u)s^2_0(u-x)P_i(x-y)B(x)p_j(x,y)-
\Big(\mathrm{A}(\sigma)P_j(u-y)+\hat{P}_j(u-y)\Big)B^2(u)p_j(u,y).
\end{equation*}
Then the sum from (\ref{29-2-45}) is rewritten as three terms
\begin{align}\nonumber
	2\mathrm{A}(\sigma)\int_{\mathbb{R}^{4\times3}}\mathrm{d}^4x\mathrm{d}^4y\mathrm{d}^4u&\,
	B^2(y)P_i(y-u)p_i(y,u)B(u)s^2_0(u-x)B(x)P_i(x-y)p_j(x,y)+
	\\-
	\mathrm{A}^2(\sigma)&\int_{\mathbb{R}^{4\times2}}\mathrm{d}^4y\mathrm{d}^4u\,\nonumber
	B^2(y)P_i(y-u)p_i(y,u)B^2(u)P_j(u-y)p_j(u,y)+
	\\+
	&\int_{\mathbb{R}^{4\times2}}\mathrm{d}^4y\mathrm{d}^4u\,\nonumber
	B^2(y)\hat{P}_i(y-u)p_i(y,u)B^2(u)\hat{P}_j(u-y)p_j(u,y).
\end{align}
Considering the fact that the last term gives a singularity with respect to the parameter $\Lambda$ only in the case when $i=j=0$, the answer for (\ref{29-2-40}) is as follows
\begin{equation}\label{29-2-48}
\mathrm{I}_{11}^{11}\big(s_0^2,P_ip_i,s_0^2,P_jp_j\big)\stackrel{\mathrm{s.p.}}{=}
2\mathrm{A}(\sigma)\mathrm{I}^{11}_{\, 2}\big(P_ip_i,P_jp_j,s_0^2\big)-\mathrm{A}^2(\sigma)\mathrm{I}^{2}_{2}\big(P_ip_iP_jp_j\big)+\delta_{i0}\delta_{j0}\mathrm{I}^{2}_{2}(\hat{P}_0^2).
\end{equation}
Finally, adding up (with corresponding coefficients) formulas (\ref{29-2-5})--(\ref{29-2-7}), (\ref{29-2-12}), (\ref{29-2-18}), (\ref{29-2-22}), (\ref{29-2-29}), (\ref{29-2-30}), (\ref{29-2-35}), (\ref{29-2-52})--(\ref{29-2-38}), and  (\ref{29-2-48}), we get the following statement.
\begin{figure}[h]
	\center{\includegraphics[width=0.085\linewidth]{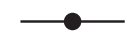}}
	\caption{The additional vertex for the functional from (\ref{29-2-54}).}
	\label{pic:29-7}
\end{figure}
\begin{figure}[h]
	\center{\includegraphics[width=0.56\linewidth]{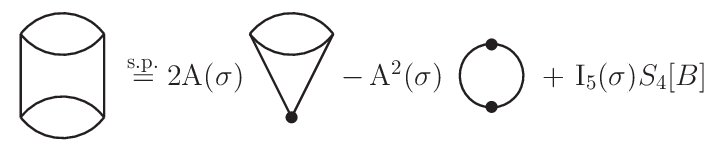}}
	\caption{Diagrammatic representation for equality (\ref{29-2-49}).}
	\label{pic:29-8}
\end{figure}
\begin{lemma}\label{29-l-2}
Taking into account all of the above, the following relation holds
\begin{equation}\label{29-2-49}
	\mathrm{I}_{11}^{11}\big((G^\Lambda)^2,G^\Lambda,(G^\Lambda)^2,G^\Lambda\big)\stackrel{\mathrm{s.p.}}{=}
	2\mathrm{A}(\sigma)\mathrm{I}^{11}_{\,2}\big(G^\Lambda,G^\Lambda,(G^\Lambda)^2\big)
	-\mathrm{A}^2(\sigma)\mathrm{I}^{2}_{2}\big((G^\Lambda)^2\big)
	+\mathrm{I}_{5}(\sigma)S_4[B],
\end{equation}
where the integral $\mathrm{I}_{5}(\sigma)$ is defined by the equality from (\ref{29-8-5}).
\end{lemma}
Note that the latter relation allows for a very elegant diagrammatic representation, shown in Fig. \ref{pic:29-8}. To do this, we need to enter one additional vertex, see Fig. \ref{pic:29-7}, for the functional
\begin{equation}\label{29-2-54}
\int_{\mathbb{R}^4}\mathrm{d}^4x\,B^2(x)\phi(x)
\end{equation}
in addition to the vertices in Fig. \ref{pic:29-1} and \ref{pic:29-2}. Such diagrammatic equality is a clear example of the fairness of the $\mathcal{R}$-operation. Indeed, subtracting from the diagram on the left hand side a twice-occurring subdiagram with "loop" and adding a subdiagram with "loop squared", we get a part of the classical action with a singular coefficient.

\subsection{Diagram $d_4$}
\label{29:sec:tri:d-4}

Let us move on to the study of the $d_4$ diagram shown in Fig. \ref{pic:29-5}. It contains six Green's functions $G^\Lambda$, but they are arranged in such a way that, unlike $d_3$, the diagram $d_4$ contains only a logarithmic singularity of the first order. Let us move on to the explicit form
\begin{align}\label{29-6-1}
d_4=
\int_{\mathbb{R}^{4\times4}}\mathrm{d}^4x\mathrm{d}^4y\mathrm{d}^4u\mathrm{d}^4z\,&
B(x)B(y)B(u)B(z)\times\\\nonumber\times&
G^{\Lambda}(x,y)G^{\Lambda}(x,u)G^{\Lambda}(x,z)
G^{\Lambda}(y,u)G^{\Lambda}(y,z)G^{\Lambda}(u,z).
\end{align}
The analysis of this diagram mainly uses the method proposed in the previous section, with the division of the Green's function into in- and out-parts. Therefore, we will pay attention only to the key feature. Namely, the fact that the following integral
\begin{equation}\label{29-6-2}
\Upsilon(R_0;z)=
\int_{\mathbb{R}^{4\times2}}\mathrm{d}^4x\mathrm{d}^4y\,
R_0(x-y)R_0(x)R_0(x-z)
R_0(y)R_0(y-z)=\frac{3\zeta(3)}{2^7\pi^4}R_0(z),
\end{equation}
constructed using the functions from (\ref{29-3-15}) without regularization, converges and is explicitly calculated, see \cite{29-5,29-6}. Therefore, in the diagram from (\ref{29-6-1}), only the contribution constructed from the main part of the asymptotics of the Green function can be considered. It also follows from relation (\ref{29-6-2}) that the singularity is proportional to the logarithm in the first power.

Let us define the following auxiliary function
\begin{equation}\label{29-6-3}
R_0^{\Lambda,\sigma}(x)=R_0^\Lambda(x)\chi\big(|x|\leqslant1/\sigma\big),
\end{equation}
then, taking into account the last remarks, the singular part is written out in the form
\begin{align}\nonumber
d_4\stackrel{\mathrm{s.p.}}{=}
\int_{\mathbb{R}^{4\times4}}\mathrm{d}^4x\mathrm{d}^4y\mathrm{d}^4u\mathrm{d}^4z\,&
B(x)B(y)B(u)B(z)\times\\\nonumber\times&
R_0^{\Lambda,\sigma}(x-y)R_0^{\Lambda,\sigma}(x-u)R_0^{\Lambda,\sigma}(x-z)
R_0^{\Lambda,\sigma}(y-u)R_0^{\Lambda,\sigma}(y-z)R_0^{\Lambda,\sigma}(u-z).
\end{align}
It is convenient to use the factorization here. To do this, select\footnote{Because of the symmetry, choosing a specific point is not a limitation.} the point $u$ and re-decompose the background fields near it. This is possible because the distance between all arguments is limited by the characteristic functions. Thus, keeping only the main order and making the following series of substitutions
\begin{equation*}\label{29-6-6}
x\to x+u,\,\,\,
y\to y+u,\,\,\,
z\to z+u,
\end{equation*}
we can make sure that the equality holds
\begin{equation}\label{29-6-5}
d_4\stackrel{\mathrm{s.p.}}{=}S_4[B]
\alpha(\Lambda,\sigma),
\end{equation}
where
\begin{equation*}\label{29-6-9}
\alpha(\Lambda,\sigma)=
\int_{\mathbb{R}^{4\times3}}\mathrm{d}^4x\mathrm{d}^4y\mathrm{d}^4z\,
R_0^{\Lambda,\sigma}(x-y)R_0^{\Lambda,\sigma}(x)
R_0^{\Lambda,\sigma}(x-z)
R_0^{\Lambda,\sigma}(y)R_0^{\Lambda,\sigma}(y-z)R_0^{\Lambda,\sigma}(z),
\end{equation*}
which, taking into account the definition from (\ref{29-6-2}), is rewritten as
\begin{equation}\label{29-6-9-1}
\alpha(\Lambda,\sigma)=\int_{\mathrm{B}_{1/\sigma}}\mathrm{d}^4z\,R_0^\Lambda(z)\Upsilon(R_0^{\Lambda,\sigma};z).
\end{equation}
Next, the basic idea is to perform two transitions
\begin{equation}\label{29-6-9-2}
\Upsilon(R_0^{\Lambda,\sigma};z)\to\Upsilon(R_0^{\Lambda};z)\to\Upsilon(R_0^{\phantom{1}};z)
\end{equation}
in the integrand from (\ref{29-6-9-1}). Such transformations do not affect the singular part of the integral under study. Let us look at these transitions in more detail.

In the first case, the support of the function is expanded (\ref{29-6-3}) by moving to the point $\sigma=0$. Using  formula (\ref{29-6-2}), it can be argued that both functions, $\Upsilon(R_0^{\Lambda,\sigma};z)$ and $\Upsilon(R_0^{\Lambda};z)$, are determined by convergent integrals and, when removing the regularization $\Lambda\to+\infty$, they can have a singularity only at the point $z=0$. Moreover, the additional term with the difference
\begin{equation*}\label{29-6-9-3}
\Upsilon(R_0^{\Lambda,\sigma};z)-\Upsilon(R_0^{\Lambda};z)
\end{equation*}
is a limited function before and after the removal of regularization. This is due to the combination of the corresponding supports. For example, consider one of the contributions
\begin{equation*}\label{29-6-9-4}
\int_{\mathbb{R}^{4\times2}}
\mathrm{d}^4x\mathrm{d}^4y
\Big[\Big(R_0^{\Lambda,\sigma}(x-y)-R_0^{\Lambda}(x-y)\Big)R_0^{\Lambda,\sigma}(x)
R_0^{\Lambda,\sigma}(x-z)
R_0^{\Lambda,\sigma}(y)R_0^{\Lambda,\sigma}(y-z)\Big]\Big|_{\Lambda\to+\infty}.
\end{equation*}
It is clear that with the tendency $|z|\to0$, one would expect divergence due to the appearance of a non-integrable density in the integrand. Indeed, the combination $R_0(y)R_0(y-z)$ tends to the behavior of $|y|^{-4}$, which is a non-integrable function. However, due to the combination of supports $\chi\big(|x-y|>1/\sigma\big)\chi\big(|x|\leqslant 1/\sigma\big)$ the domain of integration is reduced and, thus, the bad behavior is neutralized.

In the second transition (\ref{29-6-9-2}), we add the functions whose supports are included into $\mathrm{B}_{1/\Lambda}$. This, taking into account the convergence of the integral from (\ref{29-6-2}), leads to the fact that the additional term with the corresponding difference as a function of $z/\Lambda$ is rewritten as
\begin{align}\nonumber
\Upsilon(R_0^{\Lambda};z/\Lambda)-\Upsilon(R_0^{\phantom{1}};z/\Lambda)=\,&
\Lambda^2\Big(\Upsilon(R_0^{1};z)-\Upsilon(R_0^{\phantom{1}};z)\Big)\\\nonumber
=\,&
\Lambda^2|z|^{-2}\Big(\Upsilon(R_0^{|z|};z/|z|)-\Upsilon(R_0^{\phantom{1}};z/|z|)\Big).
\end{align}
Moreover, it can be checked that for large values of $|z|$, the last combination behaves like $\Lambda^2|z|^{-2-\delta}$, where $\delta>0$. This is due to the fact that the value $\Upsilon(R_0^{\phantom{1}};z/|z|)=q$ does not really depend on the variable $z$, while the function $\Upsilon(R_0^{|z|};z/|z|)$ tends to the number $q$ at $|z|\to+\infty$ fast enough due to the tendency of the characteristic function $\chi\big(|\cdot|\leqslant1/|z|\big)$ to the unit. At the same time, the behavior near zero, as before, is no worse than $\Lambda^2|z|^{-2}$. The latter reasoning means that the integral of the difference is finite at $\Lambda\to+\infty$
\begin{equation*}\label{29-6-9-6}
\bigg|\int_{\mathrm{B}_{1/\sigma}}\mathrm{d}^4z\,R_0^\Lambda(z)\Big(\Upsilon(R_0^{\Lambda};z)-\Upsilon(R_0^{\phantom{1}};z)\Big)\bigg|\leqslant
\int_{\mathbb{R}^4}\mathrm{d}^4z\Big|R_0^1(z)\Big(\Upsilon(R_0^{1};z)-\Upsilon(R_0^{\phantom{1}};z)\Big)\Big|<\infty.
\end{equation*}

Thus, the double transition from (\ref{29-6-9-2}) changes the integral from (\ref{29-6-9-1}) only by a finite value. Therefore, the following chain of equalities is valid
\begin{equation}\label{29-6-9-7}
\alpha(\Lambda,\sigma)\stackrel{\mathrm{s.p.}}{=}\int_{\mathrm{B}_{1/\sigma}}\mathrm{d}^4z\,R_0^\Lambda(z)\Upsilon(R_0;z)\stackrel{\mathrm{s.p.}}{=}\frac{3\zeta(3)L}{16(4\pi^2)^3}.
\end{equation}
Finally, using relations (\ref{29-6-5}) and (\ref{29-6-9-7}), we can formulate one more statement.

\begin{lemma}\label{29-l-3}
Taking into account all of the above, the equality holds
\begin{equation*}\label{29-6-15}
d_4\stackrel{\mathrm{s.p.}}{=}\frac{3\zeta(3)L}{16(4\pi^2)^3}S_4[B].
\end{equation*}
\end{lemma}

\subsection{Diagram $d_5$}
\label{29:sec:tri:d-5}

Let us move on to the analysis of the third three-loop diagram $d_5$, see Fig. \ref{pic:29-5}. The study of this diagram largely repeats the steps of studying $d_3$, so some calculations will be omitted or replaced with brief comments. First, we write out the necessary auxiliary decompositions:
\begin{equation*}\label{29-5-1}
G^\Lambda(x,y)=R_0^\Lambda(x-y)+PS_3^\Lambda(x,y),
\end{equation*}
where $R_0^\Lambda(x)$ is defined in (\ref{29-3-17}), and
\begin{equation}\label{29-5-18}
PS_3^\Lambda(x,y)=-R_{10}^\Lambda(x-y)m^2-R_{11}^\Lambda(x-y)\frac{\lambda \big(B^2(x)+B^2(y)\big)}{4}
	+PS^\Lambda(x,y),
\end{equation}
and also
\begin{equation*}\label{29-5-2}
\Big(G^\Lambda(x,y)\Big)^2=b_1(x,y)+2s_0(x-y)PS_3^\Lambda(x,y)+s_0^2(x-y),
\end{equation*}
where the function $b_1(x,y)$ is defined by the last equality.

Then the diagram $d_5$ is represented as the following sum of subdiagrams
\begin{equation*}\label{29-5-3}
\mathrm{I}^{11}_{\,0}\big((G^\Lambda)^2,(G^\Lambda)^2,G^\Lambda\big)=
\sum_{g_1^{\phantom{1}},g_2^{\phantom{1}}\in
	\{b_1^{\phantom{1}},2s_0^{\phantom{1}}PS_3^\Lambda,s_0^2\}\,\,}
\sum_{g_3^{\phantom{1}}=PS_3^\Lambda,R_0^\Lambda}\mathrm{I}^{11}_{\,0}(g_1,g_2,g_3).
\end{equation*}
According to the general idea, we will consider all the contributions separately. First, let us highlight the terms that obviously (taking into account the considerations from Section \ref{29:sec:tri:d-3}) do not contain singularities with respect to the regularizing parameter $\Lambda$. They are as follows:
\begin{equation*}\label{29-5-4}
\mathrm{I}^{11}_{\,0}\big(b_1^{\phantom{1}},b_1^{\phantom{1}},PS_3^\Lambda\big)\stackrel{\mathrm{s.p.}}{=}0,\,\,\,
2\mathrm{I}^{11}_{\,0}\big(2s_0^{\phantom{1}}PS_3^\Lambda,b_1^{\phantom{1}},PS_3^\Lambda\big)\stackrel{\mathrm{s.p.}}{=}0,\,\,\,
\mathrm{I}^{11}_{\,0}\big(b_1^{\phantom{1}},b_1^{\phantom{1}},R_0^\Lambda\big)\stackrel{\mathrm{s.p.}}{=}0,
\end{equation*}
\begin{equation*}\label{29-5-5}
\mathrm{I}^{11}_{\,0}\big(2s_0^{\phantom{1}}PS_3^\Lambda,2s_0^{\phantom{1}}PS_3^\Lambda,PS_3^\Lambda\big)\stackrel{\mathrm{s.p.}}{=}0,\,\,\,
2\mathrm{I}^{11}_{\,0}\big(2s_0^{\phantom{1}}PS_3^\Lambda,b_1^{\phantom{1}},R_0^\Lambda\big)\stackrel{\mathrm{s.p.}}{=}0,\,\,\,
\mathrm{I}^{11}_{\,0}\big(2s_0^{\phantom{1}}PS_3^\Lambda,2s_0^{\phantom{1}}PS_3^\Lambda,R_0^\Lambda\big)\stackrel{\mathrm{s.p.}}{=}0.
\end{equation*}
For example, let us comment on the last most non-trivial contribution
\begin{equation*}\label{29-5-6}
\int_{\mathbb{R}^{4\times3}}\mathrm{d}^4x\mathrm{d}^4y\mathrm{d}^4z\,
2s_0^{\phantom{1}}(z-x)PS_3^\Lambda(z,x)2s_0^{\phantom{1}}(x-y)PS_3^\Lambda(x,y)B(y)R_0^\Lambda(y-z)B(z).
\end{equation*}
All its component parts contain the function $s_0$. However, they are integrable after the regularization is removed. Indeed, consider the function
\begin{equation*}\label{29-5-7}
\rho_8(x,z)=\int_{\mathbb{R}^{4}}\mathrm{d}^4y\,
2s_0^{\phantom{1}}(x-y)PS_3^\Lambda(x,y)B(y)R_0^\Lambda(y-z)B(z).
\end{equation*}
After removing the regularization near the diagonal $x\sim z$, the main term of the asymptotics behaves like $\ln^2(|x-z|)$ or weaker. Therefore, when integrated with the remaining part, we obtain the final result, since the function $\ln^3(|x|)/|x|^2$ is integrable near zero in the four-dimensional space.

Next, consider the following three contributions
\begin{equation*}\label{29-5-8}
2\mathrm{I}^{11}_{\,0}\big(s_0^2,b_1^{\phantom{1}},PS_3^\Lambda\big)+
2\mathrm{I}^{11}_{\,0}\big(s_0^2,b_1^{\phantom{1}},R_0^\Lambda\big)+
2\mathrm{I}^{11}_{\,0}\big(s_0^2,2s_0^{\phantom{1}}PS_3^\Lambda,PS_3^\Lambda\big),
\end{equation*}
which can be analyzed with the usage of the main method from Section \ref{29:sec:tri:d-3}. Indeed, we note that the functions
\begin{equation*}\label{29-5-11}
\rho_9(x,z)=\int_{\mathbb{R}^{4}}\mathrm{d}^4y\,
b_1^{\phantom{1}}(x,y)B(y)PS_3^\Lambda(y,z)B(z),
\end{equation*}
\begin{equation*}\label{29-5-12}
\rho_{10}(x,z)=\int_{\mathbb{R}^{4}}\mathrm{d}^4y\,
b_1^{\phantom{1}}(x,y)B(y)R_0^\Lambda(y-z)B(z),
\end{equation*}
\begin{equation*}\label{29-5-13}
\rho_{11}(x,z)=\int_{\mathbb{R}^{4}}\mathrm{d}^4y\,
2s_0^{\phantom{1}}(x-y)PS_3^\Lambda(x,y)B(y)PS_3^\Lambda(y,z)B(z),
\end{equation*}
have finite limits at $x=z$ before and after the removal of regularization. Therefore, we can use the subtracting and adding of $\rho_i(z,z)$ in each of the integrals, where $i=9,10,11$. The part with the difference do not contain singularities with respect to $\Lambda$, since $\rho_i(x,z)-\rho_i(z,z)$ neutralizes the singularity in $s_0^2(x-z)$. As a result, the remaining part after shifting $x\to z+x$ is factorized, and the equalities are obtained
\begin{equation*}\label{29-5-14}
\mathrm{I}^{11}_{\,0}\big(s_0^2,b_1^{\phantom{1}},PS_3^\Lambda\big)\stackrel{\mathrm{s.p.}}{=}\mathrm{A}(\sigma)\mathrm{I}^1_1\big(b_1^{\phantom{1}}PS_3^\Lambda\big),
\end{equation*}
\begin{equation*}\label{29-5-15}
\mathrm{I}^{11}_{\,0}\big(s_0^2,b_1^{\phantom{1}},R_0^\Lambda\big)\stackrel{\mathrm{s.p.}}{=}\mathrm{A}(\sigma)\mathrm{I}^1_1\big(b_1^{\phantom{1}}R_0^\Lambda\big),
\end{equation*}
\begin{equation*}\label{29-5-10}
\mathrm{I}^{11}_{\,0}\big(s_0^2,2s_0^{\phantom{1}}PS_3^\Lambda,PS_3^\Lambda\big)\stackrel{\mathrm{s.p.}}{=}\mathrm{A}(\sigma)\mathrm{I}^1_1\big(2s_0^{\phantom{1}}PS_3^\Lambda PS_3^\Lambda\big).
\end{equation*}

The remaining contributions have the following form
\begin{equation}\label{29-5-9}
\mathrm{I}^{11}_{\,0}\big(s_0^2,s_0^2,PS_3^\Lambda\big)+
2\mathrm{I}^{11}_{\,0}\big(2s_0^{\phantom{1}}PS_3^\Lambda,s_0^2,R_0^\Lambda\big)+
\mathrm{I}^{11}_{\,0}\big(s_0^2,s_0^2,R_0^\Lambda\big).
\end{equation}
Let us start with the first term and use the additional decomposition for the function $PS_3^\Lambda$ (\ref{29-5-18}) in the equivalent symmetrized form, see (\ref{29-2-31})--(\ref{29-2-51}) and (\ref{29-2-41})--(\ref{29-2-42}),
\begin{equation*}\label{29-5-16}
PS_3^\Lambda(x,y)=P_1(x-y)p_1(x,y)+P_2(x-y)p_2(x,y)+PS^\Lambda(x,y).
\end{equation*}
Then the first contribution from (\ref{29-5-9}) is represented by the sum of three terms
\begin{equation}\label{29-5-17}
\mathrm{I}^{11}_{\,0}\big(s_0^2,s_0^2,PS_3^\Lambda\big)=
\mathrm{I}^{11}_{\,0}\big(s_0^2,s_0^2,PS^\Lambda\big)+
\mathrm{I}^{11}_{\,0}\big(s_0^2,s_0^2,P_1^{\phantom{1}}p_1^{\phantom{1}}\big)+
\mathrm{I}^{11}_{\,0}\big(s_0^2,s_0^2,P_2^{\phantom{1}}p_2^{\phantom{1}}\big).
\end{equation}
Let us consider the first one. In this case, we subtract and add $B(y)PS^\Lambda(y,x)B(x)$, then we get two integrals
\begin{multline}\label{29-5-19}
\int_{\mathbb{R}^{4\times3}}\mathrm{d}^4x\mathrm{d}^4y\mathrm{d}^4z\,
s_0^2(z-x)s_0^2(x-y)\Big(B(y)PS^\Lambda(y,z)B(z)-B(y)PS^\Lambda(y,x)B(x)\Big)+\\+
\int_{\mathbb{R}^{4\times3}}\mathrm{d}^4x\mathrm{d}^4y\mathrm{d}^4z\,
s_0^2(z-x)s_0^2(x-y)B(y)PS^\Lambda(y,x)B(x).
\end{multline}
It is clear that in the second term, after shifting $z\to z+x$, we obtain the well known factorization. In the first case, the difference in parentheses neutralizes the singularity in $s_0^2(z-x)$. Therefore, the function
\begin{equation}\label{29-5-20}
\rho_{12}(y,x)=
\int_{\mathbb{R}^{4}}\mathrm{d}^4z\,
s_0^2(z-x)\Big(B(y)PS^\Lambda(y,z)B(z)-B(y)PS^\Lambda(y,x)B(x)\Big)
\end{equation}
has a finite limit at $x\to y$, and it becomes possible to subtract and add $\rho_{12}(y,y)$. Thus, taking into account the additional factorization, after the shift, the sum of the integrals from (\ref{29-5-19}) is written as
\begin{equation*}\label{29-5-21}
\int_{\mathbb{R}^{4\times2}}\mathrm{d}^4x\mathrm{d}^4y\,
s_0^2(x-y)\Big(\rho_{12}(y,x)-\rho_{12}(y,y)\Big)+\mathrm{A}(\sigma)\int_{\mathbb{R}^{4}}\mathrm{d}^4y\,\rho_{12}(y,y)+\mathrm{A}(\sigma)\mathrm{I}_1^1\big(s_0^2PS^\Lambda\big).
\end{equation*}
It is clear that the first term does not give singularities, because the difference neutralizes the singularity remaining in $s_0^2(x-y)$. Therefore, representing (\ref{29-5-20}) as the sum of two terms, we obtain
\begin{align}\nonumber
\mathrm{I}^{11}_{\,0}\big(s_0^2,s_0^2,PS^\Lambda\big)\stackrel{\mathrm{s.p.}}{=}&
2\mathrm{A}(\sigma)\mathrm{I}_1^1\big(s_0^2PS^\Lambda\big)-\mathrm{A}^2(\sigma)\int_{\mathbb{R}^{4}}\mathrm{d}^4y\,B^2(y)PS^\Lambda(y,y)\\\nonumber
\stackrel{\mathrm{s.p.}}{=}&
2\mathrm{A}(\sigma)\mathrm{I}_1^1\big(s_0^2PS^\Lambda\big)-\mathrm{A}^2(\sigma)\mathrm{J}_2[B].
\end{align}
Let us move on to the second term from (\ref{29-5-17}). It has the following form
\begin{equation*}\label{29-5-23}
\int_{\mathbb{R}^{4\times3}}\mathrm{d}^4x\mathrm{d}^4y\mathrm{d}^4z\,
s_0^2(z-x)s_0^2(x-y)B(y)P_1^{\phantom{1}}(y-z)\,p_1^{\phantom{1}}(y,z)B(z).
\end{equation*}
We note only the main stages of the calculation. First, we subtract and add the term $B(y)p_1(y,x)B(x)$ (without the multiplier $P_1$). Note that the function
\begin{equation*}\label{29-5-24}
\rho_{13}(y,x)=
\int_{\mathbb{R}^{4}}\mathrm{d}^4z\,
P_1^{\phantom{1}}(y-z)s_0^2(z-x)\Big(B(y)p_1^{\phantom{1}}(y,z)B(z)-B(y)p_1^{\phantom{1}}(y,x)B(x)\Big)
\end{equation*}
has a finite limit at $x\to y$ before and after the removal of regularization, since the function $\ln(|x|)/|x|^3$ is integrable in the four-dimensional space. Therefore, by subtracting and adding the function $\rho_{13}(y,y)$ and ignoring the convergent part, we get
\begin{equation*}\label{29-5-25}
\mathrm{A}(\sigma)\int_{\mathbb{R}^{4}}\mathrm{d}^4y\,\rho_{13}(y,y)+\int_{\mathbb{R}^{4\times2}}\mathrm{d}^4x\mathrm{d}^4y\,
s_0^2(x-y)B(y)p_1^{\phantom{1}}(y,x)B(x)
\bigg(\int_{\mathbb{R}^{4}}\mathrm{d}^4z\,P_1^{\phantom{1}}(y-z)s_0^2(z-x)\bigg).
\end{equation*}
Next, applying relation (\ref{29-8-21}), we come to the following equality
\begin{equation}\label{29-5-26}
2\mathrm{A}(\sigma)\mathrm{I}_1^1\big(s_0^2P_1^{\phantom{1}}p_1^{\phantom{1}}\big)-
\mathrm{A}(\sigma)\bigg(\int_{\mathbb{R}^{4}}\mathrm{d}^4x\,s_0^2(x)P_1^{\phantom{1}}(x)\bigg)\bigg(
\int_{\mathbb{R}^{4}}\mathrm{d}^4y\,B^2(y)p_1^{\phantom{1}}(y,y)\bigg)
+\mathrm{I}_1^1\big(s_0^2\hat{P}_1^{\phantom{1}}p_1^{\phantom{1}}\big),
\end{equation}
which can be rewritten as
\begin{equation*}\label{29-5-261}
	2\mathrm{A}(\sigma)\mathrm{I}_1^1\big(s_0^2P_1^{\phantom{1}}p_1^{\phantom{1}}\big)
+m^2\mathrm{A}(\sigma)\mathrm{I}_2(\sigma)S_m[B]-
m^2\mathrm{I}_9(\sigma)S_m[B].
\end{equation*}
The answer for the third term from (\ref{29-5-17}) is obtained by replacing the indices $1\to2$ in $\{P_1,p_1,\hat{P}_1\}$ in the answer (\ref{29-5-26}) and is explicitly written as follows
\begin{equation*}\label{29-5-262}
	2\mathrm{A}(\sigma)\mathrm{I}_1^1\big(s_0^2P_1^{\phantom{1}}p_1^{\phantom{1}}\big)
	+\frac{\lambda}{2}\mathrm{A}(\sigma)\mathrm{I}_3(\sigma)S_4[B]-
	\frac{\lambda}{2}\mathrm{I}_{10}(\sigma)S_4[B].
\end{equation*}

Let us move on to the second contribution from (\ref{29-5-9}). From its explicit representation
\begin{equation*}\label{29-5-27}
\mathrm{I}^{11}_{\,0}(2s_0^{\phantom{1}}PS_3^\Lambda,s_0^2,R_0^\Lambda)=
\int_{\mathbb{R}^{4\times3}}\mathrm{d}^4x\mathrm{d}^4y\mathrm{d}^4z\,
2s_0^{\phantom{1}}(z-x)PS_3^\Lambda(z,x)s_0^2(x-y)B(y)R_0^\Lambda(y-z)B(z),
\end{equation*}
it follows that it is convenient to use $B(y)=B(y)\pm B(x)$ as the first subtraction, then the contribution can be rewritten as
\begin{multline}\label{29-5-28}
\int_{\mathbb{R}^{4\times3}}\mathrm{d}^4x\mathrm{d}^4y\mathrm{d}^4z\,
2s_0^{\phantom{1}}(z-x)PS_3^\Lambda(z,x)s_0^2(x-y)\Big(B(y)-B(x)\Big)R_0^\Lambda(y-z)B(z)+\\+\int_{\mathbb{R}^{4\times3}}\mathrm{d}^4x\mathrm{d}^4y\mathrm{d}^4z\,
2s_0^{\phantom{1}}(z-x)PS_3^\Lambda(z,x)s_0^2(x-y)B(x)R_0^\Lambda(y-z)B(z).
\end{multline}
It can be noted that the first term does not contain singular terms. Indeed, the integral of the form
\begin{equation*}\label{29-5-30}
\int_{\mathbb{R}^{4}}\mathrm{d}^4y\,
s_0^2(x-y)\Big(B(y)-B(x)\Big)R_0^\Lambda(y-z)B(z),
\end{equation*}
after removing the regularization, contains the function $|x-z|^{-1}$ or weaker in the main term of the asymptotics. Therefore, the remaining integral converges, since the function $\ln(|x|)/|x|^3$ is integrable in the four-dimensional space. Then, after auxiliary factorizations in the second term (\ref{29-5-28}) and using relation (\ref{29-8-21}), the answer is written as follows
\begin{align}\nonumber
\mathrm{I}^{11}_{\,0}(2s_0^{\phantom{1}}PS_3^\Lambda,s_0^2,R_0^\Lambda)
\stackrel{\mathrm{s.p.}}{=}&
\mathrm{A}(\sigma)\mathrm{I}_1^1\big(2s^2_0PS_3^\Lambda\big)+\mathrm{I}_1^1\big(2s_0PS_3^\Lambda\hat{R}_0^\Lambda\big)\\\nonumber
\stackrel{\mathrm{s.p.}}{=}&
\mathrm{A}(\sigma)\mathrm{I}_1^1\big(2s^2_0PS_3^\Lambda\big)+2\mathrm{I}_{8}(\sigma)\mathrm{J}_2[B]-2m^2\mathrm{I}_{11}(\sigma)S_m[B]-\lambda\mathrm{I}_{12}(\sigma)S_4[B].
\end{align}

Let us move on to the last term from (\ref{29-5-9})
\begin{equation}\label{29-5-31}
\mathrm{I}^{11}_{\,0}(s_0^{2},s_0^2,R_0^\Lambda)=
\int_{\mathbb{R}^{4\times3}}\mathrm{d}^4x\mathrm{d}^4y\mathrm{d}^4z\,
s_0^2(z-x)s_0^2(x-y)B(y)R_0^\Lambda(y-z)B(z).
\end{equation}
At first glance, this contribution is considered very easy. Indeed, this would be a good idea to directly substitute a result for the integral
\begin{equation*}\label{29-5-32}
\int_{\mathbb{R}^{4}}\mathrm{d}^4x\,s_0^2(z-x)s_0^2(x-y)
\end{equation*}
and write out the answer wanted. However, finding the asymptotics for the last integral is a laborious task. Therefore, it is easier to use adding and subtracting. To do this, we represent $B(z)$ as
\begin{equation}\label{29-5-33}
B(z)=B(x)+(z-x)^\mu\partial_{x^\mu}B(x)+
\frac{1}{2}(z-x)^{\mu\nu}\partial_{x^\mu}\partial_{x^\nu}B(x)+B_3(x,z).
\end{equation}
It is clear that using equality (\ref{29-8-21}), the answer for the first term is written out instantly and has the form
\begin{align}\nonumber
\int_{\mathbb{R}^{4\times3}}\mathrm{d}^4x\mathrm{d}^4y\mathrm{d}^4z\,
s_0^2(z-x)s_0^2(x-y)&B(y)R_0^\Lambda(y-z)B(x)=
\mathrm{A}(\sigma)\mathrm{I}_1^1\big(s^3_0\big)+\mathrm{I}_1^1\big(s_0^2\hat{R}_0^\Lambda\big)\stackrel{\mathrm{s.p.}}{=}
\\\label{29-5-341}
\stackrel{\mathrm{s.p.}}{=}&
\mathrm{A}(\sigma)\mathrm{I}_1^1\big(s^3_0\big)
+\mathrm{I}_{7}(\sigma)S_m[B]-
\frac{S_2[B]}{8}\int_{\mathbb{R}^{4}}\mathrm{d}^4x\,s_0^2(x)\hat{R}_0^\Lambda(x)|x|^2.
\end{align}

Further, in the second term, we decompose the function $\partial_{x^\mu}B(x)$ near the point $y$
\begin{equation*}\label{29-5-35}
\partial_{x^\mu}B(x)=\partial_{y^\mu}B(y)+(x-y)^\nu\partial_{y^\mu}\partial_{y^\nu}B(y)+\ldots,
\end{equation*}
where the ellipsis marks the terms neutralizing the singularity. Then, after the variable shifts
$x\to x+y$ and $z\to z+y$, we get
\begin{align}\nonumber
\int_{\mathbb{R}^{4\times3}}\mathrm{d}^4x&\mathrm{d}^4y\mathrm{d}^4z\,
s_0^2(z-x)s_0^2(x-y)B(y)R_0^\Lambda(y-z)(z-x)^\mu\partial_{x^\mu}B(x)\stackrel{\mathrm{s.p.}}{=}\\
\nonumber
\stackrel{\mathrm{s.p.}}{=}&
\bigg(\int_{\mathbb{R}^{4}}\mathrm{d}^4y\,B(y)\partial_{y^\mu}B(y)\bigg)
\bigg(\int_{\mathbb{R}^{4\times2}}\mathrm{d}^4x\mathrm{d}^4z\,
s_0^2(z-x)s_0^2(x)R_0^\Lambda(z)(z-x)^\mu\bigg)+
\\\nonumber+&
\bigg(\int_{\mathbb{R}^{4}}\mathrm{d}^4y\,
B(y)\partial_{y^\mu}\partial_{y^\nu}B(y)\bigg)
\bigg(\int_{\mathbb{R}^{4\times2}}\mathrm{d}^4x\mathrm{d}^4z\,
s_0^2(z-x)s_0^2(x)R_0^\Lambda(z)(z-x)^\mu x^\nu
\bigg).
\end{align}
Note that the first term does not give a singular contribution, since the integral in parentheses is zero. Indeed, by making one more substitution of the form $x\to-x$ and $z\to-z$, we get the same integral with the minus sign. The second term can be rewritten as
\begin{equation}\label{29-5-37}
-\frac{S_2[B]}{4}
\int_{\mathbb{R}^{4\times2}}\mathrm{d}^4x\mathrm{d}^4z\,
s_0^2(x)x^\mu
R_0^\Lambda(z+x)s_0^2(z)z_\mu,
\end{equation}
where the substitution $x^\nu z^\mu\to x^\sigma z_\sigma\delta^{\mu\nu}/4$ was used due to spherical symmetry.

Next, the third term of the decomposition from (\ref{29-5-33}) must be decomposed at the point $y$. At the same time, it is important to preserve only the zero order, since the following corrections obviously neutralize the divergences completely. Thus, we get
\begin{multline}\label{29-5-38}
\frac{1}{2}
\int_{\mathbb{R}^{4\times3}}\mathrm{d}^4x\mathrm{d}^4y\mathrm{d}^4z\,
s_0^2(z-x)s_0^2(x-y)B(y)R_0^\Lambda(y-z)
(z-x)^{\mu\nu}\partial_{y^\mu}\partial_{y^\nu}B(y)
\stackrel{\mathrm{s.p.}}{=}\\
\stackrel{\mathrm{s.p.}}{=}-\frac{S_2[B]}{8}
\int_{\mathbb{R}^{4\times2}}\mathrm{d}^4x\mathrm{d}^4z\,
s_0^2(x)R_0^\Lambda(x-z)
s_0^2(z)|z|^2.
\end{multline}
Returning to the last fourth term in (\ref{29-5-33}), it should be noted that the function
\begin{equation*}\label{29-5-39}
\rho_{14}(y,x)=
\int_{\mathbb{R}^{4}}\mathrm{d}^4z\,
B(y)R_0^\Lambda(y-z)B_3(x,z)s_0^2(z-x)
\end{equation*}
has a finite limit at $x\to y$ before and after removing the regularization. Following the general logic, subtracting and adding the function $\rho_{14}(y,y)$ and performing factorization using the well known shift, we get the answer in the form
\begin{multline}\label{29-5-40}
\int_{\mathbb{R}^{4\times3}}\mathrm{d}^4x\mathrm{d}^4y\mathrm{d}^4z\,
s_0^2(z-x)s_0^2(x-y)B(y)R_0^\Lambda(y-z)B_3^{\phantom{1}}(x,z)\stackrel{\mathrm{s.p.}}{=}
\\\stackrel{\mathrm{s.p.}}{=}
\mathrm{A}(\sigma)
\int_{\mathbb{R}^{4}}\mathrm{d}^4y\,\rho_{14}(y,y)=
\mathrm{A}(\sigma)
\int_{\mathbb{R}^{4\times2}}\mathrm{d}^4y\mathrm{d}^4z\,
B(y)s_0^3(y-z)B_3^{\phantom{1}}(y,z).
\end{multline}
Note that we can substitute the definition of $B_3(x,z)$ from (\ref{29-5-33}) into the last integral
\begin{equation}\label{29-5-41}
\int_{\mathbb{R}^{4\times2}}\mathrm{d}^4y\mathrm{d}^4z\,
B(y)s_0^3(y-z)B_3(y,z)=\mathrm{I}_1^1\big(s_0^3\big)-S_m[B]
\bigg(\int_{\mathbb{R}^{4}}\mathrm{d}^4x\,s_0^3(x)\bigg)+
\frac{S_2[B]}{8}\bigg(\int_{\mathbb{R}^{4}}\mathrm{d}^4x\,s_0^3(x)|x|^2\bigg)
\end{equation}
and get the singularity in standard terms. Finally, summing up the results (\ref{29-5-341}), (\ref{29-5-37}), (\ref{29-5-38}), (\ref{29-5-40}), and (\ref{29-5-41}), for the third contribution from (\ref{29-5-31}) we get the following answer
\begin{align}\nonumber
\mathrm{I}^{11}_{\,0}(s_0^{2},s_0^2,s_0^{\phantom{1}})&\stackrel{\mathrm{s.p.}}{=}
2\mathrm{A}(\sigma)\mathrm{I}_1^1\big(s_0^3\big)+S_m[B]\bigg(\int_{\mathbb{R}^{4}}\mathrm{d}^4x\,\Big(s_0^2(x)\hat{R}_0^\Lambda(x)-
\mathrm{A}(\sigma)s_0^3(x)\Big)\bigg)+
\\\nonumber
+&
S_2[B]\bigg(
-\frac{1}{8}
\int_{\mathbb{R}^{4\times2}}\mathrm{d}^4x\mathrm{d}^4z\,
s_0^2(x)
R_0^\Lambda(x-z)|x-z|^2s_0^2(z)
+\frac{\mathrm{A}(\sigma)}{4}
\int_{\mathbb{R}^{4}}\mathrm{d}^4z\,
s_0^3(z)|z|^2
\bigg).
\end{align}

As a result, putting together all the contributions studied in this section, we can formulate the final statement.

\begin{lemma}\label{29-l-4}
Taking into account all of the above, the following equality holds
\begin{align*}
\mathrm{I}^{11}_{\,0}\big((G^\Lambda)^2,(G^\Lambda)^2,G^\Lambda\big)
\stackrel{\mathrm{s.p.}}{=}&2\mathrm{A}(\sigma)
\mathrm{I}^{1}_{1}\big((G^\Lambda)^3\big)
+\mathrm{J}_2[B]\Big(-\mathrm{A}^2(\sigma)+4
\mathrm{I}_{8}(\sigma)\Big)+\\
+&S_2[B]\bigg(-\frac{1}{8}
\mathrm{I}_{6}(\sigma)
+\frac{1}{4}\mathrm{A}(\sigma)
\mathrm{I}_{1}(\sigma)\bigg)+\\
+&
S_m[B]\Big(\mathrm{I}_{7}(\sigma)-\mathrm{A}(\sigma)\mathrm{I}_{4}(\sigma)
+m^2\mathrm{A}(\sigma)\mathrm{I}_{2}(\sigma)-
m^2\mathrm{I}_{9}(\sigma)-4
m^2\mathrm{I}_{11}(\sigma)\Big)+\\
+&
S_4[B]\bigg(\frac{\lambda}{2}\mathrm{A}(\sigma)\mathrm{I}_{3}(\sigma)-
\frac{\lambda}{2}\mathrm{I}_{10}(\sigma)-2\lambda
\mathrm{I}_{12}(\sigma)\bigg),
\end{align*}
where the corresponding definitions for the integrals $\mathrm{I}_i(\sigma)$ are presented in Section \ref{29:sec:app:vs}.
\end{lemma}

\subsection{Relations for diagrams}
\label{29:sec:tri:sootn}

In this section, we will study the relations for linear combinations of diagrams arising on the right hand side of equation (\ref{29-1-32}). In this case, firstly we calculate the combination with the usage of the condition $\tilde{z}_{2,1}=0$, and then we separately compute the rest part. Let us start with the diagram $d_8$ and the counter diagrams $cd_3$ and $cd_4$. They appear with the following numerical coefficients
\begin{equation}\label{29-7-1}
Cb_1=\frac{\lambda^2d_8}{16}+\frac{\lambda cd_3}{4}+\frac{cd_4}{4}.
\end{equation}
It can be noted that such linear combination is represented as the following single term
\begin{equation*}\label{29-7-2}
\bigg(Cb_1-Cb_1\Big|_{B=0}\bigg)\bigg|_{\tilde{z}_{2,1}=0}=
\frac{1}{16}
\int_{\mathbb{R}^{4\times2}}\mathrm{d}^4x\mathrm{d}^4y\Big[
\mathrm{V}_1(x)\big(G^\Lambda(x,y)\big)^2\mathrm{V}_1(y)-\tilde{\kappa}_1(x,y)\Big],
\end{equation*}
where, using (\ref{29-1-11}) and (\ref{29-1-27}), we have
\begin{equation*}\label{29-7-3}
\mathrm{V}_1(x)=\lambda G^\Lambda(x,x)+2m^2z_1+\lambda z_{4,1}B^2(x).
\end{equation*}
Also, the function $\tilde{\kappa}_1(x,y)$ subtracts a singular density independent of the background field $B$. The explicit form of such density is not important in further study.
Further, substituting the asymptotic decomposition (\ref{29-3-18}) and the results from (\ref{29-4-6}) and (\ref{29-4-7}) for the coefficients of the renormalization constants, we obtain
\begin{equation*}\label{29-7-4}
\mathrm{V}_1(x)=\lambda^2 B^2(x)\mathrm{A}(\sigma)+
2m^2\tilde{z}_1+\lambda\tilde{z}_{4,1}B^2(x)+\lambda PS^\Lambda(x,x).
\end{equation*}
Note that the power singularity $R_0^\Lambda(0)$ does not appear in the equality. Also, the relation $\mathrm{A}(\sigma)=R_{11}^\Lambda(0)$ was used, which is valid by construction. Only logarithmic singularities and finite parts remain.

Let us now move on to the next linear combination containing the diagrams $d_3$ and $d_6$ and the counter diagram $cd_2$. According to (\ref{29-1-32}), it has the form
\begin{equation}\label{29-7-5}
Cb_2=\frac{\lambda^4d_3}{16}-\frac{\lambda^3d_6}{8}-\frac{\lambda^2cd_2}{4}.
\end{equation}
Again, using the definitions for the elements of the diagram technique from Section \ref{29:sec:1:state}, the decomposition of the Green's function (\ref{29-3-18}) and Lemma \ref{29-l-2}\footnote{Here it should be noted that the result of Lemma \ref{29-l-2} can be substituted with another $\sigma_1\neq\sigma$. However, in Section \ref{29:sec:tri:d-3} it was mentioned that the singular part does not depend on the choice of the parameter. Therefore, for convenience reasons, $\sigma_1=\sigma$ was taken.}, the divergent part of the combination from (\ref{29-7-5}) can be represented as
\begin{equation*}\label{29-7-6}
\bigg(Cb_2-Cb_2\Big|_{B=0}\bigg)\bigg|_{\tilde{z}_{2,1}=0}=
\frac{\lambda^2}{16}\int_{\mathbb{R}^{4}}\mathrm{d}^4x\,\rho_{15}(x)
\mathrm{V}_2(x)
-\frac{\lambda^4}{16}\mathrm{A}^2(\sigma)\mathrm{I}_2^2\big((G^\Lambda)^2\big)+
\frac{\lambda^4}{16}\mathrm{I}_5(\sigma)S_4[B],
\end{equation*}
where
\begin{equation*}\label{29-7-7}
\rho_{15}(x)=\int_{\mathbb{R}^{4\times2}}\mathrm{d}^4z\mathrm{d}^4y\,
G^\Lambda(x,z)B(z)\Big(G^\Lambda(z,y)\Big)^2B(y)G^\Lambda(y,x),
\end{equation*}
\begin{align}\label{29-7-8}
\mathrm{V}_2(x)=&2\lambda^2\mathrm{A}(\sigma)B^2(x)-2\mathrm{V}_1(x)\\\nonumber
=&-4m^2\tilde{z}_1-2\lambda\tilde{z}_{4,1}B^2(x)-2\lambda PS^\Lambda(x,x).
\end{align}
Note that the function $\mathrm{V}_2(x)$ does not contain singularities. It is $\mathcal{O}(1)$ with respect to the parameter $\Lambda$. For such function, we can use the decomposition
\begin{multline}\label{29-7-11}
\int_{\mathbb{R}^{4}}\mathrm{d}^4x\,\rho_{15}(x)
\mathrm{V}_2(x)\stackrel{\mathrm{s.p.}}{=}\mathrm{A}(\sigma)
\int_{\mathbb{R}^{4\times2}}\mathrm{d}^4x\mathrm{d}^4y\,
B^2(x)\big(G^\Lambda(x,y)\big)^2\mathrm{V}_2(y)+\\+
\bigg(\int_{\mathbb{R}^{4}}\mathrm{d}^4y\,B^2(y)\mathrm{V}_2(y)\bigg)\bigg(
\int_{\mathrm{B}_{1/\sigma}}\mathrm{d}^4x\,R_0^\Lambda(x)\hat{R}_0^\Lambda(x)\bigg),
\end{multline}
where the considerations from Section \ref{29:sec:tri:d-3} where used. Also, given the fact that the linear combinations under study do not contain power singularities, it is possible to omit terms in $PS^\Lambda(x,x)$, which are decreasing as a power of $\Lambda$, and replace it with $PS_1^\Lambda(x,x)$. Therefore, using relation (\ref{29-7-8}), we can write out the singular part, depending on the background field, for the sum of linear combinations (\ref{29-7-1}) and (\ref{29-7-5}) with the condition $\tilde{z}_{2,1}=0$ in the form
\begin{equation*}\label{29-7-31}
	\frac{\mathrm{A}(\sigma)}{16}\int_{\mathbb{R}^{4}}\mathrm{d}^4y\Big(\mathrm{V}_3^2(y)-
	4m^4\tilde{z}_1^2\Big)
	-\frac{\lambda^2}{8}\bigg(\int_{\mathbb{R}^{4}}\mathrm{d}^4y\,B^2(y)\mathrm{V}_3(y)\bigg)\mathrm{I}_8(\sigma)+
	\frac{\lambda^4}{16}\mathrm{I}_5(\sigma)S_4[B],
\end{equation*}
where
\begin{equation*}\label{29-7-12}
	\mathrm{V}_3(x)=2m^2\tilde{z}_1+\lambda\tilde{z}_{4,1}B^2(x)+\lambda PS_1^\Lambda(x,x).
\end{equation*}
Decomposing the last relation, we can formulate one more statement.
\begin{lemma}\label{29-l-5}
Taking into account all of the above, the following equality is true
\begin{align}\label{29-7-10}
	\bigg(\frac{\lambda^4d_3}{16}-\frac{\lambda^3d_6}{8}+
	\frac{\lambda^2d_8}{16}-\frac{\lambda^2cd_2}{4}+\frac{\lambda cd_3}{4}+&\frac{cd_4}{4}-\tilde{\kappa}\bigg)\bigg|_{\tilde{z}_{2,1}=0}
	\stackrel{\mathrm{s.p.}}{=}\\\stackrel{\mathrm{s.p.}}{=}&\nonumber
	S_m[B]\bigg(\frac{m^2\lambda\tilde{z}_1\tilde{z}_{4,1}}{4}\mathrm{A}(\sigma)-\frac{m^2\lambda^2\tilde{z}_1}{4}\mathrm{I}_8(\sigma)\bigg)+
	\\+&\nonumber
	S_4[B]\bigg(\frac{\lambda^2\tilde{z}_{4,1}^2}{16}\mathrm{A}(\sigma)+\frac{\lambda^4}{16}\mathrm{I}_5(\sigma)-\frac{\lambda^3\tilde{z}_{4,1}}{8}\mathrm{I}_8(\sigma)\bigg)+
	\\+&\nonumber
	\mathrm{J}_1[B]\bigg(\frac{m^2\lambda\tilde{z}_1}{4}\mathrm{A}(\sigma)\bigg)+
	\\+&\nonumber
	\mathrm{J}_2[B]\bigg(\frac{\lambda^2\tilde{z}_{4,1}}{8}\mathrm{A}(\sigma)-\frac{\lambda^3}{8}\mathrm{I}_8(\sigma)\bigg)+
	\\+&\nonumber
	\mathrm{J}_4[B]\bigg(\frac{\lambda^2}{16}\mathrm{A}(\sigma)\bigg),
\end{align}
where $\tilde{\kappa}$ subtracts a singular density independent of the background field.
\end{lemma}
The next linear combination contains the diagrams $d_1$ and $d_5$ and, according to formula (\ref{29-1-32}), is written out as follows
\begin{equation}\label{29-7-13}
\frac{\lambda^2 z_{4,1}d_1}{6}-\frac{\lambda^3d_5}{8}.
\end{equation}
Then, applying Lemma \ref{29-l-4} to the diagram $d_5$, taking into account relation (\ref{29-4-7}) in the form
\begin{equation*}\label{29-7-14}
\frac{\lambda^2 z_{4,1}}{6}-\frac{\lambda^3\mathrm{A}(\sigma)}{4}=\frac{\lambda^2\tilde{z}_{4,1}}{6},
\end{equation*}
we get the explicit expression for the singular part
\begin{align}\label{29-7-15}
\frac{\lambda^2 z_{4,1}d_1}{6}&-\frac{\lambda^3d_5}{8}\stackrel{\mathrm{s.p.}}{=}
S_2[B]\bigg(-\frac{\lambda^2\tilde{z}_{4,1}}{48}\mathrm{I}_{1}(\sigma)
-\frac{\lambda^3}{32}\mathrm{A}(\sigma)
\mathrm{I}_{1}(\sigma)
+\frac{\lambda^3}{64}
\mathrm{I}_{6}(\sigma)
\bigg)+\\\label{29-7-15-1}+&
S_m[B]\bigg(\frac{\lambda^2\tilde{z}_{4,1}}{6}\mathrm{I}_{4}(\sigma)+\frac{\lambda^3}{8}\mathrm{A}(\sigma)\mathrm{I}_{4}(\sigma)-\frac{\lambda^2\tilde{z}_{4,1}m^2}{2}\mathrm{I}_{2}(\sigma)-\frac{m^2\lambda^3}{8}\mathrm{A}(\sigma)\mathrm{I}_{2}(\sigma)-\\\nonumber&
~~~~~~~~~~~~~~~~~~~~~~~~~~~~~~~~~~~~~~~~~~~~~~~~~~~~
-\frac{\lambda^3}{8}\mathrm{I}_{7}(\sigma)+
\frac{m^2\lambda^3}{8}\mathrm{I}_{9}(\sigma)+
\frac{m^2\lambda^3}{2}\mathrm{I}_{11}(\sigma)
\bigg)+\\\label{29-7-15-2}+
&S_4[B]\bigg(-\frac{\lambda^3\tilde{z}_{4,1}}{4}\mathrm{I}_{3}(\sigma)
-\frac{\lambda^4}{16}\mathrm{A}(\sigma)\mathrm{I}_{3}(\sigma)+
\frac{\lambda^4}{16}\mathrm{I}_{10}(\sigma)+\frac{\lambda^4}{4}
\mathrm{I}_{12}(\sigma)
\bigg)+\\\label{29-7-15-3}+&
\mathrm{J}_2[B]
\bigg(\frac{\lambda^2\tilde{z}_{4,1}}{2}\mathrm{A}(\sigma)
+\frac{\lambda^3}{8}\mathrm{A}^2(\sigma)-\frac{\lambda^3}{2}
\mathrm{I}_{8}(\sigma)
\bigg).
\end{align}

Next, we present the results for the diagram $d_7$ and two counter diagrams $z_{4,1}d_2$ and $cd_5$. Since such contributions contain no more than two integration operators over $\mathbb{R}^4$, we can use the mentioned above calculation method for the two-loop contribution, that is, the usual limitation of the integration domain and variable shift. The result for $d_7$ has the form
\begin{align}\label{29-7-16}
\frac{\lambda^2\big(d_7-\tilde{\kappa}_7\big)}{48}\stackrel{\mathrm{s.p.}}{=}&
S_m[B]
\bigg(
-\frac{\lambda^3}{24}\mathrm{I}_{14}(\sigma)
+\frac{m^2\lambda^3}{8}\mathrm{I}_{16}(\sigma)+
\frac{m^2\lambda^3}{24}\mathrm{I}_{19}(\sigma)
\bigg)+
\\+&\label{29-7-16-4}
S_4[B]
\bigg(
\frac{3\lambda^4}{96}\mathrm{I}_{17}(\sigma)+\frac{\lambda^4}{96}\mathrm{I}_{20}(\sigma)
\bigg)+
\\\label{29-7-16-1}+&
\mathrm{J}_1[B]
\bigg(\frac{\lambda^2}{12}\mathrm{I}_{4}(\sigma)-\frac{m^2\lambda^2}{4}\mathrm{I}_{2}(\sigma)\bigg)+\\\label{29-7-16-2}+
&\mathrm{J}_2[B]
\bigg(-\frac{\lambda^3}{8}\mathrm{I}_{3}(\sigma)\bigg)+\mathrm{J}_3[B]
\bigg(\frac{\lambda^2}{96}\mathrm{I}_{1}(\sigma)\bigg)+\mathrm{J}_4[B]
\bigg(\frac{\lambda^2}{8}\mathrm{A}(\sigma)\bigg),
\end{align}
where $\tilde{\kappa}_7$ subtracts a singularity independent of the background field.

Continuing, the answer for the counter diagram $-\lambda z_{4,1}d_2/8$ is instantly written out using the result for the two-loop contribution from (\ref{29-4-12})--(\ref{29-4-15}). Then, the singular part of the counter diagram $-cd_5/2$ has the form
\begin{align}\label{29-7-19}
-\frac{\big(cd_5-\tilde{\kappa}_5\big)}{2}\stackrel{\mathrm{s.p.}}{=}&
\mathrm{A}_1(\sigma)S_m[B]+\mathrm{A}_2(\sigma)S_4[B]+\\\label{29-7-19-1}+&
\mathrm{J}_{1}[B]
\bigg(-\frac{m^2z_2}{2}\bigg)
+\mathrm{J}_{2}[B]
\bigg(-\frac{\lambda z_{4,2}}{4}\bigg)
+\mathrm{J}_{3}[B]
\bigg(\frac{z_{2,2}}{2}\bigg),
\end{align}
where $\tilde{\kappa}_5$ subtracts a singularity independent of the background field, and
\begin{multline}\label{29-7-20}
\mathrm{A}_1(\sigma)=
\frac{\lambda z_{2,2}}{4}A_0^\Lambda(x)R_{11}^\Lambda(x)\big|_{x=0}
-\frac{m^2\lambda z_{2,2}}{4}A_0^\Lambda(x)R_{21}^\Lambda(x)\big|_{x=0}+\\
+\frac{m^2\lambda z_2}{4}R_{11}^\Lambda(0)
-\frac{m^4\lambda z_2}{4}R_{21}^\Lambda(0)
-\frac{\lambda z_{4,2}}{4}R_{0}^\Lambda(0)
+\frac{m^2\lambda z_{4,2}}{4}R_{10}^\Lambda(0)
-\frac{m^4\lambda z_{4,2}}{8}R_{20}^\Lambda(0),
\end{multline}
\begin{equation}\label{29-7-21}
\mathrm{A}_2(\sigma)=
-\frac{\lambda^2 z_{2,2}}{16}A_0^\Lambda(x)R_{22}^\Lambda(x)\big|_{x=0}
-\frac{m^2\lambda^2 z_2}{16}R_{22}^\Lambda(0)
+\frac{\lambda^2 z_{4,2}}{8}R_{11}^\Lambda(0)
-\frac{m^2\lambda^2 z_{4,2}}{8}R_{21}^\Lambda(0).
\end{equation}
Note that in (\ref{29-7-19-1}), to obtain $\mathrm{J}_3[B]$, the deformed operator $A_0^\Lambda(\cdot)$ was replaced with the limiting one $A_0^{\phantom{1}}(\cdot)$, which is possible when acting on the component $PS_1^\Lambda$, namely
\begin{equation}\label{29-7-19-2}
L\int_{\mathbb{R}^4}\mathrm{d}^4x
\Big(\big(A_0^\Lambda(y)-A_0^{\phantom{1}}(y)\big)PS_1^\Lambda(y,x)-\kappa_{s}(x)\Big)\Big|_{y=x}
\stackrel{\mathrm{s.p.}}{=}0,
\end{equation}
where $\kappa_s(x)$ is a singular density independent of the background field. The proof\footnote{Note that equality is proved for the regularization under study, see Section \ref{29:sec:reg-4}. When expanding the type of regularization, for example, with additional deformation of the mass parameter, such equality must be investigated separately. Moreover, in this case, additional counter diagrams may appear.} of this equality is in Section \ref{29:sec:app:doc}.

Similarly, using the formula (\ref{29-21-1}), we calculate the discrepancy in the combination of Lemma \ref{29-l-5}, depending on $\tilde{z}_{2,1}$. The answer for it can be formulated as follows
\begin{align}\label{29-21-2}
	Cb_1+Cb_2-\big(Cb_1+Cb_2\big)\big|_{\tilde{z}_{2,1}=0}+\kappa_{Cb}\stackrel{\mathrm{s.p.}}{=}&
	\frac{\lambda^2\tilde{z}_{2,1}}{32}\mathrm{I}_1(\sigma)S_2[B]+
	\mathrm{A}_3(\sigma)S_m[B]+\mathrm{A}_4(\sigma)S_4[B]+\\\label{29-21-3}+&
	\mathrm{J}_{1}[B]
	\bigg(\frac{\lambda \tilde{z}_{2,1}}{4}\Big(R_{0}^\Lambda(0)-m^2R_{10}^\Lambda(0)-m^2R_{11}^\Lambda(0)\Big)\bigg)
	+\\\label{29-21-6}+&\mathrm{J}_{2}[B]
	\bigg(-\frac{3\lambda^2\tilde{z}_{2,1}}{4}\mathrm{A}(\sigma)\bigg),
\end{align}
where
\begin{multline}\label{29-21-4}
	\mathrm{A}_3(\sigma)=-\frac{\lambda^2\tilde{z}_{2,1}}{4}\big(\mathrm{I}_4(\sigma)-3m^2\mathrm{I}_2(\sigma)-m^2\mathrm{I}_8(\sigma)\big)+\frac{\tilde{z}_{2,1}^2}{4}\bigg(m^2\lambda\mathrm{A}(\sigma)-\lambda R_{0}^\Lambda(0)+m^2\lambda R_{10}^\Lambda(0)+m^2\lambda R_{11}^\Lambda(0)\bigg)+\\+\tilde{z}_{2,1}\bigg(-\frac{m^2\lambda}{2}\mathrm{A}(\sigma)z_1+\frac{m^2\lambda^2}{8}R_{0}^\Lambda(0)R_{21}^\Lambda(0)+\frac{\lambda\tilde{z}_{4,1}}{4}\Big(R_{0}^\Lambda(0)-m^2R_{10}^\Lambda(0)-m^2R_{11}^\Lambda(0)\Big)\bigg)
,
\end{multline}
\begin{equation}\label{29-21-5}
	\mathrm{A}_4(\sigma)=
\frac{\lambda^3\tilde{z}_{2,1}}{8}\big(3\mathrm{I}_3(\sigma)+\mathrm{I}_8(\sigma)\big)+\frac{3\tilde{z}_{2,1}^2\lambda^2}{16}\mathrm{A}(\sigma)+\tilde{z}_{2,1}\bigg(
\frac{\lambda^3}{32}R_{0}^\Lambda(0)R_{22}^\Lambda(0)-\frac{\lambda^3}{8}\mathrm{A}^2(\sigma)-
\frac{\lambda^2}{4}\mathrm{A}(\sigma)\tilde{z}_{4,1}
\bigg).
\end{equation}

Thus, for all the individual diagrams from (\ref{29-1-32}) and some of their linear combinations, the singular parts were written out in general terms. That is, in the notation for the asymptotic decomposition (\ref{29-3-18})--(\ref{29-3-21}) for the Green's function. The further task is divided into two parts. First, we need to show that all non-local contributions are successfully cancelled. This will demonstrate that the general considerations of renormalization theory are correct for our regularization. Secondly, we need to write out an explicit answer for the three-loop coefficients $\{z_{2,3},z_3,z_{4,3}\}$ of the renormalization constants.

Let us start with the first point. To do this, write out the coefficients for the functionals $\mathrm{J}_i[B]$, where $i=1,2,3,4$, and make sure that they do not contain singular components. In increasing order of the index, the relations are as follows
\begin{multline*}
	\mathrm{J}_1[B]:\,\,\,
\frac{m^2\lambda\tilde{z}_1}{4}\mathrm{A}(\sigma)+
\frac{\lambda \tilde{z}_{2,1}}{4}\Big(R_{0}^\Lambda(0)-m^2R_{10}^\Lambda(0)-m^2R_{11}^\Lambda(0)\Big)+
\frac{\lambda^2}{12}\mathrm{I}_4(\sigma)-\\-
\frac{m^2\lambda^2}{4}\mathrm{I}_2(\sigma)-
\frac{m^2z_2}{2}-
\frac{\lambda z_{4,1}}{8}\Big(2R_0^\Lambda(0)-2m^2R_{10}^\Lambda(0)\Big)=
-\frac{m^2\tilde{z}_2}{2}\stackrel{\mathrm{s.p.}}{=}0,
\end{multline*}
\begin{equation*}
	\mathrm{J}_2[B]:\,\,\,
\frac{5\lambda^2\tilde{z}_{4,1}}{8}\mathrm{A}(\sigma)-\frac{3\lambda^2\tilde{z}_{2,1}}{4}\mathrm{A}(\sigma)-\frac{5\lambda^3}{8}\mathrm{I}_8(\sigma)+\frac{\lambda^3}{8}\mathrm{A}^2(\sigma)-\frac{\lambda^3}{8}\mathrm{I}_3(\sigma)-
\frac{\lambda z_{4,2}}{4}+\frac{\lambda z_{4,1}}{8}\lambda R_{11}^\Lambda(0)=
-\frac{\lambda\tilde{z}_{4,2}}{4}\stackrel{\mathrm{s.p.}}{=}0,
\end{equation*}
\begin{equation*}\label{29-7-24}
	\mathrm{J}_3[B]:\,\,\,
\frac{\lambda^2}{96}\mathrm{I}_1(\sigma)+\frac{z_{2,2}}{2}=
\frac{\tilde{z}_{2,2}}{2}
\stackrel{\mathrm{s.p.}}{=}0,
\end{equation*}
\begin{equation*}\label{29-7-25}
	\mathrm{J}_4[B]:\,\,\,
\frac{3\lambda^2}{16}\mathrm{A}(\sigma)-\frac{\lambda z_{4,1}}{8}=-\frac{\lambda\tilde{z}_{4,1}}{8}
\stackrel{\mathrm{s.p.}}{=}0,
\end{equation*}
where we have used the equalities
\begin{equation*}\label{29-7-26}
R_{11}^\Lambda(0)=\mathrm{A}(\sigma)\,\,\,\,\,\,\mbox{and}\,\,\,\,\,\,
\mathrm{I}_8(\sigma)=\mathrm{I}_3(\sigma)-\mathrm{A}^2(\sigma).
\end{equation*}
Thus, it was verified by direct calculations that non-local terms do not contain singular coefficients. The expressions for the renormalization coefficients follow from the summation of the contributions from Lemmas \ref{29-l-3} and \ref{29-l-5}, and formulas (\ref{29-7-15})--(\ref{29-7-21}) and (\ref{29-21-2})--(\ref{29-21-5}).
After some arithmetic calculations, the following statement can be formulated.
\begin{lemma}\label{29-l-6}
Taking into account all the above, the singular component of the right hand side of equality (\ref{29-1-32}), that is, the three-loop contribution with corresponding counterterms, is equal to the linear combination of the functionals $S_2[B]$, $S_m[B]$, and $S_4[B]$. Non-local functionals $\{\mathrm{J}_i[B]\}_{i=1}^4$ from (\ref{29-3-28})--(\ref{29-3-29}), which appear in separate diagrams with singular coefficients, enter with finite  multipliers with respect to the parameter $\Lambda$. The three-loop coefficients $\{z_{2,3},z_3,z_{4,3}\}$ for renormalization constants are written out in general terms as follows
\begin{equation*}\label{29-7-27}
z_{2,3}=-\frac{\lambda^2\tilde{z}_{4,1}}{24}\mathrm{I}_{1}(\sigma)
-\frac{\lambda^3}{16}\mathrm{A}(\sigma)
\mathrm{I}_{1}(\sigma)
+\frac{\lambda^3}{32}
\mathrm{I}_{6}(\sigma)+\frac{\lambda^2\tilde{z}_{2,1}}{16}
\mathrm{I}_{1}(\sigma)+\tilde{z}_{2,3},
\end{equation*}
\begin{multline*}
z_3=\frac{\lambda\tilde{z}_1\tilde{z}_{4,1}}{2}\mathrm{A}(\sigma)
-\frac{\lambda^2\tilde{z}_1}{2}\mathrm{I}_8(\sigma)+
\frac{\lambda^2\tilde{z}_{4,1}}{3m^2}\mathrm{I}_{4}(\sigma)+\frac{\lambda^3}{4m^2}\mathrm{A}(\sigma)\mathrm{I}_{4}(\sigma)-\\-\lambda^2\tilde{z}_{4,1}\mathrm{I}_{2}(\sigma)-\frac{\lambda^3}{4}\mathrm{A}(\sigma)\mathrm{I}_{2}(\sigma)
-\frac{\lambda^3}{4m^2}\mathrm{I}_{7}(\sigma)+
\frac{\lambda^3}{4}\mathrm{I}_{9}(\sigma)+
\lambda^3\mathrm{I}_{11}(\sigma)
-\frac{\lambda^3}{12m^2}\mathrm{I}_{14}(\sigma)+\\
+\frac{\lambda^3}{4}\mathrm{I}_{16}(\sigma)+
\frac{\lambda^3}{12}\mathrm{I}_{19}(\sigma)
-\frac{\lambda z_{4,1}}{4}\bigg(-\frac{\lambda}{m^2} R_0^\Lambda(0)R_{11}^\Lambda(0)+\\+\lambda R_{10}^\Lambda(0)R_{11}^\Lambda(0)+\lambda R_0^\Lambda(0)R_{21}^\Lambda(0)\bigg)
+\frac{2}{m^2}\Big(\mathrm{A}_1(\sigma)+\mathrm{A}_3(\sigma)\Big)+\tilde{z}_3,
\end{multline*}
\begin{multline*}
z_{4,3}=\frac{3\lambda\tilde{z}_{4,1}^2}{2}\mathrm{A}(\sigma)
+\frac{3\lambda^3}{2}\mathrm{I}_{5}(\sigma)-3\lambda^2\tilde{z}_{4,1}\mathrm{I}_{8}(\sigma)
-6\lambda^2\tilde{z}_{4,1}\mathrm{I}_{3}(\sigma)-\\
-\frac{3\lambda^3}{2}\mathrm{A}(\sigma)\mathrm{I}_{3}(\sigma)
+\frac{3\lambda^3}{2}\mathrm{I}_{10}(\sigma)
+6\lambda^3\mathrm{I}_{12}(\sigma)
+\frac{3\lambda^3}{4}\mathrm{I}_{17}(\sigma)+\frac{\lambda^3}{4}\mathrm{I}_{20}(\sigma)+\\
+\frac{3\zeta(3)L\lambda^3}{16(4\pi^2)^3}-
\frac{3\lambda^2z_{4,1}}{4}\Big(\mathrm{A}^2(\sigma)+R_0^\Lambda(0)R_{22}^\Lambda(0)\Big)+
\frac{24}{\lambda}\Big(\mathrm{A}_2(\sigma)+\mathrm{A}_4(\sigma)\Big)+\tilde{z}_{4,3}.
\end{multline*}
\end{lemma}
Finally, moving from arbitrary constants $\{\tilde{z}_{2,3},\tilde{z}_{3},\tilde{z}_{4,3}\}$ to
	$\{\bar{z}_{2,3},\bar{z}_{3},\bar{z}_{4,3}\}$ and using the explicit form of asymptotic expansions from Section \ref{29:sec:app:vch}, we get the answer in the form
\begin{align*}
z_{2,3}=~&L^2a_2+La_1+\bar{z}_{2,3},
\\
z_{3}=~&\Lambda^2\Big(L^2b_6+Lb_5+b_4\Big)+L^3b_3+L^2b_2+Lb_1+\bar{z}_{3},
\\
z_{4,3}=~&L^3c_3+L^2c_2+Lc_1
+\bar{z}_{4,3},
\end{align*}
where
\begin{align}\label{29-7-40}
a_1=&\,\frac{\lambda^3\big(\alpha_2(\mathbf{f})-\alpha_6(\mathbf{f})\big)}{64(4\pi^2)^3}-
\frac{\lambda^2\bar{z}_{4,1}}{48(4\pi^2)^2}+\frac{\lambda^2\bar{z}_{2,1}}{32(4\pi^2)^2},
\\\label{29-7-41}
a_2=&-\frac{\lambda^3}{128(4\pi^2)^3},
\end{align}
\begin{align}
\label{29-7-42}
b_1=&\,\frac{\lambda^3}{4(4\pi^2)^3}
\bigg(\frac{47}{192}
+\frac{\alpha_2^{\phantom{1}}(\mathbf{f})}{48}
-\frac{(4\pi^2)^2\alpha_3(\mathbf{f})}{12}
+\frac{(4\pi^2)^2\alpha_{11}(\mathbf{f})}{3}
\bigg)+\\\nonumber&-
\frac{\lambda^2(\bar{z}_1-4\bar{z}_{2,1}+2\bar{z}_{4,1})}{16(4\pi^2)^2}
+
\frac{\lambda\big(\bar{z}_1^{\phantom{1}}\bar{z}_{4,1}^{\phantom{1}}+\bar{z}_2^{\phantom{1}}-
	2\bar{z}_{2,2}^{\phantom{1}}+\bar{z}_{4,2}^{\phantom{1}}-2(\bar{z}_1^{\phantom{1}}+\bar{z}_{4,1}^{\phantom{1}})\bar{z}_{2,1}^{\phantom{1}}+3\bar{z}_{2,1}^2\big)}{4(4\pi^2)},
\\\label{29-7-43}
b_2=&-\frac{41\lambda^3}{384(4\pi^2)^3}
+
\frac{\lambda^2(\bar{z}_1-4\bar{z}_{2,1}+2\bar{z}_{4,1})}{8(4\pi^2)^2},
\\\label{29-7-44}
b_3=&\,\frac{7\lambda^3}{96(4\pi^2)^3},
\\\label{29-7-45}
b_4=&-\frac{\lambda^3}{12m^2}
\bigg(
\frac{3+\alpha_1(\mathbf{f})+6\alpha_2(\mathbf{f})}{8(4\pi^2)^3}-
\frac{\alpha_2(\mathbf{f})\alpha_3(\mathbf{f})}{2(4\pi^2)}+
3\alpha_7(\mathbf{f})+3\alpha_8(\mathbf{f})+\alpha_9(\mathbf{f})+\alpha_{10}(\mathbf{f})
\bigg)+
\\\nonumber&+
\frac{\lambda^2\alpha_3(\mathbf{f})\bar{z}_{4,1}}{3m^2}-\frac{\lambda^2\alpha_3(\mathbf{f})\bar{z}_{2,1}}{2m^2}
+\frac{\lambda(1+\mathbf{f}(0))\big(\bar{z}_{2,1}^{\phantom{1}}\bar{z}_{4,1}^{\phantom{1}}-\bar{z}_{2,1}^2-\lambda\bar{z}_{2,1}^{\phantom{1}}\alpha_2(\mathbf{f})/(8\pi^2)+\bar{z}_{2,2}^{\phantom{1}}-\bar{z}_{4,2}^{\phantom{1}}\big)}{2(4\pi^2)m^2},
\\\label{29-7-37}
b_5=&
-\frac{\lambda^2(1+\mathbf{f}(0))}{16(4\pi^2)^2m^2}
\bigg(
-\frac{35\lambda}{12(4\pi^2)}+12\bar{z}_{4,1}-18\bar{z}_{2,1}
\bigg)+\frac{\lambda^3\alpha_3(\mathbf{f})}{4(4\pi^2)m^2},
\\\label{29-7-38}
b_6=&-\frac{9\lambda^3(1+\mathbf{f}(0))}{32(4\pi^2)^3m^2},
\end{align}

\begin{align}\label{29-7-39}
c_1=&-\frac{3\lambda^3}{2(4\pi^2)^3}
\bigg(
-\frac{77}{384}+\frac{3\alpha_1(\mathbf{f})}{8}-\frac{19\alpha_2(\mathbf{f})}{96}-\frac{3\alpha_2^2(\mathbf{f})}{16}
\bigg)
+\frac{3\lambda^3\zeta(3)}{16(4\pi^2)^3}
-\\\nonumber&-\frac{3\lambda^3}{2(4\pi^2)}\bigg(
\frac{\alpha_3(\mathbf{f})}{24}+
\frac{3\alpha_4(\mathbf{f})}{2}+\frac{3\alpha_5(\mathbf{f})}{2}-
\frac{\alpha_{11}(\mathbf{f})}{6}
\bigg)-\\\nonumber&
-\frac{9\lambda^2\bar{z}_{4,1}}{8(4\pi^2)^2}+\frac{3\lambda^2\bar{z}_{2,1}}{2(4\pi^2)^2}
-\frac{3\lambda}{2(4\pi^2)}\bigg(-\frac{3\bar{z}_{2,1}^2}{2}
-\frac{\bar{z}_{4,1}^2}{2}+2\bar{z}_{2,1}\bar{z}_{4,1}+\bar{z}_{2,2}-\bar{z}_{4,2}
\bigg),
\\\label{29-7-391}
c_2=&-\frac{89\lambda^3}{128(4\pi^2)^3}+\frac{27\lambda^2\bar{z}_{4,1}}{16(4\pi^2)^2}-\frac{9\lambda^2\bar{z}_{2,1}}{4(4\pi^2)^2},
\\\label{29-7-392}
c_3=&\,\frac{27\lambda^3}{64(4\pi^2)^3}.
\end{align}

\section{General problem statement}
\label{29:sec:gen}
\subsection{On transition from basic to general model}
\label{29:sec:gen:op}
Consider a generalization for the model with the action (\ref{29-1-1}) by moving from the real field $\phi(\cdot)$ to a set of real fields $\{\phi_i(\cdot)\}_{i=1}^n$, where $n\in\mathbb{N}$. Moreover, the interaction of the fourth power can be supplemented by a cubic dependence. Then the classical action is represented as
\begin{equation}\label{29-16-1}
\int_{\mathbb{R}^4}\mathrm{d}^4x\bigg[
t^i\phi_i(x)+
\frac{1}{2}\phi_i(x)A^{ij}(x)\phi_j(x)+\frac{g^{ijk}}{3!}\phi_i(x)\phi_j(x)\phi_k(x)+
\frac{\lambda^{ijkn}}{4!}\phi_i(x)\phi_j(x)\phi_k(x)\phi_n(x)
\bigg],
\end{equation}
where
\begin{equation*}\label{29-16-2}
A^{ij}(x)=\delta^{ij}A_0(x)+M^{ij},\,\,\,
A_0(x)=-\partial_{x_\mu}\partial_{x^\mu},
\end{equation*}
and the coefficients $t^i$, $M^{ij}$, $g^{ijk}$, and $\lambda^{ijkn}$, are real and completely symmetric with respect to permutations of the indices. Note that when choosing
\begin{equation}\label{29-16-222}
n=1,\,\,\,t^1=0,\,\,\, M^{11}=m^2,\,\,\,
g^{111}=0,\,\,\,\lambda^{1111}=\lambda,
\end{equation}
the action from (\ref{29-16-1}) transit to the previously studied in (\ref{29-1-1}).
The background field method can be applied to such action within the framework of the functional integral representation mentioned above. To do this, according to the general logic, it is necessary to make the shift
\begin{equation*}\label{29-16-3}
\phi_i(x)\to B_i(x)+\sqrt{\hbar}\phi_i(x).
\end{equation*}
Then the quantum action reduces to a perturbative decomposition, in which the following objects must be taken as vertices instead of (\ref{29-1-9})--(\ref{29-1-11})
\begin{equation}\label{29-16-21}
\Gamma_1\to
\int_{\mathbb{R}^4}\mathrm{d}^4x\,\phi_i(x)\bigg(t^i+A^{ij}(x)B_j(x)+
\frac{g^{ikj}}{2}B_k(x)B_j(x)
+\frac{\lambda^{iknj}}{3!}B_k(x)B_n(x)B_j(x)\bigg),
\end{equation}
\begin{equation*}\label{29-16-22}
\Gamma_2\to
\int_{\mathbb{R}^4}\mathrm{d}^4x\,\phi_i(x)\bigg(A^{ij}(x)+g^{ikj}B_k(x)
+\frac{\lambda^{iknj}}{2}B_k(x)B_n(x)\bigg)\phi_j(x),
\end{equation*}
\begin{equation*}\label{29-16-23}
\Gamma_3\to\int_{\mathbb{R}^4}\mathrm{d}^4x
\Big(g^{ijk}+\lambda^{ijkn}B_n(x)\Big)
\phi_i(x)\phi_j(x)\phi_k(x),
\end{equation*}
\begin{equation*}\label{29-16-4}
\Gamma_4\to\int_{\mathbb{R}^4}\mathrm{d}^4x\,\lambda^{ijkn}\phi_i(x)\phi_j(x)\phi_k(x)\phi_n(x).
\end{equation*}
Such set of elements for the diagrammatic technique needs to be supplemented by the Green's function, which in this case has two indices and is a solution to the equation
\begin{equation*}\label{29-16-5}
\bigg(\delta^{im}A_0(x)+V^{mj}(x)\bigg)
G_{mj}(x,y)=\delta^{ij}\delta(x-y),
\end{equation*}
where the potential is the matrix
\begin{equation}\label{29-16-6}
V^{ij}(x)=M^{ij}+g^{ikj}B_k(x)
+\frac{\lambda^{iknj}}{2}B_k(x)B_n(x).
\end{equation}
With this transition, the general appearance of Feynman diagrams will not change. The next step is the introduction of regularization. It can be introduced, following the general logic from Section \ref{29:sec:reg-4}, by adding a regularizing additive to the classical action. In this case, it has the form\footnote{Note that in a more general case, for each field $\phi_i$, a separate operator deformation can be introduced, that is, we have $A_0^{\phantom{1}}\delta^{ij}\to A_{0,i}^{\Lambda}\delta^{ij}$.}
\begin{equation}\label{29-16-7}
\frac{1}{2}\int_{\mathbb{R}^4}\mathrm{d}^4x\,\phi_i(x)\Big(A_0^\Lambda(x)-A_0^{\phantom{1}}(x)\Big)\phi_i(x).
\end{equation}
As it was shown, in this case, the Green's function acquires the deformation $G_{ij}^{\phantom{1}}(x,y)\to G_{ij}^\Lambda(x,y)$ and has near the diagonal, when $x\sim y$, the following asymptotic decomposition\footnote{The functions $R_{11}^\Lambda$ and $R_{22}^\Lambda$ are used here instead of $R_{10}^\Lambda$, $R_{20}^\Lambda$, and $R_{21}^\Lambda$, with explicit regularization.}
\begin{multline*}
G_{ij}^\Lambda(x,y)=R_0^\Lambda(x-y)\delta^{ij}-\frac{1}{2}R_{11}^\Lambda(x-y)\big(V^{ij}(x)+V^{ij}(y)\big)+\\+\frac{1}{4}R_{22}^\Lambda(x-y)\big(V^{ik}(x)V^{kj}(x)+V^{ik}(y)V^{kj}(y)\big)+PS_{1,ij}^\Lambda(x,y),
\end{multline*}
where the latest functions are defined and described in Section \ref{29:sec:reg-3}. 

Let us move on to the final step -- renormalization. Note that in this case it can be carried out by the following substitutions in the classical deformed action
\begin{equation}\label{29-16-9}
t^i\to t_r^i=t^i+\sum_{m=1}^{+\infty}\hbar^mt_{r,m}^i,\,\,\,
M^{ij}\to M_r^{ij}=M^{ij}+\sum_{m=1}^{+\infty}\hbar^mM_{r,m}^{ij},
\end{equation}
\begin{equation}\label{29-16-10}
g^{ijk}\to g_r^{ijk}=g^{ijk}+\sum_{m=1}^{+\infty}\hbar^mg_{r,m}^{ijk},\,\,\,
\lambda^{ijkn}\to \lambda_r^{ijkn}=\lambda^{ijkn}+\sum_{m=1}^{+\infty}\hbar^m\lambda_{r,m}^{ijkn},
\end{equation}
\begin{equation}\label{29-16-11}
A_0^\Lambda\delta^{ij},A_0^{\phantom{1}}\delta^{ij}\to
A_0^\Lambda\delta^{ij}_r,A_0^{\phantom{1}}\delta^{ij}_r,\,\,\,\mbox{where}\,\,\,
\delta_r^{ij}=\delta^{ij}+\sum_{m=1}^{+\infty}\hbar^m\delta_{r,m}^{ij}.
\end{equation}
In this case, in addition to the three renormalization constants, two more are added. This is due to the appearance in the classical action (\ref{29-16-1}) of terms with the first and third powers of the fields.

Thus, to obtain the three coefficients for the renormalization constants, we do not need to perform all the calculations again. It is only necessary to correct the calculations by replacing the coupling constants and background fields with new coefficients when calculating (in key relations). This can be done according to the following set of rules:
\begin{itemize}
	\item when using the decomposition of the Green's function, replace the existing potential with the matrix analog from (\ref{29-16-6})
	\begin{equation*}\label{29-16-12}
	m^2+B^2(x)/2\to V^{ij}(x);
	\end{equation*}
\item replace the coefficient at the vertex with four outer lines as follows
\begin{equation*}\label{29-16-13}
	\lambda\to\lambda^{ijkn};
\end{equation*}
\item replace the coefficient at the vertex with three outer lines as follows
\begin{equation*}\label{29-16-14}
	\lambda B(x)\to g^{ijk}+\lambda^{ijkn}B_n(x);
\end{equation*}
\item "connect" free indexes according to the diagram form.
\end{itemize}
It is important to note that substantially new vertices do not appear with this generalization. Terms with the first power (\ref{29-16-21}) do not participate in the quantum action, since they violate strong connectivity. However, they are still part of the quantum equation of motion. However, at each renormalization step, terms proportional to the first power of the background field appear, and thus the constant $t^i$ is renormalized according to the general scheme (\ref{29-16-9}).

\subsection{Coefficient values}
\label{29:sec:gen:ot}
Let us show a variant of the transition from the case $n=1$ to an arbitrary $n>1$ using the example of the first loop. For the rest of the cases, we will formulate only the final answers, since the calculations are simple and monotonous. Let us introduce a few auxiliary functions to begin with
\begin{equation}\label{29-16-16}
S_{2,ij}[B]=\int_{\mathbb{R}^4}\mathrm{d}^4x\big(\partial_{x_\mu}B_i(x)\big)\big(\partial_{x^\mu}B_j(x)\big),\,\,\,
S_{i_1\ldots i_k}[B]=\int_{\mathbb{R}^4}\mathrm{d}^4x\,B_{i_1}(x)\cdot\ldots\cdot B_{i_k}(x).
\end{equation}
Next, we write out an analog for formula (\ref{29-4-1}), taking into account the latest comments, as follows
\begin{multline*}
-\frac{R_0^\Lambda(0)}{2}\int_{\mathbb{R}^4}\mathrm{d}^4x\bigg(
g^{iik}B_k(x)+\frac{\lambda^{iikj}}{2}B_k(x)B_j(x)
\bigg)+\frac{\mathrm{A}(\sigma)}{4}\int_{\mathbb{R}^4}\mathrm{d}^4x
\bigg(
2M^{ij}g^{ijk}B_k(x)+\\+g^{ijk_1}g^{ijk_2}B_{k_1}(x)B_{k_2}(x)+
M^{ij}\lambda^{ijk_1k_2}B_{k_1}(x)B_{k_2}(x)+
g^{k_1ij}\lambda^{k_2k_3ij}B_{k_1}(x)B_{k_2}(x)B_{k_3}(x)+\\+
\frac{1}{4}\lambda^{k_1k_2ij}\lambda^{k_3k_4ij}
B_{k_1}(x)B_{k_2}(x)B_{k_3}(x)B_{k_4}(x)
\bigg).
\end{multline*}
Given the definitions from (\ref{29-16-16}) and the auxiliary notation from Section \ref{29:sec:app:kom}, the last decomposition can be represented in the following form
\begin{align*}
&S_{i}[B]\bigg(-\frac{R_0^\Lambda(0)}{2}\mathbf{T}^i_{1,1,1}+\frac{\mathrm{A}(\sigma)}{2}
\mathbf{T}^i_{1,1,2}\bigg)+\\
+&
S_{i_1i_2}[B]\bigg(-\frac{R_0^\Lambda(0)}{4}\mathbf{T}^{i_1i_2}_{1,2,1}+
\frac{\mathrm{A}(\sigma)}{4}\mathbf{T}^{i_1i_2}_{1,2,2}
+\frac{\mathrm{A}(\sigma)}{4}\mathbf{T}^{i_1i_2}_{1,2,3}\bigg)+\\
+&
S_{i_1i_2i_3}[B]\frac{\mathrm{A}(\sigma)}{4}\mathbf{T}^{i_1i_2i_3}_{1,3,1}+
S_{i_1i_2i_3i_4}[B]\frac{\mathrm{A}(\sigma)}{16}\mathbf{T}^{i_1i_2i_3i_4}_{1,4,1}.
\end{align*}
Then, using the symbol $\mathbf{S}$ to denote index symmetrization divided by the corresponding factorial\footnote{For example, for a coefficient with two indexes $\mathbf{T}^{ij}$ we have $\mathbf{ST}^{ij}=\big(\mathbf{T}^{ij}+\mathbf{T}^{ji}\big)/2!$.}, the renormalization in the first loop looks like this
\begin{equation}\label{29-16-18}
t^i_{r,1}=-\frac{1}{2}R_0^\Lambda(0)\mathbf{T}^i_{1,1,1}+\frac{1}{2}\mathrm{A}(\sigma)
\mathbf{T}^i_{1,1,2}
+\tilde{t}^i_{r,1},
\end{equation}
\begin{equation}\label{29-16-25}
\delta^{i_1i_2}_{r,1}=\tilde{\delta}^{i_1i_2}_{r,1},
\end{equation}
\begin{equation}\label{29-16-19}
M^{i_1i_2}_{r,1}=-\frac{1}{2}R_0^\Lambda(0)\mathbf{T}^{i_1i_2}_{1,2,1}+
\frac{1}{2}\mathrm{A}(\sigma)\mathbf{T}^{i_1i_2}_{1,2,2}
+\frac{1}{2}\mathrm{A}(\sigma)\mathbf{T}^{i_1i_2}_{1,2,3}+\tilde{M}^{i_1i_2}_{r,1},
\end{equation}
\begin{equation}\label{29-16-33}
g^{i_1i_2i_3}_{r,1}=\frac{3}{2}\mathrm{A}(\sigma)\mathbf{ST}^{i_1i_2i_3}_{1,3,1}
+\tilde{g}^{i_1i_2i_3},
\end{equation}
\begin{equation}\label{29-16-24}
\lambda^{i_1i_2i_3i_4}_{r,1}=
\frac{3}{2}\mathrm{A}(\sigma)\mathbf{ST}^{i_1i_2i_3i_4}_{1,4,1}+
\tilde{\lambda}^{i_1i_2i_3i_4}_{r,1},
\end{equation}
where $\tilde{t}^i_{r,1}$, $\tilde{\delta}^{i_1i_2}_{r,1}$, $\tilde{M}^{i_1i_2}_{r,1}$, $\tilde{g}^{i_1i_2i_3}$, and $\tilde{\lambda}^{i_1i_2i_3i_4}_{r,1}$ are fully symmetric constant coefficients having the asymptotics $\mathcal{O}(1)$ with respect to the regularization parameter $\Lambda$.

Similarly, the coefficients for the second loop can be obtained. To do this, it is convenient to write out local singular parts for individual diagrams $d_1$, $d_2$, and $cd_1$. Explicit equalities are presented in Section \ref{29:sec:app:kom-1}. The answers for the coefficients of the renormalization constants, taking into account the notation from Section \ref{29:sec:app:kom}, are written out as follows
\begin{align}\label{29-18-4}
	t^i_{r,2}=&\frac{1}{6}\mathrm{I}_4(\sigma)\mathbf{T}^i_{2,1,1}-
	\frac{1}{2}\mathrm{I}_2(\sigma)\mathbf{T}^i_{2,1,2}-
	\frac{1}{4}\mathrm{I}_3(\sigma)\mathbf{T}^i_{2,1,3}
	+\frac{1}{4}R_0^\Lambda(0)\mathrm{A}(\sigma)\mathbf{T}^i_{2,1,4}
	-\frac{1}{4}\mathrm{A}^2(\sigma)\mathbf{T}^i_{2,1,5}+\\\label{29-18-42}+&\frac{1}{2}\mathrm{A}(\sigma)\mathbf{T}^i_{2,1,6}-\frac{1}{2}
	R_0^\Lambda(0)\mathbf{T}^i_{2,1,7}
	+\frac{1}{2}\mathrm{A}(\sigma)\mathbf{T}^i_{2,1,8}-\mathrm{A}(\sigma)\mathbf{T}^i_{2,1,9}+
	\frac{1}{2}R_0^\Lambda(0)\mathbf{T}^i_{2,1,10}+\tilde{t}^i_{r,2},
\end{align}
\begin{equation}\label{29-16-26}
	\delta^{i_1i_2}_{r,2}=-\frac{1}{48}\mathrm{I}_1(\sigma)\mathbf{T}^{i_1i_2}_{2,2,1}
	+\tilde{\delta}^{i_1i_2}_{r,2},
\end{equation}
\begin{align}\label{29-16-27}
	M^{i_1i_2}_{r,2}=&
	\frac{1}{6}\mathrm{I}_4(\sigma)\mathbf{T}^{i_1i_2}_{2,2,1}-
	\frac{1}{2}\mathrm{I}_2(\sigma)\mathbf{T}^{i_1i_2}_{2,2,2}-
	\frac{1}{4}\mathrm{I}_3(\sigma)\mathbf{T}^{i_1i_2}_{2,2,3}-
	\mathrm{I}_3(\sigma)\mathbf{ST}^{i_1i_2}_{2,2,4}
	+\frac{1}{4}R_0^\Lambda(0)\mathrm{A}(\sigma)\mathbf{T}^{i_1i_2}_{2,2,5}
	-\\\label{29-16-271}
	-&\frac{1}{4}\mathrm{A}^2(\sigma)\mathbf{T}^{i_1i_2}_{2,2,6}
	-\frac{1}{4}\mathrm{A}^2(\sigma)\mathbf{T}^{i_1i_2}_{2,2,7}
	+\frac{1}{2}\mathrm{A}(\sigma)\mathbf{T}^{i_1i_2}_{2,2,8}
	+\mathrm{A}(\sigma)\mathbf{ST}^{i_1i_2}_{2,2,9}-
	\frac{1}{2}R_0^\Lambda(0)\mathbf{T}^{i_1i_2}_{2,2,10}+
	\\\label{29-16-272}+&
	\frac{1}{2}\mathrm{A}(\sigma)\mathbf{T}^{i_1i_2}_{2,2,11}
	-\mathrm{A}(\sigma)\mathbf{T}^{i_1i_2}_{2,2,12}
	-\mathrm{A}(\sigma)\mathbf{T}^{i_1i_2}_{2,2,13}
	+\frac{1}{2}R_0^\Lambda(0)\mathbf{T}^{i_1i_2}_{2,2,14}
	+\tilde{M}^{i_1i_2}_{r,2},
\end{align}
\begin{align}\label{29-16-28}
	g^{i_1i_2i_3}_{r,2}=&-\frac{3}{2}\mathrm{I}_3(\sigma)\mathbf{ST}^{i_1i_2i_3}_{2,3,1}-
	\frac{3}{2}\mathrm{I}_3(\sigma)\mathbf{ST}^{i_1i_2i_3}_{2,3,2}
	-\frac{3}{4}\mathrm{A}^2(\sigma)\mathbf{ST}^{i_1i_2i_3}_{2,3,3}+
	\\\label{29-16-281}
	&+\frac{3}{2}\mathrm{A}(\sigma)\mathbf{ST}^{i_1i_2i_3}_{2,3,4}+\frac{3}{2}\mathrm{A}(\sigma)\mathbf{ST}^{i_1i_2i_3}_{2,3,5}-3\mathrm{A}(\sigma)\mathbf{ST}^{i_1i_2i_3}_{2,3,6}+\tilde{g}^{i_1i_2i_3}_{r,2},
\end{align}
\begin{equation}\label{29-16-29}
	\lambda^{i_1i_2i_3i_4}_{r,2}=
	-3\mathrm{I}_3(\sigma)\mathbf{ST}^{i_1i_2i_3i_4}_{2,4,1}
	-\frac{3}{4}\mathrm{A}^2(\sigma)\mathbf{ST}^{i_1i_2i_3i_4}_{2,4,2}
	+3\mathrm{A}(\sigma)\mathbf{ST}^{i_1i_2i_3i_4}_{2,4,3}
	-3\mathrm{A}(\sigma)\mathbf{ST}^{i_1i_2i_3i_4}_{2,4,4}
	+\tilde{\lambda}^{i_1i_2i_3i_4}_{r,2},
\end{equation}
where arbitrary constant coefficients were introduced as it was done above. Note that the non-local parts are not written out in Section \ref{29:sec:app:kom-1}. It is not difficult to show that they are cancelled in the final combination.

Continuing the calculations, it can be shown that the coefficients for the third loop have the following form
\begin{equation}\label{29-16-30}
t^i_{r,3}=\sum_{k=1}^6\mathbf{\overline{T}}^{i}_{3,1,k}+\tilde{t}^i_{r,3},
\end{equation}
\begin{equation}\label{29-16-31}
\delta^{i_1i_2}_{r,3}=-\frac{1}{48}\mathrm{I}_1(\sigma)\Big(\mathbf{T}^{i_1i_2}_{3,2,9}+\mathbf{T}^{i_2i_1}_{3,2,9}\Big)+
\frac{1}{32}\mathrm{I}_6(\sigma)\mathbf{T}^{i_1i_2}_{3,2,14}-
\frac{1}{16}\mathrm{A}(\sigma)
\mathrm{I}_1(\sigma)\mathbf{T}^{i_1i_2}_{3,2,14}+\frac{1}{16}\mathrm{I}_1(\sigma)\mathbf{T}^{i_1i_2}_{3,2,42}+\tilde{\delta}^{i_1i_2}_{r,3},
\end{equation}
\begin{equation}\label{29-16-32}
M^{i_1i_2}_{r,3}=2\sum_{k=1}^6\mathbf{S\overline{T}}^{i_1i_2}_{3,1,k}+\tilde{M}^{i_1i_2}_{r,3},
\end{equation}
\begin{equation}\label{29-16-34}
g^{i_1i_2i_3}_{r,3}=6\sum_{k=1}^6\mathbf{S\overline{T}}^{i_1i_2i_3}_{3,1,k}+\tilde{g}^{i_1i_2i_3}_{r,3},
\end{equation}
\begin{equation}\label{29-16-35}
\lambda^{i_1i_2i_3i_4}_{r,3}=24
\sum_{k=1}^6\mathbf{S\overline{T}}^{i_1i_2i_3i_4}_{3,1,k}
+\tilde{\lambda}^{i_1i_2i_3i_4}_{r,3},
\end{equation}
where definitions and relations from Sections \ref{29:sec:app:kom} and \ref{29:sec:app:kom-1} were used. Thus, three-loop coefficients for the renormalization constants in the generalized model were found. In the special case (\ref{29-16-222}), they pass into the previously studied ones in Theorem \ref{29-th-1}.

\section{Conclusion}
\label{29:sec:conc}
This section is mostly devoted to the comments accumulated during the calculations. However, it would be appropriate here to list once again the main results set out in Section \ref{29:sec:res}.
\begin{itemize}
	\item The cutoff regularization in the coordinate representation has been implemented and studied. It preserves the connection between the quantum equation of motion and the quantum action.
	\item The first three coefficients for the renormalization constants for the scalar model with the quartic interaction have been calculated both in the basic formulation and in the generalized one.
	\item The consistency of the coefficients with previously known results has been shown.
	\item The asymptotic expansion for the Green's function near the diagonal has been studied.
	\item Asymptotic expansions for the regularized integrals have been calculated.
	\item The rules of the renormalization $\mathcal{R}$-operation have been tested using the example of three-loop diagrams.
	\item The first three coefficients for the $\beta$-function have been calculated (they are consistent with the general theory).
\end{itemize}

\subsection{Some remarks}

\noindent{\textbf{On the quantum equation of motion.}} Section \ref{29:sec:1:state} summarizes the derivation of the formula for the quantum action $W$, which was presented as a sum of strongly connected diagrams. At the same time, as it was noted, the quantum action is actually a functional $W[B]$, depending on the background field $B$, which, in turn, is a solution to the quantum equation of motion $Q[B](x)=0$ with a set of boundary conditions. The functional $Q[B](x)$ itself is determined by the sum of all strongly connected diagrams with one external free line.

The definition of the background field using the quantum equation of motion in the above formulation is fundamental, since such field allows us to zero out all not strongly connected diagrams when deriving representation (\ref{29-1-16}) for the quantum action. At the same time, there is the relation from (\ref{29-3-7}), which says that the quantum equation of motion is obtained by varying the quantum action with respect to the background field.

The proposed and successfully implemented regularization, see Section \ref{29:sec:reg}, allows us to preserve the connection between the quantum equation of motion and the quantum action after the corresponding deformation, see (\ref{29-3-8}), and after the renormalization procedure, see (\ref{29-3-24}). This approach ensures that after the renormalization of the quantum action, the quantum equation of motion is automatically renormalized. From this point of view, the coefficients of the renormalization constants found are universal.

\vspace{2mm}
\noindent{\textbf{On the singular components in $PS_1^\Lambda(x,y)$.}} It is important to note that the function $PS_1^\Lambda(x,y)$, which depends on the background field $B$ and is presented in Section \ref{29:sec:app:doc}, contains not only smooth non-local parts and logarithmic contributions with "sufficient" smoothness, but also functions which, when calculating divergences, lead to singularities, but vanish due to the presence of a "complete" derivative. They can be represented as derivative of a function.

Such contributions include the terms containing $\partial_{x^\nu}B(x)$ and $\partial_{x^\nu}\partial_{x^\mu}B(x)$. During the calculation process, such components of the Green's function were not discussed and were in the definition of $PS_1^\Lambda(x,y)$. Now, we emphasize that they do not contribute. This fact was verified separately by explicit calculations.

\vspace{2mm}
\noindent{\textbf{On the invariance of coefficients for the renormalization constants.}} In the recent work, see \cite{Kharuk-2021}, when studying the two-loop contribution to the effective action of the Yang--Mills theory, it was noticed that the renormalization coefficients are invariant with respect to the shift of the Green's function by local smooth zero modes of a special kind. Note that the same property holds in this paper. This can be verified either by direct calculations, or by referring to the shift of the function $PS_1^\Lambda(x,y)$, which enters the renormalized effective action without singular coefficients.

\vspace{2mm}
\noindent{\textbf{On changing the basis for singular components.}} It is also important to pay attention to the fact that the choice of functions (\ref{29-3-19})--(\ref{29-3-21}) is not the only one. They can be replaced by other functions that have the same singular behavior. In other words, we can change the "enough smooth" part by sending the difference to $PS_1^\Lambda(x,y)$. Taking into account that the function $PS_1^\Lambda(x,y)$ is excluded from the calculations (without singular coefficients), such replacement does not affect the final answer for the renormalization coefficients. For example, another interesting choice might be
\begin{equation*}\label{29-20-1}
R_0^\Lambda(x-y)\to R_0^{\Lambda,\sigma}(x-y),
\end{equation*}
\begin{equation*}\label{29-20-2}
R_{11}^\Lambda(x-y)\to
\int_{\mathbb{R}^4}\mathrm{d}^4z\,R_0^{\Lambda,\sigma}(x-z)R_0^{\Lambda,\sigma}(z-y),
\end{equation*}
\begin{multline*}
R_{22}^\Lambda(x-y)\to
2\int_{\mathbb{R}^4\times\mathbb{R}^4}\mathrm{d}^4z\mathrm{d}^4u\,R_0^{\Lambda,\sigma}(x-z)
R_0^{\Lambda,\sigma}(z-u)R_0^{\Lambda,\sigma}(u-y)-\\-2\int_{\mathrm{B}_{1/\sigma}\times\mathrm{B}_{1/\sigma}}\mathrm{d}^4z\mathrm{d}^4u\,R_0^{\phantom{1}}(z)
R_0^{+\infty,\sigma}(z-u)R_0^{\phantom{1}}(u),
\end{multline*}
where the function $R_0^{\Lambda,\sigma}(\cdot)$ is defined in (\ref{29-6-3}).
Such functions lead to a more symmetrical appearance of singular divergent diagrams, although less pleasant from the computational point of view compared to those used in this work.

\vspace{2mm}
\noindent{\textbf{About a simpler calculation.}} It should be noted that in Section \ref{29:sec:tri:d-3}, we can directly apply the expression (\ref{29-8-21}) and get the decomposition for the singular part. However, despite the fact that the result will be correct, such a procedure is not proof. Indeed, in this case it is not clear why there are no other singularities, and why such a mechanism should work. It is for this reason that a neat proof of the applicability of the decomposition has been proposed in (\ref{29-8-21}). A similar situation was faced in Section \ref{29:sec:tri:d-5}, where higher-order divergences appear.

\vspace{2mm}
\noindent{\textbf{On the matrix coefficients.}} In the generalized formulation of the problem (\ref{29-16-1}), the coefficients have several indices. This leads to the fact that sets of bulky expressions with various combinations appear in the final answers, see Section \ref{29:sec:app:kom}. Note that when calculating, it is more convenient to work not with the analytical representation, but with a diagrammatic one. Indeed, if we compare a vertex with four lines to the coefficient $\lambda^{i_1i_2i_3i_4}$, a vertex with three lines to the coefficient $g^{i_1i_2i_3}$, etc., then as a result all combinations from Section \ref{29:sec:app:kom} will take a quite friendly visual form. The analytical answer is written out in the paper, since it is comparable to the diagrammatic one. A similar idea with a transition to "Feynman diagrams" is used in group theory, see \cite{cvi}.

\vspace{2mm}
\noindent{\textbf{On the deformation of the classical action.}} As it was mentioned above, the regularization is introduced by deforming the classical action. In the case of the quartic interaction, the deformation is performed by adding the regularizing additive $S[\phi,\Lambda]$, see formulas (\ref{29-10-3}) and (\ref{29-16-7}), which depends on the field $\phi(\cdot)$. At the same time, it was assumed that the field $\phi(\cdot)$ belongs to the functional space after applying the background field method.

Note that in fact, we can add the regularizing term even before applying the background field method, that is, add $S[B+\sqrt{\hbar}\phi,\Lambda]/\hbar$ instead of $S[\phi,\Lambda]$. This does not affect the renormalization procedure and the coefficients of the renormalization constants. In this case, changes (replacing $A_0^{\phantom{1}}$ with $A_0^\Lambda$) will occur only in the classical action $S[B]$ and the vertex with one outer line $\Gamma_1[B,\phi]$.

\vspace{2mm}
\noindent{\textbf{On an additional verification of the coefficients obtained.}} Of course, special attention should be paid to additional verification of the calculated coefficients. We emphasize that divergences of the form $\Lambda^2L^k$ and $L^{k+1}$ for $k\geqslant 1$ can be found without laborious calculation of the main diagrams.

To verify this, we apply the operator $-\sigma\partial_{\sigma}$ to both sides of the equalities from (\ref{29-1-30})--(\ref{29-1-32}). Then, on both sides, only those parts that depend on the renormalization coefficients will be differentiated. For example, in  formula (\ref{29-1-32}), the terms with diagrams $d_3$--$d_8$ will vanish, since they do not depend on the auxiliary parameter $\sigma$. There it is generated in the process of finding the singular part when dividing integrals and Green's function into parts. Thus, to calculate the mentioned divergences, it is sufficient to study only the counter diagrams, which is a much simpler task.

It should be noted that such check has been successfully passed.

\subsection{Acknowledgements}

The work is supported by the Ministry of Science and Higher Education of the Russian
Federation, grant 075-15-2022-289, and by the Foundation for the Advancement of Theoretical Physics and Mathematics "BASIS", grant "Young Russian Mathematics".

\vspace{2mm}
The author expresses gratitude to N.V.Kharuk for a careful reading of the text, numerous comments, criticism and suggestions. Additionally, A.V.Ivanov expresses special gratefulness to N.V.Kharuk and K.A.Ivanov for creating comfortable and stimulating conditions for writing the work. The author is also grateful to M.A.Russkikh for discussing sections \ref{29:sec:contr-2} and \ref{29:sec:tri:d-5} as part of his master's degree education.

\vspace{2mm}
\textbf{Data availability statement.} Data sharing not applicable to this article as no datasets were generated or analysed during the current study.

\vspace{2mm}
\textbf{Conflict of interest statement.} The author states that there is no conflict of interest.

\section{Appendix}
\label{29:sec:app}

\subsection{On the coefficients of heat kernel}
\label{29:sec:app:hk}
This section shows the relationship of the decomposition for the Green's function from Section \ref{29:sec:reg-3} with the representation near the diagonal, where coefficients are used for the asymptotic decomposition of the heat kernel \cite{28,vas1,vas2} for small values of proper time. For certainty, we introduce the Laplace operator in the four-dimensional Euclidean space of the form
\begin{equation*}\label{29-19-2}
\mathcal{A}(x)=-\partial_{x_\mu}\partial_{x^\mu}-v(x),
\end{equation*}
where $v(x)$ is a smooth real scalar potential. Then, repeating the general structure of the decomposition of the Green's function $\mathcal{G}(x,y)$ near the diagonal $x\sim y$ from works \cite{29,psi}, we obtain
\begin{equation}\label{29-19-1}
\mathcal{G}(x,y)=\sum_{k=0}^{+\infty}R_{k}(x-y)a_k(x,y)+\mbox{smooth non-local part.}
\end{equation}
In the last formula, the functions $R_{k}(\cdot)$ include singular components, including logarithmic behavior. The coefficients $a_k(x,y)$ are called the Seeley--DeWitt coefficients \cite{110,10-10}. Hadamard, Minakshisundaram \cite{111}, and Gilkey \cite{8} are also often mentioned in the name. These coefficients are smooth functions and for close arguments $x\sim y$ can be decomposed into the Taylor series in terms of the potential and its derivatives. The last part of decomposition (\ref{29-19-1}) also has smooth behavior and, actually, is a potential-dependent functional.

It is known that there is no closed formula for the Seeley--DeWitt coefficients outside the diagonal, but they can be found near the diagonal as a series by degrees of the vector $x^\mu-y^\mu$. There are a number of methods \cite{30,f6,31,31-1,33} to calculate them. The first orders are as follows
\begin{equation}\label{29-19-3}
a_0(x,y)=1,\,\,\,
a_1(x,y)=v(y)+o(1),\,\,\,
a_2(x,y)=\frac{1}{6}\partial_{y_\mu}\partial_{y^\mu}v(y)+\frac{1}{2}v^2(y)+o(1),
\end{equation}
where the smallness of the correction is considered relative to the difference of the vectors $x$ and $y$. 

We show that the decomposition from (\ref{29-19-1}) is consistent with that used in (\ref{29-3-18}) when removing the regularization. Indeed, note that in the scalar model, the potential is
\begin{equation*}\label{29-19-4}
v(x)=m^2+\frac{\lambda}{2}B^2(x).
\end{equation*}
Next, select $R_0(\cdot)$ as in (\ref{29-3-15}), and also using (\ref{29-3-19}) and (\ref{29-3-21}), we take
\begin{equation*}\label{29-19-5}
R_1^{\phantom{1}}(\cdot)=R_{11}^\Lambda(\cdot)\Big|_{\Lambda\to+\infty},\,\,\,
R_2^{\phantom{1}}(\cdot)=R_{22}^\Lambda(\cdot)\Big|_{\Lambda\to+\infty}.
\end{equation*}
After that, we define the function in terms of decomposition (\ref{29-19-1}) as follows
\begin{align}\nonumber
PS_1^{+\infty}(x,y)=&R_{1}(x-y)\Big(a_1(x,y)-\big(v(x)+v(y)\big)/2\Big)+\\\nonumber
+&
R_{2}(x-y)\Big(a_2(x,y)-\big(v^2(x)+v^2(y)\big)/4\Big)+
\\\nonumber
+&\sum_{k=3}^{+\infty}R_{k}(x-y)a_k(x,y)+\mbox{smooth non-local part.}
\end{align}
Then the series from (\ref{29-19-3}) is rewritten as follows
\begin{equation*}\label{29-19-7}
\mathcal{G}(x,y)=
R_0(x-y)+R_{1}(x-y)\frac{\big(v(x)+v(y)\big)}{2}+
R_{2}(x-y)\frac{\big(v^2(x)+v^2(y)\big)}{4}+PS_1^{+\infty}(x,y),
\end{equation*}
which is completely consistent with decomposition (\ref{29-3-18}) and the explicit form of the $PS_1$-part presented in Section \ref{29:sec:app:doc}.

\subsection{Auxiliary integrals}
\label{29:sec:app:vs}
This section contains a list of auxiliary integrals that arise during calculations. They are based on regularized functions mentioned above and depend on the dimensional parameter $\Lambda$. Such integrals are singular functions depending on the regularizing parameter, and tend to infinity (diverge) as $\Lambda\to+\infty$.

Taking into account the definition from (\ref{29-8-21}), the set of integrals can be written as follows
\begin{equation}\label{29-8-1}
\mathrm{I}_1(\sigma)=\int_{\mathrm{B}_{1/\sigma}}\mathrm{d}^4x\Big(R_0^\Lambda(x)\Big)^3|x|^2,
\end{equation}
\begin{equation}\label{29-8-2}
\mathrm{I}_{2}(\sigma)=\int_{\mathrm{B}_{1/\sigma}}\mathrm{d}^4x\Big(R_0^\Lambda(x)\Big)^2R_{10}^\Lambda(x),
\end{equation}
\begin{equation}\label{29-8-3}
\mathrm{I}_{3}(\sigma)=\int_{\mathrm{B}_{1/\sigma}}\mathrm{d}^4x\Big(R_0^\Lambda(x)\Big)^2R_{11}^\Lambda(x),
\end{equation}
\begin{equation}\label{29-8-4}
\mathrm{I}_{4}(\sigma)=\int_{\mathrm{B}_{1/\sigma}}\mathrm{d}^4x\Big(R_0^\Lambda(x)\Big)^3,
\end{equation}
\begin{equation}\label{29-8-5}
\mathrm{I}_{5}(\sigma)=\int_{\mathrm{B}_{1/\sigma}}\mathrm{d}^4x
\Big(\hat{R}_0^\Lambda(x)\Big)^2,
\end{equation}
\begin{equation}\label{29-8-6}
\mathrm{I}_{6}(\sigma)=\int_{\mathrm{B}_{1/\sigma}\times\mathrm{B}_{1/\sigma}}\mathrm{d}^4x\mathrm{d}^4z\,
\Big(R_0^\Lambda(x)\Big)^2
R_0^\Lambda(x-z)|x-z|^2\Big(R_0^\Lambda(x)\Big)^2,
\end{equation}
\begin{equation}\label{29-8-7}
\mathrm{I}_{7}(\sigma)=\int_{\mathrm{B}_{1/\sigma}}\mathrm{d}^4x\,
\Big(R_0^\Lambda(x)\Big)^2\hat{R}_0^\Lambda(x),
\end{equation}
\begin{equation}\label{29-8-8}
\mathrm{I}_{8}(\sigma)=\int_{\mathrm{B}_{1/\sigma}}\mathrm{d}^4x\,R_0^\Lambda(x)\hat{R}_0^\Lambda(x),
\end{equation}
\begin{equation}\label{29-8-9}
\mathrm{I}_{9}(\sigma)=\int_{\mathrm{B}_{1/\sigma}}\mathrm{d}^4x\Big(R_0^\Lambda(x)\Big)^2\hat{R}_{10}^\Lambda(x),
\end{equation}
\begin{equation}\label{29-8-10}
\mathrm{I}_{10}(\sigma)=\int_{\mathrm{B}_{1/\sigma}}\mathrm{d}^4x\Big(R_0^\Lambda(x)\Big)^2\hat{R}_{11}^\Lambda(x),
\end{equation}
\begin{equation}\label{29-8-11}
\mathrm{I}_{11}(\sigma)=\int_{\mathrm{B}_{1/\sigma}}\mathrm{d}^4x\,R_0^\Lambda(x)\hat{R}_0^\Lambda(x)R_{10}^\Lambda(x),
\end{equation}
\begin{equation}\label{29-8-12}
\mathrm{I}_{12}(\sigma)=\int_{\mathrm{B}_{1/\sigma}}\mathrm{d}^4x\,R_0^\Lambda(x)\hat{R}_0^\Lambda(x)R_{11}^\Lambda(x),
\end{equation}
\begin{equation}\label{29-8-13}
\mathrm{I}_{13}(\sigma)=\int_{\mathrm{B}_{1/\sigma}}\mathrm{d}^4x\Big(R_0^\Lambda(x)\Big)^3R_{10}^\Lambda(x),
\end{equation}
\begin{equation}\label{29-8-14}
\mathrm{I}_{14}(\sigma)=\int_{\mathrm{B}_{1/\sigma}}\mathrm{d}^4x\Big(R_0^\Lambda(x)\Big)^3R_{11}^\Lambda(x),
\end{equation}
\begin{equation}\label{29-8-15}
\mathrm{I}_{15}(\sigma)=\int_{\mathrm{B}_{1/\sigma}}\mathrm{d}^4x\Big(R_0^\Lambda(x)\Big)^2R_{10}^\Lambda(x)R_{10}^\Lambda(x),
\end{equation}
\begin{equation}\label{29-8-19}
\mathrm{I}_{16}(\sigma)=\int_{\mathrm{B}_{1/\sigma}}\mathrm{d}^4x\Big(R_0^\Lambda(x)\Big)^2R_{10}^\Lambda(x)R_{11}^\Lambda(x),
\end{equation}
\begin{equation}\label{29-8-20}
\mathrm{I}_{17}(\sigma)=\int_{\mathrm{B}_{1/\sigma}}\mathrm{d}^4x\Big(R_0^\Lambda(x)\Big)^2R_{11}^\Lambda(x)R_{11}^\Lambda(x),
\end{equation}
\begin{equation}\label{29-8-16}
\mathrm{I}_{18}(\sigma)=\int_{\mathrm{B}_{1/\sigma}}\mathrm{d}^4x\Big(R_0^\Lambda(x)\Big)^3R_{20}^\Lambda(x),
\end{equation}
\begin{equation}\label{29-8-17}
\mathrm{I}_{19}(\sigma)=\int_{\mathrm{B}_{1/\sigma}}\mathrm{d}^4x\Big(R_0^\Lambda(x)\Big)^3R_{21}^\Lambda(x),
\end{equation}
\begin{equation}\label{29-8-18}
\mathrm{I}_{20}(\sigma)=\int_{\mathrm{B}_{1/\sigma}}\mathrm{d}^4x\Big(R_0^\Lambda(x)\Big)^3R_{22}^\Lambda(x).
\end{equation}
\subsection{Asymptotics of integrals}
\label{29:sec:app:vch}
\noindent\textbf{Auxiliary notations.}
To begin with, let us write out the key relation obtained in \cite{Ivanov-2022}, which connects the usual function $R_0$ from (\ref{29-3-15}) with a deformed analog. For $s>0$, it has the form
\begin{equation*}\label{29-9-3}
\frac{1}{2\pi^2}
\int_{\mathrm{S}^3}\mathrm{d}^3\hat{y}\,R_0(s\hat{y}+x)=
\frac{1}{4\pi^2}
\begin{cases}
\,\,s^{-2}, &\mbox{if $|x|\leqslant s$}\\
|x|^{-2}, & \mbox{if $|x|>s$}
\end{cases}
=R_0^{1/s}(x)-\frac{\mathbf{f}\big(|x|^2/s^2\big)}{4\pi^2s^2},
\end{equation*}
where the integration was performed with the usage of the  standard measure on the three-dimensional sphere $\mathrm{S}^3$ with unit radius centered at the origin. When $s=1/\Lambda$, the already known function from (\ref{29-3-17}) is obtained. Next, we define the following set of additional functions
\begin{equation*}\label{29-9-11}
f_2^\Lambda(x)=\frac{\Lambda^2}{4\pi^2}\Big(1+\mathbf{f}\big(|x|^2\Lambda^2\big)\Big)
\chi\big(|x|\leqslant1/\Lambda\big),
\end{equation*}
\begin{equation}\label{29-9-12}
f_1^\Lambda(x)=f_2^\Lambda(x)-
\frac{\chi\big(|x|\leqslant1/\Lambda\big)}{4\pi^2|x|^2},
\end{equation}
\begin{equation}\label{29-9-5}
h_i(\Lambda y)=\frac{1}{4\pi^2\Lambda^2|y|^2}\int_{\mathrm{B}_{\Lambda |y|}}\mathrm{d}^4x\,
\Big(f_2^1(x)\Big)^i+
\frac{1}{4\pi^2}\int_{\mathrm{B}_{1}\setminus\mathrm{B}_{\Lambda |y|}}\mathrm{d}^4x\,
\Big(f_2^1(x)\Big)^i\frac{1}{|x|^2},
\end{equation}
where $i\in\mathbb{N}\cup\{0\}$, and also three numbers
\begin{equation*}\label{29-9-8}
\alpha_1(\mathbf{f})=-\frac{1}{4}+
\int_{\mathrm{B}_{1}}\mathrm{d}^4x\,f_2^1(x)=
\int_{\mathrm{B}_{1}}\mathrm{d}^4x\,f_1^1(x),
\end{equation*}
\begin{equation}\label{29-9-10}
\alpha_2(\mathbf{f})=2(4\pi^2)
\int_{\mathrm{B}_1}\mathrm{d}^4x\Big(f_2^1(x)\Big)^2,
\end{equation}
\begin{equation*}\label{29-9-9}
\alpha_3(\mathbf{f})=\frac{1}{4(4\pi^2)^2}+
\int_{\mathrm{B}_1}\mathrm{d}^4x\Big(f_2^1(x)\Big)^3
=\int_{\mathbb{R}^4}\mathrm{d}^4x\Big(R_0^1(x)\Big)^3,
\end{equation*}
which are functionals that depend on the function $\mathbf{f}$. It can be noted that on the sphere $|y|=1/\Lambda$, the following relations are valid between the latter functions
\begin{equation}\label{29-9-13}
\alpha_1(\mathbf{f})=-\frac{1}{4}+4\pi^2h_1(\Lambda y)\Big|_{|y|=1/\Lambda},
\end{equation}
\begin{equation}\label{29-9-14}
\alpha_2(\mathbf{f})=2(4\pi^2)^2
h_2(\Lambda y)\Big|_{|y|=1/\Lambda},
\end{equation}
\begin{equation}\label{29-9-15}
\alpha_3(\mathbf{f})=\frac{1}{4(4\pi^2)^2}+4\pi^2
h_3(\Lambda y)\Big|_{|y|=1/\Lambda}.
\end{equation}
Further, we introduce the integral for the product of the function $R_0$ and its deformed analogue as follows
\begin{equation}\label{29-9-16}
t_i^\Lambda(y)=\int_{\mathrm{B}_{1/\sigma}}\mathrm{d}^4x\,
\Big(R_0^\Lambda(x)\Big)^iR_0^{\phantom{1}}(x+y),
\end{equation}
\begin{equation}\label{29-9-17}
p_i^\Lambda(y)=\int_{\mathrm{B}_{1/\sigma}}\mathrm{d}^4x\,
\Big(R_0^\Lambda(x)\Big)^if_1^{\Lambda}(x+y),
\end{equation}
where $i\in\mathbb{N}\cup\{0\}$. Taking into account the definitions from (\ref{29-3-17}) and (\ref{29-9-12}), the relation follows immediately
\begin{equation*}\label{29-9-18}
\int_{\mathrm{B}_{1/\sigma}}\mathrm{d}^4x\,
\Big(R_0^\Lambda(x)\Big)^iR_0^{\Lambda}(x+y)=t_i^\Lambda(y)+p_i^\Lambda(y),
\end{equation*}
which transforms the composite block for most basic integrals.
Note that in three-loop calculations, only the cases with $i\in\{1,2,3\}$ occur and are important. Using the above-mentioned equalities, we can verify the validity of the following relations
\begin{equation}\label{29-9-6}
t_1^\Lambda(y)=
\begin{cases}
	~~~~~~
	h_1(\Lambda y)+\frac{L}{2(4\pi^2)},&\mbox{if $|y|\leqslant 1/\Lambda$};\\
	\frac{1}{4(4\pi^2)}
	-\frac{\ln(|y|\sigma)}{2(4\pi^2)}+\frac{\alpha_1(\mathbf{f})}{4\pi^2\Lambda^2|y|^2}
	,&\mbox{if $1/\Lambda<|y|\leqslant 1/\sigma$};\\
	~~~
	\frac{1}{16\pi^2|y|^2\sigma^2}+\frac{\alpha_1(\mathbf{f})}{4\pi^2|y|^2\Lambda^2},&\mbox{if $1/\sigma<|y|$},
\end{cases}
\end{equation}
\begin{equation}\label{29-9-4}
t_2^\Lambda(y)=
\begin{cases}
	~~~~~~~~~~~~~~~~~
\Lambda^2h_2(\Lambda y)-\frac{\sigma^2-\Lambda^2}{4(4\pi^2)^2},&\mbox{if $|y|\leqslant 1/\Lambda$};\\
-\frac{1}{4(4\pi^2)^2}\Big(\sigma^2-\frac{1}{|y|^2}\Big)+\frac{\ln(|y|\sigma)}{2(4\pi^2)^2|y|^2}+\mathrm{A}(\sigma)R_0(y)
,&\mbox{if $1/\Lambda<|y|\leqslant 1/\sigma$};\\
~~~~~~~~~~~~~~~~~~~~~
\mathrm{A}(\sigma)R_0(y),&\mbox{if $1/\sigma<|y|$},
\end{cases}
\end{equation}
\begin{equation}\label{29-9-7}
t_3^\Lambda(y)=
\begin{cases}
	~~~~~~~~
	\Lambda^4h_3(\Lambda y)-\frac{\sigma^4-\Lambda^4}{8(4\pi^2)^3},&\mbox{if $|y|\leqslant 1/\Lambda$};\\
	-\frac{\sigma^4}{8(4\pi^2)^3}-\frac{1}{8(4\pi^2)^3|y|^4}+\frac{\Lambda^2\alpha_3(\mathbf{f})}{4\pi^2|y|^2}
	,&\mbox{if $1/\Lambda<|y|\leqslant 1/\sigma$};\\
	~~~~~
	-\frac{\sigma^2}{4(4\pi^2)^3|y|^2}+\frac{\Lambda^2\alpha_3(\mathbf{f})}{4\pi^2|y|^2},&\mbox{if $1/\sigma<|y|$}.
\end{cases}
\end{equation}
Note that all functions are continuous and depend on the module of the variable, in this case on $|y|$. The latter property is a consequence of averaging over the sphere $\mathrm{S}^3$. Moreover, the mentioned functions can withstand the use of the Laplace operator. Indeed, using the relation
\begin{equation*}\label{29-9-19}
-\partial_{y_\mu}\partial_{y^\mu}R_0(y+x)=\delta(y+x),
\end{equation*}
we obtain
\begin{equation*}\label{29-9-20}
-\partial_{y_\mu}\partial_{y^\mu}t_i^\Lambda(y)=\Big(R_0^\Lambda(y)\Big)^i\chi\big(|y|\leqslant1/\sigma\big).
\end{equation*}
At the end of the introductory paragraph with auxiliary notations, we define one more additional set of auxiliary numbers according to the following formulas
\begin{equation}\label{29-9-25}
\alpha_4(\mathbf{f})=\int_{\mathrm{B}_1}\mathrm{d}^4y\,h_2^{\phantom{1}}(y)f_2^1(y),
\end{equation}
\begin{equation}\label{29-9-26}
\alpha_5(\mathbf{f})=\int_{\mathbb{R}^{4\times2}}\mathrm{d}^4x\mathrm{d}^4y\,
\Big(R_0^1(x)\Big)^2f_1^1(x+y)R_0^1(y),
\end{equation}
\begin{equation}\label{29-9-44}
\alpha_6(\mathbf{f})=2(4\pi^2)^2\int_{\mathrm{B}_1}\mathrm{d}^4y\Big(f_2^1(y)\Big)^3|y|^2,
\end{equation}
\begin{equation}\label{29-9-61}
	\alpha_7(\mathbf{f})=\int_{\mathbb{R}^{4\times2}}\mathrm{d}^4x\mathrm{d}^4y\,
	\Big(R_0^1(x)\Big)^2f_1^1(x+y)\Big(R_0^1(y)\Big)^2,
\end{equation}
\begin{equation}\label{29-9-62}
	\alpha_8(\mathbf{f})=\int_{\mathrm{B}_1}\mathrm{d}^4y\,h_2^{\phantom{1}}(y)\Big(f_2^1(y)\Big)^2,
\end{equation}
\begin{equation}\label{29-9-91}
	\alpha_9(\mathbf{f})=\int_{\mathbb{R}^{4\times2}}\mathrm{d}^4y\mathrm{d}^4x
	\Big(R_0^1(y)\Big)^3f_1^1(y+x)R_0^1(x),
\end{equation}
\begin{equation}\label{29-9-92}
	\alpha_{10}(\mathbf{f})=\int_{\mathrm{B}_1}\mathrm{d}^4y\,h_1^{\phantom{1}}(y)\Big(f_2^1(y)\Big)^3,
\end{equation}
\begin{equation}\label{29-9-97}
	\alpha_{11}(\mathbf{f})=\int_{\mathrm{B}_1}\mathrm{d}^4y\,h_3^{\phantom{1}}(y),
\end{equation}
\begin{equation}\label{29-9-120}
	\alpha_{12}(\mathbf{f})=\int_{\mathrm{B}_{1}}\mathrm{d}^4x\,
	f_2^1(x)h_1^{\phantom{1}}(x),
\end{equation}
\begin{equation}\label{29-9-123}
	\alpha_{13}(\mathbf{f})=
	\int_{\mathrm{B}_{1}\times\mathrm{B}_{1}}\mathrm{d}^4x\mathrm{d}^4y\,f_1^1(x)R_0^1(x+y)f_1^1(y)+\int_{\mathrm{B}_{1}}\mathrm{d}^4x\,f_1^1(x)h_1^{\phantom{1}}(x).
\end{equation}

\noindent\textbf{Calculating the integral $\mathrm{I}_{1}(\sigma)$.} Let us use the definition from (\ref{29-3-17}) and the fact that the area of the unit sphere in the four-dimensional space is $2\pi^2$. Then we can write out
\begin{equation*}\label{29-9-1}
\mathrm{I}_{1}(\sigma)=\frac{\alpha_6(\mathbf{f})}{2(4\pi^2)^2}+\frac{2\pi^2}{(4\pi^2)^3}\int_{1/\Lambda}^{1/\sigma}
\frac{dr}{r}=\frac{L+\alpha_6(\mathbf{f})}{2(4\pi^2)^2}\stackrel{\mathrm{s.p.}}{=}\frac{L}{2(4\pi^2)^2}.
\end{equation*}

\noindent\textbf{Calculating the integrals $\mathrm{I}_{2}(\sigma)$ and $\mathrm{I}_{3}(\sigma)$.} Let us start with the second integral, since the first one is a generalization. Let us use the definition from (\ref{29-3-19}) and auxiliary functions (\ref{29-9-16}) and (\ref{29-9-17}), then we represent the integral as
\begin{equation}\label{29-9-2}
\mathrm{I}_{3}(\sigma)=\int_{\mathrm{B}_{1/\sigma}}\mathrm{d}^4y\,t_2^\Lambda(y)R_0^\Lambda(y)+
\int_{\mathrm{B}_{1/\sigma}}\mathrm{d}^4y\,p_2^\Lambda(y)R_0^\Lambda(y).
\end{equation}
In this case, the first integral can be explicitly calculated using formulas (\ref{29-3-17}) and (\ref{29-9-4}). At the same time, it is convenient to divide the integration area into $\mathrm{B}_{1/\Lambda}$ and $\mathrm{B}_{1/\sigma}\setminus\mathrm{B}_{1/\Lambda}$. In the first case, the contribution has the form
\begin{equation*}\label{29-9-21}
\int_{\mathrm{B}_1}\mathrm{d}^4y\,h_2^{\phantom{1}}(y)f_2^1(y)+
\frac{4\alpha_1(\mathbf{f})+1}{16(4\pi^2)^2}+\mathcal{O}\big(1/\Lambda^2\big),
\end{equation*}
where the correction includes the term with $\sigma^2$ from the first line of (\ref{29-9-4}). Further, the following contribution is obtained from the second integration area
\begin{align}\nonumber
2\pi^2\int_{1/\Lambda}^{1/\sigma}ds\,\frac{s}{4\pi^2}
\bigg(-\frac{\big(\sigma^2-s^{-2}\big)}{4(4\pi^2)^2}+\frac{\ln(s\sigma)}{2(4\pi^2)^2s^2}&+\frac{\mathrm{A}(\sigma)}{4\pi^2s^2}\bigg)=\\\nonumber
&=-\frac{L^2}{8(4\pi^2)^2}+\frac{L+16\pi^2L\mathrm{A}(\sigma)-1/2}{8(4\pi^2)^2}+\mathcal{O}\big(1/\Lambda^2\big)
\\\nonumber&=\frac{2L^2+2L+4L\alpha_2(\mathbf{f})-1}{16(4\pi^2)^2}+
\mathcal{O}\big(1/\Lambda^2\big).
\end{align}
Moving on to the second term of formula (\ref{29-9-2}), it should be noted that $p_2^\Lambda(y)$ contains in the definition (\ref{29-9-17}) a function with a compact support. Indeed, $\mathrm{supp}\big(f_1^\Lambda(\cdot)\big)\subset\mathrm{B}_{1/\Lambda}$, so the integral
\begin{equation*}\label{29-9-23}
\int_{\mathrm{B}_{1/\sigma}}\mathrm{d}^4x\int_{\mathrm{B}_{1/\sigma}}\mathrm{d}^4y\,
\Big(R_0^\Lambda(x)\Big)^2f_1^{\Lambda}(x+y)R_0^\Lambda(y)
\end{equation*}
converges when removing the regularization, and after the changes $x\to x/\Lambda$ and $y\to y/\Lambda$ is equal to
\begin{equation}\label{29-9-27}
\alpha_5(\mathbf{f})-
\int_{\mathbb{R}^4}\mathrm{d}^4x\int_{\mathrm{B}_{1}}\mathrm{d}^4y\,
\Big(R_0^1(x)\Big)^2R_0^1(x+y)f_1^{1}(y)\Big(1-\chi\big(|x|\leqslant\Lambda/\sigma\big)
\chi\big(|x+y|\leqslant\Lambda/\sigma\big)\Big).
\end{equation}
Further, taking into account the inequalities $|y|\leqslant1$ and
\begin{equation*}\label{29-9-28}
\Big|1-\chi\big(|x|\leqslant\Lambda/\sigma\big)
\chi\big(|x+y|\leqslant\Lambda/\sigma\big)\Big|\leqslant
\Big|1-\chi\big(|x|\leqslant\Lambda/\sigma-1\big)\Big|,
\end{equation*}
the absolute value of the second term from (\ref{29-9-27}) can be estimated by the integral\footnote{It is implied here that the parameter $\Lambda$ is so large that $\Lambda>3\sigma$. It is for this reason that we can write $R_0^1(\cdot)$ instead of $\big|R_0^1(\cdot)\big|$.}
\begin{equation*}\label{29-9-29}
\int_{\mathrm{B}_{\Lambda/\sigma-1}}\mathrm{d}^4x\int_{\mathrm{B}_{1}}\mathrm{d}^4y\,
\Big(R_0^1(x)\Big)^2R_0^1(x+y)\Big|f_1^{1}(y)\Big|\leqslant
\frac{1}{2(4\pi^2)^2}
\bigg(\int_{\mathrm{B}_{1}}\mathrm{d}^4y\Big|f_1^{1}(y)\Big|\bigg)
\int_{\Lambda/\sigma-1}^{+\infty}\frac{\mathrm{d}s}{s(s-1)^2},
\end{equation*}
where we have used the relation $|x+y|^{-1}\leqslant(|x|-1)^{-1}$ taking into account the integration area.
Finally, using the inequality $\ln(1+s)\leqslant s$ for $s>-1$ and the chain of equalities
\begin{equation*}\label{29-9-30}
\int_{\Lambda/\sigma-1}^{+\infty}\frac{\mathrm{d}s}{s(s-1)^2}=
\frac{1}{\Lambda/\sigma-2}+\ln\bigg(1-\frac{1}{\Lambda/\sigma-1}\bigg)\leqslant
\frac{1}{\Lambda/\sigma-2}-\frac{1}{\Lambda/\sigma-1}=
\frac{1}{(\Lambda/\sigma-2)(\Lambda/\sigma-1)},
\end{equation*}
we conclude that the second term from (\ref{29-9-27}) can be considered as the correction $\mathcal{O}\big(1/\Lambda^2\big)$.

Thus, up to a constant term, the answer for the integral $\mathrm{I}_{3}(\sigma)$ is written as
\begin{equation*}\label{29-9-24}
\mathrm{I}_{3}(\sigma)=
\frac{L^2+L+2L\alpha_2(\mathbf{f})+2\alpha_1(\mathbf{f})}{8(4\pi^2)^2}+
\alpha_4(\mathbf{f})+\alpha_5(\mathbf{f})+
\mathcal{O}\big(1/\Lambda^2\big),
\end{equation*}
or, taking into account only the equation for singular parts,
\begin{equation}\label{29-9-31}
\mathrm{I}_{3}(\sigma)\stackrel{\mathrm{s.p.}}{=}
\frac{L^2+L+2L\alpha_2(\mathbf{f})}{8(4\pi^2)^2}.
\end{equation}
In the assumptions of Section \ref{29:sec:reg-4}, the following equality holds
\begin{equation*}\label{29-9-321}
	\mathrm{I}_{2}(\sigma)=\mathrm{I}_{3}(\sigma).
\end{equation*}

\noindent\textbf{Calculating the integral $\mathrm{I}_{4}(\sigma)$.} Consider the explicit form of the integral from (\ref{29-8-4}). Using the additional scaling of the integration variable $x\to x/\Lambda$, we obtain
\begin{equation}\label{29-9-33}
\mathrm{I}_{4}(\sigma)=\Lambda^2\int_{\mathrm{B}_{\Lambda/\sigma}}\mathrm{d}^4x\Big(R_0^1(x)\Big)^3
=\Lambda^2\alpha_3(\mathbf{f})-\frac{\sigma^2}{4(4\pi^2)^2}
\stackrel{\mathrm{s.p.}}{=}\Lambda^2\alpha_3(\mathbf{f}),
\end{equation}
where we have used the representation from (\ref{29-3-17}) and the limitation of the parameter $\sigma$.

\vspace{4mm}
\noindent\textbf{Calculating the integral $\mathrm{I}_{5}(\sigma)$.} When analyzing the integral (\ref{29-8-5}), it is convenient to transform taking into account the functions from (\ref{29-9-16}) and (\ref{29-9-17}). Indeed, the following relation is correct
\begin{equation*}\label{29-9-34}
\hat{R}_0^\Lambda(x)=p_2^\Lambda(x)+\big(t_2^\Lambda(x)-\mathrm{A}(\sigma)R_0^\Lambda(x)\big).
\end{equation*}
Then the main integral is rewritten as four terms
\begin{multline}\label{29-9-35}
\mathrm{I}_{5}(\sigma)=\int_{\mathrm{B}_{1/\sigma}}\mathrm{d}^4x\Big(p_2^\Lambda(x)\Big)^2+
2\int_{\mathrm{B}_{1/\sigma}}\mathrm{d}^4x\,p_2^\Lambda(x)t_2^\Lambda(x)-\\-
2\mathrm{A}(\sigma)\int_{\mathrm{B}_{1/\sigma}}\mathrm{d}^4x\,p_2^\Lambda(x)R_0^\Lambda(x)+
\int_{\mathrm{B}_{1/\sigma}}\mathrm{d}^4x\Big(t_2^\Lambda(x)-\mathrm{A}(\sigma)R_0^\Lambda(x)\Big)^2.
\end{multline}
This division is convenient because it allows us to divide a non-trivial task into simple parts. Indeed, after substituting $p_2^\Lambda(x)$ into the first term and scaling the variables, we get the integral
\begin{equation*}\label{29-9-36}
\int_{\mathrm{B}_{\Lambda/\sigma}}\mathrm{d}^4x\int_{\mathrm{B}_{\Lambda/\sigma}}\mathrm{d}^4y
\int_{\mathrm{B}_{\Lambda/\sigma}}\mathrm{d}^4z
\Big(R_0^1(x)\Big)^2
f_1^1(x+y)f_1^1(y+z)
\Big(R_0^1(z)\Big)^2,
\end{equation*}
which is finite after the regularization is removed, since $\mathrm{supp}\big(f_1^1(\cdot)\big)\subset\mathrm{B}_1$. To clearly see the convergence, we write out the limit integral with the absolute value of the integrand, which obviously is more that the absolute value of the last expression
\begin{equation*}\label{29-9-37}
\int_{\mathrm{B}_{1}}\mathrm{d}^4x\int_{\mathrm{B}_{2}}\mathrm{d}^4y
\int_{\mathbb{R}^4}\mathrm{d}^4z
\Big|f_1^1(x)f_1^1(x+y)\Big|
\Big(R_0^1(y+z)\Big)^2
\Big(R_0^1(z)\Big)^2.
\end{equation*}
Here, we have used the permutation of multipliers after the limit transition $\mathrm{B}_{\Lambda/\sigma}\to\mathbb{R}^4$ and the reduction of the integration domain taking into account the support of the function $f_1^1(\cdot)$. Thus, we get
\begin{equation*}\label{29-9-38}
\int_{\mathrm{B}_{1/\sigma}}\mathrm{d}^4x\Big(p_2^\Lambda(x)\Big)^2
\stackrel{\mathrm{s.p.}}{=}0.
\end{equation*}
Let us move on to the second term from (\ref{29-9-35}). In fact, it resembles the second term from (\ref{29-9-2}), so the basic estimation procedure will be the same. The key difference is that after scaling $x\to x/\Lambda$ in the domain $1\leqslant|x|<\Lambda/\sigma$, in addition to the function $|x|^{-2}$, also $\ln(|x|)|x|^{-2}$ and $\sigma^2/\Lambda^2$ appear, for which the estimates can be written in a similar\footnote{In the case of $\sigma^2/\Lambda^2$, the contribution tends to zero at $\Lambda\to+\infty$, since decreasing of $\Lambda^{-2}$ is stronger than the logarithmic singularity $L$, which follows from the integration of $\Big(R_0^1(x)\Big)^2$.} way. Therefore, it can be argued that
\begin{equation*}\label{29-9-39}
2\int_{\mathrm{B}_{1/\sigma}}\mathrm{d}^4x\,p_2^\Lambda(x)t_2^\Lambda(x)=
2\int_{\mathbb{R}^4}\mathrm{d}^4x
\int_{\mathrm{B}_{1}}\mathrm{d}^4y
\Big(R_0^1(x)\Big)^2\tilde{t}_2(x+y)f_1^1(y)+o\big(1/\Lambda\big)\stackrel{\mathrm{s.p.}}{=}0,
\end{equation*}
where
\begin{equation*}\label{29-9-40}
\tilde{t}_2(x)=
\begin{cases}
	~~~~~
	h_2(x)+\frac{1}{4(4\pi^2)^2},&\mbox{if $|x|\leqslant 1$};\\
	\frac{1+2\alpha_2(\mathbf{f})}{4(4\pi^2)^2|x|^2}+\frac{\ln(|x|)}{2(4\pi^2)^2|x|^2}
	,&\mbox{if $|x|>1$}.
\end{cases}
\end{equation*}
Next, moving on to the third term from (\ref{29-9-35}), we immediately note that the integral is equal to the second term from (\ref{29-9-2}), which has already been discussed. Therefore, we will straightway write out the chain of relations
\begin{equation*}\label{29-9-41}
-2\mathrm{A}(\sigma)\int_{\mathrm{B}_{1/\sigma}}\mathrm{d}^4x\,p_2^\Lambda(x)R_0^\Lambda(x)
=-2\mathrm{A}(\sigma)\alpha_5(\mathbf{f})+\mathcal{O}\big(L/\Lambda^2\big)
\stackrel{\mathrm{s.p.}}{=}-2\mathrm{A}(\sigma)\alpha_5(\mathbf{f})
\stackrel{\mathrm{s.p.}}{=}-\frac{L\alpha_5(\mathbf{f})}{4\pi^2}.
\end{equation*}
The last contribution from (\ref{29-9-35}) after splitting the integration domain and the corresponding scaling is rewritten as
\begin{equation*}\label{29-9-42}
\int_{\mathrm{B}_{1}}\mathrm{d}^4x
\bigg(h_2(y)-\frac{\sigma^2/\Lambda^2-1}{4(4\pi^2)^2}-\mathrm{A}(\sigma)f_2^1(y)
\bigg)^2+\frac{1}{2(4\pi^2)^3}
\int_{1/\Lambda}^{1/\sigma}\mathrm{d}s\,s^3
\bigg(-\frac{\sigma^2}{4}+\frac{1}{4s^2}+\frac{\ln(s\sigma)}{2s^2}
\bigg)^2.
\end{equation*}
It is clear that the terms with $\sigma^2$ can be deleted, since they will be included in the correction of the form $\mathcal{O}\big(L/\Lambda^2\big)$. The remaining contributions (after discarding the correction term $\mathcal{O}(1)$) have the form
\begin{multline*}
\frac{\mathrm{A}^2(\sigma)\alpha_2(\mathbf{f})}{2(4\pi^2)}-
2\mathrm{A}(\sigma)\alpha_4(\mathbf{f})-\frac{\mathrm{A}(\sigma)\big(1+4\alpha_1(\mathbf{f})\big)}{8(4\pi^2)^2}+\frac{1}{2(4\pi^2)^3}\bigg(\frac{L^3}{12}-\frac{L^2}{8}+\frac{L}{16}
\bigg)\stackrel{\mathrm{s.p.}}{=}\\\stackrel{\mathrm{s.p.}}{=}
\frac{L^3}{24(4\pi^2)^3}-\frac{L^2\big(1-2\alpha_2(\mathbf{f})\big)}{16(4\pi^2)^3}+\frac{L}{4\pi^2}
\bigg(\frac{-1-8\alpha_1^{\phantom{1}}(\mathbf{f})+8\alpha_2^2(\mathbf{f})}{32(4\pi^2)^2}-\alpha_4(\mathbf{f})\bigg).
\end{multline*}
Eventually, collecting all the terms together, we obtain the following result
\begin{equation*}\label{29-9-45}
\mathrm{I}_5(\sigma)
\stackrel{\mathrm{s.p.}}{=}
\frac{L^3}{24(4\pi^2)^3}-\frac{L^2\big(1-2\alpha_2(\mathbf{f})\big)}{16(4\pi^2)^3}+\frac{L}{4\pi^2}
\bigg(\frac{-1-8\alpha_1^{\phantom{1}}(\mathbf{f})+8\alpha_2^2(\mathbf{f})}{32(4\pi^2)^2}-\alpha_4(\mathbf{f})-\alpha_5(\mathbf{f})\bigg).
\end{equation*}
\\
\noindent\textbf{Calculating the integral $\mathrm{I}_6(\sigma)$.} When calculating this integral, note that the function $R_0^\Lambda(x)|x|^2$ can be rewritten as follows
\begin{equation*}\label{29-9-46}
R_0^\Lambda(x)|x|^2=\frac{1}{4\pi^2}+r^\Lambda_0(x),\,\,\,
r^\Lambda_0(x)=\bigg(R_0^\Lambda(x)|x|^2-\frac{1}{4\pi^2}\bigg)\chi\big(|x|\leqslant1/\Lambda\big).
\end{equation*}
Then we represent the integral in the form
\begin{equation*}\label{29-9-47}
\mathrm{I}_6(\sigma)=\frac{\mathrm{A}^2(\sigma)}{4\pi^2}+
\int_{\mathrm{B}_{1/\sigma}}\mathrm{d}^4x\int_{\mathrm{B}_{1/\sigma}}\mathrm{d}^4y\,
\Big(R_0^\Lambda(x)\Big)^2
r^\Lambda_0(x+y)
\Big(R_0^\Lambda(y)\Big)^2.
\end{equation*}
The last term has a finite limit when the regularization is removed, and for all $\Lambda$ its absolute value is less than the following number
\begin{equation*}\label{29-9-48}
\int_{\mathbb{R}^4}\mathrm{d}^4x\int_{\mathrm{B}_{1}}\mathrm{d}^4y\,
\Big(R_0^1(x)\Big)^2
\Big(R_0^1(x+y)\Big)^2
\Big|r^1_0(y)\Big|.
\end{equation*}
Thus, the final answer has the form
\begin{equation*}\label{29-9-49}
\mathrm{I}_6(\sigma)\stackrel{\mathrm{s.p.}}{=}\frac{\mathrm{A}^2(\sigma)}{4\pi^2}
\stackrel{\mathrm{s.p.}}{=}
\frac{L^2}{4(4\pi^2)^3}+
\frac{L\alpha_2(\mathbf{f})}{2(4\pi^2)^3}.
\end{equation*}
\\
\noindent\textbf{Calculating the integral $\mathrm{I}_7(\sigma)$.} First of all, we use formulas (\ref{29-9-16}) and (\ref{29-9-17}), and then we represent the integral in the form of two parts
\begin{equation}\label{29-9-50}
\mathrm{I}_7(\sigma)=
\int_{\mathrm{B}_{1/\sigma}}\mathrm{d}^4y\Big(R_0^\Lambda(y)\Big)^2p_2^\Lambda(y)+
\int_{\mathrm{B}_{1/\sigma}}\mathrm{d}^4y
\Big(R_0^\Lambda(y)\Big)^2\Big(t_2^\Lambda(y)-\mathrm{A}(\sigma)R_0^\Lambda(y)\Big).
\end{equation}
Let us start with the first term. After the standard scaling of the integration variables, we get
\begin{equation*}\label{29-9-51}
\Lambda^2
\int_{\mathrm{B}_{\Lambda/\sigma}}\mathrm{d}^4x
\int_{\mathrm{B}_{\Lambda/\sigma}}\mathrm{d}^4y
\Big(R_0^1(x)\Big)^2
f_1^1(x+y)
\Big(R_0^1(y)\Big)^2.
\end{equation*}
Note that the integral (without $\Lambda^2$) is convergent at $\Lambda\to+\infty$ and is equal to $\alpha_7(\mathbf{f})$. Moreover, the correction to the limit value decreases faster than $1/\Lambda^2$. Indeed, with the usage of the method used in calculating $\mathrm{I}_3(\sigma)$ can be verified that the difference is estimated as follows
\begin{multline*}
\bigg|\int_{\mathrm{B}_{\Lambda/\sigma}}\mathrm{d}^4x
\int_{\mathrm{B}_{\Lambda/\sigma}}\mathrm{d}^4y
\Big(R_0^1(x)\Big)^2
f_1^1(x+y)
\Big(R_0^1(y)\Big)^2-\alpha_7(\mathbf{f})\bigg|\leqslant\\\leqslant
\frac{1}{2(4\pi^2)^3}
\bigg(\int_{\mathrm{B}_{1}}\mathrm{d}^4x\Big|f_1^1(x)\Big|\bigg)
\int_{\Lambda/\sigma-1}^{+\infty}\frac{\mathrm{d}s}{s(s-1)^4},
\end{multline*}
which is a correction of the form $\mathcal{O}\big(1/\Lambda^4\big)$. Therefore, the relation holds
\begin{equation*}\label{29-9-53}
\int_{\mathrm{B}_{1/\sigma}}\mathrm{d}^4y\Big(R_0^\Lambda(y)\Big)^2p_2^\Lambda(y)=
\Lambda^2\alpha_7(\mathbf{f})+\mathcal{O}\big(1/\Lambda^2\big).
\end{equation*}
Let us move on to the second term from (\ref{29-9-50}). Substituting the explicit form (\ref{29-9-4}) into it and dividing the integration area into two parts, we get
\begin{equation*}\label{29-9-54}
\Lambda^2\int_{\mathrm{B}_{1}}\mathrm{d}^4y\Big(f_2^1(y)\Big)^2
\bigg(h_2(y)-\frac{\sigma^2/\Lambda^2-1}{4(4\pi^2)^2}-\mathrm{A}(\sigma)f_2^1(y)\bigg)
+
\frac{1}{2(4\pi^2)^3}
\int_{1/\Lambda}^{1/\sigma}\frac{\mathrm{d}s}{s}
\bigg(-\frac{\sigma^2}{4}+\frac{1}{4s^2}+\frac{\ln(s\sigma)}{2s^2}\bigg),
\end{equation*}
the singular part of which is equal to
\begin{equation*}\label{29-9-55}
\Lambda^2\alpha_8(\mathbf{f})+\frac{\Lambda^2\alpha_2(\mathbf{f})}{8(4\pi^2)^3}
-\Lambda^2\mathrm{A}(\sigma)\bigg(\alpha_3(\mathbf{f})-\frac{1}{4(4\pi^2)^2}\bigg)
-\frac{\sigma^2L}{8(4\pi^2)^3}
+\frac{\Lambda^2}{16(4\pi^2)^3}
+\frac{\Lambda^2-2L\Lambda^2}{16(4\pi^2)^3}.
\end{equation*}
Therefore, the final expression has the following form
\begin{equation*}\label{29-9-56}
\mathrm{I}_7(\sigma)\stackrel{\mathrm{s.p.}}{=}
-\frac{L\Lambda^2\alpha_3(\mathbf{f})}{2(4\pi^2)}+
\Lambda^2\bigg(\frac{1+2\alpha_2(\mathbf{f})}{8(4\pi^2)^3}
-\frac{\alpha_2(\mathbf{f})\alpha_3(\mathbf{f})}{2(4\pi^2)}+\alpha_7(\mathbf{f})+\alpha_8(\mathbf{f})\bigg)
-\frac{\sigma^2L}{8(4\pi^2)^3}.
\end{equation*}
\\
\noindent\textbf{Calculating the integral $\mathrm{I}_8(\sigma)$.} In this case, it is convenient to reduce the calculation of the integral to what has already been studied. Indeed, it can be noted that the following relation is true
\begin{equation*}\label{29-9-57}
\mathrm{I}_8(\sigma)
=\mathrm{I}_3(\sigma)-\mathrm{A}(\sigma)
\int_{\mathrm{B}_{1/\sigma}}\mathrm{d}^4y\Big(R_0^\Lambda(y)\Big)^2
=\mathrm{I}_3(\sigma)-\mathrm{A}^2(\sigma).
\end{equation*}
Therefore, using formula (\ref{29-9-31}), we get
\begin{equation*}\label{29-9-58}
\mathrm{I}_8(\sigma)\stackrel{\mathrm{s.p.}}{=}	
\frac{-L^2+L-2L\alpha_2(\mathbf{f})}{8(4\pi^2)^2}.
\end{equation*}
\\
\noindent\textbf{Calculating the integrals $\mathrm{I}_9(\sigma)$ and $\mathrm{I}_{10}(\sigma)$.} Let us start with the second integral, since the first one is a generalization. We describe the explicit form using formula (\ref{29-8-10}) and the representation from (\ref{29-3-19})
\begin{equation*}\label{29-9-59}
\mathrm{I}_{10}(\sigma)=
\int_{\mathrm{B}_{1/\sigma}\times\mathrm{B}_{1/\sigma}}\mathrm{d}^4x\mathrm{d}^4y
\Big(R_0^\Lambda(x)\Big)^2
R_{11}^\Lambda(x-y)
\Big(R_0^\Lambda(y)\Big)^2-\mathrm{A}(\sigma)\mathrm{I}_{3}(\sigma).
\end{equation*}
Next, we rewrite the representation for the function $R_{11}^\Lambda(x-y)$ as follows
\begin{align}\nonumber
\int_{\mathrm{B}_{1/\sigma}}\mathrm{d}^4z\,
R_{0}^\Lambda(x-y+z)R_{0}^\Lambda(z)=&
\int_{\mathbb{R}^4}\mathrm{d}^4z\,
R_{0}^\Lambda(x+z)R_{0}^\Lambda(y+z)\chi\big(|y+z|\leqslant1/\sigma\big)\\\nonumber
=&
\int_{\mathrm{B}_{1/\sigma}}\mathrm{d}^4z\,
R_{0}^\Lambda(x+z)R_{0}^\Lambda(y+z)+\\\nonumber
+&
\int_{\mathbb{R}^4\setminus\mathrm{B}_{1/\sigma}}\mathrm{d}^4z\,
R_{0}^\Lambda(x+z)R_{0}^\Lambda(y+z)\chi\big(|y+z|\leqslant1/\sigma\big)-\\\nonumber
-&
\int_{\mathrm{B}_{1/\sigma}}\mathrm{d}^4z\,
R_{0}^\Lambda(x+z)R_{0}^\Lambda(y+z)\chi\big(|y+z|>1/\sigma\big).
\end{align}
Note that the contribution with the first term can be reduced to the integral $\mathrm{I}_{5}(\sigma)$, then
\begin{multline}\label{29-9-64}
\mathrm{I}_{10}(\sigma)=\mathrm{I}_{5}(\sigma)+\mathrm{A}(\sigma)\mathrm{I}_{3}(\sigma)-
\mathrm{A}^3(\sigma)+\\+
\int_{\mathbb{R}^4\setminus\mathrm{B}_{1/\sigma}}\mathrm{d}^4z
\Big(t_2^\Lambda(z)+p_2^\Lambda(z)\Big)r_{<}^\Lambda(z)-
\int_{\mathrm{B}_{1/\sigma}}\mathrm{d}^4z
\Big(t_2^\Lambda(z)+p_2^\Lambda(z)\Big)r_{>}^\Lambda(z),
\end{multline}
where
\begin{equation*}\label{29-9-65}
r_{<}^\Lambda(z)=
\int_{\mathrm{B}_{1/\sigma}}\mathrm{d}^4y\,
R_{0}^\Lambda(z+y)\chi\big(|z+y|\leqslant1/\sigma\big)\Big(R_0^\Lambda(y)\Big)^2,
\end{equation*}
\begin{equation*}\label{29-9-66}
r_{>}^\Lambda(z)=
\int_{\mathrm{B}_{1/\sigma}}\mathrm{d}^4y\,
R_{0}^\Lambda(z+y)\chi\big(|z+y|>1/\sigma\big)\Big(R_0^\Lambda(y)\Big)^2.
\end{equation*}
Let us take a closer look at the last two integrals from (\ref{29-9-64}). Given the explicit form of the function $t_2^\Lambda(z)$ in the domain $\mathbb{R}^4\setminus\mathrm{B}_{1/\sigma}$, it can be replaced with $\mathrm{A}(\sigma)R_0^\Lambda(z)$. Thus, we have
\begin{align}\label{29-9-67}
\int_{\mathbb{R}^4\setminus\mathrm{B}_{1/\sigma}}\mathrm{d}^4z\,
t_2^\Lambda(z)r_{<}^\Lambda(z)&=\mathrm{A}(\sigma)
\int_{\mathbb{R}^4\setminus\mathrm{B}_{1/\sigma}}\mathrm{d}^4z\,
R_0^\Lambda(z)r_{<}^\Lambda(z)\\\nonumber&\stackrel{\mathrm{s.p.}}{=}
\mathrm{A}(\sigma)
\int_{\mathbb{R}^4\setminus\mathrm{B}_{1/\sigma}}\mathrm{d}^4z
\int_{\mathrm{B}_{1/\sigma}}\mathrm{d}^4y\,
R_0(z)
R_{0}(z+y)\chi\big(|z+y|\leqslant1/\sigma\big)\Big(R_0(y)\Big)^2,
\end{align}
where the fact that the limit integral exists was used\footnote{Here we need to pay attention to the fact that as $y$ tends to zero, the integration area is reduced, so the singularity in $R_0^2(y)$ is neutralized, and the integral converges.}, and the correction when multiplied by $L$ does not give a contribution to the singular part. Continuing, we note that $\mathrm{supp}\big(p_2^\Lambda(\cdot)\big)\subset\mathrm{B}_{1/\sigma+1/\Lambda}$, hence
\begin{equation*}\label{29-9-68}
\int_{\mathrm{B}_{1/\sigma+1/\Lambda}\setminus\mathrm{B}_{1/\sigma}}\mathrm{d}^4z\,
p_2^\Lambda(z)r_{<}^\Lambda(z)\stackrel{\mathrm{s.p.}}{=}0.
\end{equation*}
Performing similar reasoning, we can formulate the following statements about the last integral in (\ref{29-9-64}). They have the form
\begin{align}\label{29-9-69}
\int_{\mathrm{B}_{1/\sigma}}\mathrm{d}^4z\,
t_2^\Lambda(z)r_{<}^\Lambda(z)&\stackrel{\mathrm{s.p.}}{=}\mathrm{A}(\sigma)
\int_{\mathrm{B}_{1/\sigma}}\mathrm{d}^4z\,
R_0^\Lambda(z)r_{>}^\Lambda(z)\\\nonumber&\stackrel{\mathrm{s.p.}}{=}
\mathrm{A}(\sigma)
\int_{\mathrm{B}_{1/\sigma}}\mathrm{d}^4z
\int_{\mathrm{B}_{1/\sigma}}\mathrm{d}^4y\,
R_0(z)
R_{0}(z+y)\chi\big(|z+y|>1/\sigma\big)\Big(R_0(y)\Big)^2,
\end{align}
\begin{equation*}\label{29-9-70}
\int_{\mathrm{B}_{1/\sigma}}\mathrm{d}^4z\,
p_2^\Lambda(z)r_{>}^\Lambda(z)\stackrel{\mathrm{s.p.}}{=}0.
\end{equation*}
It remains to be noted that the singular parts in (\ref{29-9-67}) and (\ref{29-9-69}) are equal to each other. Indeed, the first integral passes into the second after the permutation $R_{0}(z+y)\chi\big(|z+y|>1/\sigma\big)$ and $R_{0}(z+y)\chi\big(|z+y|\leqslant1/\sigma\big)$ and a shift of the variable. Consequently, the singular parts in the last two terms of (\ref{29-9-64}) cancel each other, and the final answer is presented as
\begin{align*}
\mathrm{I}_{10}(\sigma)\stackrel{\mathrm{s.p.}}{=}&\mathrm{I}_{5}(\sigma)+\mathrm{A}(\sigma)\mathrm{I}_{3}(\sigma)-
\mathrm{A}^3(\sigma)\\\nonumber\stackrel{\mathrm{s.p.}}{=}&-
\frac{L^3}{48(4\pi^2)^3}-\frac{L^2\alpha_2(\mathbf{f})}{16(4\pi^2)^3}+\\\nonumber&+\frac{L}{4\pi^2}
\bigg(\frac{-1-4\alpha_1^{\phantom{1}}(\mathbf{f})+2\alpha_2^{\phantom{1}}(\mathbf{f})}{32(4\pi^2)^2}-
\frac{\alpha_4(\mathbf{f})+\alpha_5(\mathbf{f})}{2}\bigg).
\end{align*}
Note that in the assumptions of Section \ref{29:sec:reg-4}, the following equality is true
\begin{equation*}\label{29-9-721}
	\mathrm{I}_{9}(\sigma)=\mathrm{I}_{10}(\sigma).
\end{equation*}

\noindent\textbf{Calculating the integrals $\mathrm{I}_ {11}(\sigma)$ and $\mathrm{I}_{12}(\sigma)$.} Following the general logic, let us start with the last case. We use the representations from (\ref{29-9-16}) and (\ref{29-9-17}), then the integral can be rewritten as
\begin{equation*}\label{29-9-73}
\mathrm{I}_{12}(\sigma)=\int_{\mathrm{B}_{1/\sigma}}\mathrm{d}^4y
\,R_0^\Lambda(y)\Big(p_2^\Lambda(y)+t_2^\Lambda(y)-\mathrm{A}(\sigma)R_0^\Lambda(y)\Big)
\Big(p_1^\Lambda(y)+t_1^\Lambda(y)\Big).
\end{equation*}
Most of the terms from the decomposition are analyzed by the methods mentioned above. Therefore, omitting unnecessary details, we present some relations for singular terms
\begin{equation*}\label{29-9-74}
\int_{\mathrm{B}_{1/\sigma}}\mathrm{d}^4y
\,R_0^\Lambda(y)p_2^\Lambda(y)p_1^\Lambda(y)\stackrel{\mathrm{s.p.}}{=}0,
\end{equation*}
\begin{equation*}\label{29-9-75}
\int_{\mathrm{B}_{1/\sigma}}\mathrm{d}^4y
\,R_0^\Lambda(y)p_2^\Lambda(y)
t_1^\Lambda(y)
\stackrel{\mathrm{s.p.}}{=}\frac{L}{8\pi^2}
\int_{\mathrm{B}_{1/\sigma}}\mathrm{d}^4y
\,R_0^\Lambda(y)p_2^\Lambda(y)
\stackrel{\mathrm{s.p.}}{=}\frac{L\alpha_5(\mathbf{f})}{8\pi^2},
\end{equation*}
\begin{equation*}\label{29-9-76}
\int_{\mathrm{B}_{1/\sigma}}\mathrm{d}^4y
\,R_0^\Lambda(y)t_2^\Lambda(y)p_1^\Lambda(y)
\stackrel{\mathrm{s.p.}}{=}0,
\end{equation*}
\begin{equation*}\label{29-9-77}
\int_{\mathrm{B}_{1/\sigma}}\mathrm{d}^4y
\,R_0^\Lambda(y)\Big(-\mathrm{A}(\sigma)R_0^\Lambda(y)\Big)
p_1^\Lambda(y)
\stackrel{\mathrm{s.p.}}{=}-\frac{L}{8\pi^2}
\int_{\mathrm{B}_{1/\sigma}}\mathrm{d}^4y\Big(R_0^\Lambda(y)\Big)^2
p_1^\Lambda(y)\stackrel{\mathrm{s.p.}}{=}-\frac{L\alpha_5(\mathbf{f})}{8\pi^2}.
\end{equation*}
There is only one term left, which is convenient to calculate using the explicit form of the functions from (\ref{29-9-6}) and (\ref{29-9-4}) and dividing the integration domain into two parts. After changing the variable in the domain $|y|\leqslant 1/\Lambda$, we have
\begin{multline*}
\int_{\mathrm{B}_{1}}\mathrm{d}^4y\,f_2^1(y)
\bigg(h_2^{\phantom{1}}(y)-\frac{\sigma^2/\Lambda^2-1}{4(4\pi^2)^2}-\mathrm{A}(\sigma)f_2^1(y)\bigg)
\bigg(h_1^{\phantom{1}}(y)+\frac{L}{2(4\pi^2)}\bigg)\stackrel{\mathrm{s.p.}}{=}\\\stackrel{\mathrm{s.p.}}{=}
-\frac{\mathrm{A}(\sigma)L\alpha_2(\mathbf{f})}{4(4\pi^2)^2}+
\frac{L\big(1+4\alpha_1(\mathbf{f})\big)}{32(4\pi^2)^3},
\end{multline*}
where we have used the relation
\begin{equation}\label{29-9-80}
\int_{\mathrm{B}_{1}}\mathrm{d}^4y\,f_2^1(y)h_2^{\phantom{1}}(y)=
\int_{\mathrm{B}_{1}}\mathrm{d}^4y\,\Big(f_2^1(y)\Big)^2h_1^{\phantom{1}}(y).
\end{equation}
Next, in the domain $1/\Lambda<|y|\leqslant 1/\sigma$ we get
\begin{align*}
\frac{1}{4(4\pi^2)^3}
\frac{}{}\int_{1/\Lambda}^{1/\sigma}\mathrm{d}s\,s
\bigg(-\frac{\sigma^2}{2}+\frac{1}{2s^2}
+\frac{\ln(s\sigma)}{s^2}\bigg)
\bigg(\frac{1}{4}
-\frac{\ln(s\sigma)}{2}&+\frac{\alpha_1(\mathbf{f})}{\Lambda^2s^2}\bigg)\stackrel{\mathrm{s.p.}}{=}
\\\nonumber\stackrel{\mathrm{s.p.}}{=}&-
\frac{L^3}{24(4\pi^2)^3}+
\frac{L\big(1-4\alpha_1(\mathbf{f})\big)}{32(4\pi^2)^3}.
\end{align*}
Therefore, summing up all the terms, we get the final answer in the form
\begin{equation*}\label{29-9-81}
\mathrm{I}_{12}(\sigma)\stackrel{\mathrm{s.p.}}{=}-
\frac{L^3}{24(4\pi^2)^3}-
\frac{L^2\alpha_2(\mathbf{f})}{8(4\pi^2)^3}+
\frac{L\big(1-2\alpha_2^2(\mathbf{f})\big)}{16(4\pi^2)^3}.
\end{equation*}
Also, in the assumptions of Section \ref{29:sec:reg-4}, the following equality is true
\begin{equation*}\label{29-9-821}
	\mathrm{I}_{11}(\sigma)=\mathrm{I}_{12}(\sigma).
\end{equation*}

\noindent\textbf{Calculating the integrals $\mathrm{I}_{13}(\sigma)$ and $\mathrm{I}_{14}(\sigma)$.} Let us consider the explicit form of the last integral and use the representations from (\ref{29-9-16}) and (\ref{29-9-17}) for $i=1$. Then we get
\begin{equation}\label{29-9-83}
\mathrm{I}_{14}(\sigma)=
\int_{\mathrm{B}_{1/\sigma}}\mathrm{d}^4y\Big(R_0^\Lambda(y)\Big)^3\Big(p_1^\Lambda(y)+t_1^\Lambda(y)\Big).
\end{equation}
Let us start with the first term, in which we scale the variables,
\begin{equation*}\label{29-9-84}
\int_{\mathrm{B}_{1/\sigma}}\mathrm{d}^4y\Big(R_0^\Lambda(y)\Big)^3p_1^\Lambda(y)
=\Lambda^2
\int_{\mathrm{B}_{\Lambda/\sigma}\times\mathrm{B}_{\Lambda/\sigma}}\mathrm{d}^4y\mathrm{d}^4x
\Big(R_0^1(y)\Big)^3f_1^1(y+x)R_0^1(x).
\end{equation*}
Next, replace the integration domain $\mathrm{B}_{\Lambda/\sigma}\times\mathrm{B}_{\Lambda/\sigma}$ with $\mathbb{R}^{4\times2}$. Indeed, this can be done, since the inequality is true
\begin{equation*}\label{29-9-85}
\bigg|\int_{\mathrm{B}_{\Lambda/\sigma}\times\mathrm{B}_{\Lambda/\sigma}}\mathrm{d}^4y\mathrm{d}^4x
\Big(R_0^1(y)\Big)^3f_1^1(y+x)R_0^1(x)-\alpha_9(\mathbf{f})\bigg|\leqslant
\frac{1}{2(4\pi^2)^3}\bigg(\int_{\mathrm{B}_{1}}\mathrm{d}^4y\Big|f_1^1(y)\Big|\bigg)
\int_{\Lambda/\sigma-1}^{+\infty}\frac{\mathrm{d}s}{s^3(s-1)^2},
\end{equation*}
therefore, the correction can be included into $\mathcal{O}\big(1/\Lambda^4\big)$. So, we get
\begin{equation*}\label{29-9-86}
\int_{\mathrm{B}_{1/\sigma}}\mathrm{d}^4y\Big(R_0^\Lambda(y)\Big)^3p_1^\Lambda(y)=
\Lambda^2\alpha_9(\mathbf{f})+\mathcal{O}\big(1/\Lambda^2\big)
\stackrel{\mathrm{s.p.}}{=}\Lambda^2\alpha_9(\mathbf{f}).
\end{equation*}
The second term from (\ref{29-9-83}) is analyzed using direct substitution of (\ref{29-9-6}) and splitting the integration domain into two regions. In the range $|y|\leqslant 1/\Lambda$ after an appropriate scaling we have
\begin{equation*}\label{29-9-87}
\Lambda^2L\bigg(\frac{\alpha_3(\mathbf{f})}{2(4\pi^2)}-\frac{1}{8(4\pi^2)^3}\bigg)	
+\Lambda^2\alpha_{10}(\mathbf{f}).
\end{equation*}
Further, in the domain $1/\Lambda<|y|\leqslant 1/\sigma$ we get
\begin{equation*}\label{29-9-88}
\frac{1}{2(4\pi^2)^3}\int_{1/\Lambda}^{1/\sigma}\frac{\mathrm{d}s}{s^3}
\bigg(\frac{1}{4}-\frac{\ln(s\sigma)}{2}+\frac{\alpha_1(\mathbf{f})}{\Lambda^2s^2}\bigg)
\stackrel{\mathrm{s.p.}}{=}\frac{\Lambda^2}{2(4\pi^2)^3}
\bigg(\frac{L}{4}+\frac{\alpha_1(\mathbf{f})}{4}
\bigg).
\end{equation*}
Finally, summing up all the contributions, we write out the asymptotics in the form
\begin{equation*}\label{29-9-89}
\mathrm{I}_{14}(\sigma)\stackrel{\mathrm{s.p.}}{=}
\frac{\Lambda^2L\alpha_3(\mathbf{f})}{2(4\pi^2)}+
\Lambda^2\bigg(\frac{\alpha_1(\mathbf{f})}{8(4\pi^2)^3}+\alpha_{9}(\mathbf{f})+\alpha_{10}(\mathbf{f})\bigg).
\end{equation*}
Also, in the assumptions of Section \ref{29:sec:reg-4}, the following equality is true
\begin{equation*}\label{29-9-901}
	\mathrm{I}_{13}(\sigma)=
	\mathrm{I}_{14}(\sigma).
\end{equation*}

\noindent\textbf{Calculating the integrals $\mathrm{I}_{15}(\sigma)$, $\mathrm{I}_{16}(\sigma)$, and $\mathrm{I}_{17}(\sigma)$.} Following the general logic, we substitute the functions from (\ref{29-9-16}) and (\ref{29-9-17}) into the last integral, then the contribution is rewritten as
\begin{equation}\label{29-9-93}
\mathrm{I}_{17}(\sigma)=
\int_{\mathrm{B}_{1/\sigma}}\mathrm{d}^4y\Big(R_0^\Lambda(y)\Big)^2
\Big(p_1^\Lambda(y)p_1^\Lambda(y)+2p_1^\Lambda(y)t_1^\Lambda(y)+t_1^\Lambda(y)t_1^\Lambda(y)\Big).
\end{equation}
Repeating the basic steps of processing similar integrals that were studied above, we immediately write out the results for the first two contributions
\begin{equation*}\label{29-9-94}
\int_{\mathrm{B}_{1/\sigma}}\mathrm{d}^4y\Big(R_0^\Lambda(y)\Big)^2\Big(p_1^\Lambda(y)\Big)^2
\stackrel{\mathrm{s.p.}}{=}0,
\end{equation*}
\begin{equation*}\label{29-9-95}
2\int_{\mathrm{B}_{1/\sigma}}\mathrm{d}^4y\Big(R_0^\Lambda(y)\Big)^2p_1^\Lambda(y)t_1^\Lambda(y)
\stackrel{\mathrm{s.p.}}{=}\frac{L}{4\pi^2}
\int_{\mathrm{B}_{1/\sigma}}\mathrm{d}^4y\Big(R_0^\Lambda(y)\Big)^2p_1^\Lambda(y)
\stackrel{\mathrm{s.p.}}{=}\frac{L\alpha_5(\mathbf{f})}{4\pi^2}.
\end{equation*}
Note that in the latter case, we have used the fact that the combination $t_1^\Lambda(y)-L/(8\pi^2)$ in the region $|y|\leqslant 1/\sigma$ is expressed in terms of functions depending on $|y|\Lambda$.
Continuing, the third term from (\ref{29-9-93}) is analyzed by dividing the integration domain. So, for $|y|\leqslant1/\Lambda$ we have
\begin{equation*}\label{29-9-96}
\frac{L^2}{4(4\pi^2)^2}\int_{\mathrm{B}_{1}}\mathrm{d}^4y\Big(f_2^1(y)\Big)^2+
\frac{L}{4\pi^2}\int_{\mathrm{B}_{1}}\mathrm{d}^4y\Big(f_2^1(y)\Big)^2h_1^{\phantom{1}}(y)=
\frac{L^2\alpha_2(\mathbf{f})}{8(4\pi^2)^3}+\frac{L\alpha_4(\mathbf{f})}{4\pi^2},
\end{equation*}
where we have used the relation (\ref{29-9-80}).
Next, in the region $1/\Lambda<|y|\leqslant1/\sigma$ we get
\begin{equation*}\label{29-9-98}
\frac{1}{2(4\pi^2)^3}\int_{1/\Lambda}^{1/\sigma}\frac{\mathrm{d}s}{s}
\bigg(
\frac{1}{4}-\frac{\ln(s\sigma)}{2}+\frac{\alpha_1(\mathbf{f})}{\Lambda^2s^2}
\bigg)^2\stackrel{\mathrm{s.p.}}{=}
\frac{L^3}{24(4\pi^2)^3}+
\frac{L^2}{16(4\pi^2)^3}+
\frac{L\big(1+8\alpha_1(\mathbf{f})\big)}{32(4\pi^2)^3}.
\end{equation*}
Finally, after summation, the answer is presented in the form
\begin{equation*}\label{29-9-99}
\mathrm{I}_{17}(\sigma)\stackrel{\mathrm{s.p.}}{=}
\frac{L^3}{24(4\pi^2)^3}+
\frac{L^2\big(1+2\alpha_2(\mathbf{f})\big)}{16(4\pi^2)^3}+
\frac{L}{4\pi^2}
\bigg(
\frac{1+8\alpha_1(\mathbf{f})}{32(4\pi^2)^2}+\alpha_4(\mathbf{f})+\alpha_5(\mathbf{f})
\bigg).
\end{equation*}
Also, in the assumptions of Section \ref{29:sec:reg-4}, the following equality is true
\begin{equation*}\label{29-9-1011}
	\mathrm{I}_{15}(\sigma)=\mathrm{I}_{16}(\sigma)=\mathrm{I}_{17}(\sigma).
\end{equation*}

\noindent\textbf{Calculating the integrals $\mathrm{I}_{18}(\sigma)$, $\mathrm{I}_{19}(\sigma)$, and $\mathrm{I}_{20}(\sigma)$.} Let us start with the last integral, see (\ref{29-8-18}). We use the definition (\ref{29-3-21}) and the representations from (\ref{29-9-16}) and (\ref{29-9-17}), then the contribution can be rewritten as
\begin{equation}\label{29-9-102}
\mathrm{I}_{20}(\sigma)=2
\int_{\mathrm{B}_{1/\sigma}}\mathrm{d}^4y\Big(p_3^\Lambda(y)p_1^\Lambda(y)+
p_3^\Lambda(y)t_1^\Lambda(y)+t_3^\Lambda(y)p_1^\Lambda(y)
+t_3^\Lambda(y)t_1^\Lambda(y)\Big)-2\tilde{c}_2\mathrm{I}_{4}(\sigma).
\end{equation}
The first two terms are analyzed taking into account the general idea outlined above, as follows
\begin{equation*}\label{29-9-103}
2\int_{\mathrm{B}_{1/\sigma}}\mathrm{d}^4y\,p_3^\Lambda(y)p_1^\Lambda(y)\stackrel{\mathrm{s.p.}}{=}0,
\end{equation*}
\begin{equation*}\label{29-9-104}
2\int_{\mathrm{B}_{1/\sigma}}\mathrm{d}^4y\,p_3^\Lambda(y)t_1^\Lambda(y)\stackrel{\mathrm{s.p.}}{=}
\frac{L}{4\pi^2}
\int_{\mathrm{B}_{\Lambda/\sigma}\times\mathrm{B}_{\Lambda/\sigma}}\mathrm{d}^4x\mathrm{d}^4y
\Big(R_0^1(x)\Big)^3f_1^1(x+y)
\stackrel{\mathrm{s.p.}}{=}
\frac{L\alpha_1(\mathbf{f})\alpha_3(\mathbf{f})}{4\pi^2},
\end{equation*}
where, in the last transition, we have used the fact  that the difference is estimated from above by the inequality
\begin{equation*}\label{29-9-105}
\bigg|\int_{\mathrm{B}_{\Lambda/\sigma}\times\mathrm{B}_{\Lambda/\sigma}}\mathrm{d}^4x\mathrm{d}^4y
\Big(R_0^1(x)\Big)^3f_1^1(x+y)-\alpha_1(\mathbf{f})\alpha_3(\mathbf{f})\bigg|\leqslant
\frac{1}{2(4\pi^2)^2}
\bigg(\int_{\mathrm{B}_{1}}\mathrm{d}^4y\Big|f_1^1(y)\Big|\bigg)
\int_{\Lambda/\sigma-1}^{+\infty}\frac{\mathrm{d}s}{s^3}
\end{equation*}
and, taking into account the total multiplier, it can be included into the correction $\mathcal{O}\big(L/\Lambda^2\big)$. Let us now consider the third term from (\ref{29-9-102})
\begin{equation*}\label{29-9-106}
2\int_{\mathrm{B}_{1/\sigma}}\mathrm{d}^4y\,t_3^\Lambda(y)p_1^\Lambda(y).
\end{equation*}
Here, it should immediately be noted that the region $|y|\leqslant 1/\Lambda$ does not give a singular contribution, since in this case two functions with compact supports are obtained. Next, in the region $1/\Lambda<|y|\leqslant 1/\sigma$, the function $t_3^\Lambda(y)$ consists of three parts
\begin{equation*}\label{29-9-107}
-\frac{\sigma^4}{8(4\pi^2)^3}-\frac{1}{8(4\pi^2)^3|y|^4}+\frac{\Lambda^2\alpha_3(\mathbf{f})}{4\pi^2|y|^2}.
\end{equation*}
It is clear that the first two terms do not give a singular contribution. Indeed, after scaling the variables, they look like
\begin{equation*}\label{29-9-108}
-\frac{\sigma^4}{4(4\pi^2)^3\Lambda^4}\int_{(\mathrm{B}_{\Lambda/\sigma}\setminus\mathrm{B}_{1})\times\mathrm{B}_{\Lambda/\sigma}}\mathrm{d}^4y\mathrm{d}^4x\,f_1^1(y+x)R_0^1(x)=\mathcal{O}\big(1/\Lambda^2\big),
\end{equation*}
\begin{equation*}\label{29-9-109}
-\frac{1}{4(4\pi^2)^3}\int_{(\mathrm{B}_{\Lambda/\sigma}\setminus\mathrm{B}_{1})\times\mathrm{B}_{\Lambda/\sigma}}\mathrm{d}^4y\mathrm{d}^4x\,\frac{1}{|y|^4}f_1^1(y+x)R_0^1(x)=\mathcal{O}(1),
\end{equation*}
where an explicit estimate has been omitted, as it is similar to those mentioned above.
The third term has a different form, and the following chain of relations is valid for it
\begin{align*}
\frac{\alpha_3(\mathbf{f})}{2\pi^2}\int_{(\mathrm{B}_{\Lambda/\sigma}\setminus\mathrm{B}_{1})\times\mathrm{B}_{\Lambda/\sigma}}\mathrm{d}^4y\mathrm{d}^4x&\,\frac{1}{|y|^2}f_1^1(y+x)R_0^1(x)=\\=&
\frac{\alpha_3(\mathbf{f})}{2\pi^2}\int_{(\mathrm{B}_{\Lambda/\sigma}\setminus\mathrm{B}_{1})\times\mathrm{B}_{1}}\mathrm{d}^4y\mathrm{d}^4x\,\frac{R_0^1(y+x)}{|y|^2}\chi\big(|y+x|\leqslant\Lambda/\sigma\big)f_1^1(x)\stackrel{\mathrm{s.p.}}{=}\\\stackrel{\mathrm{s.p.}}{=}&
\frac{\alpha_3(\mathbf{f})}{2\pi^2}\int_{(\mathrm{B}_{\Lambda/\sigma}\setminus\mathrm{B}_{1})\times\mathrm{B}_{1}}\mathrm{d}^4y\mathrm{d}^4x\,\frac{R_0^1(y)}{|y|^2}f_1^1(x)
\stackrel{\mathrm{s.p.}}{=}\frac{L\alpha_1(\mathbf{f})\alpha_3(\mathbf{f})}{4\pi^2},
\end{align*}
where, in the second step, a correction of the form $\mathcal{O}(1)$ was discarded. Indeed, it can be verified that integrals with
\begin{equation*}\label{29-9-113}
\Big(R_0^1(y+x)\chi\big(|y+x|\leqslant\Lambda/\sigma\big)-R_0^1(y+x)\Big)\,\,\,\mbox{and}\,\,\,
\Big(R_0^1(y+x)-R_0^1(y)\Big)
\end{equation*}
have finite limits when removing the regularization $\Lambda\to+\infty$.
Therefore, we get
\begin{equation*}\label{29-9-111}
2\int_{\mathrm{B}_{1/\sigma}}\mathrm{d}^4y\,t_3^\Lambda(y)p_1^\Lambda(y)
\stackrel{\mathrm{s.p.}}{=}\frac{L\alpha_1(\mathbf{f})\alpha_3(\mathbf{f})}{4\pi^2}.
\end{equation*}
Moving on to the fourth term from (\ref{29-9-102}), let us use the explicit form of the functions from (\ref{29-9-6}) and (\ref{29-9-7}) and divide the integration domain into two parts. For $|y|\leqslant1/\Lambda$ we have
\begin{equation*}\label{29-9-112}
2\int_{\mathrm{B}_{1}}\mathrm{d}^4y
\bigg(h_3(y)-\frac{\sigma^4/\Lambda^4-1}{8(4\pi^2)^3}\bigg)
\bigg(h_1(y)+\frac{L}{2(4\pi^2)}\bigg)\stackrel{\mathrm{s.p.}}{=}\frac{L}{64(4\pi^2)^3}+
\frac{L\alpha_{11}(\mathbf{f})}{4\pi^2}.
\end{equation*}
In the domain $1/\Lambda<|y|\leqslant1/\sigma$ we get
\begin{multline*}
\frac{1}{4\pi^2}\int_{1/\Lambda}^{1/\sigma}\mathrm{d}s\,s^3
\bigg(-\frac{\sigma^4}{8(4\pi^2)^2}-\frac{1}{8(4\pi^2)^2s^4}+\frac{\Lambda^2\alpha_3(\mathbf{f})}{s^2}\bigg)
\bigg(\frac{1}{4}
-\frac{\ln(s\sigma)}{2}+\frac{\alpha_1(\mathbf{f})}{\Lambda^2s^2}\bigg)\stackrel{\mathrm{s.p.}}{=}\\\stackrel{\mathrm{s.p.}}{=}-\frac{L^2}{32(4\pi^2)^3}-\frac{L}{32(4\pi^2)^3}
-\frac{L\alpha_3(\mathbf{f})}{4(4\pi^2)}+\frac{L\alpha_1(\mathbf{f})\alpha_3(\mathbf{f})}{4\pi^2}+\frac{\Lambda^2\alpha_3(\mathbf{f})}{4(4\pi^2)\sigma^2}.
\end{multline*}
In the fifth term, see (\ref{29-9-102}), we can use the asymptotics (\ref{29-9-33}) and the following chain of relations to find $\tilde{c}_2$ from (\ref{29-3-27})
\begin{equation}\label{29-9-115}
\tilde{c}_2=\int_{\mathrm{B}_{1/\sigma}}\mathrm{d}^4y\,R_0^{\phantom{1}}(y)t_1^{+\infty}(y)=
\frac{1}{8(4\pi^2)}\int_{0}^{1/\sigma}\mathrm{d}s\,s
\Big(1-2\ln(s\sigma)\Big)=\frac{1}{8(4\pi^2)\sigma^2}.
\end{equation}
Then it can be rewritten as
\begin{equation*}\label{29-9-116}
-2\tilde{c}_2\mathrm{I}_{4}(\sigma)\stackrel{\mathrm{s.p.}}{=}
-\frac{\Lambda^2\alpha_3(\mathbf{f})}{4(4\pi^2)\sigma^2}.
\end{equation*}
Summing up all the studied parts, we come to the answer
\begin{equation*}\label{29-9-117}
\mathrm{I}_{20}(\sigma)\stackrel{\mathrm{s.p.}}{=}
-\frac{L^2}{32(4\pi^2)^3}
+\frac{L}{4\pi^2}\bigg(
-\frac{1}{64(4\pi^2)^2}
-\frac{\alpha_3(\mathbf{f})}{4}+3\alpha_1(\mathbf{f})\alpha_3(\mathbf{f})
+\alpha_{11}(\mathbf{f})\bigg).
\end{equation*}
In addition, in the assumptions of Section \ref{29:sec:reg-4}, the following equality is true
\begin{equation*}\label{29-9-1191}
	\mathrm{I}_{18}(\sigma)=\mathrm{I}_{19}(\sigma)=\mathrm{I}_{20}(\sigma).
\end{equation*}

\noindent\textbf{Calculating $R_{22}^\Lambda(0)$.} In this case, it is necessary to study the asymptotics of the integral, including terms proportional to $1/\Lambda^2$. Let us use the functions from (\ref{29-9-16}) and (\ref{29-9-17}), then
\begin{equation}\label{29-9-118}
R_{22}^\Lambda(0)=2
\int_{\mathrm{B}_{1/\sigma}}\mathrm{d}^4x\Big(R_0^\Lambda(x)t_1^\Lambda(x)+
R_0^\Lambda(x)p_1^\Lambda(x)
\Big)-\frac{1}{16\pi^2\sigma^2}.
\end{equation}
The first term is considered with the usage of the explicit substitution of the expression from (\ref{29-9-6}) and splitting the integration domain into two parts
\begin{equation*}\label{29-9-119}
\frac{L\alpha_1(\mathbf{f})}{2\pi^2\Lambda^2}+
\frac{2\alpha_{12}(\mathbf{f})}{\Lambda^2}+\frac{\sigma^{-2}-\Lambda^{-2}}{16\pi^2},
\end{equation*}
where $\alpha_{12}(\mathbf{f})$ is defined by formula (\ref{29-9-120}). The second term in (\ref{29-9-118}), taking into account formula (\ref{29-6-3}), is explicitly written as follows
\begin{multline*}
2\int_{\mathrm{B}_{1/\sigma}}\mathrm{d}^4x\,R_0^\Lambda(x)p_1^\Lambda(x)=-2
\int_{\mathrm{B}_{1/\Lambda}\times\mathrm{B}_{1/\sigma}}\mathrm{d}^4x\mathrm{d}^4y\,f_1^\Lambda(x)\Big(R_0^\Lambda(x+y)-R_0^{\Lambda,\sigma}(x+y)\Big)R_0^\Lambda(y)+\\+2
\int_{\mathrm{B}_{1/\Lambda}}\mathrm{d}^4x\,f_1^\Lambda(x)\Big(t_1^\Lambda(x)+p_1^\Lambda(x)\Big).
\end{multline*}
In the last sum, the absolute value of the first term  is estimated from above by the expression
\begin{equation*}\label{29-9-122}
\frac{\sigma^2}{2\pi^2\Lambda^2}
\bigg(\int_{\mathrm{B}_{1}}\mathrm{d}^4x\Big|f_1^1(x)\Big|\bigg)\int_{1/\sigma-1/\Lambda}^{1/\sigma}\mathrm{d}s\,s,
\end{equation*}
therefore, it is included into the correction $\mathcal{O}\big(1/\Lambda^3\big)$ and can be omitted. Then, the second term, taking into account the definition from (\ref{29-9-123}) is equal to
\begin{equation*}\label{29-9-124}
\frac{L\alpha_1(\mathbf{f})}{4\pi^2\Lambda^2}+
\frac{2\alpha_{13}(\mathbf{f})}{\Lambda^2}.
\end{equation*}
Finally, summing up all the terms, we obtain the relation
\begin{equation}\label{29-9-126}
R_{22}^\Lambda(0)=
\frac{3L\alpha_1(\mathbf{f})}{4\pi^2\Lambda^2}+
\frac{2\alpha_{12}(\mathbf{f})+2\alpha_{13}(\mathbf{f})}{\Lambda^2}-
\frac{1}{16\pi^2\Lambda^2}+\mathcal{O}\big(1/\Lambda^3\big).
\end{equation}

\subsection{One auxiliary proof}
\label{29:sec:app:doc}
In this part of the appendix, we consider the proof of one special property (\ref{29-7-19-2}) for the non-local component $PS_1^\Lambda(x,y)$ of the Green's function $G^\Lambda(x,y)$. First, using the decompositions from (\ref{29-3-6}) and (\ref{29-3-18}), we write out its explicit formula as the sum of several terms
\begin{equation*}\label{29-12-1}
PS_1^\Lambda(x,y)=\sum_{k=0}^3g_k(x,y),
\end{equation*}
where
\begin{equation*}\label{29-12-5}
g_0(x,y)=G_0^\Lambda(x-y)-R_0^\Lambda(x-y)+R_{10}^\Lambda(x-y)m^2-\frac{1}{2}
R_{20}^\Lambda(x-y)m^4,
\end{equation*}
\begin{multline*}
g_1(x,y)=-\frac{\lambda}{2}\int_{\mathbb{R}^4}\mathrm{d}^4x_1\,
G_0^\Lambda(x-x_1)B^2(x_1)G_0^\Lambda(x_1-y)+\\+\frac{\lambda}{4}R_{11}^\Lambda(x-y)\big(B^2(x)+B^2(y)\big)-
\frac{m^2\lambda}{4}
R_{21}^\Lambda(x-y)\big(B^2(x)+B^2(y)\big),
\end{multline*}
\begin{multline*}
g_2(x,y)=
\frac{\lambda^2}{4}\int_{\mathbb{R}^{4\times2}}\mathrm{d}^4x_1\mathrm{d}^4x_2\,
G_0^\Lambda(x-x_1)B^2(x_1)G_0^\Lambda(x_1-x_2)B^2(x_2)G_0^\Lambda(x_2-y)
-\\-\frac{\lambda^2}{16}R_{22}^\Lambda(x-y)\big(B^4(x)+B^4(y)\big),
\end{multline*}
as well as the part that depends on higher powers of the background field
\begin{multline*}
g_3(x,y)=\sum_{k=3}^{+\infty}\frac{(-\lambda)^k}{2^k}
\int_{\mathbb{R}^{4\times k}}\mathrm{d}^4x_1\ldots\mathrm{d}^4x_k\,
G_0^\Lambda(x-x_1)B^2(x_1)G_0^\Lambda(x_1-x_2)B^2(x_2)\times\ldots\\\ldots\times G_0^\Lambda(x_{k-1}-x_k)B^2(x_k)G_0^\Lambda(x_k-y).
\end{multline*}
Such splitting is convenient because it shows not only an obvious dependence on a power of the background field, but also an increasing smoothness. Separately, we note that in the further the conditions from (\ref{29-10-6}) are used for the introduced regularization. Note that $g_0(x,y)$ does not depend on the background field and is subtracted in formula (\ref{29-7-19-2}) by the density $\kappa_s(\cdot)$, so it is not required to study it. Next, adding and subtracting $m^2$ and describing the difference of operators in the form
\begin{equation*}\label{29-12-6}
A_0^\Lambda(x)-A_0^{\phantom{1}}(x)=
A^\Lambda(x)-A(x),
\end{equation*}
we can make sure that the equality holds
\begin{equation*}\label{29-12-7}
\big(A^\Lambda(x)-A(x)\big)g_3(x,y)\xrightarrow{\Lambda\to+\infty}0
\end{equation*}
for all argument values, including the diagonal $y=x$.
Such transition is obtained due to the sufficient smoothness of the function and properties (\ref{29-3-25}). Let us move on to the remaining functions $g_2(x,y)$ and $g_1(x,y)$, first applying the operator $A^\Lambda(x)$, and then the operator $A(x)$. In the first case, we get
\begin{equation}\label{29-12-9}
	\int_{\mathbb{R}^4}\mathrm{d}^4x\Big(A^\Lambda(x)g_2(x,y)\Big)\Big|_{y=x}=\frac{\lambda^2}{4}\mathrm{I}_2^2\big((G_0^\Lambda)^2\big)-\frac{\lambda^2}{4}R_{11}^\Lambda(0)S_4[B]+\mathcal{O}\big(L/\Lambda^2\big),
\end{equation}
\begin{align}\nonumber
\int_{\mathbb{R}^4}\mathrm{d}^4x\Big(A^\Lambda(x)g_1(x,y)\Big)\Big|_{y=x}=&~S_m[B]\bigg(
-\frac{\lambda}{2}G_0^\Lambda(0)+\frac{\lambda}{2}R_0^\Lambda(0)-\frac{m^2\lambda}{2}R_{11}^\Lambda(0)\bigg)
+\mathcal{O}\big(L/\Lambda^2\big)\\\label{29-12-81}=&-\frac{\lambda}{2}
S_m[B]PS_0^\Lambda(0)+\mathcal{O}\big(L/\Lambda^2\big),
\end{align}
where the terms with $R_{21}^\Lambda(0)$ and $R_{22}^\Lambda(0)$ are included in the correction. 

Now let us apply the limit operator $A(x)$ to the same functions. First, we note that the auxiliary integrals
\begin{equation*}\label{29-12-12}
\rho_{16}(z,y)=\int_{\mathbb{R}^{4}}\mathrm{d}^4x_2
\Big(B^2(z)G_0^\Lambda(z-x_2)B^2(x_2)G_0^\Lambda(x_2-y)-
B^2(z)R_0^{\Lambda,\sigma}(z-x_2)B^2(x_2)R_0^{\Lambda,\sigma}(x_2-y)\Big),
\end{equation*}
\begin{equation*}\label{29-12-14}
\rho_{17}(z,y)=\int_{\mathbb{R}^{4}}\mathrm{d}^4x_2
\Big(
B^2(z)R_0^{\Lambda,\sigma}(z-x_2)B^2(x_2)R_0^{\Lambda,\sigma}(x_2-y)-
R_0^{\Lambda}(z-x_2)R_0^{\Lambda,\sigma}(x_2-y)B^4(y)\Big),
\end{equation*}
\begin{multline*}
\rho_{18}(z,y)=\int_{\mathbb{R}^{4\times2}}\mathrm{d}^4x_1\mathrm{d}^4x_2
\Big(G_0^\Lambda(z-x_1)R_0^{\Lambda}(x_1-x_2)R_0^{\Lambda,\sigma}(x_2-y)B^4(y)\Big)-\\-\frac{1}{4}R_{22}^\Lambda(x-y)\big(B^4(x)+B^4(y)\big),
\end{multline*}
are convergent before and after the removal of regularization and have finite limits on the diagonal $z=y$.
Here we have used the definition from (\ref{29-6-3}). The following relations are valid for such functions
\begin{equation*}\label{29-12-13}
\bigg(A(x)\int_{\mathbb{R}^{4}}\mathrm{d}^4z\,G_0^\Lambda(x-z)\rho_{j}(z,y)\bigg)\bigg|_{y=x}=\rho_{j}(x,x)+o(1),\,\,\,j=16,17,
\end{equation*}
\begin{equation*}\label{29-12-131}
A(x)\rho_{18}(x,y)\Big|_{y=x}=
o(1).
\end{equation*}
Thus, summing up all the terms, we are convinced that the following equality holds
\begin{align*}
\int_{\mathbb{R}^4}\mathrm{d}^4x\Big(A(x)g_2(x,y)\Big)\Big|_{y=x}=&~\frac{\lambda^2}{4}
\int_{\mathbb{R}^4}\mathrm{d}^4x\Big(\rho_{16}(x,x)+\rho_{17}(x,x)\Big)+o(1)\\
=&~
\frac{\lambda^2}{4}\mathrm{I}_2^2\big((G_0^\Lambda)^2\big)-\frac{\lambda^2}{4}R_{11}^\Lambda(0)S_4[B]+o(1),
\end{align*}
which in the main part coincides with (\ref{29-12-9}). Moving on to the function $g_1(x,y)$, it is convenient to use the definitions from (\ref{29-3-19}) and (\ref{29-3-21}) and rewrite it as
\begin{multline*}
\int_{\mathbb{R}^4}\mathrm{d}^4x\Big(A(x)g_1(x,y)\Big)\Big|_{y=x}=-\frac{\lambda S_m[B]}{2}
\int_{\mathbb{R}^4}\mathrm{d}^4x
\Big(G_0^\Lambda(x)A(x)G_0^\Lambda(x)-\\-
R_0^{\Lambda,\sigma}(x)A(x)R_0^{\Lambda}(x)+2m^2R_{11}^{\Lambda}(x)A_0^{\phantom{1}}(x)R_0^{\Lambda}(x)
\Big)+\mathcal{O}\big(L/\Lambda^2\big),
\end{multline*}
where the correction includes the term $R_{22}^\Lambda(0)$ without derivatives.
Note that in the last term, integration must occur over the ball $\mathrm{B}_{1/\sigma}$. Replacing $\mathrm{B}_{1/\sigma}\to\mathbb{R}^4$ is possible due to the reason that 
\begin{equation*}\label{29-12-19}
	\mathrm{supp}\big(A_0(\cdot)R_0^\Lambda(\cdot)\big)\subset\mathrm{B}_{1/\Lambda}
\end{equation*}
by construction, see (\ref{29-3-17}).
Next, applying the convergence of $A(x)G_0^\Lambda(x-y)\to\delta(x-y)$ in the sense of generalized functions, we rewrite the individual terms as follows
\begin{equation}\label{29-12-11}
\int_{\mathbb{R}^4}\mathrm{d}^4x\,G_0^\Lambda(x)A(x)G_0^\Lambda(x)=
PS_0^\Lambda(0)+\int_{\mathbb{R}^4}\mathrm{d}^4x\Big(R_0^\Lambda(x)-m^2R_{11}^\Lambda(x)\Big)A(x)G_0^\Lambda(x)+o(1),
\end{equation}
\begin{equation}\label{29-12-17}
\int_{\mathbb{R}^4}\mathrm{d}^4x
\Big(R_0^\Lambda(x)-m^2R_{11}^\Lambda(x)\Big)\Big(A(x)-A^\Lambda(x)\Big)
\Big(G_0^\Lambda(x)-R_0^\Lambda(x)+m^2R_{11}^\Lambda(x)\Big)=o(1),
\end{equation}
\begin{equation}\label{29-12-18}
\int_{\mathbb{R}^4}\mathrm{d}^4x\,R_{11}^\Lambda(x)
\Big(A(x)-A^\Lambda(x)\Big)R_{11}^\Lambda(x)=o(1),
\end{equation}
\begin{equation}\label{29-12-21}
\int_{\mathbb{R}^4}\mathrm{d}^4x\Big(
R_{0}^{\Lambda}(x)A_0^{\phantom{1}}(x)R_{11}^{\Lambda}(x)-
R_{11}^{\Lambda}(x)A_0^{\phantom{1}}(x)R_0^{\Lambda}(x)
\Big)=0.
\end{equation}
Therefore, using  formula (\ref{29-12-17}) in the integral (\ref{29-12-11}), it is possible to make the following chain of replacements
\begin{multline*}
A(x)G_0^\Lambda(x)=\Big(A(x)-A^\Lambda(x)\Big)G_0^\Lambda(x)+\delta(x)\to\\\to
\Big(A(x)-A^\Lambda(x)\Big)\Big(R_0^\Lambda(x)-m^2R_{11}^\Lambda(x)\Big)+\delta(x)=\\=
A_0(x)\Big(R_0^\Lambda(x)-m^2R_{11}^\Lambda(x)\Big)+m^2R_0^\Lambda(x)\chi\big(|x|\leqslant1/\sigma\big),
\end{multline*}
then, taking into account formulas (\ref{29-12-18}) and (\ref{29-12-21}), the equality holds
\begin{equation*}\label{29-12-22}
\int_{\mathbb{R}^4}\mathrm{d}^4x\Big(A(x)g_1(x,y)\Big)\Big|_{y=x}=
-\frac{\lambda}{2}S_m[B]PS_0^\Lambda(0)+o(1).
\end{equation*}
It is clear that the final part coincides with the one obtained in (\ref{29-12-81}). By explicitly writing out estimates for the correction contributions, we can make sure that all the terms $o(1)$ above can be included into $\mathcal{O}(1/L)$, from which the final statement, see (\ref{29-7-19-2}), follows.

\subsection{Special case of auxiliary integrals}
In Section \ref{29:sec:app:vch}, we have obtained the first terms of the asymptotic expansion with respect to the regularizing parameter $\Lambda$ for the integrals from Section \ref{29:sec:app:vs}. It has been shown that some terms depend on the function $\mathbf{f}(\cdot)$ from (\ref{29-3-17}). This section discusses a special case of asymptotic expansions for $\mathbf{f}\equiv0$. To do this, we need to calculate the numbers $\alpha_1(0)$--$\alpha_{13}(0)$, see (\ref{29-9-13})--(\ref{29-9-15}) and (\ref{29-9-25})--(\ref{29-9-123}). 

Let us start with their components and calculate the function from formula (\ref{29-9-5})
\begin{equation*}\label{29-13-1}
h_k(y)\Big|_{\mathbf{f}=0}=\frac{2-|y|^2}{8(4\pi^2)^k},
\end{equation*}
where $k\in\mathbb{N}\cup\{0\}$ and $|y|\leqslant1$. Then most of the numbers are written out instantly
\begin{equation*}\label{29-13-2}
\alpha_1(0)=-\frac{1}{8},\,\,\,
\alpha_2(0)=\frac{1}{4},\,\,\,
\alpha_3(0)=\frac{3}{8(4\pi^2)^2},\,\,\,
\alpha_4(0)=\frac{1}{48(4\pi^2)^2,}
\end{equation*}
\begin{equation*}\label{29-13-3}
\alpha_6(0)=\frac{1}{6},\,\,\,
\alpha_8(0)=\frac{1}{48(4\pi^2)^3},\,\,\,
\alpha_{10}(0)=\frac{1}{48(4\pi^2)^3},\,\,\,
\alpha_{11}(0)=\frac{1}{48(4\pi^2)^2},\,\,\,
\alpha_{12}(0)=\frac{1}{48(4\pi^2)}.
\end{equation*}
Four more numbers need to be calculated separately. Let us start with $\alpha_{13}(0)$. It is convenient to find such integral using the preliminary transition to the momentum representation. Let us recall a few basic formulas, see \cite{Ivanov-2022}, which we will use
\begin{equation*}\label{29-13-5}
\frac{1}{2\pi^2}\int_{\mathrm{S}^3}\mathrm{d}^3\hat{z}\,e^{ix_\mu\hat{z}^\mu}
=
\frac{2J_1(|x|)}{|x|},
\end{equation*}
\begin{equation*}\label{29-13-4}
R_0^{\phantom{1}}(x)=\frac{1}{(2\pi)^4}\int_{\mathbb{R}^4}\mathrm{d}^4z\,\frac{e^{ix_\mu z^\mu}}{|z|^2}
,\,\,\,
R_0^{1}(x)=\frac{1}{(2\pi)^4}\int_{\mathbb{R}^4}\mathrm{d}^4z\,\frac{e^{ix_\mu z^\mu}}{|z|^2}
\frac{2J_1(|z|)}{|z|},
\end{equation*}
where $J_n(\cdot)$ is the Bessel function of the first kind. Then, passing into the momentum representation in the first integral from (\ref{29-9-123}) and explicitly calculating the second one, we obtain the relation
\begin{equation*}\label{29-13-6}
\alpha_{13}(0)=\frac{\Theta_{13}}{4\pi^2},
\end{equation*}
where the integral
\begin{equation*}\label{29-13-7}
\Theta_{13}=\int_{0}^{+\infty}\frac{\mathrm{d}s}{s^4}\bigg(\frac{2}{s}J_1(s)-1\bigg)^2J_1(s)-\frac{5}{192}=-0,01122021\pm10^{-8}
\end{equation*}
can be found numerically with a given accuracy. By performing similar reasoning, the following set of equalities can be obtained
\begin{equation*}\label{29-13-8}
\alpha_{5}(0)=\frac{\Theta_{5}}{(4\pi^2)^2},\,\,\,
\alpha_{7}(0)=\frac{\Theta_{7}}{(4\pi^2)^3},\,\,\,
\alpha_{9}(0)=\frac{\Theta_{9}}{(4\pi^2)^3},
\end{equation*}
where
\begin{equation*}\label{29-13-9}
\Theta_{5}=\int_{0}^{+\infty}\frac{\mathrm{d}s}{s^3}\bigg(\frac{2}{s}J_1(s)-1\bigg)J_1(s)\big(e_1(s)+e_2(s)\big),
\end{equation*}
\begin{equation*}\label{29-13-10}
\Theta_{7}=\int_{0}^{+\infty}\frac{\mathrm{d}s}{2s}\bigg(\frac{2}{s}J_1(s)-1\bigg)\big(e_1(s)+e_2(s)\big)^2,
\end{equation*}
\begin{equation*}\label{29-13-11}
\Theta_{9}=\int_{0}^{+\infty}\frac{\mathrm{d}s}{s^3}\bigg(\frac{2}{s}J_1(s)-1\bigg)J_1(s)\big(e_1(s)+e_3(s)\big),
\end{equation*}
\begin{equation*}\label{29-13-12}
e_1(s)=\int_0^1\mathrm{d}t\,t^2J_1(st)=\frac{J_2(s)}{s},
\end{equation*}
\begin{equation*}\label{29-13-13}
e_2(s)=\int_1^{+\infty}\frac{\mathrm{d}t}{t^2}J_1(st)=\frac{s^3{}_2F_3\big(1,1;2,2,3;-s^2/4\big)-16s\ln(s/2)-8s(2\gamma-1)}{32},
\end{equation*}
\begin{equation*}\label{29-13-14}
e_3(s)=\int_1^{+\infty}\frac{\mathrm{d}t}{t^4}J_1(st)=\frac{-s^5{}_2F_3\big(1,1;2,3,4;-s^2/4\big)+48s^3\ln(s/2)+s^3(48\gamma-60)+192s}{768}.
\end{equation*}
Here $\gamma$ is the Euler--Mascheroni constant, and ${}_2F_3$ is a hypergeometric function. The required numbers can be calculated numerically and have the form
\begin{align}\label{29-13-15}
\Theta_{5}=&-0,04298947\pm10^{-8},
\\\label{29-13-151}
\Theta_{7}=&-0,02667863\pm10^{-8},
\\\label{29-13-152}
\Theta_{9}=&-0,02771715\pm10^{-8}.
\end{align}
Finally, under the assumption $\mathbf{f}\equiv 0$, the asymptotic expansions for the integrals from Section \ref{29:sec:app:vs} can be rewritten in the following form
\begin{equation*}\label{29-14-1}
	\mathrm{I}_1(\sigma)=\frac{1}{(4\pi^2)^2}
	\bigg(\frac{L}{2}+\frac{1}{12}\bigg)+o(1),
\end{equation*}
\begin{equation*}\label{29-14-2}
	\mathrm{I}_{2}(\sigma)=\mathrm{I}_{3}(\sigma)=\frac{1}{(4\pi^2)^2}
	\bigg(\frac{L^2}{8}+\frac{3L}{16}+\frac{96\Theta_5-1}{96}\bigg)+o(1),
\end{equation*}
\begin{equation*}\label{29-14-4}
	\mathrm{I}_{4}(\sigma)=\frac{1}{(4\pi^2)^2}
	\bigg(\frac{3L\Lambda^2}{8}-\frac{\sigma^2}{4}\bigg)+o(1),
\end{equation*}
\begin{equation*}\label{29-14-5}
	\mathrm{I}_{5}(\sigma)=\frac{1}{(4\pi^2)^2}
	\bigg(\frac{L^3}{24}-\frac{L^2}{32}-\frac{L(192\Theta_5+1)}{192}\bigg)+\mathcal{O}(1),
\end{equation*}
\begin{equation*}\label{29-14-6}
	\mathrm{I}_{6}(\sigma)=\frac{1}{(4\pi^2)^3}
	\bigg(\frac{L^2}{4}+\frac{L}{8}\bigg)+\mathcal{O}(1),
\end{equation*}
\begin{equation*}\label{29-14-7}
	\mathrm{I}_{7}(\sigma)=\frac{1}{(4\pi^2)^3}
	\bigg(-\frac{3L\Lambda^2}{16}+\frac{\Lambda^2(31+192\Theta_7)}{192}-\frac{L\sigma^2}{8}\bigg)+\mathcal{O}(1),
\end{equation*}
\begin{equation*}\label{29-14-8}
	\mathrm{I}_{8}(\sigma)=\frac{1}{(4\pi^2)^2}
	\bigg(-\frac{L^2}{8}+\frac{L}{16}\bigg)+\mathcal{O}(1),
\end{equation*}
\begin{equation*}\label{29-14-9}
	\mathrm{I}_{9}(\sigma)=\mathrm{I}_{10}(\sigma)=\frac{1}{(4\pi^2)^3}
	\bigg(-\frac{L^3}{48}-\frac{L^2}{64}-\frac{L(1+48\Theta_5)}{96}\bigg)+\mathcal{O}(1),
\end{equation*}
\begin{equation*}\label{29-14-11}
	\mathrm{I}_{11}(\sigma)=\mathrm{I}_{12}(\sigma)=\frac{1}{(4\pi^2)^3}
	\bigg(-\frac{L^3}{24}-\frac{L^2}{32}+\frac{7L}{128}\bigg)+\mathcal{O}(1),
\end{equation*}
\begin{equation*}\label{29-14-13}
	\mathrm{I}_{13}(\sigma)=\mathrm{I}_{14}(\sigma)=\frac{1}{(4\pi^2)^3}
	\bigg(\frac{3L\Lambda^2}{16}+\frac{\Lambda^2(1+192\Theta_9)}{192}\bigg)+\mathcal{O}(1),
\end{equation*}
\begin{equation*}\label{29-14-15}
	\mathrm{I}_{15}(\sigma)=\mathrm{I}_{16}(\sigma)=\mathrm{I}_{17}(\sigma)=\frac{1}{(4\pi^2)^3}
	\bigg(\frac{L^3}{24}+\frac{3L^2}{32}+\frac{L(1+48\Theta_5)}{48}\bigg)+\mathcal{O}(1),
\end{equation*}
\begin{equation*}\label{29-14-16}
	\mathrm{I}_{18}(\sigma)=\mathrm{I}_{19}(\sigma)=\mathrm{I}_{20}(\sigma)=\frac{1}{(4\pi^2)^3}
	\bigg(-\frac{L^2}{32}-\frac{11L}{48}\bigg)+\mathcal{O}(1).
\end{equation*}
Also, the last set can be supplemented with the following asymptotic expansion, see formula (\ref{29-9-126}),
\begin{equation*}\label{29-14-17}
R_{20}^\Lambda(0)=R_{21}^\Lambda(0)=R_{22}^\Lambda(0)=
-\frac{3L}{8(4\pi^2)\Lambda^2}+
\frac{-5+48\Theta_{13}}{24(4\pi^2)\Lambda^2}+\mathcal{O}\big(1/\Lambda^3\big).
\end{equation*}

\subsection{Combinations of coupling constants}
\label{29:sec:app:kom}
\noindent{\textbf{Notations for the first loop.}}

\begin{equation*}\label{29-17-1}
\mathbf{T}_{1,1,1}^{i}=g^{ijj},\,\,\,
\mathbf{T}_{1,1,2}^{i}=g^{ij_1j_2}M^{j_1j_2},
\end{equation*}
\begin{equation*}\label{29-17-2}
\mathbf{T}_{1,2,1}^{i_1i_2}=\lambda^{i_1i_2jj},\,\,\,
\mathbf{T}_{1,2,2}^{i_1i_2}=g^{i_1j_1j_2}g^{j_1j_2i_2},\,\,\,
\mathbf{T}_{1,2,3}^{i_1i_2}=\lambda^{i_1i_2j_1j_2}M^{j_1j_2},
\end{equation*}
\begin{equation*}\label{29-17-3}
\mathbf{T}_{1,3,1}^{i_1i_2i_3}=\lambda^{i_1i_2j_1j_2}g^{j_1j_2i_3},\,\,\,
\mathbf{T}_{1,4,1}^{i_1i_2i_3i_4}=\lambda^{i_1i_2j_1j_2}\lambda^{j_1j_2i_3i_4}.
\end{equation*}

\noindent{\textbf{Notations for the second loop.}}
\begin{equation*}\label{29-17-51}
	\mbox{\textbf{Combinations for}}\,\, B_i:
\end{equation*}
\begin{equation*}\label{29-17-4}
\mathbf{T}_{2,1,1}^{i}=\lambda^{ij_1j_2j_3}g^{j_1j_2j_3},\,\,\,
\mathbf{T}_{2,1,2}^{i}=\lambda^{ij_1j_2j_3}g^{j_1j_2j_4}M^{j_3j_4},\,\,\,
\mathbf{T}_{2,1,3}^{i}=g^{ij_1j_2}g^{j_1j_3j_4}g^{j_2j_3j_4},
\end{equation*}
\begin{equation*}\label{29-17-5}
\mathbf{T}_{2,1,4}^{i}=g^{ij_1j_2}\lambda^{j_1j_2j_3j_3},\,\,\,
\mathbf{T}_{2,1,5}^{i}=g^{ij_1j_2}\lambda^{j_1j_2j_3j_4}M^{j_3j_4},
\end{equation*}
\begin{equation*}\label{29-17-6}
\mathbf{T}_{2,1,6}^{i}=g^{ij_1j_2}M^{j_1j_2}_{r,1},\,\,\,
\mathbf{T}_{2,1,7}^{i}=g^{ijj}_{r,1},\,\,\,
\mathbf{T}_{2,1,8}^{i}=g^{ij_1j_2}_{r,1}M^{j_1j_2},
\end{equation*}
\begin{equation*}\label{29-17-6-1}
	\mathbf{T}_{2,1,9}^{i}=g^{ij_1j_2}M^{j_1j_3}\tilde{\delta}_{r,1}^{j_3j_2},\,\,\,
	\mathbf{T}_{2,1,10}^{i}=g^{ij_1j_2}\tilde{\delta}_{r,1}^{j_1j_2}.
\end{equation*}

\begin{equation*}\label{29-17-52}
	\mbox{\textbf{Combinations for}}\,\, B_{i_1}B_{i_2}:
\end{equation*}
\begin{equation*}\label{29-17-7}
\mathbf{T}_{2,2,1}^{i_1i_2}=\lambda^{i_1j_1j_2j_3}\lambda^{j_1j_2j_3i_2},\,\,\,
\mathbf{T}_{2,2,2}^{i_1i_2}=\lambda^{i_1j_1j_2j_3}\lambda^{j_1j_2j_4i_2}M^{j_3j_4},
\end{equation*}
\begin{equation*}\label{29-17-8}
\mathbf{T}_{2,2,3}^{i_1i_2}=\lambda^{i_1i_2j_1j_2}g^{j_1j_3j_4}g^{j_2j_3j_4},\,\,\,
\mathbf{T}_{2,2,4}^{i_1i_2}=g^{i_1j_1j_2}g^{j_1j_3j_4}\lambda^{j_2j_3j_4i_2},
\end{equation*}
\begin{equation*}\label{29-17-9}
\mathbf{T}_{2,2,5}^{i_1i_2}=\lambda^{i_1i_2j_1j_2}\lambda^{j_1j_2j_3j_3},\,\,\,
\mathbf{T}_{2,2,6}^{i_1i_2}=g^{i_1j_1j_2}\lambda^{j_1j_2j_3j_4}g^{j_3j_4i_2},
\end{equation*}
\begin{equation*}\label{29-17-10}
\mathbf{T}_{2,2,7}^{i_1i_2}=\lambda^{i_1i_2j_1j_2}\lambda^{j_1j_2j_3j_4}M^{j_3j_4},\,\,\,
\mathbf{T}_{2,2,8}^{i_1i_2}=\lambda^{i_1i_2j_1j_2}M^{j_1j_2}_{r,1},
\end{equation*}
\begin{equation*}\label{29-17-11}
\mathbf{T}_{2,2,9}^{i_1i_2}=g^{i_1j_1j_2}_{r,1}g^{j_1j_2i_2},\,\,\,
\mathbf{T}_{2,2,10}^{i_1i_2}=\lambda^{i_1i_2jj}_{r,1},\,\,\,
\mathbf{T}_{2,2,11}^{i_1i_2}=\lambda^{i_1i_2j_1j_2}_{r,1}M^{j_1j_2},
\end{equation*}
\begin{equation*}\label{29-17-11-1}
	\mathbf{T}_{2,2,12}^{i_1i_2}=\lambda^{i_1i_2j_1j_2}M^{j_1j_3}\tilde{\delta}_{r,1}^{j_3j_2},\,\,\,
	\mathbf{T}_{2,2,13}^{i_1i_2}=g^{i_1j_1j_3}g^{i_2j_3j_2}\tilde{\delta}_{r,1}^{j_1j_2},\,\,\,
	\mathbf{T}_{2,2,14}^{i_1i_2}=\lambda^{i_1i_2j_1j_2}\tilde{\delta}_{r,1}^{j_1j_2}.
\end{equation*}

\begin{equation*}\label{29-17-53}
	\mbox{\textbf{Combinations for}}\,\, B_{i_1}B_{i_2}B_{i_3}:
\end{equation*}
\begin{equation*}\label{29-17-13}
\mathbf{T}_{2,3,1}^{i_1i_2i_3}=g^{i_1j_1j_2}\lambda^{j_1j_3j_4i_2}\lambda^{j_2j_3j_4i_3},\,\,\,
\mathbf{T}_{2,3,2}^{i_1i_2i_3}=\lambda^{i_1i_2j_1j_2}g^{j_1j_3j_4}\lambda^{j_2j_3j_4i_3},
\end{equation*}
\begin{equation*}\label{29-17-14}
\mathbf{T}_{2,3,3}^{i_1i_2i_3}=\lambda^{i_1i_2j_1j_2}\lambda^{j_1j_2j_3j_4}g^{j_3j_4i_3},
\end{equation*}
\begin{equation*}\label{29-17-15}
\mathbf{T}_{2,3,4}^{i_1i_2i_3}=\lambda^{i_1i_2j_1j_2}g^{j_1j_2i_3}_{r,1},\,\,\,
\mathbf{T}_{2,3,5}^{i_1i_2i_3}=\lambda^{i_1i_2j_1j_2}_{r,1}g^{j_1j_2i_3},
\end{equation*}
\begin{equation*}\label{29-17-15-1}
	\mathbf{T}_{2,3,6}^{i_1i_2i_3}=\lambda^{i_1i_2j_1j_2}g^{j_2j_3i_3}\tilde{\delta}_{r,1}^{j_1j_3}.
\end{equation*}

\begin{equation*}\label{29-17-54}
	\mbox{\textbf{Combinations for}}\,\, B_{i_1}B_{i_2}B_{i_3}B_{i_4}:
\end{equation*}
\begin{equation*}\label{29-17-16}
\mathbf{T}_{2,4,1}^{i_1i_2i_3i_4}=\lambda^{i_1i_2j_1j_2}\lambda^{j_1j_3j_4i_3}\lambda^{j_2j_3j_4i_4},
\end{equation*}
\begin{equation*}\label{29-17-17}
\mathbf{T}_{2,4,2}^{i_1i_2i_3i_4}=\lambda^{i_1i_2j_1j_2}\lambda^{j_1j_2j_3j_4}\lambda^{j_3j_4i_3i_4},\,\,\,
\mathbf{T}_{2,4,3}^{i_1i_2i_3i_4}=\lambda^{i_1i_2j_1j_2}_{r,1}\lambda^{j_1j_2i_3i_4},
\end{equation*}
\begin{equation*}\label{29-17-17-1}
	\mathbf{T}_{2,4,4}^{i_1i_2i_3i_4}=\lambda^{i_1i_2j_1j_2}\lambda^{i_3i_4j_3j_2}\tilde{\delta}_{r,1}^{j_1j_3}.
\end{equation*}

\noindent{\textbf{Notations for the third loop.}}
\begin{equation*}\label{29-17-55}
	\mbox{\textbf{Combinations for}}\,\, B_{i}:
\end{equation*}
\begin{equation*}\label{29-17-18}
\mathbf{T}_{3,1,1}^{i}=\tilde{g}^{ij_1j_2}_{r,1}\tilde{M}^{j_1j_2}_{r,1},\,\,\,
\mathbf{T}_{3,1,2}^{i}=\lambda^{ij_1j_2j_3}g^{j_1j_4j_5}g^{j_4j_5j_6}g^{j_6j_2j_3},
\end{equation*}
\begin{equation*}\label{29-17-12}
\mathbf{T}_{3,1,3}^{i}=\tilde{g}^{ij_1j_2}_{r,1}g^{j_1j_3j_4}g^{j_2j_3j_4},\,\,\,
\mathbf{T}_{3,1,4}^{i}=\lambda^{ij_1j_2j_3}g^{j_1j_2j_4}\tilde{M}^{j_3j_4}_{r,1},
\end{equation*}
\begin{equation*}\label{29-17-19}
\mathbf{T}_{3,1,5}^{i}=\lambda^{ij_1j_2j_3}g^{j_1j_2j_3}_{r,1},\,\,\,
\mathbf{T}_{3,1,6}^{i}=\lambda^{ij_1j_2j_3}_{r,1}g^{j_1j_2j_3},
\end{equation*}
\begin{equation*}\label{29-17-20}
\mathbf{T}_{3,1,7}^{i}=\lambda^{ij_1j_2j_3}g^{j_4j_2j_3}_{r,1}M^{j_1j_4},\,\,\,
\mathbf{T}_{3,1,8}^{i}=\lambda^{ij_1j_2j_3}_{r,1}g^{j_4j_2j_3}M^{j_1j_4},
\end{equation*}
\begin{equation*}\label{29-17-21}
\mathbf{T}_{3,1,9}^{i}=g^{ij_1j_2}g^{j_1j_3j_4}g^{j_2j_3j_4}_{r,1},\,\,\,
\mathbf{T}_{3,1,10}^{i}=\lambda^{ij_1j_2j_3}\lambda^{j_2j_3j_4j_5}g^{j_1j_4j_5},
\end{equation*}
\begin{equation*}\label{29-17-22}
\mathbf{T}_{3,1,11}^{i}=\lambda^{ij_1j_2j_3}\lambda^{j_2j_3j_4j_5}g^{j_6j_4j_5}M^{j_1j_6},\,\,\,
\mathbf{T}_{3,1,12}^{i}=g^{ij_1j_6}g^{j_1j_2j_3}\lambda^{j_2j_3j_4j_5}g^{j_6j_4j_5},
\end{equation*}
\begin{equation*}\label{29-17-23}
\mathbf{T}_{3,1,13}^{i}=g^{ij_3j_6}g^{j_1j_2j_3}\lambda^{j_2j_6j_4j_5}g^{j_1j_4j_5},\,\,\,
\mathbf{T}_{3,1,14}^{i}=\lambda^{ij_1j_2j_3}\lambda^{j_2j_6j_4j_5}g^{j_1j_4j_5}M^{j_3j_6},
\end{equation*}
\begin{equation*}\label{29-17-24}
\mathbf{T}_{3,1,15}^{i}=\lambda^{ij_1j_2j_3}\lambda^{j_2j_3j_4j_5}g^{j_1j_4j_6}M^{j_5j_6},\,\,\,
\mathbf{T}_{3,1,16}^{i}=g^{ij_1j_2}\lambda^{j_1j_3j_4j_5}\lambda^{j_2j_3j_4j_5},
\end{equation*}
\begin{equation*}\label{29-17-25}
\mathbf{T}_{3,1,17}^{i}=g^{ij_1j_2}\lambda^{j_1j_3j_4j_5}\lambda^{j_2j_3j_4j_6}M^{j_5j_6},\,\,\,
\mathbf{T}_{3,1,18}^{i}=g^{ij_1j_2}\lambda^{j_1j_3j_4j_5}\lambda^{j_6j_3j_4j_5}M^{j_2j_6},
\end{equation*}
\begin{equation*}\label{29-17-26}
\mathbf{T}_{3,1,19}^{i}=\lambda^{ij_1j_2j_3}g^{j_1j_4j_5}g^{j_5j_6j_3}g^{j_2j_4j_6},\,\,\,
\mathbf{T}_{3,1,20}^{i}=g^{ij_1j_2}\lambda^{j_1j_2j_3j_3}_{r,1},
\end{equation*}
\begin{equation*}\label{29-17-27}
\mathbf{T}_{3,1,21}^{i}=g^{ij_1j_2}\lambda^{j_1j_2j_3j_4}_{r,1}M^{j_3j_4},\,\,\,
\mathbf{T}_{3,1,22}^{i}=g^{ij_1j_2}\lambda^{j_1j_4j_3j_3}_{r,1}M^{j_2j_4},
\end{equation*}
\begin{equation*}\label{29-17-28}
\mathbf{T}_{3,1,23}^{i}=g^{ij_1j_2}M^{j_1j_2}_{r,2},\,\,\,
\mathbf{T}_{3,1,24}^{i}=g^{ijj}_{r,2},\,\,\,
\mathbf{T}_{3,1,25}^{i}=g^{ij_1j_2}_{r,2}M^{j_1j_2},
\end{equation*}
\begin{equation*}\label{29-17-29}
\mathbf{T}_{3,1,26}^{i}=g^{ij_1j_2}\delta^{j_1j_2}_{r,2},\,\,\,
\mathbf{T}_{3,1,27}^{i}=g^{ij_1j_2}M^{j_2j_3}M^{j_3j_1}_{r,2},\,\,\,
\mathbf{T}_{3,1,28}^{i}=g^{ij_1j_2}\delta^{j_1j_3}_{r,2}M^{j_2j_3},
\end{equation*}
\begin{equation*}\label{29-17-29-1}
	\mathbf{T}_{3,1,29}^{i}=\lambda^{ij_1j_2j_3}g^{j_2j_3j_4}\tilde{\delta}_{r,1}^{j_1j_4},\,\,\,
	\mathbf{T}_{3,1,30}^{i}=\lambda^{ij_1j_2j_3}g^{j_2j_3j_4}\big(M^{j_1j_5}\tilde{\delta}_{r,1}^{j_5j_4}+M^{j_4j_5}\tilde{\delta}_{r,1}^{j_5j_1}\big),
\end{equation*}
\begin{equation*}\label{29-17-29-2}
	\mathbf{T}_{3,1,31}^{i}=\lambda^{ij_1j_2j_3}g^{j_5j_3j_4}M^{j_2j_5}\tilde{\delta}_{r,1}^{j_1j_4},\,\,\,
	\mathbf{T}_{3,1,32}^{i}=g^{ij_1j_2}g^{j_2j_3j_4}g^{j_1j_3j_5}\tilde{\delta}_{r,1}^{j_5j_4},
\end{equation*}
\begin{equation*}\label{29-17-29-3}
	\mathbf{T}_{3,1,33}^{i}=g^{ij_1j_2}g^{j_2j_3j_4}g^{j_5j_3j_4}\tilde{\delta}_{r,1}^{j_5j_1},\,\,\,
	\mathbf{T}_{3,1,34}^{i}=g^{ij_1j_2}\tilde{\delta}_{r,1}^{j_1j_3}M^{j_3j_4}\tilde{\delta}_{r,1}^{j_4j_2},
\end{equation*}
\begin{equation*}\label{29-17-29-4}
	\mathbf{T}_{3,1,35}^{i}=g^{ij_1j_2}\tilde{\delta}_{r,1}^{j_1j_3}\tilde{\delta}_{r,1}^{j_3j_2},\,\,\,
	\mathbf{T}_{3,1,36}^{i}=g^{ij_1j_2}M^{j_1j_3}\tilde{\delta}_{r,1}^{j_3j_4}\tilde{\delta}_{r,1}^{j_4j_2},
\end{equation*}
\begin{equation*}\label{29-17-29-5}
	\mathbf{T}_{3,1,37}^{i}=g^{ij_1j_2}M^{j_2j_3}\lambda^{j_1j_3j_4j_5}
	\tilde{\delta}_{r,1}^{j_4j_5},\,\,\,
	\mathbf{T}_{3,1,38}^{i}=\tilde{g}^{ij_1j_2}_{r,1}\tilde{\delta}_{r,1}^{j_1j_2},
\end{equation*}
\begin{equation*}\label{29-17-29-6}
	\mathbf{T}_{3,1,39}^{i}=\tilde{g}^{ij_1j_2}_{r,1}M^{j_2j_3}\tilde{\delta}_{r,1}^{j_1j_3},\,\,\,
	\mathbf{T}_{3,1,40}^{i}=g^{ij_1j_2}\tilde{M}^{j_2j_3}_{r,1}\tilde{\delta}_{r,1}^{j_1j_3}.
\end{equation*}

\begin{equation*}\label{29-17-56}
	\mbox{\textbf{Combinations for}}\,\, B_{i_1}B_{i_2}:
\end{equation*}
\begin{equation*}\label{29-17-30}
\mathbf{T}_{3,2,1}^{i_1i_2}=\tilde{\lambda}^{i_1i_2j_1j_2}_{r,1}\tilde{M}^{j_1j_2}_{r,1},\,\,\,
\mathbf{T}_{3,2,2}^{i_1i_2}=\tilde{g}^{i_1j_1j_2}_{r,1}\tilde{g}^{j_1j_2i_2}_{r,1},
\end{equation*}
\begin{equation*}\label{29-17-32}
\mathbf{T}_{3,2,3}^{i_1i_2}=\lambda^{i_1j_1j_2j_3}g^{j_1j_4j_5}g^{j_4j_5j_6}\lambda^{i_2j_2j_3j_6},\,\,\,
\mathbf{T}_{3,2,4}^{i_1i_2}=\lambda^{i_1j_1j_2j_3}\lambda^{i_2j_1j_4j_5}g^{j_4j_5j_6}g^{j_2j_3j_6},
\end{equation*}
\begin{equation*}\label{29-17-33}
\mathbf{T}_{3,2,5}^{i_1i_2}=\lambda^{i_1j_1j_2j_3}g^{j_1j_4j_5}\lambda^{i_2j_4j_5j_6}g^{j_2j_3j_6},\,\,\,
\mathbf{T}_{3,2,6}^{i_1i_2}=\lambda^{i_1j_1j_2j_3}\lambda^{i_2j_1j_2j_4}\tilde{M}^{j_3j_4}_{r,1},
\end{equation*}
\begin{equation*}\label{29-17-34}
\mathbf{T}_{3,2,7}^{i_1i_2}=\lambda^{i_1j_1j_2j_3}g^{j_1j_2j_4}\tilde{g}^{i_2j_3j_4}_{r,1},\,\,\,
\mathbf{T}_{3,2,8}^{i_1i_2}=g^{j_1j_2j_3}g^{j_1j_2j_4}\tilde{\lambda}^{i_1i_2j_3j_4}_{r,1},
\end{equation*}
\begin{equation*}\label{29-17-35}
\mathbf{T}_{3,2,9}^{i_1i_2}=\lambda^{i_1j_1j_2j_3}\lambda^{i_2j_1j_2j_3}_{r,1},\,\,\,
\mathbf{T}_{3,2,10}^{i_1i_2}=\lambda^{i_1j_1j_2j_3}\lambda^{i_2j_1j_2j_4}_{r,1}M^{j_3j_4},\,\,\,
\mathbf{T}_{3,2,11}^{i_1i_2}=\lambda^{i_1i_2j_1j_2}g^{j_1j_3j_4}g^{j_2j_3j_4}_{r,1},
\end{equation*}
\begin{equation*}\label{29-17-36}
\mathbf{T}_{3,2,12}^{i_1i_2}=g^{i_1j_1j_2}g^{j_1j_3j_4}_{r,1}\lambda^{i_2j_2j_3j_4},\,\,\,
\mathbf{T}_{3,2,13}^{i_1i_2}=g^{i_1j_1j_2}g^{j_1j_3j_4}\lambda^{i_2j_2j_3j_4}_{r,1},
\end{equation*}
\begin{equation*}\label{29-17-37}
\mathbf{T}_{3,2,14}^{i_1i_2}=\lambda^{i_1j_1j_2j_3}\lambda^{j_2j_3j_4j_5}\lambda^{i_2j_1j_4j_5},\,\,\,
\mathbf{T}_{3,2,15}^{i_1i_2}=\lambda^{i_1j_1j_2j_3}\lambda^{j_2j_3j_4j_5}\lambda^{i_2j_6j_4j_5}M^{j_1j_6},
\end{equation*}
\begin{equation*}\label{29-17-38}
\mathbf{T}_{3,2,16}^{i_1i_2}=g^{i_1j_1j_6}g^{j_1j_2j_3}\lambda^{j_2j_3j_4j_5}\lambda^{i_2j_6j_4j_5},\,\,\,
\mathbf{T}_{3,2,17}^{i_1i_2}=\lambda^{i_1i_2j_1j_6}g^{j_1j_2j_3}\lambda^{j_2j_3j_4j_5}g^{j_6j_4j_5},
\end{equation*}
\begin{equation*}\label{29-17-39}
\mathbf{T}_{3,2,18}^{i_1i_2}=\lambda^{i_1i_2j_5j_6}g^{j_1j_2j_3}\lambda^{j_2j_3j_4j_5}g^{j_1j_4j_6},\,\,\,
\mathbf{T}_{3,2,19}^{i_1i_2}=g^{i_1j_5j_6}g^{j_1j_2j_3}\lambda^{j_2j_3j_4j_5}\lambda^{i_2j_1j_4j_6},
\end{equation*}
\begin{equation*}\label{29-17-40}
\mathbf{T}_{3,2,20}^{i_1i_2}=\lambda^{i_1j_1j_2j_3}\lambda^{j_2j_3j_4j_5}g^{j_1j_4j_6}g^{i_2j_5j_6},\,\,\,
\mathbf{T}_{3,2,21}^{i_1i_2}=\lambda^{i_1j_1j_2j_3}\lambda^{j_2j_3j_4j_5}\lambda^{i_2j_1j_4j_6}M^{j_5j_6},
\end{equation*}
\begin{equation*}\label{29-17-41}
\mathbf{T}_{3,2,22}^{i_1i_2}=\lambda^{i_1i_2j_1j_2}\lambda^{j_1j_3j_4j_5}
\lambda^{j_2j_3j_4j_5},\,\,\,
\mathbf{T}_{3,2,23}^{i_1i_2}=\lambda^{i_1i_2j_1j_2}\lambda^{j_1j_3j_4j_5}
\lambda^{j_2j_3j_4j_6}M^{j_5j_6},
\end{equation*}
\begin{equation*}\label{29-17-42}
\mathbf{T}_{3,2,24}^{i_1i_2}=g^{i_1j_1j_2}\lambda^{j_1j_3j_4j_5}
\lambda^{j_2j_3j_4j_6}g^{i_2j_5j_6},\,\,\,
\mathbf{T}_{3,2,25}^{i_1i_2}=\lambda^{i_1j_1j_2j_3}\lambda^{i_2j_1j_4j_5}g^{j_4j_6j_2}g^{j_3j_5j_6},
\end{equation*}
\begin{equation*}\label{29-17-43}
\mathbf{T}_{3,2,26}^{i_1i_2}=\lambda^{i_1i_2j_1j_2}\lambda^{j_1j_2j_3j_3}_{r,1},\,\,\,
\mathbf{T}_{3,2,27}^{i_1i_2}=\lambda^{i_1i_2j_1j_2}\lambda^{j_1j_2j_3j_4}_{r,1}M^{j_3j_4},\,\,\,
\mathbf{T}_{3,2,28}^{i_1i_2}=g^{i_1j_1j_2}\lambda^{j_1j_2j_3j_4}_{r,1}g^{i_2j_3j_4},
\end{equation*}
\begin{equation*}\label{29-17-44}
\mathbf{T}_{3,2,29}^{i_1i_2}=\lambda^{i_1i_2j_1j_2}\lambda^{j_1j_4j_3j_3}_{r,1}M^{j_2j_4},\,\,\,
\mathbf{T}_{3,2,30}^{i_1i_2}=g^{i_1j_1j_4}g^{i_2j_2j_4}\lambda^{j_1j_2j_3j_3}_{r,1},
\end{equation*}
\begin{equation*}\label{29-17-45}
\mathbf{T}_{3,2,31}^{i_1i_2}=\lambda^{i_1i_2j_1j_2}M^{j_1j_2}_{r,2},\,\,\,
\mathbf{T}_{3,2,32}^{i_1i_2}=g^{i_1j_1j_2}g^{i_2j_1j_2}_{r,2},\,\,\,
\mathbf{T}_{3,2,33}^{i_1i_2}=\lambda^{i_1i_2jj}_{r,2},
\end{equation*}
\begin{equation*}\label{29-17-46}
\mathbf{T}_{3,2,34}^{i_1i_2}=\lambda^{i_1i_2j_1j_2}_{r,2}M^{j_1j_2},\,\,\,
\mathbf{T}_{3,2,35}^{i_1i_2}=\lambda^{i_1i_2j_1j_2}\delta^{j_1j_2}_{r,2},\,\,\,
\mathbf{T}_{3,2,36}^{i_1i_2}=\lambda^{i_1i_2j_1j_2}M^{j_2j_3}M^{j_3j_1}_{r,2},
\end{equation*}
\begin{equation*}\label{29-17-47}
\mathbf{T}_{3,2,37}^{i_1i_2}=g^{i_1j_1j_2}g^{i_2j_2j_3}M^{j_3j_1}_{r,2},\,\,\,
\mathbf{T}_{3,2,38}^{i_1i_2}=\lambda^{i_1i_2j_1j_2}M^{j_2j_3}\delta^{j_3j_1}_{r,2},\,\,\,
\mathbf{T}_{3,2,39}^{i_1i_2}=g^{i_1j_1j_2}g^{i_2j_2j_3}\delta^{j_3j_1}_{r,2},
\end{equation*}
\begin{equation*}\label{29-17-471}
	\mathbf{T}_{3,2,40}^{i_1i_2}=g^{j_1i_1j_2}g^{j_2i_2j_3}\lambda^{j_3j_4j_5j_6}\lambda^{j_1j_4j_5j_6},\,\,\,
	\mathbf{T}_{3,2,41}^{i_1i_2}=\lambda^{j_1i_1i_2j_2}M^{j_2j_3}\lambda^{j_3j_4j_5j_6}\lambda^{j_1j_4j_5j_6},
\end{equation*}
\begin{equation*}\label{29-17-47-1}
	\mathbf{T}_{3,2,42}^{i_1i_2}=\lambda^{i_1j_1j_2j_3}\tilde{\delta}_{r,1}^{j_3j_4}\lambda^{j_1j_2j_4i_2},\,\,\,
	\mathbf{T}_{3,2,43}^{i_1i_2}=\lambda^{i_1j_1j_2j_3}\lambda^{j_1j_2j_4i_2}\tilde{\delta}_{r,1}^{j_3j_5}M^{j_5j_4},
\end{equation*}
\begin{equation*}\label{29-17-47-2}
\mathbf{T}_{3,2,44}^{i_1i_2}=\lambda^{i_1j_1j_2j_3}\lambda^{j_1j_5j_4i_2}\tilde{\delta}_{r,1}^{j_2j_5}M^{j_3j_4},\,\,\,
\mathbf{T}_{3,2,45}^{i_1i_2}=\lambda^{i_1i_2j_1j_2}\tilde{\delta}_{r,1}^{j_1j_5}g^{j_5j_3j_4}g^{j_2j_3j_4},
\end{equation*}
\begin{equation*}\label{29-17-47-3}
	\mathbf{T}_{3,2,46}^{i_1i_2}=\lambda^{i_1i_2j_1j_2}\tilde{\delta}_{r,1}^{j_3j_5}g^{j_1j_3j_4}g^{j_2j_5j_4},\,\,\,
	\mathbf{T}_{3,2,47}^{i_1i_2}=g^{i_1j_1j_2}\tilde{\delta}_{r,1}^{j_1j_5}g^{j_5j_3j_4}\lambda^{j_2j_3j_4i_2},
\end{equation*}
\begin{equation*}\label{29-17-47-4}
	\mathbf{T}_{3,2,48}^{i_1i_2}=g^{i_1j_1j_2}\tilde{\delta}_{r,1}^{j_2j_5}g^{j_1j_3j_4}\lambda^{j_5j_3j_4i_2},\,\,\,
	\mathbf{T}_{3,2,49}^{i_1i_2}=g^{i_1j_1j_2}\tilde{\delta}_{r,1}^{j_3j_5}g^{j_1j_3j_4}\lambda^{j_2j_5j_4i_2},
\end{equation*}
\begin{equation*}\label{29-17-47-7}
	\mathbf{T}_{3,2,50}^{i_1i_2}=g^{i_1j_1j_2}\tilde{\delta}_{r,1}^{j_2j_3}\tilde{\delta}_{r,1}^{j_1j_4}g^{i_2j_3j_4},\,\,\,
	\mathbf{T}_{3,2,51}^{i_1i_2}=\lambda^{i_1i_2j_1j_2}\tilde{\delta}_{r,1}^{j_2j_4}M^{j_4j_3}\tilde{\delta}_{r,1}^{j_1j_3},
\end{equation*}
\begin{equation*}\label{29-17-47-5}
	\mathbf{T}_{3,2,52}^{i_1i_2}=\lambda^{i_1i_2j_1j_2}\tilde{\delta}_{r,1}^{j_2j_3}\tilde{\delta}_{r,1}^{j_1j_3},\,\,\,
	\mathbf{T}_{3,2,53}^{i_1i_2}=\lambda^{i_1i_2j_1j_2}\tilde{\delta}_{r,1}^{j_4j_3}M^{j_2j_4}\tilde{\delta}_{r,1}^{j_1j_3},
\end{equation*}
\begin{equation*}\label{29-17-47-6}
	\mathbf{T}_{3,2,54}^{i_1i_2}=g^{i_1j_1j_2}\tilde{\delta}_{r,1}^{j_2j_3}\tilde{\delta}_{r,1}^{j_3j_4}g^{i_2j_1j_4},\,\,\,
	\mathbf{T}_{3,2,55}^{i_1i_2}=\lambda^{i_1i_2j_1j_2}M^{j_2j_3}\lambda^{j_1j_3j_4j_5}
	\tilde{\delta}_{r,1}^{j_4j_5},
\end{equation*}
\begin{equation*}\label{29-17-47-8}
	\mathbf{T}_{3,2,56}^{i_1i_2}=g^{i_1j_1j_2}g^{i_2j_2j_3}\lambda^{j_1j_3j_4j_5}
	\tilde{\delta}_{r,1}^{j_4j_5},\,\,\,
	\mathbf{T}_{3,2,57}^{i_1i_2}=\tilde{\lambda}^{i_1i_2j_1j_2}_{r,1}
	\tilde{\delta}_{r,1}^{j_1j_2},
\end{equation*}
\begin{equation*}\label{29-17-47-9}
	\mathbf{T}_{3,2,58}^{i_1i_2}=\lambda^{i_1i_2j_1j_2}_{r,1}
	\tilde{\delta}_{r,1}^{j_1j_3}\tilde{M}^{j_3j_2}_{r,1},\,\,\,
	\mathbf{T}_{3,2,59}^{i_1i_2}=g^{i_1j_1j_2}_{r,1}
	\tilde{\delta}_{r,1}^{j_1j_3}\tilde{g}^{i_2j_3j_2}_{r,1},\,\,\,
	\mathbf{T}_{3,2,60}^{i_1i_2}=\tilde{\lambda}^{i_1i_2j_1j_2}_{r,1}
	\tilde{\delta}_{r,1}^{j_1j_3}M^{j_3j_4}.
\end{equation*}

\begin{equation*}\label{29-17-57}
	\mbox{\textbf{Combinations for}}\,\, B_{i_1}B_{i_2}B_{i_3}:
\end{equation*}
\begin{equation*}\label{29-17-48}
\mathbf{T}_{3,3,1}^{i_1i_2i_3}=\tilde{\lambda}^{i_1i_2j_1j_2}_{r,1}\tilde{g}^{i_3j_1j_2}_{r,1},\,\,\,
\mathbf{T}_{3,3,2}^{i_1i_2i_3}=\lambda^{i_1j_1j_2j_3}g^{j_1j_4j_5}\lambda^{j_4j_5j_6i_2}\lambda^{j_6j_2j_3i_3},
\end{equation*}
\begin{equation*}\label{29-17-49}
\mathbf{T}_{3,3,3}^{i_1i_2i_3}=\lambda^{i_1j_1j_2j_3}\lambda^{i_2j_2j_3j_4}
\tilde{g}^{i_3j_4j_1}_{r,1},\,\,\,
\mathbf{T}_{3,3,4}^{i_1i_2i_3}=g^{j_1j_2j_3}\lambda^{i_3j_2j_3j_4}
\tilde{\lambda}^{i_1i_2j_4j_1}_{r,1},
\end{equation*}
\begin{equation*}\label{29-17-50}
\mathbf{T}_{3,3,5}^{i_1i_2i_3}=\lambda^{i_1j_1j_2j_3}\lambda^{j_2j_3j_4i_2}_{r,1}g^{j_1j_4i_3},\,\,\,
\mathbf{T}_{3,3,6}^{i_1i_2i_3}=\lambda^{i_1i_2j_1j_2}g^{j_1j_3j_4}_{r,1}\lambda^{j_3j_4j_2i_3},
\end{equation*}
\begin{equation*}\label{29-17-58}
\mathbf{T}_{3,3,7}^{i_1i_2i_3}=\lambda^{i_1i_2j_1j_2}g^{j_1j_3j_4}\lambda^{j_3j_4j_2i_3}_{r,1},\,\,\,
\mathbf{T}_{3,3,8}^{i_1i_2i_3}=\lambda^{i_1j_1j_2j_3}g^{i_2j_3j_4}\lambda^{j_4j_5j_6i_3}
\lambda^{j_5j_6j_1j_2},
\end{equation*}
\begin{equation*}\label{29-17-59}
\mathbf{T}_{3,3,9}^{i_1i_2i_3}=g^{j_1j_2j_3}\lambda^{i_1i_2j_3j_4}\lambda^{j_4j_5j_6i_3}
\lambda^{j_5j_6j_1j_2},\,\,\,
\mathbf{T}_{3,3,10}^{i_1i_2i_3}=g^{j_1j_2j_3}\lambda^{j_3j_4j_6i_3}\lambda^{j_4j_5i_1i_2}
\lambda^{j_5j_6j_1j_2},
\end{equation*}
\begin{equation*}\label{29-17-60}
\mathbf{T}_{3,3,11}^{i_1i_2i_3}=\lambda^{i_3j_1j_2j_3}g^{j_3j_4j_6}\lambda^{j_4j_5i_1i_2}
\lambda^{j_5j_6j_1j_2},\,\,\,
\mathbf{T}_{3,3,12}^{i_1i_2i_3}=\lambda^{i_1j_1j_2j_3}\lambda^{j_3j_4j_6i_3}g^{j_4j_5i_2}
\lambda^{j_5j_6j_1j_2},
\end{equation*}
\begin{equation*}\label{29-17-61}
\mathbf{T}_{3,3,13}^{i_1i_2i_3}=\lambda^{i_1i_2j_1j_2}\lambda^{j_1j_3j_4j_5}g^{j_5j_6i_3}
\lambda^{j_6j_4j_3j_2},\,\,\,
\mathbf{T}_{3,3,14}^{i_1i_2i_3}=\lambda^{i_1i_2j_1j_2}\lambda^{j_1j_3j_4j_5}
\lambda^{j_5j_4j_3j_6}g^{j_2j_6i_3},
\end{equation*}
\begin{equation*}\label{29-17-62}
\mathbf{T}_{3,3,15}^{i_1i_2i_3}=\lambda^{i_1j_1j_2j_3}\lambda^{i_2j_2j_4j_5}\lambda^{i_3j_4j_1j_6}g^{j_3j_5j_6},\,\,\,
\mathbf{T}_{3,3,16}^{i_1i_2i_3}=\lambda^{i_1i_2j_1j_2}\lambda^{j_1j_2j_3j_4}_{r,1}g^{j_3j_4i_3},
\end{equation*}
\begin{equation*}\label{29-17-63}
\mathbf{T}_{3,3,17}^{i_1i_2i_3}=\lambda^{i_1i_2j_1j_2}g^{j_2j_4i_3}\lambda^{j_1j_4j_3j_3}_{r,1},\,\,\,
\mathbf{T}_{3,3,18}^{i_1i_2i_3}=\lambda^{i_1i_2j_1j_2}g^{j_1j_2i_3}_{r,2},
\end{equation*}
\begin{equation*}\label{29-17-64}
\mathbf{T}_{3,3,19}^{i_1i_2i_3}=\lambda^{i_1i_2j_1j_2}_{r,2}g^{j_1j_2i_3},\,\,\,
\mathbf{T}_{3,3,20}^{i_1i_2i_3}=\lambda^{i_1i_2j_1j_2}g^{j_2j_3i_3}M^{j_3j_1}_{r,2},\,\,\,
\mathbf{T}_{3,3,21}^{i_1i_2i_3}=\lambda^{i_1i_2j_1j_2}g^{j_2j_3i_3}\delta^{j_3j_1}_{r,2},
\end{equation*}
\begin{equation*}\label{29-17-64-1}
\mathbf{T}_{3,3,22}^{i_1i_2i_3}=g^{i_1j_1j_2}\lambda^{j_5j_3j_4i_2}
\tilde{\delta}_{r,1}^{j_1j_5}\lambda^{j_2j_3j_4i_3}
,\,\,\,
\mathbf{T}_{3,3,23}^{i_1i_2i_3}=g^{i_1j_1j_2}\lambda^{j_1j_3j_4i_2}
\tilde{\delta}_{r,1}^{j_4j_5}\lambda^{j_2j_3j_5i_3},
\end{equation*}
\begin{equation*}\label{29-17-64-2}
	\mathbf{T}_{3,3,24}^{i_1i_2i_3}=\lambda^{i_1i_2j_1j_2}g^{j_5j_3j_4}
	\tilde{\delta}_{r,1}^{j_1j_5}\lambda^{j_2j_3j_4i_3}
	,\,\,\,
	\mathbf{T}_{3,3,25}^{i_1i_2i_3}=\lambda^{i_1i_2j_1j_2}g^{j_1j_3j_4}
	\tilde{\delta}_{r,1}^{j_2j_5}\lambda^{j_5j_3j_4i_3},
\end{equation*}
\begin{equation*}\label{29-17-64-3}
	\mathbf{T}_{3,3,26}^{i_1i_2i_3}=\lambda^{i_1i_2j_1j_2}g^{j_1j_3j_4}
	\tilde{\delta}_{r,1}^{j_4j_5}\lambda^{j_2j_3j_5i_3},\,\,\,
	\mathbf{T}_{3,3,27}^{i_1i_2i_3}=\lambda^{i_1i_2j_1j_2}g^{i_3j_3j_4}
	\tilde{\delta}_{r,1}^{j_1j_3}\tilde{\delta}_{r,1}^{j_2j_4},
\end{equation*}
\begin{equation*}\label{29-17-64-4}
	\mathbf{T}_{3,3,28}^{i_1i_2i_3}=\lambda^{i_1i_2j_1j_2}g^{i_3j_1j_4}
	\tilde{\delta}_{r,1}^{j_2j_3}\tilde{\delta}_{r,1}^{j_3j_4}
	,\,\,\,
	\mathbf{T}_{3,3,29}^{i_1i_2i_3}=\lambda^{i_1i_2j_1j_2}g^{i_3j_2j_3}\lambda^{j_1j_3j_4j_5}
	\tilde{\delta}_{r,1}^{j_4j_5},
\end{equation*}
\begin{equation*}\label{29-17-64-5}
	\mathbf{T}_{3,3,30}^{i_1i_2i_3}=\lambda^{i_1i_2j_1j_2}\tilde{g}^{i_3j_2j_3}_{r,1}
	\tilde{\delta}_{r,1}^{j_3j_1}
	,\,\,\,
	\mathbf{T}_{3,3,31}^{i_1i_2i_3}=\tilde{\lambda}^{i_1i_2j_1j_2}_{r,1}g^{i_3j_2j_3}
	\tilde{\delta}_{r,1}^{j_3j_1}.
\end{equation*}

\begin{equation*}\label{29-17-71}
	\mbox{\textbf{Combinations for}}\,\, B_{i_1}B_{i_2}B_{i_3}B_{i_4}:
\end{equation*}
\begin{equation*}\label{29-17-65}
\mathbf{T}_{3,4,1}^{i_1i_2i_3i_4}=\tilde{\lambda}^{i_1i_2j_1j_2}_{r,1}\tilde{\lambda}^{i_3i_4j_1j_2}_{r,1},\,\,\,
\mathbf{T}_{3,4,2}^{i_1i_2i_3i_4}=\lambda^{i_1j_1j_2j_3}\lambda^{i_4j_1j_4j_5}\lambda^{j_4j_5j_6i_2}\lambda^{j_6j_2j_3i_3},
\end{equation*}
\begin{equation*}\label{29-17-66}
\mathbf{T}_{3,4,3}^{i_1i_2i_3i_4}=\lambda^{i_1j_1j_2j_3}\lambda^{i_2j_2j_3j_4}
\tilde{\lambda}^{i_3i_4j_4j_1}_{r,1},\,\,\,
\mathbf{T}_{3,4,4}^{i_1i_2i_3i_4}=\lambda^{i_1j_1j_2j_3}\lambda^{j_2j_3j_4i_2}_{r,1}\lambda^{i_4j_1j_4i_3},
\end{equation*}
\begin{equation*}\label{29-17-67}
\mathbf{T}_{3,4,5}^{i_1i_2i_3i_4}=\lambda^{i_1j_1j_2j_3}\lambda^{i_2i_4j_3j_4}\lambda^{j_4j_5j_6i_3}
\lambda^{j_5j_6j_1j_2},\,\,\,
\mathbf{T}_{3,4,6}^{i_1i_2i_3i_4}=\lambda^{i_4j_1j_2j_3}\lambda^{j_3j_4j_6i_3}\lambda^{j_4j_5i_1i_2}
\lambda^{j_5j_6j_1j_2},
\end{equation*}
\begin{equation*}\label{29-17-68}
\mathbf{T}_{3,4,7}^{i_1i_2i_3i_4}=\lambda^{i_1i_2j_1j_2}\lambda^{j_1j_3j_4j_5}\lambda^{j_5j_6i_3i_4}
\lambda^{j_6j_4j_3j_2},\,\,\,
\mathbf{T}_{3,4,8}^{i_1i_2i_3i_4}=\lambda^{i_1i_2j_1j_2}\lambda^{j_1j_3j_4j_5}
\lambda^{j_5j_4j_3j_6}\lambda^{j_2j_6i_3i_4},
\end{equation*}
\begin{equation*}\label{29-17-69}
\mathbf{T}_{3,4,9}^{i_1i_2i_3i_4}=\lambda^{i_1j_1j_2j_3}\lambda^{i_2j_2j_4j_5}\lambda^{i_3j_4j_1j_6}\lambda^{j_3j_5j_6i_4},\,\,\,
\mathbf{T}_{3,4,10}^{i_1i_2i_3i_4}=\lambda^{i_1i_2j_1j_2}\lambda^{j_1j_2j_3j_4}_{r,1}\lambda^{j_3j_4i_3i_4},
\end{equation*}
\begin{equation*}\label{29-17-70}
\mathbf{T}_{3,4,11}^{i_1i_2i_3i_4}=\lambda^{i_1i_2j_1j_2}\lambda^{j_2j_4i_3i_4}\lambda^{j_1j_4j_3j_3}_{r,1},\,\,\,
\mathbf{T}_{3,4,12}^{i_1i_2i_3i_4}=\lambda^{i_1i_2j_1j_2}\lambda^{j_1j_2i_3i_4}_{r,2},
\end{equation*}
\begin{equation*}\label{29-17-72}
	\mathbf{T}_{3,4,13}^{i_1i_2i_3i_4}=\lambda^{i_1i_2j_1j_2}\lambda^{j_2j_3i_3i_4}M^{j_3j_1}_{r,2},\,\,\,
	\mathbf{T}_{3,4,14}^{i_1i_2i_3i_4}=\lambda^{i_1i_2j_1j_2}\lambda^{j_2j_3i_3i_4}\delta^{j_3j_1}_{r,2},
\end{equation*}
\begin{equation*}\label{29-17-72-1}
	\mathbf{T}_{3,4,15}^{i_1i_2i_3i_4}=\lambda^{i_1i_2j_1j_2}\tilde{\delta}_{r,1}^{j_1j_5}\lambda^{j_5j_3j_4i_3}\lambda^{j_2j_3j_4i_4}
	,\,\,\,
	\mathbf{T}_{3,4,16}^{i_1i_2i_3i_4}=\lambda^{i_1i_2j_1j_2}\lambda^{j_1j_3j_4i_3}\tilde{\delta}_{r,1}^{j_4j_5}\lambda^{j_2j_3j_5i_4}
	,
\end{equation*}
\begin{equation*}\label{29-17-72-2}
	\mathbf{T}_{3,4,17}^{i_1i_2i_3i_4}=\lambda^{i_1i_2j_1j_2}\lambda^{j_3j_4i_3i_4}\tilde{\delta}_{r,1}^{j_1j_3}\tilde{\delta}_{r,1}^{j_2j_4}
	,\,\,\,
	\mathbf{T}_{3,4,18}^{i_1i_2i_3i_4}=\lambda^{i_1i_2j_1j_2}\lambda^{j_1j_4i_3i_4}\tilde{\delta}_{r,1}^{j_2j_3}\tilde{\delta}_{r,1}^{j_3j_4}
	,
\end{equation*}
\begin{equation*}\label{29-17-72-3}
	\mathbf{T}_{3,4,19}^{i_1i_2i_3i_4}=\lambda^{i_1i_2j_1j_2}\lambda^{i_3i_4j_2j_3}\lambda^{j_1j_3j_4j_5}
	\tilde{\delta}_{r,1}^{j_4j_5}
	,\,\,\,
	\mathbf{T}_{3,4,20}^{i_1i_2i_3i_4}=\lambda^{i_1i_2j_1j_2}\tilde{\lambda}^{j_1j_3i_3i_4}_{r,1}\tilde{\delta}_{r,1}^{j_2j_3}
	.
\end{equation*}

\subsection{Local parts for the diagrams}
\label{29:sec:app:kom-1}

\noindent{\textbf{Generalization for $\lambda^2d_1$ from (\ref{29-4-8})--(\ref{29-4-10}):}}
\begin{align*}
&S_i[B]\Big(
2\mathrm{I}_4(\sigma)\mathbf{T}^i_{2,1,1}-
6\mathrm{I}_2(\sigma)\mathbf{T}^i_{2,1,2}-
3\mathrm{I}_3(\sigma)\mathbf{T}^i_{2,1,3}
\Big)-\frac{1}{8}S_{2,i_1i_2}[B]\mathrm{I}_8(\sigma)\mathbf{T}^{i_1i_2}_{2,2,1}+
\\
+
&S_{i_1i_2}[B]\Big(
\mathrm{I}_4(\sigma)\mathbf{T}^{i_1i_2}_{2,2,1}-
3\mathrm{I}_2(\sigma)\mathbf{T}^{i_1i_2}_{2,2,2}-
\frac{3}{2}\mathrm{I}_3(\sigma)\mathbf{T}^{i_1i_2}_{2,2,3}-
6\mathrm{I}_3(\sigma)\mathbf{T}^{i_1i_2}_{2,2,4}
\Big)+
\\
+&S_{i_1i_2i_3}[B]\Big(
-3\mathrm{I}_3(\sigma)\mathbf{T}^{i_1i_2i_3}_{2,3,1}-
3\mathrm{I}_3(\sigma)\mathbf{T}^{i_1i_2i_3}_{2,3,2}
\Big)-\frac{3}{2}S_{i_1i_2i_3i_4}[B]\mathrm{I}_3(\sigma)\mathbf{T}^{i_1i_2i_3i_4}_{2,4,1}.
\end{align*}

\noindent{\textbf{Generalization for $\lambda d_2$ from (\ref{29-4-12})--(\ref{29-4-15}):}}
\begin{align*}
	&S_i[B]\Big(
	-2R_0^\Lambda(0)\mathrm{A}(\sigma)\mathbf{T}^i_{2,1,4}
+2\mathrm{A}^2(\sigma)\mathbf{T}^i_{2,1,5}
	\Big)+
	\\
	+
	&S_{i_1i_2}[B]\Big(
-R_0^\Lambda(0)\mathrm{A}(\sigma)\mathbf{T}^{i_1i_2}_{2,2,5}
+\mathrm{A}^2(\sigma)\mathbf{T}^{i_1i_2}_{2,2,6}
+\mathrm{A}^2(\sigma)\mathbf{T}^{i_1i_2}_{2,2,7}
	\Big)+
	\\
	+&S_{i_1i_2i_3}[B]\mathrm{A}^2(\sigma)\mathbf{T}^{i_1i_2i_3}_{2,3,3}
	+\frac{1}{4}S_{i_1i_2i_3i_4}[B]\mathrm{A}^2(\sigma)\mathbf{T}^{i_1i_2i_3i_4}_{2,4,2}+\\\nonumber+&
	R_0^\Lambda(0)R_{22}^\Lambda(0)\int_{\mathbb{R}^4}\mathrm{d}^4x\Big(V^{i_1j}(x)V^{ji_2}(x)-M^{i_1j}M^{ji_2}\Big)\lambda^{i_1i_2kk}.
\end{align*}

\noindent{\textbf{Generalization for $cd_1$ from (\ref{29-4-16})--(\ref{29-4-18}):}}
\begin{align*}
	&S_i[B]\Big(
	-\mathrm{A}(\sigma)\mathbf{T}^i_{2,1,6}+
	R_0^\Lambda(0)\mathbf{T}^i_{2,1,7}
	-\mathrm{A}(\sigma)\mathbf{T}^i_{2,1,8}
	\Big)+
	\\
	+
	&S_{i_1i_2}[B]\bigg(
	-\frac{1}{2}\mathrm{A}(\sigma)\mathbf{T}^{i_1i_2}_{2,2,8}
	-\mathrm{A}(\sigma)\mathbf{T}^{i_1i_2}_{2,2,9}+
	\frac{1}{2}R_0^\Lambda(0)\mathbf{T}^{i_1i_2}_{2,2,10}-
	\frac{1}{2}\mathrm{A}(\sigma)\mathbf{T}^{i_1i_2}_{2,2,11}
	\bigg)+
	\\
	+&S_{i_1i_2i_3}[B]
	\bigg(-\frac{1}{2}\mathrm{A}(\sigma)\mathbf{T}^{i_1i_2i_3}_{2,3,4}-\frac{1}{2}\mathrm{A}(\sigma)\mathbf{T}^{i_1i_2i_3}_{2,3,5}\bigg)
	-\frac{1}{4}S_{i_1i_2i_3i_4}[B]\mathrm{A}(\sigma)\mathbf{T}^{i_1i_2i_3i_4}_{2,4,3}-\\\nonumber-&
	\frac{1}{4}
	R_0^\Lambda(0)R_{22}^\Lambda(0)\int_{\mathbb{R}^4}\mathrm{d}^4x\Big(V^{i_1j}(x)V^{ji_2}(x)-M^{i_1j}M^{ji_2}\Big)\lambda^{i_1i_2kk}.
\end{align*}

\noindent{\textbf{Generalization for the linear combination from (\ref{29-7-10}):}}
\begin{equation*}\label{29-18-5}
S_{i}[B]\mathbf{\overline{T}}^{i}_{3,1,1}+
S_{i_1i_2}[B]\mathbf{\overline{T}}^{i_1i_2}_{3,2,1}+
S_{i_1i_2i_3}[B]\mathbf{\overline{T}}^{i_1i_2i_3}_{3,3,1}+
S_{i_1i_2i_3i_4}[B]\mathbf{\overline{T}}^{i_1i_2i_3i_4}_{3,4,1},
\end{equation*}
where
\begin{equation*}\label{29-18-6}
\mathbf{\overline{T}}^{i}_{3,1,1}=
\frac{1}{2}\mathrm{A}(\sigma)\mathbf{T}^{i}_{3,1,1}+
\frac{1}{4}\mathrm{I}_5(\sigma)\mathbf{T}^{i}_{3,1,2}-
\frac{1}{4}\mathrm{I}_8(\sigma)\Big(\mathbf{T}^{i}_{3,1,3}+2\mathbf{T}^{i}_{3,1,4}\Big),
\end{equation*}
\begin{equation*}\label{29-18-7}
\mathbf{\overline{T}}^{i_1i_2}_{3,2,1}=
\frac{1}{4}\mathrm{A}(\sigma)\Big(\mathbf{T}^{i_1i_2}_{3,2,1}+\mathbf{T}^{i_1i_2}_{3,2,2}\Big)+
\frac{1}{8}\mathrm{I}_5(\sigma)
\Big(\mathbf{T}^{i_1i_2}_{3,2,3}+\mathbf{T}^{i_1i_2}_{3,2,4}+\mathbf{T}^{i_1i_2}_{3,2,5}\Big)-
\frac{1}{8}\mathrm{I}_8(\sigma)\Big(
2\mathbf{T}^{i_1i_2}_{3,2,6}+4\mathbf{T}^{i_1i_2}_{3,2,7}+\mathbf{T}^{i_1i_2}_{3,2,8}
\Big),
\end{equation*}
\begin{equation*}\label{29-18-8}
\mathbf{\overline{T}}^{i_1i_2i_3}_{3,3,1}=
\frac{1}{4}\mathrm{A}(\sigma)\mathbf{T}^{i_1i_2i_3}_{3,3,1}+
\frac{1}{4}\mathrm{I}_5(\sigma)\mathbf{T}^{i_1i_2i_3}_{3,3,2}-
\frac{1}{4}\mathrm{I}_8(\sigma)\Big(\mathbf{T}^{i_1i_2i_3}_{3,3,3}+
\mathbf{T}^{i_1i_2i_3}_{3,3,4}\Big),
\end{equation*}
\begin{equation*}\label{29-18-9}
\mathbf{\overline{T}}^{i_1i_2i_3i_4}_{3,4,1}=\frac{1}{16}\mathrm{A}(\sigma)
\mathbf{T}^{i_1i_2i_3i_4}_{3,4,1}+\frac{1}{16}\mathrm{I}_5(\sigma)\mathbf{T}^{i_1i_2i_3i_4}_{3,4,2}-
\frac{1}{8}\mathrm{I}_8(\sigma)\mathbf{T}^{i_1i_2i_3i_4}_{3,4,3}.
\end{equation*}

\noindent{\textbf{Generalization for the linear combination from (\ref{29-7-13}):}}
\begin{equation*}\label{29-18-51}
	S_{i}[B]\mathbf{\overline{T}}^{i}_{3,1,2}+
	S_{i_1i_2}[B]\mathbf{\overline{T}}^{i_1i_2}_{3,2,2}+
	S_{i_1i_2i_3}[B]\mathbf{\overline{T}}^{i_1i_2i_3}_{3,3,2}+
	S_{i_1i_2i_3i_4}[B]\mathbf{\overline{T}}^{i_1i_2i_3i_4}_{3,4,2}+
\end{equation*}
\begin{equation*}\label{29-18-10}
+S_{2,i_1i_2}[B]\bigg(-\frac{1}{48}\mathrm{I}_1(\sigma)\mathbf{T}^{i_1i_2}_{3,2,9}+
\frac{1}{64}\mathrm{I}_6(\sigma)\mathbf{T}^{i_1i_2}_{3,2,14}-
\frac{1}{32}\mathrm{A}(\sigma)
\mathrm{I}_1(\sigma)\mathbf{T}^{i_1i_2}_{3,2,14}\bigg),
\end{equation*}
where
\begin{multline*}
	\mathbf{\overline{T}}^{i}_{3,1,2}=
	\frac{1}{6}\mathrm{I}_4(\sigma)\mathbf{T}^{i}_{3,1,5}+
	\frac{1}{6}\mathrm{I}_4(\sigma)\mathbf{T}^{i}_{3,1,6}-
	\frac{1}{2}\mathrm{I}_2(\sigma)\mathbf{T}^{i}_{3,1,7}-
	\frac{1}{2}\mathrm{I}_2(\sigma)\mathbf{T}^{i}_{3,1,8}-
	\frac{1}{2}\mathrm{I}_3(\sigma)\mathbf{T}^{i}_{3,1,9}-\\
	-\frac{1}{4}\big(\mathrm{I}_7(\sigma)-\mathrm{A}(\sigma)\mathrm{I}_4(\sigma)\big)
	\mathbf{T}^{i}_{3,1,10}-
	\frac{\mathrm{A}(\sigma)\mathrm{I}_3(\sigma)-\mathrm{I}_{10}(\sigma)}{8}
	\Big(2\mathbf{T}^{i}_{3,1,11}+\mathbf{T}^{i}_{3,1,12}\Big)+\\+
	\frac{1}{2}\mathrm{I}_{12}(\sigma)
	\Big(\mathbf{T}^{i}_{3,1,13}+\mathbf{T}^{i}_{3,1,14}+\mathbf{T}^{i}_{3,1,15}\Big),
\end{multline*}
\begin{multline*}
	\mathbf{\overline{T}}^{i_1i_2}_{3,2,2}=
	\frac{1}{6}\mathrm{I}_4(\sigma)\mathbf{T}^{i_1i_2}_{3,2,9}
	-\frac{1}{2}\mathrm{I}_2(\sigma)\mathbf{T}^{i_1i_2}_{3,2,10}
	-\frac{1}{4}\mathrm{I}_3(\sigma)\mathbf{T}^{i_1i_2}_{3,2,11}
	-\frac{1}{2}\mathrm{I}_3(\sigma)\mathbf{T}^{i_1i_2}_{3,2,12}
	-\frac{1}{2}\mathrm{I}_3(\sigma)\mathbf{T}^{i_1i_2}_{3,2,13}-\\
	-\frac{1}{8}\big(\mathrm{I}_7(\sigma)-\mathrm{A}(\sigma)\mathrm{I}_4(\sigma)\big)\mathbf{T}^{i_1i_2}_{3,2,14}
	-\frac{\mathrm{A}(\sigma)\mathrm{I}_3(\sigma)-\mathrm{I}_{10}(\sigma)}{16}
	\Big(2\mathbf{T}^{i_1i_2}_{3,2,15}+4\mathbf{T}^{i_1i_2}_{3,2,16}+\mathbf{T}^{i_1i_2}_{3,2,17}\Big)+\\+
	\frac{1}{4}\mathrm{I}_{12}(\sigma)\Big(\mathbf{T}^{i_1i_2}_{3,2,18}+2\mathbf{T}^{i_1i_2}_{3,2,19}+2\mathbf{T}^{i_1i_2}_{3,2,20}+2\mathbf{T}^{i_1i_2}_{3,2,21}\Big),
\end{multline*}
\begin{multline*}
	\mathbf{\overline{T}}^{i_1i_2i_3}_{3,3,2}=
	-\frac{1}{2}\mathrm{I}_3(\sigma)\mathbf{T}^{i_1i_2i_3}_{3,3,5}
	-\frac{1}{4}\mathrm{I}_3(\sigma)\mathbf{T}^{i_1i_2i_3}_{3,3,6}
	-\frac{1}{4}\mathrm{I}_3(\sigma)\mathbf{T}^{i_1i_2i_3}_{3,3,7}-\\
	-\frac{\mathrm{A}(\sigma)\mathrm{I}_3(\sigma)-\mathrm{I}_{10}(\sigma)}{8}
	\Big(\mathbf{T}^{i_1i_2i_3}_{3,3,8}+\mathbf{T}^{i_1i_2i_3}_{3,3,9}\Big)+
	\frac{1}{4}\mathrm{I}_{12}(\sigma)
	\Big(\mathbf{T}^{i_1i_2i_3}_{3,3,10}+\mathbf{T}^{i_1i_2i_3}_{3,3,11}+2\mathbf{T}^{i_1i_2i_3}_{3,3,12}\Big),
\end{multline*}
\begin{equation*}\label{29-18-91}
	\mathbf{\overline{T}}^{i_1i_2i_3i_4}_{3,4,2}=
	-\frac{1}{4}\mathrm{I}_3(\sigma)
	\mathbf{T}^{i_1i_2i_3i_4}_{3,4,4}
	-\frac{1}{16}\big(\mathrm{A}(\sigma)\mathrm{I}_3(\sigma)-\mathrm{I}_{10}(\sigma)\big)\mathbf{T}^{i_1i_2i_3i_4}_{3,4,5}+
	\frac{1}{4}\mathrm{I}_{12}(\sigma)\mathbf{T}^{i_1i_2i_3i_4}_{3,4,6}.
\end{equation*}

\noindent{\textbf{Generalization for the sum of (\ref{29-7-16}) and $\lambda^4d_4/24$ from Lemma \ref{29-l-3}:}}
\begin{equation*}\label{29-18-52}
	S_{i}[B]\mathbf{\overline{T}}^{i}_{3,1,3}+
	S_{i_1i_2}[B]\mathbf{\overline{T}}^{i_1i_2}_{3,2,3}+
	S_{i_1i_2i_3}[B]\mathbf{\overline{T}}^{i_1i_2i_3}_{3,3,3}+
	S_{i_1i_2i_3i_4}[B]\mathbf{\overline{T}}^{i_1i_2i_3i_4}_{3,4,3},
\end{equation*}
where
\begin{equation*}\label{29-18-62}
	\mathbf{\overline{T}}^{i}_{3,1,3}=
	-\frac{1}{12}\mathrm{I}_{14}(\sigma)\mathbf{T}^{i}_{3,1,16}
	+\frac{1}{4}\mathrm{I}_{17}(\sigma)\mathbf{T}^{i}_{3,1,17}
	+\frac{1}{12}\mathrm{I}_{20}(\sigma)\mathbf{T}^{i}_{3,1,18}
	+\frac{\zeta(3)L}{32(4\pi^2)^3}\mathbf{T}^{i}_{3,1,19},
\end{equation*}
\begin{equation*}\label{29-18-72}
	\mathbf{\overline{T}}^{i_1i_2}_{3,2,3}=
	-\frac{1}{24}\mathrm{I}_{14}(\sigma)\mathbf{T}^{i_1i_2}_{3,2,22}
	+\frac{1}{8}\mathrm{I}_{17}(\sigma)\Big(\mathbf{T}^{i_1i_2}_{3,2,23}+\mathbf{T}^{i_1i_2}_{3,2,24}\Big)
	+\frac{1}{24}\mathrm{I}_{20}(\sigma)\Big(\mathbf{T}^{i_1i_2}_{3,2,40}+\mathbf{T}^{i_1i_2}_{3,2,41}\Big)
	+\frac{3\zeta(3)L}{64(4\pi^2)^3}\mathbf{T}^{i_1i_2}_{3,2,25},
\end{equation*}
\begin{equation*}\label{29-18-82}
	\mathbf{\overline{T}}^{i_1i_2i_3}_{3,3,3}=
	\frac{1}{8}\mathrm{I}_{17}(\sigma)\mathbf{T}^{i_1i_2i_3}_{3,3,13}
	+\frac{1}{24}\mathrm{I}_{20}(\sigma)\mathbf{T}^{i_1i_2i_3}_{3,3,14}
	+\frac{\zeta(3)L}{32(4\pi^2)^3}\mathbf{T}^{i_1i_2i_3}_{3,3,15},
\end{equation*}
\begin{equation*}\label{29-18-92}
	\mathbf{\overline{T}}^{i_1i_2i_3i_4}_{3,4,3}=
	\frac{1}{32}\mathrm{I}_{17}(\sigma)\mathbf{T}^{i_1i_2i_3i_4}_{3,4,7}+
	\frac{1}{96}\mathrm{I}_{20}(\sigma)\mathbf{T}^{i_1i_2i_3i_4}_{3,4,8}+
	\frac{\zeta(3)L}{128(4\pi^2)^3}\mathbf{T}^{i_1i_2i_3i_4}_{3,4,9}.
\end{equation*}

\noindent{\textbf{Generalization for the diagram $-\lambda z_{4,1}d_2/8$:}}
\begin{equation*}\label{29-18-53}
	S_{i}[B]\mathbf{\overline{T}}^{i}_{3,1,4}+
	S_{i_1i_2}[B]\mathbf{\overline{T}}^{i_1i_2}_{3,2,4}+
	S_{i_1i_2i_3}[B]\mathbf{\overline{T}}^{i_1i_2i_3}_{3,3,4}+
	S_{i_1i_2i_3i_4}[B]\mathbf{\overline{T}}^{i_1i_2i_3i_4}_{3,4,4},
\end{equation*}
where
\begin{equation*}\label{29-18-63}
	\mathbf{\overline{T}}^{i}_{3,1,4}=
	\frac{1}{4}R_0^\Lambda(0)\mathrm{A}(\sigma)\mathbf{T}^{i}_{3,1,20}-
	\frac{1}{4}\mathrm{A}^2(\sigma)\mathbf{T}^{i}_{3,1,21}-
	\frac{1}{4}R_0^\Lambda(0)R_{22}^\Lambda(0)\mathbf{T}^{i}_{3,1,22},
\end{equation*}
\begin{equation*}\label{29-18-73}
	\mathbf{\overline{T}}^{i_1i_2}_{3,2,4}=
	\frac{1}{8}R_0^\Lambda(0)\mathrm{A}(\sigma)\mathbf{T}^{i_1i_2}_{3,2,26}-
	\frac{1}{8}\mathrm{A}^2(\sigma)\Big(\mathbf{T}^{i_1i_2}_{3,2,27}+\mathbf{T}^{i_1i_2}_{3,2,28}\Big)-
	\frac{1}{8}R_0^\Lambda(0)R_{22}^\Lambda(0)\Big(\mathbf{T}^{i_1i_2}_{3,2,29}+\mathbf{T}^{i_1i_2}_{3,2,30}\Big),
\end{equation*}
\begin{equation*}\label{29-18-83}
	\mathbf{\overline{T}}^{i_1i_2i_3}_{3,3,4}=
	-\frac{1}{8}\mathrm{A}^2(\sigma)\mathbf{T}^{i_1i_2i_3}_{3,3,16}-
	\frac{1}{8}R_0^\Lambda(0)R_{22}^\Lambda(0)\mathbf{T}^{i_1i_2i_3}_{3,3,17},
\end{equation*}
\begin{equation*}\label{29-18-93}
	\mathbf{\overline{T}}^{i_1i_2i_3i_4}_{3,4,4}=
	-\frac{1}{32}\mathrm{A}^2(\sigma)\mathbf{T}^{i_1i_2i_3i_4}_{3,4,10}-
	\frac{1}{32}R_0^\Lambda(0)R_{22}^\Lambda(0)\mathbf{T}^{i_1i_2i_3i_4}_{3,4,11}.
\end{equation*}

\noindent{\textbf{Generalization for the diagram from formula (\ref{29-7-19}):}}
\begin{equation*}\label{29-18-54}
	S_{i}[B]\mathbf{\overline{T}}^{i}_{3,1,5}+
	S_{i_1i_2}[B]\mathbf{\overline{T}}^{i_1i_2}_{3,2,5}+
	S_{i_1i_2i_3}[B]\mathbf{\overline{T}}^{i_1i_2i_3}_{3,3,5}+
	S_{i_1i_2i_3i_4}[B]\mathbf{\overline{T}}^{i_1i_2i_3i_4}_{3,4,5},
\end{equation*}
where
\begin{multline*}
	\mathbf{\overline{T}}^{i}_{3,1,5}=
	\frac{1}{2}\mathrm{A}(\sigma)\mathbf{T}^{i}_{3,1,23}
	-\frac{1}{2}R_0^\Lambda(0)\mathbf{T}^{i}_{3,1,24}+
	\frac{1}{2}\mathrm{A}(\sigma)\mathbf{T}^{i}_{3,1,25}+
	\frac{1}{2}R_0^\Lambda(0)\mathbf{T}^{i}_{3,1,26}-\\-
	\frac{1}{2}R_{22}^\Lambda(0)\mathbf{T}^{i}_{3,1,27}
	-\mathrm{A}(\sigma)\mathbf{T}^{i}_{3,1,28},
\end{multline*}
\begin{multline*}
	\mathbf{\overline{T}}^{i_1i_2}_{3,2,5}=
	\frac{1}{4}\mathrm{A}(\sigma)\mathbf{T}^{i_1i_2}_{3,2,31}+
	\frac{1}{2}\mathrm{A}(\sigma)\mathbf{T}^{i_1i_2}_{3,2,32}-
	\frac{1}{4}R_{0}^\Lambda(0)\mathbf{T}^{i_1i_2}_{3,2,33}+
	\frac{1}{4}\mathrm{A}(\sigma)\mathbf{T}^{i_1i_2}_{3,2,34}+
	\frac{1}{4}R_{0}^\Lambda(0)\mathbf{T}^{i_1i_2}_{3,2,35}-\\-\frac{1}{4}R_{22}^\Lambda(0)
	\Big(\mathbf{T}^{i_1i_2}_{3,2,36}+\mathbf{T}^{i_1i_2}_{3,2,37}\Big)-
	\frac{1}{2}\mathrm{A}(\sigma)\Big(\mathbf{T}^{i_1i_2}_{3,2,38}+\mathbf{T}^{i_1i_2}_{3,2,39}\Big),
\end{multline*}
\begin{equation*}\label{29-18-84}
	\mathbf{\overline{T}}^{i_1i_2i_3}_{3,3,5}=
	\frac{1}{4}\mathrm{A}(\sigma)\mathbf{T}^{i_1i_2i_3}_{3,3,18}+
	\frac{1}{4}\mathrm{A}(\sigma)\mathbf{T}^{i_1i_2i_3}_{3,3,19}-
	\frac{1}{4}R_{22}^\Lambda(0)\mathbf{T}^{i_1i_2i_3}_{3,3,20}-
	\frac{1}{2}\mathrm{A}(\sigma)\mathbf{T}^{i_1i_2i_3}_{3,3,21},
\end{equation*}
\begin{equation*}\label{29-18-94}
	\mathbf{\overline{T}}^{i_1i_2i_3i_4}_{3,4,5}=
	\frac{1}{8}\mathrm{A}(\sigma)\mathbf{T}^{i_1i_2i_3i_4}_{3,4,12}-
	\frac{1}{16}R_{22}^\Lambda(0)\mathbf{T}^{i_1i_2i_3i_4}_{3,4,13}-
	\frac{1}{8}\mathrm{A}(\sigma)\mathbf{T}^{i_1i_2i_3i_4}_{3,4,14}.
\end{equation*}

\noindent{\textbf{Generalization for the diagram from formula (\ref{29-21-2}):}}
\begin{equation*}\label{29-18-51-1}
	S_{i}[B]\mathbf{\overline{T}}^{i}_{3,1,6}+
	S_{i_1i_2}[B]\mathbf{\overline{T}}^{i_1i_2}_{3,2,6}+
	S_{i_1i_2i_3}[B]\mathbf{\overline{T}}^{i_1i_2i_3}_{3,3,6}+
	S_{i_1i_2i_3i_4}[B]\mathbf{\overline{T}}^{i_1i_2i_3i_4}_{3,4,6}+
\end{equation*}
\begin{equation*}\label{29-18-10-1}
	+S_{2,i_1i_2}[B]\bigg(\frac{1}{32}\mathrm{I}_1(\sigma)\mathbf{T}^{i_1i_2}_{3,2,42}\bigg),
\end{equation*}
where
\begin{multline*}
	\mathbf{\overline{T}}^{i}_{3,1,6}=
	\bigg(-\frac{1}{2}\mathrm{I}_4(\sigma)+\frac{1}{2}R_0^\Lambda(0)\mathrm{A}(\sigma)\bigg)\mathbf{T}^{i}_{3,1,29}
	+\bigg(\frac{1}{4}\mathrm{I}_2(\sigma)+\frac{1}{4}\mathrm{I}_8(\sigma)-\frac{1}{4}\mathrm{A}^2(\sigma)\bigg)\mathbf{T}^{i}_{3,1,30}+\\
	+\mathrm{I}_2(\sigma)\mathbf{T}^{i}_{3,1,31}
	+\frac{1}{2}\mathrm{I}_3(\sigma)\mathbf{T}^{i}_{3,1,32}
	+\bigg(\frac{1}{4}\mathrm{I}_3(\sigma)+\frac{1}{4}\mathrm{I}_8(\sigma)-\frac{1}{4}\mathrm{A}^2(\sigma)\bigg)\mathbf{T}^{i}_{3,1,33}
	+~~~~~\\+\frac{1}{2}\mathrm{A}(\sigma)\mathbf{T}^{i}_{3,1,34}
	-\frac{1}{2}R_0^\Lambda(0)\mathbf{T}^{i}_{3,1,35}
	+\mathrm{A}(\sigma)\mathbf{T}^{i}_{3,1,36}
	+\frac{1}{4}R_0^\Lambda(0)R_{22}^\Lambda(0)\mathbf{T}^{i}_{3,1,37}
	+\\+\frac{1}{2}R_0^\Lambda(0)\mathbf{T}^{i}_{3,1,38}
	-\mathrm{A}(\sigma)\mathbf{T}^{i}_{3,1,39}
	-\mathrm{A}(\sigma)\mathbf{T}^{i}_{3,1,40},
\end{multline*}
\begin{multline*}
	\mathbf{\overline{T}}^{i_1i_2}_{3,2,6}=
	\bigg(-\frac{1}{4}\mathrm{I}_4(\sigma)+\frac{1}{4}R_0^\Lambda(0)\mathrm{A}(\sigma)\bigg)\mathbf{T}^{i_1i_2}_{3,2,42}
	+\bigg(\frac{1}{4}\mathrm{I}_2(\sigma)+\frac{1}{4}\mathrm{I}_8(\sigma)-\frac{1}{4}\mathrm{A}^2(\sigma)\bigg)\mathbf{T}^{i_1i_2}_{3,2,43}+\\
	+\frac{1}{2}\mathrm{I}_2(\sigma)\mathbf{T}^{i_1i_2}_{3,2,44}
	+\bigg(\frac{1}{8}\mathrm{I}_3(\sigma)+\frac{1}{8}\mathrm{I}_8(\sigma)-\frac{1}{8}\mathrm{A}^2(\sigma)\bigg)\mathbf{T}^{i_1i_2}_{3,2,45}
	+\frac{1}{4}\mathrm{I}_3(\sigma)\mathbf{T}^{i_1i_2}_{3,2,46}
	+~~~~~~~~~~~~~~~\\+\bigg(\frac{1}{4}\mathrm{I}_3(\sigma)+\frac{1}{4}\mathrm{I}_8(\sigma)-\frac{1}{4}\mathrm{A}^2(\sigma)\bigg)\mathbf{T}^{i_1i_2}_{3,2,47}
	+\bigg(\frac{1}{4}\mathrm{I}_3(\sigma)+\frac{1}{4}\mathrm{I}_8(\sigma)-\frac{1}{4}\mathrm{A}^2(\sigma)\bigg)\mathbf{T}^{i_1i_2}_{3,2,48}
	+~~~\\+\mathrm{I}_3(\sigma)\mathbf{T}^{i_1i_2}_{3,2,49}
	+\frac{1}{4}\mathrm{A}(\sigma)\mathbf{T}^{i_1i_2}_{3,2,50}
	+\frac{1}{4}\mathrm{A}(\sigma)\mathbf{T}^{i_1i_2}_{3,2,51}
	-\frac{1}{4}R_0^\Lambda(0)\mathbf{T}^{i_1i_2}_{3,2,52}
	+\frac{1}{2}\mathrm{A}(\sigma)\mathbf{T}^{i_1i_2}_{3,2,53}
	+\\+\frac{1}{2}\mathrm{A}(\sigma)\mathbf{T}^{i_1i_2}_{3,2,54}
	+\frac{1}{8}R_0^\Lambda(0)R_{22}^\Lambda(0)\mathbf{T}^{i_1i_2}_{3,2,55}
	+\frac{1}{8}R_0^\Lambda(0)R_{22}^\Lambda(0)\mathbf{T}^{i_1i_2}_{3,2,56}
	+\\+\frac{1}{4}R_0^\Lambda(0)\mathbf{T}^{i_1i_2}_{3,2,57}
	-\frac{1}{2}\mathrm{A}(\sigma)\mathbf{T}^{i_1i_2}_{3,2,58}
	-\mathrm{A}(\sigma)\mathbf{T}^{i_1i_2}_{3,2,59}
	-\frac{1}{2}\mathrm{A}(\sigma)\mathbf{T}^{i_1i_2}_{3,2,60},
\end{multline*}
\begin{multline*}
	\mathbf{\overline{T}}^{i_1i_2i_3}_{3,3,6}=
	\bigg(\frac{1}{4}\mathrm{I}_3(\sigma)+\frac{1}{4}\mathrm{I}_8(\sigma)-\frac{1}{4}\mathrm{A}^2(\sigma)\bigg)\mathbf{T}^{i_1i_2i_3}_{3,3,22}
	+\frac{1}{2}\mathrm{I}_3(\sigma)\mathbf{T}^{i_1i_2i_3}_{3,3,23}
	+\\+\bigg(\frac{1}{8}\mathrm{I}_3(\sigma)+\frac{1}{8}\mathrm{I}_8(\sigma)-\frac{1}{8}\mathrm{A}^2(\sigma)\bigg)\mathbf{T}^{i_1i_2i_3}_{3,3,24}
	+\bigg(\frac{1}{8}\mathrm{I}_3(\sigma)+\frac{1}{8}\mathrm{I}_8(\sigma)-\frac{1}{8}\mathrm{A}^2(\sigma)\bigg)\mathbf{T}^{i_1i_2i_3}_{3,3,25}
	+\\+\frac{1}{2}\mathrm{I}_3(\sigma)\mathbf{T}^{i_1i_2i_3}_{3,3,26}
	+\frac{1}{4}\mathrm{A}(\sigma)\mathbf{T}^{i_1i_2i_3}_{3,3,27}
	+\frac{1}{2}\mathrm{A}(\sigma)\mathbf{T}^{i_1i_2i_3}_{3,3,28}
	+\frac{1}{8}R_0^\Lambda(0)R_{22}^\Lambda(0)\mathbf{T}^{i_1i_2i_3}_{3,3,29}
	-\\-\frac{1}{2}\mathrm{A}(\sigma)\mathbf{T}^{i_1i_2i_3}_{3,3,30}
	-\frac{1}{2}\mathrm{A}(\sigma)\mathbf{T}^{i_1i_2i_3}_{3,3,31},
\end{multline*}
\begin{multline*}
	\mathbf{\overline{T}}^{i_1i_2i_3i_4}_{3,4,6}=
	\bigg(\frac{1}{8}\mathrm{I}_3(\sigma)+\frac{1}{8}\mathrm{I}_8(\sigma)-\frac{1}{8}\mathrm{A}^2(\sigma)\bigg)\mathbf{T}^{i_1i_2i_3i_4}_{3,4,15}
	+\frac{1}{4}\mathrm{I}_3(\sigma)\mathbf{T}^{i_1i_2i_3i_4}_{3,4,16}
	+\frac{1}{16}\mathrm{A}(\sigma)\mathbf{T}^{i_1i_2i_3i_4}_{3,4,17}
	+\\+\frac{1}{8}\mathrm{A}(\sigma)\mathbf{T}^{i_1i_2i_3i_4}_{3,4,18}
	+\frac{1}{32}R_0^\Lambda(0)R_{22}^\Lambda(0)\mathbf{T}^{i_1i_2i_3i_4}_{3,4,19}
	-\frac{1}{4}\mathrm{A}(\sigma)\mathbf{T}^{i_1i_2i_3i_4}_{3,4,20}.
\end{multline*}


\begin{thebibliography}{99}
\bibitem{9}
C. Itzykson, J. B. Zuber, \textit{Quantum Field Theory}, Mcgraw-hill, New York (1980)
\bibitem{10}
M. E. Peskin, D. V. Schroeder, \textit{An Introduction to Quantum Field Theory}, Addison-Wesley, 1--868 (1995)

\bibitem{3}
L. D. Faddeev, A. A. Slavnov, \textit{Gauge Fields: An Introduction to Quantum Theory}, Frontiers in Physics \textbf{83}, Addison-Wesley, 1--236 (1991)

\bibitem{6}
J. C. Collins, \textit{Renormalization: An Introduction to Renormalization, the Renormalization Group and the Operator-Product Expansion}, Cambridge University Press (1984)
\bibitem{7}
O. I. Zavialov, \textit{Renormalized quantum field theory}, Kluwer Academic Publishers, Dodrecht, Boston, 1--524 (1990)
\bibitem{105}
D. I. Kazakov, \textit{Radiative Corrections, Divergences, Regularization, Renormalization, Renormalization Group and All That in Examples in Quantum Field Theory}, arXiv:0901.2208 [hep-ph] (2009)

\bibitem{w6}
M. Oleszczuk, \textit{A symmetry-preserving cut-off regularization},
Z. Phys. C, \textbf{64}, 533--538 (1994)
\bibitem{w7}
Sen-Ben Liao, \textit{Operator Cutoff Regularization and Renormalization Group in 
	Yang-Mills Theory}, Phys. Rev. D, \textbf{56}, 5008--5033 (1997)
\bibitem{w8}
G. Cynolter, E. Lendvai, \textit{Cutoff Regularization Method in Gauge Theories},
[arXiv:1509.07407 [hep-ph]] (2015)
\bibitem{Khar-2020}
N. V. Kharuk, \textit{Mixed type regularizations and nonlogarithmic singularities}, Questions of quantum field theory and statistical physics. Part 27, Zap. Nauchn. Sem. POMI, \textbf{494}, POMI, St. Petersburg, 2020, 242--249; J. Math. Sci. (N. Y.), \textbf{264}, 362--367 (2022) 10.1007/s10958-022-06003-7


\bibitem{19}
C. G. Bollini, J. J. Giambiagi, \textit{Dimensional Renormalization: The Number of Dimensions as a Regularizing Parameter}, Nuovo Cim. B, \textbf{12}, 20--26 (1972)
\bibitem{555}
G. 't Hooft, \textit{Dimensional regularization and the renormalization group}, Nucl. Phys. B, \textbf{61}, 455--468 (1973)


\bibitem{chi-0}
A. Brizola, O. Battistel, Marcos Sampaio, M. C. Nemes, \textit{Implicit Regularisation Technique: Calculation of the Two-loop $\phi^4_4$-theory $\beta$-function}, Mod. Phys. Lett. A, \textbf{14}, 1509--1518 (1999)
\bibitem{chi-1}
A. L. Cherchiglia, M. Sampaio, M. C. Nemes, \textit{Systematic Implementation of Implicit Regularization for Multi-Loop Feynman Diagrams}, Int. J. Mod. Phys. A, \textbf{26}, 2591--2635 (2011)
\bibitem{chi-2}
A. Cherchiglia, D. C. Arias-Perdomo, A. R. Vieira, M. Sampaio, B. Hiller, \textit{Two-loop renormalisation of gauge theories in 4D Implicit Regularisation and connections to dimensional methods}, Eur. Phys. J. C, \textbf{81}, 468 (2021)

\bibitem{Pauli-Villars}
W. Pauli, F. Villars, \textit{On the Invariant Regularization in Relativistic Quantum Theory}, Rev. Mod. Phys. \textbf{21}(3): 434--444 (1949)

\bibitem{Bakeyev-Slavnov}
T. Bakeyev, A. Slavnov, \textit{Higher covariant derivative regularization revisited}, Mod. Phys. Lett. A \textbf{11}(19), 1539--1554 (1996)

\bibitem{Gelfand-1964}
I. M. Gel'fand, G. E. Shilov, 
\textit{Generalized Functions, Volume 1: Properties and Operations},
AMS Chelsea Publishing \textbf{377}, 1--423 (1964)
\bibitem{Vladimirov-2002}
V. S. Vladimirov, \textit{Methods of the theory of generalized functions}, London,
CRC Press, 1--328 (2002)

\bibitem{Gel}
I. M. Gel'fand, M. I. Graev, N. Ya. Vilenkin, \textit{Generalized functions. Vol. 5: Integral geometry and representation theory}, Translated from the Russian by Eugene Saletan, Boston, MA: Academic Press, 1--449 (1966)
\bibitem{I-1}
S. E. Derkachev, A. V. Ivanov, \textit{Racah coefficients for the group $\mathrm{SL}(2,\mathbb{R})$}, Zap. Nauchn. Sem. POMI, \textbf{509} (2021),  99--112; J. Math. Sci. (N. Y.), \textbf{275}:3 (2023), 289--298 10.1007/s10958-023-06681-x

\bibitem{I-2}
A. V. Ivanov, \textit{On the completeness of projectors for tensor product decomposition of continuous series representations groups $\mathrm{SL}(2,\mathbb{R})$}, Zap. Nauchn. Sem. POMI, \textbf{473}, POMI, St. Petersburg, 2018, 161--173; J. Math. Sci. (N. Y.), \textbf{242}:5 (2019), 692--700 10.1007/s10958-019-04507-3
\bibitem{I-3}
S. E. Derkachev, A. V. Ivanov, L. A. Shumilov, \textit{Mellin--Barnes transformation for two-loop master-diagrams}, Zap. Nauchn. Sem. POMI, \textbf{494}, POMI, St. Petersburg, 2020, 144--167; J. Math. Sci. (N. Y.), \textbf{264}:3 (2022), 298--312 10.1007/s10958-022-05998-3


\bibitem{Soh}
Y. W. Sokhotskii, \textit{On definite integrals and functions used in series expansions}, published by M. Stasyulevich, St. Petersburg (1873)

\bibitem{29-4}
H. Kleinert, V. Schulte-Frohlinde, \textit{Critical properties of $\phi^4$-theories}, World Scientific, Singapore, 1--512 (2001)
\bibitem{29-3}
A. N. Vasil'ev, \textit{The field theoretic renormalization group in critical behavior theory and stochastic dynamics}, Boca Raton: Chapman and Hall/CRC, 1--681 (2004)

\bibitem{34}
A. V. Ivanov, N. V. Kharuk, \textit{Quantum equation of motion and two-loop cutoff renormalization for $\phi^3$ model}, Zap. Nauchn. Sem. POMI, \textbf{487}, POMI, St. Petersburg, 2019, 151--166; J. Math. Sci. (N. Y.), \textbf{257}:4 (2021), 526--536, arXiv:2203.04562, 10.1007/s10958-021-05500-5
\bibitem{Ivanov-Kharuk-2020}
A. V. Ivanov, N. V. Kharuk, \textit{Two-Loop Cutoff Renormalization of 4-D Yang--Mills Effective Action}, 2020 J. Phys. G: Nucl. Part. Phys. \textbf{48}, 015002, arXiv:2004.05999, 
10.1088/1361-6471/abb939
\bibitem{Ivanov-Kharuk-20222}
A. V. Ivanov, N. V. Kharuk, \textit{Formula for two-loop divergent part of 4-D Yang--Mills effective action},  Eur. Phys. J. C \textbf{82}, 997 (2022), arXiv:2203.07131, 10.1140/epjc/s10052-022-10921-w
\bibitem{Ivanov-Akac}
P. V. Akacevich, A. V. Ivanov, \textit{On Two-Loop Effective Action of 2D Sigma Model},  Eur. Phys. J. C \textbf{83}, 653 (2023), arXiv:2304.02374, 10.1140/epjc/s10052-023-11797-0
\bibitem{Ivanov-Kharuk-2023}
A. V. Ivanov, N. V. Kharuk, \textit{Three-loop divergences in effective action of
4-dimensional Yang--Mills theory with cutoff regularization: $\Gamma_4^2$-contribution}, Zap. Nauchn. Sem. POMI, \textbf{520}, POMI, St. Petersburg, 2023, 162--188
\bibitem{Ivanov-2022}
A. V. Ivanov, \textit{Explicit Cutoff Regularization in Coordinate Representation}, 2022 J. Phys. A: Math. Theor. \textbf{55}, 495401, arXiv:2209.01783, 10.1088/1751-8121/aca8dc

\bibitem{29-2}
E. Brezin, J. C. Le Guillou, J. Zinn-Justin, \textit{Addendum to Wilson's theory of critical phenomena and Callan-Symanzik equations in $4-\varepsilon$ dimensions}, Phys. Rev. D, \textbf{9}, 1121--1124 (1974)
\bibitem{29-1}
D. I. Kazakov, O. V. Tarasov, A. A. Vladimirov, \textit{Calculation of Critical Exponents by Quantum Field Theory Methods}, Sov. Phys. JETP \textbf{50}, 521--526 (1979)
\bibitem{12}
I. Jack, H. Osborn, \textit{Two-loop background field calculations for arbitrary background fields}, Nucl. Phys. B, \textbf{207}, 474--504 (1982)
\bibitem{2}
M. Nakahara, \textit{Geometry, topology and physics}, Second Edition, CRC Press, 1--573 (2003)

\bibitem{102}
B. S. DeWitt, \textit{Quantum Theory of Gravity. 2. The Manifestly Covariant
	Theory}, Phys. Rev. \textbf{162}, 1195--1239 (1967)
\bibitem{103}
B. S. DeWitt, \textit{Quantum Theory of Gravity. 3. Applications of the Covariant Theory}, Phys. Rev. \textbf{162}, 1239--1256 (1967)
\bibitem{24}
G. ’t Hooft, \textit{The background field method in gauge field theories}, (Karpacz, 1975), Proceedings, Acta Universitatis Wratislaviensis, \textbf{1}, Wroclaw, 345--369 (1976)
\bibitem{25}
L. F. Abbott, \textit{Introduction to the background field method}, Acta Phys. Polon. B, \textbf{13}:1--2, 33--50 (1982)
\bibitem{26}
I. Ya. Aref'eva, A. A. Slavnov, L. D. Faddeev, \textit{Generating functional for the S-matrix in gauge-invariant theories}, TMF, \textbf{21}:3, 311--321 (1974)
\bibitem{23}
L. D. Faddeev, \textit{Mass in Quantum Yang--Mills theory (comment on a Clay millenium problem)}, Bull. Braz. Math. Soc. (N. S.), \textbf{33}:2, 201--212 (2002) arXiv: 0911.1013
\bibitem{Ivanov-Russkikh}
A. V. Ivanov, M. A. Russkikh, \textit{Quantum field theory on the example of the simplest cubic model}, Zap. Nauchn. Sem. POMI, \textbf{509} (2021),  123--152; J. Math. Sci. (N. Y.), \textbf{275}:3 (2023), 306--325 arXiv:2107.14488 10.1007/s10958-023-06683-9

\bibitem{29-hp}
E. Hille, R. S. Phillips, \textit{Functional Analysis and Semi-groups}, Providence: American Mathematical Society, 1-808 (1957)

\bibitem{LR-1}
D. Benedetti, R. Gurau, S. Harribey, K. Suzuki, \textit{Long-range multi-scalar models at three loops}, 2020 J. Phys. A: Math. Theor. \textbf{53}, 445008
\bibitem{Iv-Kh-23}
A. V. Ivanov, N. V. Kharuk, \textit{Ordered Exponential and Its Features in Yang--Mills Effective Action}, 2023 Commun. Theor. Phys. \textbf{75}, 085202, 	arXiv:2301.10514, 10.1088/1572-9494/acde4e


\bibitem{29-5}
N. I. Usyukina, \textit{Calculation of many-loop diagrams of perturbation theory}, Theor Math Phys, \textbf{54}, 78--81 (1983)
\bibitem{29-6}
D. I. Kazakov, \textit{Calculation of Feynman diagrams by the “Uniqueness” method}, Theor Math Phys, \textbf{58}, 223--230 (1984)

\bibitem{Kharuk-2021}
N. V. Kharuk, \textit{Zero modes of the Laplace operator in two-loop calculations in the Yang--Mills theory}, Questions of quantum field theory and statistical physics. Part 28, Zap. Nauchn. Sem. POMI, \textbf{509} (2021), 216--226; J. Math. Sci. (N. Y.), \textbf{275}:3 (2023), 370--377 10.1007/s10958-023-06687-5

\bibitem{cvi}
P. Cvitanović, \textit{Group Theory: Birdtracks, Lie's, and Exceptional Groups}, Princeton University Press, Princeton, 1--280 (2008)

\bibitem{28}
V. Fock, \textit{Die Eigenzeit in der Klassischen- und in der Quanten- mechanik}, Sow. Phys., \textbf{12}, 404--425 (1937)
\bibitem{vas1}
D. V. Vassilevich, \textit{Heat kernel expansion: user's manual}, Phys. Rept. \textbf{388}, 279--360 (2003)
\bibitem{vas2}
D. Fursaev, D. Vassilevich, \textit{Operators, Geometry and Quanta: Methods of Spectral Geometry in Quantum Field Theory}, Springer, 1--304 (2011)

\bibitem{29}
M. Lüscher, \textit{Dimensional regularisation in the presence of large background fields}, Annals of Physics \textbf{142}, 359--392 (1982)
\bibitem{psi}
A. V. Ivanov, N. V. Kharuk, \textit{Special functions for heat kernel expansion}, Eur. Phys. J. Plus \textbf{137}: 1060 (2022), arXiv:2106.00294 10.1140/epjp/s13360-022-03176-7

\bibitem{110}
R. T. Seeley, \textit{Complex powers of an elliptic operator},
Singular Integrals, Proc. Sympos. Pure Math. \textbf{10},
Amer. Math. Soc., 288--307 (1967)
\bibitem{10-10}
B. S. DeWitt, \textit{Dynamical Theory of Groups and Fields}, Gordon and Breach, New York, 1--248 (1965)
\bibitem{111}
G. W. Gibbons, \textit{Quantum field theory in curved spacetime}, General Relativity, An Einstein Centenary Survey, 639--679 (1979)
\bibitem{8}
P. B. Gilkey, \textit{The spectral geometry of a Riemannian manifold},
J. Differ. Geom., \textbf{10}, 601--618 (1975)

\bibitem{30}
A. O. Barvinsky, G. A. Vilkovisky, \textit{The Generalized Schwinger--DeWitt Technique in Gauge Theories and Quantum Gravity},
Phys. Rept. \textbf{119}, 1--74 (1985)
\bibitem{f6}
I. G. Avramidi, \textit{Heat Kernel and Quantum Gravity}, New York: Springer, Vol. 64, 1--149 (2000)
\bibitem{31}
A. V. Ivanov, \textit{Diagram technique for the heat kernel of the covariant Laplace operator}, TMF, \textbf{198}:1 (2019), 113--132; Theoret. and Math. Phys., \textbf{198}:1 (2019), 100--117, 10.1134/S0040577919010070,  arXiv:1905.05455
\bibitem{31-1}
A. V. Ivanov, N. V. Kharuk, \textit{Non-recursive formula for trace of heat kernel}, International Conference on Days on Diffraction, DD 2019, 2019, pp. 74--77, 10.1109/DD46733.2019.9016557
\bibitem{33}
A. V. Ivanov, N. V. Kharuk, \textit{Heat kernel: Proper-time method, Fock–Schwinger gauge, path integral, and Wilson line}, TMF, \textbf{205}:2 (2020), 242--261; Theoret. and Math. Phys., \textbf{205}:2 (2020), 1456--1472, arXiv:1906.04019, 10.1134/S0040577920110057

\end{thebibliography}
\end{document}